\documentclass[12pt,a4paper,oneside]{book}
\usepackage{graphicx}
\usepackage{indentfirst}
\usepackage{amsfonts}
\usepackage{latexsym}
\usepackage{amssymb}
\usepackage{color}
\usepackage{titlesec}
\usepackage{fancyhdr}
\usepackage{amsmath,amssymb,amsfonts,amsthm}
\usepackage{graphicx}
\usepackage{enumerate}
\usepackage{mathrsfs}
\usepackage{bm}

\newcommand{\bigrule}{\titlerule[0.5mm]}

\titleformat{\chapter}[display]
{\bfseries\Huge}{
 \filleft
 \Large\chaptertitlename\
 \Large\thechapter}
{0mm} {\filleft} [\vspace{0.5mm} \bigrule]

\renewcommand{\small}{\fontsize{10pt}{14pt}\selectfont}

\title{\vspace*{-3cm}
\large{\textbf{Universidad Complutense de Madrid}\\
\vspace*{0.5cm}Facultad de Ciencias F\'{\i}sicas\\
Departamento de F\'{\i}sica Te\'orica II\\(M\'etodos Matem\'aticos de la F\'{\i}sica)\\}
\begin{figure}[h] \centering
\includegraphics[width=4.0cm]{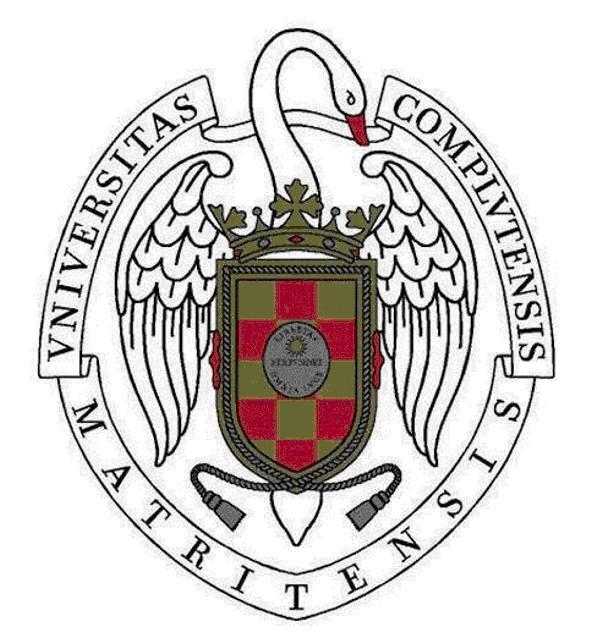}
\end{figure}
----------------------------------------------------------------- \\
\Large{Classical and Quantum Formulations of\\
$\mathbb{S}^{1}\times\mathbb{S}^{2}$ and $\mathbb{S}^{3}$ Gowdy
Models\\ Coupled with
Matter}\\---------------------------------------------------------
\\ \vspace*{0.1cm}\large{Exploring the Mathematical Aspects of\\
Gravity and Quantum Field Theory}}
\author{\Large{\textsc{Daniel G\'omez Vergel}}\vspace*{0.2cm}\\Instituto de Estructura de la Materia, CSIC\\\\\\Research Advisors:\\
Dr. Jes\'us Fernando Barbero Gonz\'alez\footnote{Instituto de Estructura de la Materia, CSIC}\\
Dr. Eduardo Jes\'us S\'anchez Villase\~{n}or\footnote{Universidad
Carlos III de Madrid}}
\date{}

\begin{document}

\newtheorem{theorem}{Theorem}[section]
\newtheorem{lema}{Lemma}[section]
\newtheorem{definit}{Definition}[section]
\newtheorem{rem}{Remark}[section]


\pagestyle{fancy} \fancyhf{}

\fancyhead[LO]{\textsc{Contents}} \fancyhead[RO,LE]{\thepage}
\renewcommand{\headrulewidth}{0.6pt}
\setlength{\headheight}{1.5\headheight}


\maketitle

\newpage
\thispagestyle{empty} \mbox{}

\tableofcontents


\chapter*{Introduction}
\addcontentsline{toc}{chapter}{Introduction}


\pagestyle{fancy} \fancyhf{}

\fancyhead[LO]{\textsc{Introduction}} \fancyhead[RO,LE]{\thepage}
\renewcommand{\headrulewidth}{0.6pt}
\setlength{\headheight}{1.5\headheight}


\begin{center}\textbf{Quantizing gravity}\\\end{center}

\indent Despite the remarkable successes of some of the main
approaches to quantum gravity, we still do not have a satisfactory
theory unifying general relativity (GR) and quantum field theory
(QFT). A suitable quantization of gravitation is, thus, the most
important unsolved problem of modern theoretical physics. The
relentless search for a final theory is motivated by the hope that
it will let us address the fundamental questions in cosmology and
astrophysics, in particular, those related to the physics of the
primitive universe and the generic problem of the appearance of
singularities in physically relevant situations such as the collapse
of compact objects and the formation process of black holes.

Both GR and QFT are incomplete on their own. On one hand, the
singularity theorems of GR state that, under certain energy
conditions satisfied by matter, singularities are expected generic
features of cosmological and collapse solutions $[1]$. The
prediction of infinite energy densities and the consequent divergent
curvature of the spacetime clearly indicates that the theory is
being applied beyond its domain of validity. On the other hand, QFT
has the problem of yielding infinities whenever the amplitudes for
multiply-connected Feynman diagrams are calculated. When possible,
these infinities are subtracted away by absorbing them into the free
parameters of the theory through a renormalization process. In this
context, quantum gravity is expected to provide a natural
ultraviolet cutoff at the Planck length, $L_{P}\sim 10^{-35}m$,
solving in part the problem of the divergencies. Concerning the
foundations of quantum theory, it is necessary to probe new
interpretations avoiding the instrumentalism of the standard
Dirac-von Neumann postulates, based upon the artificial distinction
between the quantum system under study and the external classical
observer who measures it $[2]$. This is specially so in the context
of quantum cosmology, where one faces the problem of correctly
interpreting the wave function of the universe. The
\emph{measurement problem} in quantum mechanics deserves special
attention. Recall that, given a physical system in a (possibly
mixed) state $\rho$, the \emph{result state} of an ideal measurement
of a quantum observable $A$, with respect to a set of measured
values in a Borel set $\Delta\in\mathrm{Bor}(\mathbb{R})$, is
described by the density operator
$$\rho\mapsto\rho_{A,\,\Delta}=\frac{1}{\mathrm{Tr}(\rho E^{A}(\Delta))}
\sum_{\alpha\in\Delta}E^{A}(\{\alpha\})\,\rho E^{A}(\{\alpha\})\,,$$
where $E^{A}$ is the unique spectral measure associated with $A$.
This transition is nonlocal, stochastic and irreversible, and may
come into conflict with the deterministic evolution of closed
systems. In this respect, some authors have put forward the idea
that the reduction of the wave packet may be related to GR,
concretely, to the existence of an initial singularity $[3]$.
\\
\linebreak \indent In view of the highly successful description of
electromagnetic and nuclear forces provided by the $SU(3)\times
SU(2)\times U(1)$ standard model, it is quite natural to attempt a
quantization of the gravitational field following the same strategy
that for the rest of interactions. To do this, one starts by
assuming a concrete topology for the spacetime and splitting its
metric $g_{\mu\nu}$ in the form
$$g_{\mu\nu}=g_{\mu\nu}^{\mathrm{back}}+G_{N}h_{\mu\nu}\,,$$
where $g^{\mathrm{back}}_{\mu\nu}$ is taken to be a background
metric and $h_{\mu\nu}$ is the dynamical field which measures the
deviation of the physical metric from the background. $G_{N}$
denotes the Newton constant. Quantum gravity is then seen as a
theory of small quantum fluctuations around
$g_{\mu\nu}^{\mathrm{back}}$. Note that, in this context, the use of
a background field is strictly necessary in order to apply the usual
quantum field perturbative techniques, providing a fiducial causal
structure used to discuss important items as micro-causality. In
particular, when $g^{\mathrm{back}}_{\mu\nu}$ is chosen to be the
flat metric, one can use the well-known representation theory of the
Poincar\'{e} group to show that the quanta of the $h_{\mu\nu}$ field
are massless particles with spin two. These are the so-called
\emph{gravitons}, which interact with each other and with matter
according to the Einstein-Hilbert Lagrangian or its possible
extensions. Detailed calculations lead to conclude, however, that
Einstein's GR is perturbatively non-renormalizable at two loops for
pure gravity, and at one loop for gravity coupled to matter [4].
Because of this, it is generally agreed that the quantization
procedure introduced above cannot provide a mathematically
consistent and predictive \emph{fundamental} theory valid to
arbitrarily small distances. Indeed, in a non-renormalizable theory
the number of basic parameters tends to infinity at high energies.
From this point of view, GR is rather an effective theory valid only
at low energies. There are proposals to overcome this
non-renormalizability problem by adding additional high power terms
of the Riemannian curvature to the original Lagrangian, but then
some problematic issues related to unitarity arise.
\\
\indent Nevertheless, there are examples of field theories which
\emph{do} exist as fundamental theories despite their
non-renormalizability. In the so-called \emph{asymptotic safety
scenario}, S. Weinberg pointed out the possibility that quantum
gravity could be formulated in a nonperturbative way by invoking a
non-Gaussian ultraviolet fixed point --in contrast with the standard
perturbative renormalizations based upon Gaussian fixed points at
which all couplings parameterizing the general action functional
vanish. In this case, the theory would be asymptotically safe given
the absence of unphysical singularities at high energies $[5]$. This
issue is under investigation at the present moment, although most of
the physicists focus their attention on two different research
programs, namely, (super) string theory and loop quantum gravity
(LQG). The main aim of the first approach is to perform a
unification of all known fundamental interactions, including
gravity, in terms of excitations of one-dimensional objects called
\emph{strings} that evolve in a certain background metric space.
Although the theory includes GR in its low-energy regime, so far it
has not been possible to recover univocally the correct
4-dimensional standard model. Major efforts are devoted to this
issue and also to probing the nonperturbative aspects of the theory.
Despite these serious difficulties, string theory has successfully
explained the Bekenstein-Hawking area law for a limited class of
objects, the so-called BPS-type extremal black holes, with the
correct prefactor $1/4$ relating entropy and area $[6]$. The theory
has also been successful describing the emission of Hawking
radiation. Furthermore, string theory is expected to make the
self-interacting Feynman diagrams finite order-by-order, solving in
this way the non-renormalizability problem of quantum gravity.
\\
\indent A leading alternative to string theory is LQG, a
mathematically consistent, nonperturbative, generally-covariant and
background-independent canonical quantization of GR which describes
gravity as a theory of $SU(2)$ connections and holonomies. This
formulation differs substantially from the previous approach, in the
sense that it tries to preserve the profound implications that GR
has for the notions of space, time and causality. In fact,
Einstein's GR is taken here as the basic starting point; although
high-energy corrections to Einstein's equation may appear after
quantization, they are not expected to modify the elegant
description of gravitation in terms of a curvature of the spacetime
geometry at large scales. As the main achievement of the theory, LQG
has provided the correct entropy for a wide variety of black holes,
including the Schwarzschild and Kerr types. However, there is a
quantization ambiguity due to the so-called Barbero-Immirzi
parameter appearing in the eigenvalues of the area operator. To
recover the right pre-factor $1/4$, one has to properly fix this
parameter for each class of black hole. An effective equi-spacing of
the degeneracy spectrum of microscopic black holes has also been
shown to exist $[7]$. This is in (surprising) agreement with some
solid results by Bekenstein, who inferred that the horizon area
spectra of a black hole far away from extremality must be
necessarily discrete and equally spaced in the context of any
consistent quantum theory of gravity $[8]$. Future investigations
will try to extend these results to the macroscopic limit $[9]$. In
addition, there are some relevant results concerning the resolution
of classical singularities such as the one corresponding to the
Schwarzschild metric $[10]$. Despite these remarkable successes,
however, the theory has not been able to recover the classical GR at
its low-energy limit, and a definite formulation of the dynamics
(related to the quantization of the Hamiltonian constraint) is not
yet available. Because of this, some issues such as Hawking
radiation are not well understood.
\\
\indent Finally, we must mention another promising nonperturbative
formulation of quantum gravity, the so-called causal dynamical
triangulation (CDT) approach, whose aim is to give a rigorous
mathematical meaning to the Lorentzian path integral correspon\-ding
to gravity (consisting of the usual Einstein-Hilbert action with a
cosmological term) by restricting it to geometries with a well
defined causal structure (even at the Planck length). Large
fluctuations in curvature are allowed at short scale, but in such a
way that the resulting large scale geometry is nondegenerate.
Concretely, the picture of the spacetime which emerges is as
follows: At sub-Planckian scales, spacetime is fractal with
dimension two, whereas a smooth classical geometry of effective
(spectral) dimensionality equal to four is recovered at large scales
$[11]$.
\\
\linebreak \indent It is clear that quantum gravity differs
substantially from the majority of research branches in theoretical
physics due to the absence of the experimental data necessary to
sift through the broad range of proposals. Indeed, the Planck energy
$E_{P}\sim 10^{28}eV$, at which quantum gravity effects become
relevant, is beyond the experimental range of any available (or
conceivable) particle accelerator; in fact, these effects may only
be probed in the very early age of our universe --the so-called
Planck epoch-- or in some violent astrophysical collapse processes.
Given the obvious technical difficulties to access to this energy
regime, it is not possible to subject the different tentative
theories to a rigorous validation/falsation process in the
traditional way. The feasibility of a quantum gravity formulation is
rather settled by demanding that any reasonable theory predicts some
of the well established semiclassical results already mentioned such
as the Bekenstein-Hawking entropy, the Hawking radiation, or the
quantization and equally-spacing (adiabatic invariance) of the
horizon area of black holes. The mathematical consistency of the
theory is also taken into account, as well as the successful
attainment of concrete desired objectives, as the unification of all
known fundamental forces in the case of string theory.
\\
\begin{center}\textbf{Two-Killing vector symmetry reductions}\\\end{center}

\indent Along this thesis, we will adopt a somehow modest point of
view. We will restrict ourselves to the study of some symmetry
reductions of GR which are especially useful to gain valuable
insights into the behavior of gravity in its quantum regime. In this
context, Bianchi models and two-Killing vector reductions of GR have
received a lot of attention owing to their applications in
astrophysics and cosmology. Two-Killing vector reductions, in
particular, have been widely considered as appealing testing grounds
for quantum gravity owing to the fact that they still have local
degrees of freedom\footnote{These models are usually referred to as
\emph{midi-superspace} models (see $[12]$ and references therein),
in contrast with the so-called \emph{mini-superspace} models, like
Bianchi types, which have a finite number of degrees of freedom.} as
well as (restricted) diffeomorphism invariance, two of the features
of the gravitational theory that lie at the heart of the
difficulties encountered in its quantization process. They often
admit an exact quantization, being possible to make concrete
predictions, at least in a qualitative way, about the relevant
features that a full theory of quantum gravity should have (whatever
it might be). In fact, these models have proved to be privileged
frameworks to discuss some fundamental aspects of quantum theory.
For instance, one can analyze the need for quantum evolution to be
unitary in order for physical predictions to be consistent with
causality, and how this condition can be relaxed in some definite
sense within the Heisenberg picture $[13]$. They also provide a
natural framework to apply the so-called algebraic formulation of
quantum theory, consisting in defining an appropriate $*$-algebra of
quantum observables for each system, in order to facilitate the
construction and analysis of the different Hilbert space
representations for the models.

\begin{table}
\begin{center}
\begin{tabular}{|l|l|l|l|}
\hline \hline Group $\big(G^{\scriptstyle{(2)}}\big)$ & Manifold
$\big({^{\scriptstyle{(3)}}}\Sigma\big)$ & Action &
Name of the model\\
\hline
\hline $U(1)\times U(1)$ & $\mathbb{R}^{2}\times\mathbb{S}^{1}$ & Not free & \\
& $\mathbb{R}\times\mathbb{T}^{2}$ & Free & Schmidt model\\
& $\mathbb{T}^{3}$ & Free & $\mathbb{T}^{3}$ Gowdy model\\
& $\mathbb{S}^{1}\times\mathbb{S}^{2}$ & Not free & $\mathbb{S}^{1}\times\mathbb{S}^{2}$ Gowdy model\\
& $\mathbb{S}^{3}$ & Not free & $\mathbb{S}^{3}$ Gowdy model\\
\hline
\hline $\mathbb{R}\times U(1)$ & $\mathbb{R}^{3}$ & Not free & Cylindrical gravitational waves\\
& $\mathbb{R}\times\mathbb{S}^{2}$ & Free & Cylindrical wormhole\\
& $\mathbb{R}\times\mathbb{T}^{2}$ & Free &\\
\hline \hline $\mathbb{R}^{2}$ & $\mathbb{R}^{3}$ & Free &\\
& $\mathbb{R}^{2}\times\mathbb{S}^{1}$ & Free &\\ \hline \hline
\end{tabular}
\caption{Spatial topologies compatible with the abelian biparametric
Lie group $G^{\scriptstyle{(2)}}$, whith a smooth, effective and
proper action on the spatial sections of a globally hyperbolic
spacetime
$({^{\scriptstyle{(4)}}}\mathcal{M}\simeq\mathbb{R}\times{^{\scriptstyle{(3)}}}\Sigma,{^{\scriptstyle{(4)}}}g_{ab})$.
The action of the group, unique up to automorphisms of
$G^{\scriptstyle{(2)}}$ and diffeomorphisms of
${^{\scriptstyle{(3)}}}\Sigma$, can be free or have degenerate
orbits.}\label{TableI}
\end{center}
\end{table}

\indent  When the Killing fields commute and are hypersurface
orthogonal, the resulting mo\-dels --said to be \emph{linearly
polarized}-- become specially simple and solvable. These reductions
differ from each other in the action of the isometry group and the
correspon\-ding compatible spatial topologies (see \emph{Table
\ref{TableI}} and $[14]$). The so-called linear Einstein-Rosen waves
$[15]$, which describe the propagation of linearly polarized
wave-like modes in a spacetime with noncompact spatial slices,
deserve particular attention. Here, the symmetry group is
$\mathbb{R}\times U(1)$ and the spacetime is topologically
$\mathbb{R}^4$. The quantization of this system coupled to massless
scalar fields has been rigorously analyzed recently, and has
provided several interesting features relevant for quantum gravity
$[16]$. In this context, the introduction of matter is a way to
produce quantum test particles with controllable wave functions in
order to explore the quantized spacetime geometry. A suitable gauge
fixing procedure yields a time-independent Hamiltonian which is a
nontrivial bounded function of the free Hamiltonian corresponding to
two uncoupled massless and axially symmetric scalar fields evolving
in the same fixed (1+2)-dimensional Minkowskian space. It has its
origin in the boundary terms of the Einstein-Hilbert action needed
to have a well-defined variational principle. This fact allows one
to \emph{exactly} quantize the model by using the standard
techniques of QFT in curved backgrounds, even though the system is
nonlinear and self-interacting. In particular, the quantum unitary
evolution operator can be obtained in closed form in a
straightforward way and used for a number of purposes leading to
physical applications such as the discussion of the existence of
large quantum gravity effects $[17]$ or the study of the
microcausality of the system $[16]$. Specifically, the field
commutator can be obtained from the two-point correlation functions,
interpreted here as approximate probability amplitudes for a
particle created at certain radial distance from the cylindrical
symmetry axis at certain time to be detected somewhere else at a
different instant of time. One then observes purely quantum
gravitational effects such as an enhancement of the probability of
finding the field quanta very close to the symmetry axis. The
probability amplitude is also high along lines that can be
interpreted as approximate null geodesics of an emergent axially
symmetric Minkowskian geometry, which provides a concrete example of
how classical behavior can be recovered from a quantum gravity model
$[16]$.
\\
\begin{center}\textbf{Structure of the thesis}\\\end{center}

\indent We will focus on the so-called linearly polarized Gowdy
models $[18]$ coupled to massless scalar fields. From the physical
point of view, their most salient feature is the fact that they
describe cosmological models with initial, or initial and final,
singularities. Here, the isometry group is $U(1)\times U(1)$ and the
spatial manifold is restricted to have the topology of a 3-torus
$\mathbb{T}^3$, a 3-handle $\mathbb{S}^1\times \mathbb{S}^2$, a
3-sphere $\mathbb{S}^3$, or that of the lens spaces $L(p,q)$ --that
can be studied by imposing discrete symmetries on the $\mathbb{S}^3$
case $[19]$. The exact quantization of the linearly pola\-rized
$\mathbb{T}^3$ Gowdy model in the vacuum has been profusely analyzed
in the past $[20]$. Its dynamics is governed by a quadratic
nonautonomous Hamiltonian obtained through a deparameterization
process. The gravitational local degrees of freedom can be
interpreted as those corresponding to a massless scalar field in a
fiducial background with initial singularity, so that the standard
techniques of QFT in curved spacetimes can be applied in order to
construct the quantum theory. The fact that the linear symplectic
transformations describing the classical time evolution cannot be
unitarily implemented in the physical Hilbert space when the system
is written in terms of its original variables was initially
interpreted as a serious obstacle for the feasibility of the model
$[21]$. This problem should come as no surprise, however, since a
generic feature of the quantization of infinite-dimensional linear
symplectic dynamical systems is precisely the impossibility of
defining the unitary quantum counterpart of \emph{all} linear
symplectic transformations on the phase space
$[22]$.\footnote{Consider, for example, the generic impossibility of
making sense of the unitary quantum evolution operator when dealing
with scalar fields propagating in the Minkowskian spacetime from
initial to final Cauchy surfaces that are not level surfaces of some
Minkowskian time $[23]$.} These transformations are characterized by
$*$-automorphisms defined on the corresponding abstract $*$-algebra
of quantum observables. Note that the lack of a unitary operator
implementing the quantum time evolution of the system comes into
conflict with the axiomatic structure of quantum theory itself. In
case of not rejecting the model for this reason, one must carefully
analyze the viability of a suitable probabilistic interpretation for
it, as discussed in $[24]$. Nevertheless, it is possible to overcome
this problem just by performing a suitable time-dependent
redefinition of the field $[25]$. Furthermore, by demanding the
unitarity of the dynamics and the invariance under an extra $U(1)$
symmetry generated by a residual global constraint, the existence of
a unique (up to unitary equivalence) Fock representation can be
proved for the system $[26]$.
\\
\indent The purpose of this thesis is twofold. First, we will
generalize and extend the existing literature devoted to the vacuum
3-torus model to the remaining more complicated topologies, the
3-handle and the 3-sphere, allowing also the coupling of gravity to
matter, concretely, to massless scalar fields. These topologies are
less known than the 3-torus one but equally relevant in cosmology
owing to the fact that they display both initial and final
singularities. For this reason, they become specially useful test
beds for issues related to canonical quantization in cyclic
universes. Here, as in the case of linear Einstein-Rosen waves, the
addition of matter is a useful way to probe the quantized geometry,
much in the same way as test particles are introduced in classical
GR in order to analyze the spacetime geometry. Second, concerning
the canonical quantization of the resulting gauge systems, we will
confirm and clarify several relevant results found in the literature
devoted to the vacuum 3-torus case. This will be done by placing
particular emphasis on mathematical issues such as the rigorous
application of symplectic geometry to nonautonomous Hamiltonian
systems or the algebraic formulation of quantum theory in terms of
suitable $*$-algebras of observables.
\\
\linebreak \indent  The text is structured as follows. \emph{Chapter
\ref{ChapterI}} will be devoted to the Lagrangian and Hamiltonian
formulations of the $\mathbb{S}^{1}\times\mathbb{S}^{2}$ and
$\mathbb{S}^{3}$ Gowdy models, whose treatment in previous
literature has suffered from an obvious lack of rigor.\footnote{The
Hamiltonian analysis for these models in the vacuum has only been
addressed in a partial way in [27], without providing the detailed
phase space description necessary to understand several relevant
geometrical issues.} This will be done by applying modern
differential-geometric techniques to analytical mechanics. In
contrast with the 3-torus case, the existence of degenerate orbits
under the action of the isometry group will force us to carefully
consider the regularity conditions that the dynamical variables must
verify. A Geroch symmetry reduction, and a subsequent conformal
transformation, will allow us to interpret these models as
(1+2)-dimensional gravity coupled to a set of massless scalar fields
with axial symmetry. Some details concerning this reduction will
vary depending on the topologies and will be commented separately
for each case. Among several issues, we will explain how the
topology of the spatial slices affects the definition of the
constraints, and also how the coupling of massless scalar fields is
realized in the diffe\-rent topologies. A careful application of the
Dirac-Bergmann theory of constrained systems $[28-30]$ yields a
reduced phase space description of these systems in terms of
coisotropic (or first class constrained) manifolds. This is the case
when the Poisson algebra of the constraints is a proper Lie algebra.
An appropriate partial gauge-fixing (\emph{deparameterization})
process will allow us to characterize the dynamics through
nonautonomous Hamiltonian systems, mathematically described as
cosymplectic or contact manifolds. Within this Hamiltonian setting,
we will understand in detail the mechanisms leading to the
appearance of initial and final singularities. It is important to
highlight the fact that for these models, at variance with the
3-torus case, there are no extra constraints after the
deparameterization process. This will obviously simplify the
construction of a Hilbert space representation of the canonical
commutation relations within the quantization process. In
particular, it will not be necessary to distinguish between
kinematical and physical Hilbert spaces.
\\
\linebreak \indent In \emph{Chapter \ref{ChapterII}}, we will
proceed to perform an exact Fock-type canonical quantization of the
deparameterized models. Both gravitational and matter local degrees
of freedom will be encoded in massless scalar fields evolving in the
same fixed background metric conformally equivalent to the Einstein
static (1+2)-dimensional universe, topologically
$(0,\pi)\times\mathbb{S}^{2}$. This fact will allow us to treat
these fields in a unified way in the construction of the quantum
theory by applying the usual techniques of QFT in curved spacetimes.
The appropriate starting point is the covariant formulation of the
reduced phase-space in terms of smooth real solutions to the
Klein-Gordon equation of motion. This approach is completely
equivalent to the usual canonical one, but proves specially useful
in this context in order to discuss some quantization issues such as
the existence of (in principle) many nonunitarily equivalent Fock
space representations for the canonical commutation relations, each
of them characterized by $SO(3)$-invariant complex structures on the
covariant phase space. These invariant complex structures will be
parameterized by pairs
$(\rho_\ell,\nu_\ell)\in(0,+\infty)\times\mathbb{R}$,
$\ell\in\mathbb{N}_0$. A first result will be the impossibility to
characterize the dynamics through a unitary evolution operator when
the system is described in terms of its original variables. This
feature will be proved to be insensitive to the election of the
$SO(3)$-invariant complex structure used to quantize the system.
Then, we will be forced to introduce new dynamical variables in
order to properly describe the system. Concretely, with the aim of
overcoming the non-unitarity obstruction we will perform a
re-scaling of the fields similar to the one employed in the 3-torus
case. This redefinition will be dictated \emph{precisely} by the
conformal factor that relates the Gowdy metrics to the Einstein
metric. In this way, we will provide a suitable geometrical
interpretation of the techniques previously employed in the
literature devoted to the 3-torus topology: The singular behavior
introduced by the conformal factor will be translated into the
behavior of a singular and time-dependent potential term for the
re-scaled fields. This potential term will be sufficiently
well-behaved as a function of time --in spite of being singular at
some instants-- to allow the unitary implementation of the dynamics.
Moreover, we will be able to fully characterize all
$SO(3)$-invariant complex structures for which the time evolution is
unitary in terms of the asymptotic behavior of the pairs
$(\rho_\ell,\nu_\ell)$ for large values of $\ell\,$. In addition, we
will prove that the many different $SO(3)$-invariant Fock
representations of this type are unitarily equivalent. It is
important to remark in this respect that, in absence of extra
constraints as in the 3-torus model, we will use the $SO(3)$
symmetry associated to the background metric in order to select a
preferred class of complex structures such that the Fock
quantization is unique (up to unitary equivalence). The simplicity
of the arguments used to prove these results will convince the
reader of the usefulness of the employed formalism.
\\
\linebreak \indent Finally, \emph{Chapter \ref{ChapterIII}} will be
devoted to the functional Schr\"{o}dinger representation of these
models. Here, quantum states are characterized by square integrable
functionals belonging to a $L^{2}$-space constructed from an
appropriate distributional extension of the classical configuration
space --in this case, it is given by the space of tempered
distributions on the 2-sphere $\mathbb{S}^{2}$-- endowed with a
time-dependent Gaussian measure whose support will be analyzed in
detail. By virtue of the interrelation between measure theory and
representation of canonical commutation relations, the momentum
operators will differ from the usual ones in terms of derivatives by
the appearance of linear multiplicative terms which depend on the
configuration observables. We will check that, as a consequence of
the unitary implementability of time evolution, the representations
corresponding to different values of the time parameter are
unitarily equivalent and, hence, their associated measures are
mutually absolutely continuous. We will end this chapter by
developing a general procedure to obtain the evolution operator for
the systems under study, written explicitly \emph{in closed form} in
terms of the basic field and momentum observables. This analysis
will be (implicitly) based upon the theory of adiabatic invariants
developed by Lewis in the context of the study of classical and
quantum systems with time-dependent harmonic-oscillator-type
Hamiltonians $[31]$.
\\
\linebreak \indent To conclude, we will probe the existence of
suitable semiclassical states for the Gowdy cosmologies and discuss
several possible applications of our study, as well as some open
problems to be tackled in the future.
\\
\linebreak \indent The appendices, far for providing merely
incidental details on these subjects, will give additional
information on several results attained in the main body of the
thesis, going deeply into the relevant mathematical aspects of the
text.
\\
\indent In \emph{Appendix A}, the reader will find a generalization
of the Geroch symmetry reduction procedure with respect to a
space-like hypersurface-orthogonal Killing vector field in presence
of a (symmetric) massless scalar field minimally coupled to
(1+3)-dimensional gravity. After a suitable conformal
transformation, 4-dimensional Einstein-Klein-Gordon equations turn
out to be equivalent to (1+2)-dimensional gravity coupled to two
massless scalar fields, one of them proportional to the original
scalar field, and the other being given by the logarithm of the norm
of the Killing vector field.
\\
\indent In \emph{Appendix B}, we will summarize the main proposals
and results of symplectic geometry when applied to analytical
mechanics, fixing the notation and conventions used throughout the
thesis. We will pay special attention to the nonautonomous (i.e.,
time-dependent) Hamiltonian systems, in terms of which we describe
the dynamics of the Gowdy models after deparameterization.
\\
\indent \emph{Appendix C} presents the general framework of the
mathematical description of (classical and quantum) physical
theories in terms of $C^{*}$-algebras. The main point here is to
introduce the concepts of observables and states, and how these can
be realized respectively as self-adjoint bounded operators and
vectors (or density matrices) in Hilbert spaces. The reader is
strongly advised to revisit these topics, particularly the usual
Dirac-von Neumann axiomatic structure of quantum mechanics, from
this algebraic point of view.
\\
\indent It is clear that, in the quest for a suitable quantization
of systems of infinitely many time-dependent harmonic oscillators,
like those describing the Gowdy cosmologies, the understanding of
the special features of the single quantum oscillator is
particularly advisable. In \emph{Appendix D}, we reformulate the
study of the unitary implementation of the dynamics for a single
one-dimensional harmonic oscillator with nonconstant frequency,
paying special attention to the search of semiclassical states and
closed expressions for the evolution operators.
\\
\indent Finally, \emph{Appendix E} summarizes the theory of
symme\-tric/antisymmetric Fock spaces, in particular, the definition
of the creation and annihilation operators and their canonical
commutation/anticommutation relations.
\\
\linebreak \indent Throughout the text, with the exception of the
\emph{Appendix B}, we will use the Penrose abstract index convention
with tangent space indices belonging to the beginning of the Latin
alphabet $[32]$. Lorentzian spacetime metrics will have signature
$(-+++)$ and the conventions for the curvature tensors will be those
of reference $[33]$.
\\
\linebreak \linebreak \noindent \textbf{References}
\begin{itemize}
\item [1] S. W. Hawking and G. F. R. Ellis, \emph{The Large Scale
Structure of spacetime}, Cambridge University Press (1973).
\item [2] To initiate himself into different approaches to quantum mechanics, the reader can consult, among
many others: H. Everett, Rev. Mod. Phys. \textbf{29}, 454 (1957); J.
B. Hartle, Am. J. Phys. \textbf{36}, 704 (1968); B. S. De Witt,
Physics Today \textbf{23}, 30 (1970); B. S. De Witt and N. Graham,
eds., \emph{The Many Worlds Interpretations of Quantum Mechanics},
Princeton University Press, Princeton (1973); C. Rovelli, Int. J. of
Theor. Phys. \textbf{35}, 1637 (1996).
\item [3] S. W. Hawking and R. Penrose, \emph{The Nature of Space and
Time}, Princeton University Press (1996).
\item [4] G. 't Hooft and M. Veltman, Annales de l'institut Henri Poincar\'e (A) Physique Th\'eorique \textbf{20},
69-94 (1974).
\item [5] S. Weinberg, ``Ultraviolet divergencies in quantum theories of gravitation'', in
S. W. Hawking and W. Israel, eds., \emph{General Relativity: An
Einstein Centenary Survey}, Cambridge University Press (1979); J.
Gomis and S. Weinberg, Nucl. Phys. B \textbf{469}, 473-487 (1996).
\item [6] A. Strominger and C. Vafa, Phys. Lett. B \textbf{379}, 99
(1996).
\item [7] A. Corichi, J. Diaz-Polo, and E. Fernandez-Borja, Phys. Rev. Lett. \textbf{98}, 181301 (2007); Class. Quant. Grav.
\textbf{24}, 243 (2007).
\item [8] J. D. Bekenstein, Lett. Nuovo Cimento \textbf{11}, 467 (1974). V. F. Mukhanov, JETP Letters \textbf{44}, 63
(1986); J. D. Bekenstein, ``Classical Properties, Thermodynamics and
Heuristic Quantization'', in \emph{Cosmology and Gravitation}, M.
Novello, ed., Atlantisciences, France (2000).
\item [9] J. F. Barbero G. and E. J. S. Villase\~{n}or, Physical Review D \textbf{77}, 121502(R) (2008).
\item [10] Martin Bojowald, Living Rev. Relativity \textbf{11}, 4
(2008). R Gambini and J. Pullin, Phys. Rev. Lett. \textbf{101},
161301 (2008).
\item [11]  J. Ambjorn, J. Jurkiewicz, and R. Loll, Phys. Rev. D \textbf{72} (2005)
064014; Phys. Rev. Lett. \textbf{95}, 171301 (2005).
\item [12] C. G. Torre, Int. J. Theor. Phys. \textbf{38}, 1081-1102 (1999).
\item [13] T. Jacobson, in \emph{Conceptual Problems of Quantum Gravity}, edited by A. Ashtekar and
J. Stachel, Birkh\"{a}user, Boston (1991).
\item [14] B. K. Berger, P. T. Chru\'sciel, and V. Moncrief, Ann. Phys.
\textbf{237}, 322 (1995).
\item [15] A. Einstein and N. Rosen, J. Franklin Inst. \textbf{223}, 43
(1937). A. Ashtekar and M. Pierri, J. Math. Phys. \textbf{37}, 6250
(1996).
\item [16] J. F. Barbero G., I. Garay, and E. J. S. Villase\~{n}or, Physical Review D \textbf{74}
044004 (2006); Physical Review Letters \textbf{95} 051301 (2005).
\item [17] A. Ashtekar, Phys. Rev. Lett. \textbf{77}, 4864 (1996).
M. E. Angulo and G. A. Mena Marug\'an, Int. J. Mod. Phys. D
\textbf{9}, 669-686 (2000). J. F. Barbero G., I. Garay, and E. J. S.
Villase\~{n}or, Class. Quant. Grav. \textbf{25}, 205013 (2008).
\item [18] R. H. Gowdy, Phys. Rev. Lett. \textbf{27}, 826 (1971); Ann. Phys.
\textbf{83}, 203 (1974).
\item [19] P. S. Mostert, Ann. of Math. \textbf{65}, 447 (1957); \textbf{66}, 589 (1957).
\item [20] C. W. Misner, Phys. Rev. D \textbf{8}, 3271 (1973). B. K. Berger, Ann. Phys. \textbf{83}, 458
(1974); Phys. Rev. D \textbf{11}, 2770 (1975). G. A. Mena Marug\'an,
Phys. Rev. D \textbf{56}, 908 (1997). M. Pierri, Int. J. Mod. Phys.
D \textbf{11}, 135 (2002).
\item [21] A. Corichi, J. Cortez, and H. Quevedo, Int. J. Mod. Phys. D \textbf{11},
1451 (2002).
\item [22] D. Shale, Trans. Am. Math. Soc. \textbf{103},
149 (1962).
\item [23] C. G. Torre and M. Varadarajan, Class. Quant. Grav. \textbf{16}, 2651 (1999).
\item [24] C. G. Torre, Phys. Rev. D \textbf{66}, 084017
(2002).
\item [25] A. Corichi, J. Cortez, and G. A. Mena Marug\'an,
Phys. Rev. D \textbf{73}, 041502 and 084020 (2006).
\item [26] A. Corichi, J. Cortez, G. A. Mena~Marug\'an, and J. M. Velhinho,
Class. Quant. Grav. \textbf{23}, 6301 (2006). J. Cortez, G. A.
Mena~Marug\'an, and J. M. Velhinho, Phys. Rev. D \textbf{75}, 084027
(2007).
\item [27] J. L. Hanquin and J. Demaret, J. Phys. A: Math. Gen. \textbf{16}, L5 (1983).
\item [28] M. J. Gotay, J. M. Nester, and G. Hinds, J. Math. Phys. \textbf{19}, 2388—2399 (1978).
\item [29] D. Chinea, M. de Le\'on, and J. C. Marrero, J. Math. Phys. \textbf{35}, 3410-3447 (1994).
\item [30] M. Henneaux and C. Teitelboim, \emph{Quantization of Gauge Systems}. Princeton University Press, 1994.
\item [31] H. R. Lewis Jr., Phys. Rev. Lett. \textbf{18}, 510 (1967); J.
Math. Phys. \textbf{9}, 1976-1986 (1968).
\item [32] R. Penrose and W. Rindler, \emph{Spinors and spacetime, Vol. 1: Two-Spinor Calculus and Relativistic
Fields}, Cambridge University Press (1984).
\item [33] R. M. Wald, \emph{General Relativity}, The University of Chicago Press, Chicago and London (1984).
\end{itemize}


\chapter{Hamiltonian Formulation}\label{ChapterI}
\begin{flushright}
\small{\vspace*{-0.9cm} \textbf{J. F. Barbero G., D. G. Vergel, and
E. J. S. Villase\~{n}or}\\\textbf{Classical and Quantum Gravity
\textbf{24}, 5945 (2007)}}
\end{flushright}

\newcommand{\dos}{{\scriptstyle{(2)}}}
\newcommand{\tres}{{\scriptstyle{(3)}}}
\newcommand{\cuatro}{{\scriptstyle{(4)}}}


\pagestyle{fancy} \fancyhf{}

\fancyhead[LO]{\textsc{Chapter 1. Hamiltonian Formulation}}
\fancyhead[RO,LE]{\thepage}
\renewcommand{\headrulewidth}{0.6pt}


\vspace*{0.5cm} \indent Consider a smooth, effective, and proper
action\footnote{Let $G$ be a Lie group and $\mathcal{M}$ a manifold.
The (left) action of $G$ on $\mathcal{M}$ is a differentiable map
$\sigma:G\times\mathcal{M}\rightarrow\mathcal{M}$ which satisfies
(i) $\sigma(e,p)=p$ for any $p\in\mathcal{M}$ and (ii)
$\sigma(g_1,\sigma(g_2,p))=\sigma(g_1\cdot g_2,p)$. Here, $e$
denotes the unit element of the group. The action is said to be (i)
\emph{smooth}, if the $\sigma$ mapping is $C^{\infty}$; (ii)
\emph{proper}, if the mapping
$G\times\mathcal{M}\rightarrow\mathcal{M}\times\mathcal{M}$ given by
$(g,p)\mapsto(\sigma(g,p),p)$ is proper, i.e., inverses of compact
sets are compact; (iii) \emph{effective} if the unit element $e\in
G$ is the unique element that defines the trivial action on
$\mathcal{M}$, i.e.,
$\sigma(g,p)=p\,,\,\forall\,p\in\mathcal{M}\Rightarrow g=e$ (see
$[1]$ for more details).} of the biparametric Lie Group
\begin{equation*}
G^\dos:=U(1)\times U(1)=\big\{(g_1,g_2)=(e^{ix_1},e^{ix_2})\,|
\,x_{1},x_{2}\in\mathbb{R}(\mathrm{mod}\,2\pi)\big\}
\end{equation*}
on a compact, connected, and oriented 3-manifold
${^{\scriptstyle{\tres}}}\Sigma$. This manifold is then restricted
to have the topology of a 3-torus $\mathbb{T}^3$, a 3-handle
$\mathbb{S}^1\times \mathbb{S}^2$, a 3-sphere $\mathbb{S}^3$, or
that of the lens spaces $L(p,q)$. Moreover, the action of the group
is unique up to automorphisms of $G^\dos$ and diffeomorphisms of
${^{\scriptstyle{\tres}}}\Sigma$ $[2,3]$. Next, we construct a
4-manifold ${^{\cuatro}}\mathcal{M}$, diffeomorphic to
$\mathbb{R}\times{^{\tres}}\Sigma$, such that
$({^{\cuatro}}\mathcal{M}, {^{\cuatro}}g_{ab})$ is a globally
hyperbolic spacetime endowed with a Lorentzian metric
${^{\cuatro}}g_{ab}$. We further require $G^\dos$ to act by
isometries on the spatial slices of ${^{\cuatro}}\mathcal{M}$,
obtaining in this way the so-called \emph{Gowdy models}. We will
focus on the \emph{linearly polarized cases}, where the isometry
group is generated by pairs of mutually orthogonal, commuting,
spacelike, and globally defined hypersurface-orthogonal Killing
vector fields $(\xi^{a},\sigma^{a})$.
\\
\linebreak \indent Most of the work on these models, after the
initial papers by Gowdy, has profusely analyzed the 3-torus spatial
topology; in fact, this is by far the preferred choice to discuss
quantization issues. Here, we will focus our attention on the other
possible \emph{closed} (compact and without boundary) topologies,
the 3-handle and the 3-sphere. The lens spaces $L(p,q)$ can be
studied by imposing discrete symmetries on the 3-sphere case and, in
fact, the arguments presented for $\mathbb{S}^3$ remain valid for
them. Specifically, the nonexistence of additional qualitative
phenomena in the lens spaces with respect to the $\mathbb{S}^{3}$
models has its origin in the fact that the covering
$\mathbb{S}^{3}\rightarrow L(p,q)$ is the projection map of the
quotient of $\mathbb{S}^{3}$ by a subgroup of the isometry group
$G^{\dos}$.

\section{$\mathbb{S}^1\times\mathbb{S}^2$ Gowdy models coupled to massless
scalars}{\label{handle}}

Let us start by considering the 3-handle
${^{\tres}}\Sigma=\mathbb{S}^1\times\mathbb{S}^2$, whose points can
be parameterized in the form
$(e^{i\xi},e^{i\sigma}\sin\theta,\cos\theta)$, with
$\theta\in[0,\pi]$, $\xi,\sigma\in\mathbb{R}(\mathrm{mod}\,2\pi)$.
We define the following (left) $G^\dos$-group action
$$
(g_1,g_2)\cdot(e^{i\xi},e^{i\sigma}\sin\theta,\cos\theta)=
(e^{x_1},e^{x_2})\cdot(e^{i\xi},e^{i\sigma}\sin\theta,\cos\theta)=
(e^{i(x_1+\xi)},e^{i(x_2+\sigma)}\sin\theta,\cos\theta)\,.
$$
The action of the two $U(1)$ commuting subgroup factors of $G^\dos$,
$(g_1,g_2)=(e^{ix},1)$ and $(g_1,g_2)=(1,e^{ix})$,
$x\in\mathbb{R}(\mathrm{mod}\,2\pi)$, is respectively given by
\begin{eqnarray*}
&&(e^{ix},1)\cdot(e^{i\xi},e^{i\sigma}\sin\theta,\cos\theta)=
(e^{i(x+\xi)},e^{i\sigma}\sin\theta,\cos\theta)\,,\nonumber\\
&&(1,e^{ix})\cdot(e^{i\xi},e^{i\sigma}\sin\theta,\cos\theta)=
(e^{i\xi}, e^{i(x+\sigma)}\sin\theta,\cos\theta)\,.\nonumber
\end{eqnarray*}
The corresponding tangent vectors at each point of
${^{\tres}}\Sigma$, obtained by differentiating the previous
expressions with respect to $x$ at $x=0$, are
\begin{equation}
(ie^{i\xi},0,0)\,,\quad\,(0,ie^{i\sigma}\sin\theta,0)\,.\nonumber
\end{equation}
It is straightforward to verify that both fields commute. As we can
see, the first one is never zero but the latter vanishes at
$\theta=0$ and $\theta=\pi$. This corresponds to the circles
$(e^{i\xi},0,1)$ and $(e^{i\xi},0,-1)$.
\\
\indent Consider now the 4-manifold
${^{\scriptstyle{\cuatro}}}{\mathcal{M}}\simeq\mathbb{R}\times\mathbb{S}^{1}\times\mathbb{S}^{2}$;
let $t$ be a global coordinate on $\mathbb{R}$. We want to introduce
three smooth (almost everywhere nonvanishing) vector fields
$\theta^{a}$, $\sigma^a$, and $\xi^a$, tangent to the embedded
submanifolds $\{t\}\times\mathbb{S}^{1}\times\mathbb{S}^{2}$. In
order to do this, let us fix $t_0\in\mathbb{R}$ and define on
$\{t_0\}\times\mathbb{S}^{1}\times\mathbb{S}^{2}$ the coordinate
vector fields $(\partial/\partial\theta)^{a}$,
$(\partial/\partial\sigma)^{a}$, and $(\partial/\partial\xi)^{a}$.
We then extend these fields to the entire
${^{\scriptstyle{\cuatro}}}\mathcal{M}$ space by Lie dragging them
along a smooth vector field $t^a$ defined as the tangent vector to a
smooth congruence of curves transverse to the slices
$\{t\}\times\mathbb{S}^{1}\times\mathbb{S}^{2}$. In particular, we
can simply take $t^{a}:=(\partial/\partial t)^{a}$. By definition,
the $\sigma^{a}$ field is defined to be zero in the two
aforementioned submanifolds each diffeomorphic to
$\mathbb{R}\times\mathbb{S}^{1}$; by removing them, we obtain a
4-manifold ${^{\scriptstyle{\cuatro}}}\tilde{\mathcal{M}}$ for which
the 4-tuple $(t^{a},\theta^{a},\sigma^{a},\xi^{a})$ defines a
parallelization. Once we have introduced these vector fields on
${^{\scriptstyle{\cuatro}}}\mathcal{M}$ as \textit{background}
objects, we restrict ourselves to working with 4-metrics
$^{\cuatro}g_{ab}$ satisfying the following conditions:

\begin{figure}[t] \centering
\includegraphics[width=12.5cm]{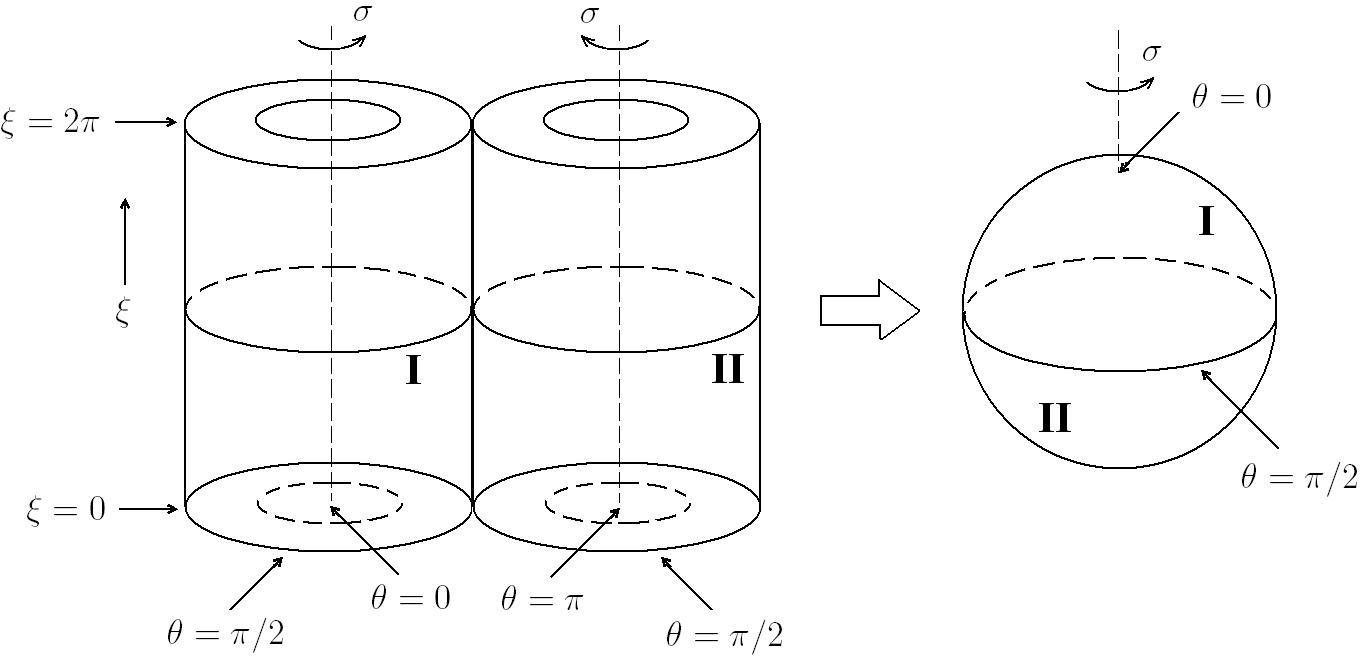}
\caption{Cylindrical coordinates-patches on the 3-handle
$\mathbb{S}^{1}\times\mathbb{S}^{2}$, which has been sliced along
the surface $\theta=\pi/2$. As a result, we obtain two solid tori
that have been further sliced at $\xi=0$ and rendered as solid
cylinders. Each section of constant value of $\xi$ consists of two
discs, I and II (one in each cylinder), being identified at their
edges to form a 2-sphere.}
\end{figure}

\begin{enumerate}
\item The action of the group $G^\dos$ on
${^{\scriptstyle{\cuatro}}}\mathcal{M}$ given by
$(g_1,g_2)\cdot(t,p)=(t,(g_1,g_2)\cdot p)$, $t\in\mathbb{R}$,
$p\in\mathbb{T}^{3}$, with $(g_1,g_2)\cdot p$ defined above, is an
action by isometries, i.e., $\xi^a$ and $\sigma^a$ are linearly
independent Killing vector fields on
${^{\scriptstyle{\cuatro}}}\tilde{\mathcal{M}}$, so that
$\mathcal{L}_{\xi}^{\cuatro}g_{ab}=0$,
$\mathcal{L}_{\sigma}^{\cuatro}g_{ab}=0$.
\item $t$ is a global time function, i.e.,
${^{\scriptstyle{\cuatro}}}g^{ab}(\mathrm{d}t)_b$ is a timelike
vector field. From now on, we will consider the manifold
${^{\scriptstyle{\cuatro}}}\mathcal{M}$ to be endowed with a time
orientation such that this vector field is past-directed.
\item $\{t\}\times\mathbb{S}^{1}\times\mathbb{S}^{2}$ are spacelike hypersurfaces for all
$t\in\mathbb{R}$.
\item $\xi^a$ and $\sigma^a$ are hypersurface orthogonal. This
defines the so called linearly polarized case. This condition means
that the twists of the two fields vanish, which will ultimately
allow us to simplify the field equations and describe the system as
a simple theory of scalar fields.
\end{enumerate}
Two simple but important results that can be proved at this point as
a consequence of the first are the following:

\bigskip

\noindent i) If $\xi^a$ and $\sigma^a$ are Killing vectors and
$[\xi,\sigma]^a=0$, then
$\mathcal{L}_\sigma(^{\cuatro}g_{ab}-\xi_a\xi_b/\lambda_{\xi})=0$,
where
$\lambda_{\xi}:={^{\cuatro}}g_{ab}\xi^{a}\xi^{b}:=\xi_{a}\xi^{a}>0$.

\bigskip

\noindent ii) Furthermore, if we define the vector $X^a$ orthogonal
to $\xi^a$ as $X^a:=\sigma^a-\xi^a(\xi^b\sigma_b)/\lambda_{\xi}$, it
satisfies $[\xi,X]^a=0$ and also
$\mathcal{L}_X(^{\cuatro}g_{ab}-\xi_a\xi_b/\lambda_{\xi})=0$. This
means that, without loss of generality, we can work with everywhere
orthogonal and commuting Killing vector fields $\xi^a$ and
$\sigma^a$. In fact, we impose:

\bigskip

\indent 5. $(\theta^{a},\sigma^{a},\xi^{a})$ are mutually
${^{\cuatro}}g$-orthogonal vector fields.

\bigskip

\indent Let us consider now the Einstein-Klein-Gordon equations
\begin{equation}\label{EKG}
{^{\cuatro}}R_{ab}=8\pi
G_{N}(\mathrm{d}\phi)_a(\mathrm{d}\phi)_b\,,\quad
{^{\cuatro}}g^{ab}\,{^{\cuatro}}\nabla_{a}{^{\cuatro}}\nabla_{b}\phi=0\,,
\end{equation}
corresponding to (1+3)-dimensional gravity minimally coupled to a
zero rest mass scalar field $\phi$ symmetric under the
diffeomorphisms generated by the Killing fields
($\mathcal{L}_{\xi}\phi=\mathcal{L}_{\sigma}\phi=0$,
$\mathcal{L}_{\xi}{^{\cuatro}}g_{ab}=\mathcal{L}_{\sigma}{^{\cuatro}}g_{ab}=0$).
Here ${^{\cuatro}}R_{ab}$ and ${^{\cuatro}}\nabla_{a}$ denote the
Ricci tensor and  the Levi-Civita connection  associated with
${^{\cuatro}}g_{ab}$, respectively. The exterior derivative of the
scalar field $\phi$ is denoted by $(\mathrm{d}\phi)_a$ and  $G_{N}$
is the Newton constant. In order to get a simplified, lower
dimensional description, we will perform a Geroch symmetry reduction
with respect to the nonvanishing Killing vector field $\xi^{a}$ on
the 3-manifold
${^{\tres}}\mathcal{M}:={^{\cuatro}}{\mathcal{M}}/U(1)\simeq\mathbb{R}\times\mathbb{S}^{2}$.
In the present situation, the hypersurface orthogonality of
$\xi^{a}$ allows us to view ${^{\tres}}\mathcal{M}$ as one of the
embedded submanifolds everywhere orthogonal to the closed orbits of
$\xi^a$, endowed with the induced metric
$${^{\tres}}g_{ab}:={^{\cuatro}}g_{ab}-\lambda_{\xi}^{-1}\xi_{a}\xi_{b}\,.$$
In the linearly polarized case, the twists of the Killing fields
vanish and the field equations can be written as those corresponding
to a set of massless scalar fields coupled to (1+2)-gravity by
performing the conformal transformation
$g_{ab}:=\lambda_{\xi}{^{\tres}}g_{ab}$. The system (\ref{EKG}) is
then equivalent to (see the \emph{Theorem \ref{GerochTheorem}} in
\emph{appendix \ref{AppendixGeroch}} for more details)
\begin{equation}
R_{ab}=\frac{1}{2}\sum_i
(\mathrm{d}\phi_i)_a(\mathrm{d}\phi_i)_b\,,\quad
g^{ab}\nabla_a\nabla_b\phi_i=0\,,\quad \mathcal{L}_\sigma
g_{ab}=0\,,\quad \mathcal{L}_\sigma\phi_i=0\,,\label{ecs}
\end{equation}
where $R_{ab}$ and $\nabla_{a}$ denote, respectively, the Ricci
tensor and the Levi-Civita connection associated with $g_{ab}$ --all
of them being three dimensional objects on
${^{\tres}}\mathcal{M}$--, and we have defined\footnote{It is
completely straightforward to couple any number $N$ of massless
scalar fields; in practice this can be done by supposing that the
index $i$ runs from $1$ to $N+1$. The subindex $i = 1$ will always
label the gravitational scalar that encodes the local gravitational
degrees of freedom.} $\phi_1:=\log\lambda_{\xi}$,
$\phi_2:=\sqrt{16\pi G_{N}}\phi$. Recall that we still have the
additional symmetry generated by the remaining Killing vector field
$\sigma^a$, which vanishes at $\theta=0,\pi$. Let us consider the
corresponding space of orbits
${^{\dos}}\mathcal{M}:={^{\tres}}\mathcal{M}/U(1)\simeq\mathbb{R}\times[0,\pi]$.
The induced 2-metric of signature $(-+)$ on $^{\dos}\mathcal{M}$ can
be written
$$s_{ab}=g_{ab}-\tau^{-2}\sigma_a\sigma_b\,,$$
where $\tau^2:=g_{ab}\sigma^{a}\sigma^{b}\ge0$ is the area density
of the symmetry $G^\dos$-group orbits, which vanishes at
$\theta=0,\pi$. In the following, we will use the notation
$\tau=+\sqrt{\tau^2}$. The global time function $t$ induces a
foliation over $^{\dos}\mathcal{M}$. Let $n^{a}$ be the $g$-unit and
future-directed ($g^{ab}n_{a}(\mathrm{d}t)_b>0$) vector field normal
to this foliation, and let $\hat{\theta}^{a}$ be the $g$-unit
spacelike vector field tangent to the slices of constant $t$, such
that
$$\theta^{a}=e^{\gamma/2}\hat{\theta}^{a}$$
for some extra field $\gamma$. If we choose the congruence of curves
with $t^{a}$ tangent to ${^{\dos}}\mathcal{M}$, then the congruence
is transverse to the foliation, and we can express
\begin{equation}\label{ta}
t^{a}=e^{\gamma/2}(Nn^{a}+N^{\theta}\hat{\theta}^{a})\,,
\end{equation}
where $N>0$ and $N^{\theta}$ are proportional to the usual lapse and
shift functions. As we will see, the unusual factor $e^{\gamma/2}$
will allow us to obtain a proper gauge algebra and simplify later
calculations. We require $N$, $N^{\theta}$, and $\gamma$ to be
smooth real-valued fields on ${^{\tres}}\mathcal{M}$. The expression
of the metric in terms of the vector fields introduced above is
$$g_{ab}=-n_{a}n_{b}+\hat{\theta}_{a}\hat{\theta}_{b}+\tau^{-2}\sigma_{a}\sigma_{b}\,,$$
where the indices are raised or lowered with the $g_{ab}$ metric.
Using equation (\ref{ta}) and taking into account the orthogonality
conditions
\begin{eqnarray}
&g_{ab}n^an^b=-1\,,&\quad g_{ab}n^a\sigma^b=0\,,\nonumber\\
&g_{ab}\sigma^a\sigma^b=\tau^2\,,&\quad
g_{ab}n^a\hat{\theta}^b=0\,,\label{orthogcond}\\
&g_{ab}\hat{\theta}^a\hat{\theta}^b=+1\,,&\quad
g_{ab}\hat{\theta}^{a}\sigma^b=0\,,\nonumber
\end{eqnarray}
the metric may be expressed as
\begin{eqnarray}\label{metric}
\hspace*{-0.3cm}g_{ab}=e^{\gamma}\hspace*{-0.08cm}\left(\left(N^{\theta2}\hspace*{-0.05cm}-\hspace*{-0.05cm}N^{2}\right)(\mathrm{d}t)_{a}(\mathrm{d}t)_{b}\hspace*{-0.05cm}+\hspace*{-0.05cm}
2N^{\theta}(\mathrm{d}t)_{(a}(\mathrm{d}\theta)_{b)}\hspace*{-0.05cm}+\hspace*{-0.05cm}
(\mathrm{d}\theta)_{a}(\mathrm{d}\theta)_{b}\right)\hspace*{-0.05cm}+\hspace*{-0.05cm}
\tau^{2}(\mathrm{d}\sigma)_{a}(\mathrm{d}\sigma)_{b}\,.
\end{eqnarray}
The fact that the vectors $(t^a,\theta^a,\sigma^a)$ commute
everywhere translates into necessary conditions that the vectors
$n^a$ and $\theta^a$ and the scalars $N$, $N^\theta$, and $\gamma$
must satisfy. As a consequence of the invariance property
$\mathcal{L}_{\sigma}g_{ab}=0$, the Lie derivative of relations
(\ref{orthogcond}) with respect to the Killing field $\sigma^{a}$
yields
\begin{eqnarray}
&g_{ab}(\mathcal{L}_{\sigma}n)^an^b=0\,,&\quad g_{ab}(\mathcal{L}_{\sigma}n)^a\sigma^b=0\,,\label{Lie_cond_A}\\
&\mathcal{L}_{\sigma}\tau^2=0\,,\hspace{1.2cm}&\quad
g_{ab}(\mathcal{L}_{\sigma}n)^a\hat{\theta}^b+
g_{ab}n^a(\mathcal{L}_{\sigma}\hat{\theta})^b=0\,,\label{Lie_cond_B}\\
&g_{ab}(\mathcal{L}_{\sigma}\hat{\theta})^a\hat{\theta}^b=0\,,&\quad
g_{ab}(\mathcal{L}_{\sigma}\hat{\theta})^a\sigma^b=0\,.\label{Lie_cond_C}
\end{eqnarray}
Equations (\ref{Lie_cond_A}) and (\ref{Lie_cond_C}) imply that the
unique nonvanishing terms of $(\mathcal{L}_{\sigma}n)^{a}$ y
$(\mathcal{L}_{\sigma}{\hat{\theta}})^{a}$ lie on the
$\hat{\theta}^{a}$ and $n^{a}$ directions, respectively. They are
related by
\begin{equation}
(\mathcal{L}_{\sigma}n)^{a}=\alpha\hat{\theta}^{a}\,,\quad
(\mathcal{L}_\sigma{\theta})^{a}=\alpha n^{a}\,,
\label{liederntheta}
\end{equation}
where $\alpha$ is an extra scalar field. The commutation relations
$[t,\sigma]^{a}=0=[\theta,\sigma]^{a}$ lead to
\begin{eqnarray}
&\mathcal{L}_{X}N+\alpha
N^{\theta}+\frac{1}{2}N\mathcal{L}_{X}\gamma=0\,,&
\mathcal{L}_{X}N^{\theta}+\alpha N+\frac{1}{2}N^{\theta}\mathcal{L}_{X}\gamma=0\,,\nonumber\\
&\alpha=0\,,\hspace{3.8cm}&\mathcal{L}_{X}\gamma=0\,,\nonumber
\end{eqnarray}
so that the fields $N$, $N^{\theta}$ and $\gamma$ are constant along
the orbits defined by the remaining Killing vector field
$\sigma^{a}$,
\begin{eqnarray}
\mathcal{L}_{\sigma}N=0\,,&\mathcal{L}_{\sigma}N^{\theta}=0\,,&\mathcal{L}_{\sigma}\gamma=0\,.\label{dersigma}
\end{eqnarray}
Note that the scalars $\phi_i$ are also constant on the orbits of
$\sigma^a$, the matter scalar $\phi_2$ because we have imposed this
from the start and the gravitational scalar $\phi_1$ due to the fact
that the two Killings $\xi^a$ and $\sigma^a$ commute:
$\mathcal{L}_\sigma\lambda_{\xi}=0$. Therefore, we will end up with
an essentially two dimensional model with fields depending only on
coordinates $t$ and $\theta$. In what follows, we will simply denote
$\mathcal{L}_t$ with a dot; however, for the moment we will keep the
notation for the Lie derivative along $\theta^{a}$,
$\mathcal{L}_\theta$, having in mind that we will later introduce a
more convenient smooth derivative for smooth axially symmetric
functions. From equation (\ref{liederntheta}), we get for $\alpha=0$
\begin{eqnarray}
&(\mathcal{L}_{\sigma}n)^{a}=0\,,&(\mathcal{L}_{\sigma}\hat{\theta})^{a}=0\,.\label{comm}
\end{eqnarray}
Finally, from the commutation relation $[\theta,t]^{a}=0$ we obtain
\begin{eqnarray}
\frac{1}{2}(\mathcal{L}_{\theta}\gamma)(Nn^a+N^\theta\hat{\theta}^a)
-\frac{1}{2}\dot{\gamma}\hat{\theta}^a+N
e^{\gamma/2}[\hat{\theta},n]^a+(\mathcal{L}_{\theta}N)n^a
+(\mathcal{L}_{\theta}N^\theta)\hat{\theta}^a=0\,.\nonumber
\end{eqnarray}
This last equation can be projected in the directions defined by the
vectors $n^{a}$, $\hat{\theta}^{a}$, and $\sigma^{a}/\tau$ to give
\begin{eqnarray}
&&\frac{1}{2}N(\mathcal{L}_{\theta}\gamma)+
\mathcal{L}_{\theta}N+Ne^{\gamma/2}n^an^b\nabla_a\hat{\theta}_b=0\,,\label{proy_1}\\
&&\frac{1}{2}N^\theta\mathcal{L}_{\theta}\gamma+\mathcal{L}_{\theta}N^\theta-\frac{1}{2}\dot{\gamma}+
Ne^{\gamma/2}\hat{\theta}^a\hat{\theta}^b\nabla_a n_b=0\,,\label{proy_2}\\
&&\hat{\theta}^a\sigma^b\nabla_an_b=0\,.\label{proy_3}
\end{eqnarray}

\subsection{Lagrangian formulation}

\indent The set of equations (\ref{ecs}) can be derived from a
(1+2)-dimensional Einstein-Hilbert action corresponding to gravity
minimally coupled to massless scalars
\begin{eqnarray}\label{act}
\displaystyle {^{\tres}}S(g_{ab},\phi_i)&=&\frac{1}{16\pi
G_{3}}\int_{(t_0,t_1)\times\mathbb{S}^{2}}{^{\tres}}\mathbf{e}\,|g|^{1/2}\left(R-
\frac{1}{2}\sum_i g^{ab}(\mathrm{d}\phi_i)_a(\mathrm{d}\phi_i)_b\right)\nonumber\\
&+& \displaystyle \frac{1}{8\pi
G_{3}}\int_{\{t_{0}\}\times\mathbb{S}^{2}}^{\{t_{1}\}\times\mathbb{S}^{2}}
{^{\dos}}\mathbf{e}\,|h|^{1/2}K\,.
\end{eqnarray}
Here, $R$ denotes the Ricci scalar associated with $g_{ab}$. $K$ and
$h_{ab}$ are, respectively, the trace of the second fundamental form
$K_{ab}$ and the induced 2-metric on the hypersurfaces diffeomorphic
to $\{t\}\times\mathbb{S}^{2}$. In terms of the $n^{a}$ vector
field, $K_{ab}={h_{a}}^{c}\nabla_{c}n_{b}$. $G_{3}$ denotes the
Newton constant per unit length in the direction of the
$\xi$-symmetric orbits. We have restricted the integration region to
a closed interval $[t_0,t_1]$. The action is written with the help
of a fiducial (i.e., non dynamical) volume form
${^{\tres}}\mathbf{e}$ compatible with the canonical volume form
${^{\tres}}\bm{\epsilon}$ defined by the 3-metric $g_{ab}$. This is
given by ${^{\tres}}\bm{\epsilon}=\sqrt{|g|}\,{^{\tres}}\mathbf{e}$.
The notation adopted here is such that in any basis where the
nonvanishing components of ${^{\tres}}\bm{\epsilon}$ have the values
$\pm 1$, the scalar $|g|^{1/2}$ coincides with the square root of
the determinant of the matrix of the metric $g_{\mu\nu}$ in that
basis. The volume form ${^{\tres}}\bm{\epsilon}$ induces a 2-form
${^{\dos}}\bm{\epsilon}_{ab}={^{\tres}}\bm{\epsilon}_{abc}n^{c}$ on
each slice $\{t\}\times \mathbb{S}^{2}$ which agrees with the volume
associated with the 2-metric $h_{ab}$. We have also introduced a
fixed volume 2-form ${^{\dos}}\mathbf{e}$ on
$\{t\}\times\mathbb{S}^{2}$ such that
${^{\dos}}\bm{\epsilon}=\sqrt{|h|}\,{^{\dos}}\mathbf{e}$. It
satisfies
$\sqrt{|g|}\,{^{\tres}}\mathbf{e}_{abc}n^{c}=\sqrt{|h|}\,{^{\dos}}\mathbf{e}_{ab}$.
We require that both ${^{\tres}}\mathbf{e}$ and
${^{\dos}}\mathbf{e}$ be time-independent, i.e.,
$\mathcal{L}_{t}{^{\tres}}\mathbf{e}=0=\mathcal{L}_{t}{^{\dos}}\mathbf{e}$.
We also demand them to be invariant under the action of the
remaining Killing vector field. In particular, given the
(1+2)-dimensional splitting of $\mathbb{R}\times\mathbb{S}^{2}$, it
is natural to choose
${^{\tres}}\mathbf{e}=\mathrm{d}t\wedge{^{\dos}}\mathbf{e}$, with
${^{\dos}}\mathbf{e}$ being the fiducial 2-form associated with a
round metric on $\mathbb{S}^2$ such that
${^{\dos}}\mathbf{e}_{ab}\theta^{a}\sigma^{b}=Ne^{\gamma}\tau/|g|^{1/2}=\sin\theta$,
i.e.,
${^{\dos}}\mathbf{e}=\sin\theta\mathrm{d}\theta\wedge\mathrm{d}\sigma$.
The last integral which appears in (\ref{act}) is notation for the
integral over the final $\{t_{1}\}\times\mathbb{S}^{2}$-hypersurface
minus the integral over the initial
$\{t_{0}\}\times\mathbb{S}^{2}$-hypersurface. This boundary term is
necessary in order to ensure that the variational principle is well
defined $[4]$. We have the Codazzi relation
$$R={^{\dos}}R+K_{ab}K^{ab}-K^{2}+2\nabla_a(n^{a}K-n^{b}\nabla_{b}n^{a})$$
where ${^{\dos}}R$ denotes the Ricci scalar associated with
$h_{ab}$. Making use of Stokes' theorem,
$$\int_{\mathrm{int}(\mathcal{U})}{^{\scriptstyle{(3)}}}\mathbf{e}|g|^{1/2}\nabla_{a}A^{a}=\int_{\partial\mathcal{U}}{^{\scriptstyle{(2)}}}\mathbf{e}|h|^{1/2}\eta_{a}A^{a}\,,
$$
with $\mathcal{U}$ being a compact and oriented manifold and
$\eta^{a}$ the exterior normal unit vector to its boundary (in our
case, $\eta^{a}=n^{a}$ on $\{t_1\}\times\mathbb{S}^{2}$ and
$\eta^{a}=-n^{a}$ on $\{t_0\}\times\mathbb{S}^{2}$), and taking into
account that $n_{a}n^{b}\nabla_{b}n^{a}=0$, we get
\begin{eqnarray}\label{actr}
\!\!\!\!\displaystyle {^{\tres}}S(g_{ab},\phi_i)\!=\!\frac{1}{16\pi
G_{3}}\int_{t_0}^{t_1}\mathrm{d}t\!\int_{\mathbb{S}^2}
{^{\dos}}\mathbf{e}\,|g|^{1/2}\Big(\,^{\dos}R+K_{ab}K^{ab}-K^2
\nonumber\\
- \frac{1}{2}\sum_i g^{ab}
(\mathrm{d}\phi_{i})_a(\mathrm{d}\phi_{i})_b\Big).&&
\end{eqnarray}
The matrix of the $h_{ab}$ metric induced on $\{t\}\times
\mathbb{S}^{2}$ takes the expression
\begin{equation}
\left[
\begin{array}{cc}
h_{\theta\theta}&h_{\theta \sigma}\\
h_{\sigma \theta}&h_{\sigma\sigma}
\end{array}
\right] =\left[
\begin{array}{cc}
e^\gamma&0\\
0& \tau^2
\end{array}
\right],
\end{equation}
so that the unique nonvanishing Christoffel symbols are
$\Gamma^{\theta}_{\theta\theta}=\frac{1}{2}\mathcal{L}_{\theta}\gamma$,
$\Gamma^{\theta}_{\sigma\sigma}=-e^{-\gamma}\tau\mathcal{L}_{\theta}\tau$,
and
$\Gamma^{\sigma}_{\theta\sigma}=\tau^{-1}\mathcal{L}_{\theta}\tau$.
Hence, the Ricci scalar is given by
\begin{equation}
{^{\scriptstyle{(2)}}}R=\tau^{-1}e^{-\gamma}\left((\mathcal{L}_{\theta}\tau)\mathcal{L}_{\theta}\gamma-2\mathcal{L}_{\theta}^{2}\tau\right),
\label{R2}
\end{equation}
where
$\mathcal{L}_{\theta}^{2}:=\mathcal{L}_{\theta}\circ\mathcal{L}_{\theta}$.
Next, we proceed to calculate the components of the extrinsic
curvature $K_{ab}={h_{a}}^{c}\nabla_{c}n_{b}$. On one hand,
\begin{eqnarray*}
\hat{\theta}^a\hat{\theta}^bK_{ab} & = &
\hat{\theta}^a\hat{\theta}^bh_a^{\,\,c}\nabla_cn_b=\hat{\theta}^a\hat{\theta}^b
\left(\hat{\theta}_a\hat{\theta}^c+\tau^{-2}\sigma_a\sigma^c\right)\nabla_cn_b
=\hat{\theta}^a\hat{\theta}^b\nabla_an_b \nonumber\\& = &
\frac{e^{-\gamma/2}}{2N}\left(\dot{\gamma}-2\mathcal{L}_{\theta}N^{\theta}-N^{\theta}\mathcal{L}_{\theta}\gamma\right),
\end{eqnarray*}
where we have used the equations (\ref{comm}) and (\ref{proy_2}).
Similarly,
\begin{eqnarray*}
\sigma^a\sigma^bK_{ab}&=&
\sigma^a\sigma^bh_a^{\,\,c}\nabla_cn_b=\sigma^a\sigma^b
\left(\hat{\theta}_a\hat{\theta}^c+\tau^{-2}\sigma_a\sigma^c\right)\nabla_cn_b
=\sigma^a\sigma^b\nabla_an_b\nonumber\\
&=&\sigma^bn^a\nabla_a\sigma_b=\tau
n^a\nabla_a\tau=\frac{\tau}{N}\left(e^{-\gamma/2}t^a-N^\theta\hat{\theta}^a\right)\nabla_a\tau\nonumber\\
&=& \frac{\tau
e^{-\gamma/2}}{N}\left(\dot{\tau}-N^\theta\mathcal{L}_{\theta}\tau\right),
\end{eqnarray*}
where we have used the relations (\ref{ta}) and (\ref{comm}), as
well as
\begin{equation*}
\hat{\theta}^aX^bK_{ab}=
\hat{\theta}^aX^bh_a^{\,\,c}\nabla_cn_b=\hat{\theta}^aX^b
\left(\hat{\theta}_a\hat{\theta}^c+\tau^{-2}X_aX^c\right)\nabla_cn_b
=\hat{\theta}^aX^b\nabla_an_b=0\,,
\end{equation*}
which follows from (\ref{comm}) and (\ref{proy_3}). Hence, we
finally obtain
\begin{equation*}
K_{ab}=\frac{e^{-\gamma/2}}{2N}\left(\dot{\gamma}-N^\theta\mathcal{L}_{\theta}\gamma-2\mathcal{L}_{\theta}N^{\theta}\right)\hat{\theta}_a\hat{\theta}_b+
\frac{e^{-\gamma/2}}{N\tau^3}\left(\dot{\tau}-N^\theta\mathcal{L}_{\theta}\tau\right)\sigma_a\sigma_b\,,
\end{equation*}
with trace $K=h^{ab}K_{ab}$. From this expression, we easily get
\begin{equation}
K_{ab}K^{ab}-K^{2}=-\frac{e^{-\gamma}}{N^{2}\tau}\left(\dot{\gamma}-N^{\theta}\mathcal{L}_{\theta}\gamma-2\mathcal{L}_{\theta}N^{\theta}\right)
\left(\dot{\tau}-N^{\theta}\mathcal{L}_{\theta}\tau\right).\label{KK-K2}
\end{equation}
For the scalar fields $\phi_{i}$, we have
\begin{eqnarray}
g^{ab}\left(\nabla_a\phi_{i}\right)\nabla_b\phi_{i}&=&
\left(-n^an^b+\hat{\theta}^a\hat{\theta}^b+\tau^{-2}X^aX^b\right)
\nabla_a\phi_{i}\nabla_b\phi_{i}\nonumber\\
&=&\Big(-\frac{e^{-\gamma}}{N^{2}}\left(t^{a}-N^{\theta}e^{\gamma/2}\hat{\theta}^{a}\right)
\left(t^{b}-N^{\theta}e^{\gamma/2}\hat{\theta}^{b}\right)+\hat{\theta}^{a}\hat{\theta}^{b}\nonumber\\
&+&\tau^{-2}X^{a}X^{b}\Big)\left(\nabla_{a}\phi_{i}\right)\nabla_{b}\phi_{i}\nonumber\\
&=&-\frac{e^{-\gamma}}{N^2}\left(\dot{\phi}_{i}^2-2N^\theta
\dot{\phi}_{i}\mathcal{L}_{\theta}\phi_{i}+\left((N^{\theta})^2-N^{2}\right)(\mathcal{L}_{\theta}\phi_{i})^{2}\right)\,.\label{phi_i}
\end{eqnarray}
Substituting (\ref{R2}), (\ref{KK-K2}) and (\ref{phi_i}) in the
action (\ref{actr}), we finally get
\begin{eqnarray}\label{act3b}
{^{\scriptstyle{(3)}}}S(N,N^{\theta},\gamma,\tau,\phi_{i})&=&\frac{1}{16\pi
G_3}\int_{t_0}^{t_1}\mathrm{d}t\!\int_{\mathbb{S}^2}
{^{\dos}}\mathbf{e}\,|g|^{1/2}e^{-\gamma}\Bigg\{
\frac{1}{\tau}\left((\mathcal{L}_{\theta}\gamma)\mathcal{L}_{\theta}\tau-2\mathcal{L}_{\theta}^{2}\tau\right)\nonumber\\
&-&\frac{1}{\tau N^2}
\left(\dot{\gamma}-2\mathcal{L}_{\theta}N^{\theta}-N^{\theta}\mathcal{L}_{\theta}\gamma
\right)\left(\dot{\tau}-N^\theta\mathcal{L}_{\theta}\tau\right)
\\
&+&\frac{1}{2N^2}\sum_{i}\hspace*{-0.05cm}\left[\dot{\phi}^2_i-2N^\theta\dot{\phi}_i\mathcal{L}_{\theta}\phi_{i}\hspace*{-0.05cm}+\hspace*{-0.05cm}
\left((N^{\theta})^2\hspace*{-0.05cm}-\hspace*{-0.05cm}N^{2}\right)(\mathcal{L}_{\theta}\phi_{i})^{2}\right]\hspace*{-0.05cm}\Bigg\}\nonumber.
\end{eqnarray}

\subsection{Regularity of the metric}

\indent From a classical point of view, the final outcome of the
Hamiltonian analysis of the system that we will discuss in the
following is a set of equations whose solutions allow us to
reconstruct a four dimensional spacetime metric and a set of scalar
fields satisfying the coupled Einstein-Klein-Gordon equations. This
means that, once we decide the functional space to which this metric
belongs, the objects that appear during the dimensional reduction,
gauge fixing and so on may be subject to some regularity conditions.
In the 3-torus case these are simple smoothness requirements but in
the present case, due to the existence of a symmetry axis, these are
more complicated. The regularity conditions that the metric
components for an axially symmetric metric must verify can be
deduced as in  $[3,5]$. Given an 4-dimensional axisymmetric
spacetime with polar coordinates $(t,z,r,\varphi)$ in a neighborhood
of the axis, any regular symmetric tensor field $M_{ab}$ must take
the form
\begin{equation}\label{Mmunu}
[M_{\mu\nu}]=\left[\begin{array}{cccc}
A&B&rD&r^{2}F\\
B&C&rE&r^{2}G\\
rD&rE&H+r^{2}J&r^{3}K\\
r^{2}F&r^{2}G&r^{3}K&r^{2}(H-r^{2}J)
\end{array}\right],
\end{equation}
where $A,B,\ldots,K$ are functions of $t$, $z$ and $r^{2}$. By using
the coordinates $(t,\theta,\sigma,\xi)$, we can write the original
4-metric ${^{\cuatro}}g_{ab}$ for the Gowdy
$\mathbb{S}^{1}\times\mathbb{S}^{2}$ models as
\begin{eqnarray}
{^{\cuatro}}g_{ab}&=&e^{(\gamma-\phi_1)}\Big((N^{\theta
2}-N^{2})(\mathrm{d}t)_a(\mathrm{d}t)_b+
2N^{\theta}(\mathrm{d}t)_{(a}(\mathrm{d}\theta)_{b)}+(\mathrm{d}\theta)_a(\mathrm{d}\theta)_b\Big)
\label{4metric}\\
&+& \tau^{2}e^{-\phi_1}(\mathrm{d}\sigma)_a(\mathrm{d}\sigma)_b+
e^{\phi_1}(\mathrm{d}\xi)_a(\mathrm{d}\xi)_b\,.\nonumber
\end{eqnarray}
Identifying $t\leftrightarrow t$, $z\leftrightarrow\xi$,
$r\leftrightarrow\theta$ and  $\varphi\leftrightarrow\sigma$, we
easily obtain the following regularity conditions for the metric
(here we impose analyticity; otherwise we need only to know the
asymptotic behavior for small values of $\sin{\theta}$)
\begin{eqnarray}
&&e^{(\gamma-\phi_1)}(N^{\theta 2}-N^2)=A(t,\cos\theta)\,,\label{reg1}\\
&&e^{(\gamma-\phi_1)}N^\theta=B(t,\cos\theta)\sin\theta\,, \label{reg2}\\
&&e^{\phi_1}=C(t,\cos\theta)\,,\label{reg3}\\
&&e^{\gamma-\phi_1}=D(t,\cos\theta)+E(t,\cos\theta)\sin^2\theta\,, \label{reg4}\\
&&\tau^2e^{-\phi_1}=\sin^2\theta\big(D(t,\cos\theta)-E(t,\cos\theta)\sin^2\theta
\big)\,,\label{reg5}
\end{eqnarray}
where $A,\,B,\,C,\,D,\,E$ are analytic in their arguments. Here,
$C>0$ and $D(t,\cos\theta)\pm E(t,\cos\theta)\sin^{2}\theta>0$, so
that $D>0$. Finally,
$0<N^{2}={N^{\theta}}^{2}-e^{-(\gamma-\phi_{g})}A(t,\cos\theta)$,
which implies
$\big(B^{2}(t,\cos\theta)-A(t,\cos\theta)E(t,\cos\theta)\big)\sin^{2}\theta>A(t,\cos\theta)D(t,\cos\theta)$.
The conditions for the fields themselves (dropping the $t$
dependence) become
\begin{eqnarray}
&&\phi_i=\hat{\phi}_{i}(\cos\theta)\,,\label{reg_1}\\
&&\gamma=\hat{\gamma}(\cos\theta)\,,\label{reg_2}\\
&&N^\theta=\hat{N}^{\theta}(\cos\theta)\sin\theta\,,\label{reg_3}\\
&&N=\hat{N}(\cos\theta)\,,\label{reg_4}\\
&&\tau=\hat{T}(\cos\theta)\sin\theta\,,\label{reg_5}\\
&&\tau^2e^{-\gamma}=\frac{D(\cos\theta)-E(\cos\theta)\sin^2\theta}{D(\cos\theta)
+E(\cos\theta)\sin^2\theta}\sin^2\theta\,,\label{reg_6}
\end{eqnarray}
where
$\hat{\phi}_{i},\hat{\gamma},\hat{N}^{\theta},\hat{N},\hat{T}:[-1,1]\rightarrow\mathbb{R}$
($\hat{N}>0$) can be written as functions of $A,\,B,\,C,\,D,\,E$.
They must be differentiable functions in $(-1,1)$ with definite
right and left derivatives at $\pm1$. This is, they must be
$C^\infty$ functions in $(-1,1)$ with bounded derivatives. Note that
the singular dependence of all relevant fields has been factored out
($\sin\theta$ is \textit{not} a smooth function on the sphere). The
functions defined on $\mathbb{S}^2$ as
$\hat{\phi}_{i}\circ\cos\theta,\hat{\gamma}\circ\cos\theta,
\hat{N}^{\theta}\circ\cos\theta,\hat{N}\circ\cos\theta,\hat{T}\circ\cos\theta$,
which are analytic on the sphere and invariant under rotations
around its symmetry axis, will be considered as the basic fields to
describe our system. In what follows, they will be simply referred
to as $\hat{\phi}_{i},\hat{\gamma},\hat{N}^{\theta},\hat{N},\hat{T}$
(without the $\circ\cos\theta$ that will only be used if the
possibility of confusion arises), and collectively as the
\emph{hat}-fields.\\
\indent Note that condition (\ref{reg_6}) implies that the values of
the fields $\hat{T}$ and $\hat{\gamma}$ at the poles of the sphere
are not independent of each other, but are related by the relations
\begin{equation}
\hat{T}(\pm1)=e^{\hat{\gamma}(\pm1)/2}\,.\label{polarS2}
\end{equation}
As we will see, these \emph{polar constraints} are necessary
ingredients to ensure the consistency of the models as they
guarantee the differentiability of the other constraints present in
them.
\\
\linebreak \indent Now, we proceed to rewrite the action
(\ref{act3b}) as the integral of a smooth function on the sphere.
This will contain essentially the $hat$-fields, some suitable smooth
derivative of them, and smooth functions of $\cos\theta$. Indeed,
given a smooth and axially symmetric function on $\mathbb{S}^2$, its
$\mathcal{L}_\theta$-derivative cannot necessarily be extended as a
smooth function on the sphere. For a concrete example, consider the
function $\cos\theta$ itself, whose derivative is given by
$\mathcal{L}_\theta\cos\theta=-\sin\theta$. We can, however, define
a smooth derivative $f^\prime$ for any smooth axially symmetric
function as the extension of
\begin{equation}\label{primederivative}
f^\prime:=-\frac{1}{\sin\theta}\partial_\theta f
\end{equation}
to $\mathbb{S}^2$. This is formally done by considering $f$ as a
function of $\cos\theta$ and differentiating. In particular,
$f^{\prime\prime}=-(\cos\theta/\sin^{3}\theta)\partial_{\theta}f+(1/\sin^{2}\theta)\partial_{\theta}^{2}f$.
In the following, the \textit{prime} symbol will always refer to
this derivative. Taking this into account, we get the action
\begin{eqnarray}
&&\frac{1}{16\pi G_3}\int_{t_0}^{t_1}
\mathrm{d}t\int_{\mathbb{S}^2}{^{\dos}}\mathbf{e}\,\bigg\{
\hat{N}[(\hat{\gamma}^\prime\hat{T}^\prime-2\hat{T}^{\prime\prime})\sin^2\theta+
(6\hat{T}^{\prime}-\hat{\gamma}^\prime\hat{T})\cos\theta+2\hat{T}]\nonumber\\
&&\hspace{19mm}+\,\,\frac{1}{\hat{N}}[\hat{N}^\theta
\hat{T}\cos\theta-\dot{\hat{T}}-\hat{N}^\theta
\hat{T}^\prime\sin^2\theta]\,[\dot{\hat{\gamma}}
+(2\hat{N}^{\theta\prime}+\hat{N}^\theta\hat{\gamma}^\prime)\sin^2\theta-2\hat{N}^\theta\cos\theta]
\nonumber\\
&&\hspace{19mm}+\,\,\frac{\hat{T}}{2\hat{N}}\sum_{i}\bigg({\dot{\hat{\phi}}}\,^2_i
+2\hat{N}^\theta\dot{\hat{\phi}}_i\,\hat{\phi}_i^\prime\sin^2\theta+(\hat{N}^{\theta2}\sin^2\theta
-\hat{N}^2)\hat{\phi}_i^{\prime2}\sin^2\theta\bigg)\bigg\}\,.\nonumber
\end{eqnarray}
From now on, we will use units such that $16\pi G_{3}=1$. In order
to express the action in its canonical form we must obtain the
corresponding canonically conjugate momenta through a Legendre
transformation,
\begin{eqnarray}
&&\Pi:=p_{\hat{N}}=0\,,\quad \Pi_{\theta}:=p_{\hat{N}^{\theta}}=0\,,\label{primaryconstr}\\
&&p_{\hat{\gamma}}=\frac{1}{\hat{N}}\left(\hat{N}^\theta
\hat{T}\cos\theta-\dot{\hat{T}}-\hat{N}^\theta
\hat{T}^\prime\sin^2\theta\right),\\
&&p_{\hat{T}}=-\frac{1}{\hat{N}}\left(\dot{\hat{\gamma}}
+(2\hat{N}^{\theta\prime}+\hat{N}^\theta\hat{\gamma}^\prime)\sin^2\theta-2\hat{N}^\theta\cos\theta\right),\\
&&p_{\hat{\phi}_{i}}=\frac{\hat{T}}{\hat{N}}\left(\dot{\hat{\phi}}_{i}+\hat{N}^{\theta}\hat{\phi}_{i}^{\prime}\sin^{2}\theta\right).
\end{eqnarray}
We see that the Lagrangian function is singular, which prevents
solving for \emph{all} the generalized velocities in terms of
momenta. This can certainly be done for $\dot{\hat{\gamma}}$,
$\dot{\hat{T}}$ and $\dot{\hat{\phi}}_{i}$, but not for
$\dot{\hat{N}}$ and $\dot{\hat{N}}^{\theta}$. The Hamiltonian
formalism of the model therefore requires the application of the
Dirac-Bergmann algorithm for constrained systems.\footnote{For more
details on the treatment of constrained Hamiltonian system, the
reader can consult the \emph{appendix \ref{AppendixSymplectic}}.}
Introducing the Lagrange multipliers $\lambda$ and $\lambda^\theta$
to enforce the primary constraints (\ref{primaryconstr}), we obtain
the action
\begin{equation}\label{actlambda}
\int_{t_0}^{t_1}\mathrm{d}t\left(\int_{\mathbb{S}^{2}}{^{\dos}}\mathbf{e}\left(
\dot{\hat{N}}p_{\hat{N}}+\dot{\hat{N}}^{\theta}p_{\hat{N}^{\theta}}
+\dot{\hat{\gamma}}p_{\hat{\gamma}}+\dot{\hat{T}}p_{\hat{T}}
+\sum_{i}\dot{\hat{\phi}}_{i}p_{\hat{\phi}_{i}}\right)-H[\lambda,\lambda^{\theta},\hat{N},\hat{N}^{\theta}]\right),
\end{equation}
with the Hamiltonian function
\begin{equation}\label{Hamiltlambda}
H[\lambda,\lambda^{\theta},\hat{N},\hat{N}^{\theta}]:=\int_{\mathbb{S}^{2}}{^{\dos}}\mathbf{e}\left(\lambda\Pi+\lambda^{\theta}\Pi_{\theta}
+\hat{N}C+\hat{N}^{\theta}C_{\theta}\right),
\end{equation}
where
\begin{eqnarray}
C&:=&-p_{\hat{\gamma}}p_{\hat{T}}+(2\hat{T}^{\prime\prime}-\hat{\gamma}^\prime\hat{T}^\prime)\sin^2\theta
+(\hat{\gamma}^\prime\hat{T}-6\hat{T}^\prime)\cos\theta-2\hat{T}\nonumber\\
&&+\frac{1}{2}\sum_{i}\bigg(\frac{p_{\hat{\phi}_i}^2}{\hat{T}}+\hat{T}\hat{\phi}_i^{\prime2}\sin^2\theta\bigg)\,,\label{C1S2}\\
C_{\theta}&:=& \Big(2p^\prime_{\hat{\gamma}}-\hat{\gamma}^\prime
p_{\hat{\gamma}}- \hat{T}^\prime
p_{\hat{T}}-\sum_{i}\hat{\phi}_i^\prime
p_{\hat{\phi}_i}\Big)\sin^2\theta +\left(\hat{T}
p_{\hat{T}}-2p_{\hat{\gamma}}\right)\cos\theta\,.\label{C2S2}
\end{eqnarray}
The variation of the action with respect to the Lagrange multipliers
$\lambda$ and $\lambda^{\theta}$ provides the primary constraints
(\ref{primaryconstr}). In order to guarantee the consistency of the
dynamics of the system, these constraints must be preserved under
the time evolution of the system (intuitively speaking, their
Poisson brackets with the Hamiltonian must vanish). This leads us to
impose the \emph{secondary constraints}
\begin{equation}\label{CandCthetazero}
C=0\,,\quad C_{\theta}=0\,.
\end{equation}
It is easy to check that, by imposing again the consistency of the
secondary constraints, one does not obtain additional (tertiary)
constraints. Therefore, the dynamical variables are restricted to
belong to a constraint surface in the canonical phase space of the
system, coordinatized by
$(\hat{N},p_{\hat{N}};\hat{N}^{\theta},p_{\hat{N}^{\theta}};\hat{\gamma},p_{\hat{\gamma}};
\hat{T},p_{\hat{T}};\hat{\phi}_{i},p_{\hat{\phi}_{i}})$. This
surface is globally defined by the constraints $\Pi=0=\Pi_{\theta}$
and $C=0=C_{\theta}$. The Hamiltonian of the system is identically
zero on it. Note, however, that the equations of motion of $\hat{N}$
and $\hat{N}^{\theta}$ are, respectively, $\dot{\hat{N}}=\lambda$
and $\dot{\hat{N}}^{\theta}=\lambda^{\theta}$, with $\lambda$ and
$\lambda^{\theta}$ being unspecified time-dependent functions. As a
consequence, the dynamical trajectories of $\hat{N}$ and
$\hat{N}^{\theta}$ are completely arbitrary. Moreover, the Hamilton
equations corresponding to the canonical pairs
$(\hat{\gamma},p_{\hat{\gamma}};\hat{T},p_{\hat{T}};\hat{\phi}_{i},p_{\hat{\phi}_i})$
are not affected by the term
$\big(\lambda\Pi+\lambda^{\theta}\Pi_{\theta}\big)$ appearing in the
Hamiltonian (\ref{Hamiltlambda}). Thus, with respect to the
dynamical variables $\hat{\gamma},\hat{T},\hat{\phi}_{i}$, the
action (\ref{actlambda}) is completely equivalent to $[6]$
\begin{equation}
S:=\int_{t_0}^{t_1}\mathrm{d}t\left(\int_{\mathbb{S}^2}{^{\dos}}\mathbf{e}\left(\dot{\hat{\gamma}}p_{\hat{\gamma}}+
\dot{\hat{T}}p_{\hat{T}}+\sum_{i}\dot{\hat{\phi}}_{i}p_{\hat{\phi}_{i}}\right)-H[\hat{N},\hat{N}^{\theta}]\right).
\end{equation}
Here, the terms proportional to $\lambda$ and $\lambda^{\theta}$
have been dropped and $\hat{N}$ and $\hat{N}^{\theta}$ are simply
treated as Lagrange multipliers. The new Hamiltonian is given by
\begin{equation}\label{HNNtheta}
H[\hat{N},\hat{N}^{\theta}]=\int_{\mathbb{S}^{2}}{^\dos}\mathbf{e}\left(\hat{N}C+\hat{N}^{\theta}C_{\theta}\right).
\end{equation}
The canonical phase space of the system, denoted by $\Gamma$, is now
coordinatized by the conjugated pairs
$(\hat{\gamma},p_{\hat{\gamma}};\hat{T},p_{\hat{T}};\hat{\phi}_{i},p_{\hat{\phi}_i})$
and endowed with the standard (weakly) symplectic form
\begin{equation}\label{omegaS2XS1}
\omega:=\int_{\mathbb{S}^2}{^{\dos}}\mathbf{e}\,\bigg(\delta\hat{\gamma}\wedge\delta
p_{\hat{\gamma}}+\delta\hat{T}\wedge\delta
p_{\hat{T}}+\sum_{i}\delta\hat{\phi}_i\wedge\delta
p_{\hat{\phi}_i}\bigg)\,.
\end{equation}
Now we proceed to analyze the gauge transformations generated by the
constraints. For this aim, we have to smear the constraints to
obtain well defined functions on the phase space,
\begin{equation}\label{CandCtheta}
C[\hat{N}_{g}]:=\int_{\mathbb{S}^{2}}{^\dos}\mathbf{e}\hat{N}_{g}C\,,\quad
C_{\theta}[\hat{N}_{g}^{\theta}]:=\int_{\mathbb{S}^{2}}{^\dos}\mathbf{e}\hat{N}_{g}^{\theta}C_{\theta}\,.
\end{equation}
It is straightforward to check that the polar constraints
(\ref{polarS2}) guarantee the differentiability of
(\ref{CandCtheta}) without further restrictions on $\hat{N}_{g}$ and
$\hat{N}_{g}^{\theta}$. Indeed, consider the exterior derivative of
$C[\hat{N}_{g}]$,
$$\delta C[\hat{N}_{g}]=\int_{\mathbb{S}^{2}}{^{\dos}}\mathbf{e}\,\sum_{\chi}\frac{\delta C[\hat{N}_{g}]}{\delta\chi(s)}\,\delta\chi(s)\,,$$
with the sum extended over all canonical coordinates of $\Gamma$ and
$\delta/\delta\chi$ denoting a functional derivative. Integrating by
parts, we encounter the surface term
$-2\hat{N}_{g}\cos\theta\delta\hat{T}+\hat{N_{g}}\hat{T}\cos\theta\delta\hat{\gamma}\big]_{\theta=0}^{\pi}$,
which vanishes by virtue of (\ref{polarS2}). The differentiability
of $C_{\theta}[\hat{N}_{g}^{\theta}]$ is trivially verified, since
the corresponding surface term is proportional to $\sin\theta$
evaluated at $\theta=0,\pi$. We then have the gauge
transformations\footnote{The following identities become very useful
in this context: Given a canonical pair $(\varphi,p_{\varphi})$ and
a smooth axially symmetric function $F$ on the sphere, we have
$\int_{\mathbb{S}^{2}}{^{\dos}}\mathbf{e}\,\{\varphi^{\prime},p_{\varphi}\}F=-F^{\prime}$
and
$\int_{\mathbb{S}^{2}}{^{\dos}}\mathbf{e}\,\{\varphi^{\prime\prime},p_{\varphi}\}F=F^{\prime\prime}$.}
\begin{eqnarray*}
&&\{\hat{\gamma},C[\hat{N}_g]\}=-\hat{N}_gp_{\hat{T}}\,,
\nonumber\\
&&\{\hat{T},C[\hat{N}_g]\}=-\hat{N}_g p_{\hat{\gamma}}\,,\nonumber\\
&&\{\hat{\phi_i},C[\hat{N}_{g}]\}=\frac{\hat{N}^g}{\hat{T}}p_{\hat{\phi}_i}\,,\nonumber\\
&&\{p_{\hat{\gamma}},C[\hat{N}_g]\}=
\hat{N}_{g}^{\prime}(\hat{T}\cos\theta-\hat{T}^\prime\sin^2\theta)+
\hat{N}_g(\hat{T}+3\hat{T}^\prime\cos\theta-\hat{T}^{\prime\prime}\sin^2\theta)\,,\nonumber\\
&&\{p_{\hat{T}},C[\hat{N}_{g}]\}=\hat{N}_{g}^{\prime}(2\cos\theta-\hat{\gamma}^\prime\sin^2\theta)+
\hat{N}_{g}(\hat{\gamma}^\prime\cos\theta-\hat{\gamma}^{\prime\prime}\sin^2\theta)-2\hat{N}_{g}^{\prime\prime}
\sin^2\theta\nonumber\\
&&\hspace{2.6cm}+\frac{\hat{N}_g}{2}\sum_{i}\bigg(\frac{p_{\hat{\phi}_i}^2}{\hat{T}^2}-
\sin^2\theta\hat{\phi}_i^{\prime2}\bigg)\,,\nonumber\\
&&\{p_{\hat{\phi}_i},C[\hat{N}_g]\}=
\hat{N}_{g}^{\prime}\hat{T}\hat{\phi}_i^{\prime}\sin^2\theta+\hat{N}_{g}[(\hat{T}^\prime\hat{\phi}_i^\prime+
\hat{T}\hat{\phi}_i^{\prime\prime})\sin^2\theta-2\hat{T}\hat{\phi_i^\prime}\cos\theta]\,,\nonumber
\end{eqnarray*}
and
\begin{eqnarray*}
&&\{\hat{\gamma},C_\theta[\hat{N}_{g}^\theta]\}=-2\hat{N}_{g}^{\theta\prime}\sin^2\theta+
\hat{N}_{g}^\theta(2\cos\theta-\hat{\gamma}^\prime\sin^2\theta)\,,\nonumber\\
&&\{\hat{T},C_\theta[\hat{N}_{g}^\theta]\}=\hat{N}_{g}^\theta(\hat{T}\cos\theta-\hat{T}^\prime\sin^2\theta)\,,\nonumber\\
&&\{\hat{\phi}_i,C_\theta[\hat{N}_{g}^\theta]\}=-\hat{N}_{g}^\theta\hat{\phi}_i^\prime\sin^2\theta\,,\nonumber\\
&&\{p_{\hat{\gamma}},C_\theta[\hat{N}_{g}^\theta]\}=
\hat{N}_{g}^\theta(2p_{\hat{\gamma}}\cos\theta-p^\prime_{\hat{\gamma}}\sin^2\theta)-
\hat{N}_{g}^{\theta\prime}p_{\hat{\gamma}}\sin^2\theta\,,\nonumber\\
&&\{p_{\hat{T}},C_\theta[\hat{N}_{g}^\theta]\}=\hat{N}_{g}^\theta(p_{\hat{T}}\cos\theta-p^\prime_{\hat{T}}\sin^2\theta)-
\hat{N}_{g}^{\theta\prime}p_{\hat{T}}\sin^2\theta\,,\nonumber\\
&&\{p_{\hat{\phi}_i},C_\theta[\hat{N}_{g}^\theta]\}=\hat{N}_{g}^\theta(2p_{\hat{\phi}_i}\cos\theta-
p^\prime_{\hat{\phi}_i}\sin^2\theta)-\hat{N}_{g}^{\theta\prime}p_{\hat{\phi}_i}\sin^2\theta\,.\nonumber
\end{eqnarray*}
Finally, we must check the stability of the polar constraints
(\ref{polarS2}), $(\hat{T}e^{-\hat{\gamma}/2})(\pm1)=1$. To this
end, we compute
\begin{eqnarray}
&&\hspace{-4mm}\{\hat{T}e^{-\hat{\gamma}/2},C[\hat{N}_{g}]\}=8\pi
G_3\hat{N}_{g}e^{-\hat{\gamma}/2}
\left(\hat{T}p_{\hat{T}}-2p_{\hat{\gamma}}\right),\nonumber\\
&&\hspace{-4mm}\{\hat{T}e^{-\hat{\gamma}/2},C_\theta[\hat{N}_{g}^\theta]\}=
e^{-\hat{\gamma}/2}\Big(\hat{T}\hat{N}_g^{\theta\prime}+
\hat{N}_{g}^{\theta}\Big(\frac{1}{2}\hat{T}\hat{\gamma}^\prime-\hat{T}^\prime\Big)\Big)\sin^2\theta\,.
\nonumber
\end{eqnarray}
The first expression vanishes at the poles as a consequence of the
constraint (\ref{C2S2}) for $\theta=0,\,\pi$ ($\sin\theta=0$ and
$|\cos\theta|=1$) whereas the second vanishes because of the
$\sin^2\theta$ factor. We then conclude that there are no secondary
constraints coming from the stability of the polar constraints.

\subsection{Deparameterization}

The dynamical variables are restricted to belong to a constraint
surface $\Gamma_{c}\subset\Gamma$ globally defined by the
constraints $C=0$, $C_{\theta}=0$. A straightforward calculation
shows that these constraints are first class in Dirac's terminology,
or equivalently that $\Gamma_{c}$ is a coisotropic submanifold of
$\Gamma$. Indeed, the Poisson algebra of the constraints is a proper
Lie algebra
\begin{eqnarray*}
&&\hspace{-2mm}\{C[\hat{N}_g],C[\hat{M}_g]\}=C_\theta[\hat{M}_g
\hat{N}_g^{\prime}-\hat{N}_g
\hat{M}_g^{\prime}]\,,\nonumber\\
&&\hspace{-2mm}\{C[\hat{N}_g],C_\theta[\hat{M}_g^\theta]\}=C[(\hat{M}_{g}^\theta
\hat{N}_{g}^{\prime}-\hat{N}_g
\hat{M}_{g}^{\theta\prime})\sin^2\theta+\hat{N}_g\hat{M}_{g}^\theta\cos\theta]\,,\nonumber\\
&&\hspace{-2mm}\{C_\theta[\hat{N}_{g}^\theta],C_\theta[\hat{M}_{g}^\theta]\}=C_\theta[(\hat{M}_{g}^\theta
\hat{N}_{g}^{\theta\prime}-\hat{N}_{g}^\theta
\hat{M}_{g}^{\theta\prime})\sin^2\theta]\,.\nonumber
\end{eqnarray*}
Note that, as a consequence of the introduction of the suitable
exponential factor $e^{\gamma/2}$ in (\ref{ta}) we have a
\textit{closed} gauge algebra $[7,8]$, i.e., with structure
\textit{constants}.
\\
\indent Motion along the directions defined by the weighted
constraints corresponds then to \emph{gauge transformations}, i.e.,
transformations that do not affect the physical state of the system.
Due to this fact, we would like to isolate the true physical degrees
of freedom of the model. As is well known, there are several
possible ways to do this. The first one is to eliminate the
variables representing the gauge degrees of freedom by introducing
the so-called \emph{reduced phase space} of the system, that is, the
(quotient) space of orbits of the gauge diffeomorphisms. Each point
on the reduced phase space is an equivalence class of points on the
constraint surface $\Gamma_{c}$, where two points are regarded as
equivalent if they differ by a symplectic transformation generated
by the (weighted) constraints. The successful implementation of the
reduction allows us not only to label gauge orbits but also provides
us with important mathematical structures (topological, symplectic,
etc) from the ones present in the initial phase space. The second
way is to fix a gauge, by choosing a global cross-section of
$\Gamma$ intersecting the gauge orbits once and only once. Here, we
will see that a \textit{partial} gauge fixing procedure
(\emph{deparameterization}) provides another interesting way to deal
with the system, allowing us to describe it as a nonautonomous
--i.e., time-dependent-- quadratic Hamiltonian system $[9,10,11]$.
\\
\indent The Hamiltonian vector fields associated with the weighted
constraints $C[\hat{N}_g]$ and $C_{\theta}[\hat{N}_{g}^{\theta}]$
are tangential to $\Gamma_c$ and define the degenerate directions of
the pull-back of $\omega$ to this submanifold. The
deparameterization procedure is based on the choice of one of these
Hamiltonian vector fields to define an evolution vector field
$E_{H_R}$ associated with some reduced Hamiltonian $H_R$ of a
generically nonautonomous system. As shown in this section, we will
be able to impose gauge fixing conditions in such a way that just
one of the first class constraints, say $\mathcal{C}$, is not fixed.
This will be used to define dynamics. Let
$\iota:\Gamma_G\rightarrow\Gamma$ denote the embedding of the gauge
fixed surface given by the first class constraints
(\ref{CandCthetazero}) and the gauge fixing conditions; the
pull-back of the symplectic form to this surface, $\iota^{*}\omega$,
has just one degenerate direction defined by a Hamiltonian vector
field $E_{H_R}$. Select then a suitable phase space variable $T$
such that $E_{H_R}(T)=1$. The level surfaces of $T$ are all
diffeomorphic to a manifold $\Gamma_{R}$ and transverse to
$E_{H_R}$, defining a foliation of $\Gamma_{G}$ with $T$ as global
time function. In this case,
$\iota^{*}\omega=-\mathrm{d}T\wedge\mathrm{d}H_R+\omega_{R}$ and
$E_{H_R}=\partial_{T}+X_{H_R}$, where $\omega_{R}$ is a weakly
nondegenerate form, and the triplet $(\Gamma_R,\omega_R,H_R(T))$
defines a nonautonomous Hamiltonian system, mathematically
characterized by a cosymplectic manifold (see the \emph{appendix
\ref{AppendixSymplectic}}).
\\
\linebreak \indent We begin by choosing gauge fixing conditions
similar to those employed in the 3-torus case $[12]$,
\begin{equation}
\hat{T}^\prime=0\,,\quad
p_{\hat{\gamma}}^\prime=0\,.\label{gauge1S2}
\end{equation}
They mean that both $\hat{T}$ and $p_{\hat{\gamma}}$ take the same
value irrespective of $\theta$, but we do not specify which one.
Note that conditions of the type $\hat{T}=T$, $p_{\hat{\gamma}}=p$,
with $T,p\in\mathbb{R}$, not only would tell us that $\hat{T}$ and
$p_{\hat{\gamma}}$ are independent of $\theta$, but also assign a
fixed value to them, thus removing additional degrees of freedom.
With our choice, there is still a dynamical mode in $\hat{T}$ which
may vary in the evolution but is constant on every spatial slice in
the (1+3)-decomposition. It will be eventually identified with a
certain function of the time parameter. A convenient way to discuss
the gauge fixing procedure is to describe the family of gauge
conditions (\ref{gauge1S2}) by introducing an orthonormal basis of
weight functions on the subspace of axially symmetric functions on
$\mathbb{S}^2$,
\begin{equation}\label{Yn}
Y_{n}:=
\left({\frac{2n+1}{4\pi}}\right)^{1/2}\mathscr{P}_n(\cos\theta)\,,\,\,\,n\in\mathbb{N}_0\,,
\end{equation}
where $\mathscr{P}_n$ are the Legendre polynomials. By expanding now
\begin{eqnarray*}
\quad\hat{T}=\sum_{n=0}^\infty \hat{T}_nY_{n}\,,\quad
p_{\hat{\gamma}}=\sum_{n=0}^\infty p_{\hat{\gamma}_n}Y_n\,,
\end{eqnarray*}
with
\begin{eqnarray*}
&&\hat{T}_n=\left({\frac{2n+1}{4\pi}}\right)^{1/2}\int_{\mathbb{S}^2}\!\!\!^{\dos}\mathbf{e}\,
\mathscr{P}_{n}(\cos\theta)\hat{T}\,,\quad
p_{\hat{\gamma}_n}=\left({\frac{2n+1}{4\pi}}\right)^{1/2}\int_{\mathbb{S}^2}\!\!\!^{\dos}\mathbf{e}\,
\mathscr{P}_n(\cos\theta)p_{\hat{\gamma}}\,,
\end{eqnarray*}
the previous gauge fixing conditions (\ref{gauge1S2}) become
\begin{equation}
\hat{T}_n=0=p_{\hat{\gamma}_n}\,,\,\,\, \forall\,n\in
\mathbb{N}\,.\label{gauge3S2}
\end{equation}
In order to see if this is a good gauge fixation --and,
alternatively, find out if some gauge freedom is left-- we compute
\begin{eqnarray}
&&\hspace{-1cm}\{\hat{T}_n,C[Y_m]\}\approx-\sqrt{\frac{(2n+1)(2m+1)}{(4\pi)^3}}p_{\hat{\gamma}0}\int_{\mathbb{S}^2}\!\!\!^{\dos}\mathbf{e}\,\mathscr{P}_n(\cos\theta)\mathscr{P}_m(\cos\theta)=\frac{1}{\sqrt{4\pi}}p_{\hat{\gamma}0}\delta_{nm}
\nonumber\\
&&\hspace{-1cm}\{\hat{T}_n,C_\theta[Y_m]\}\approx\frac{\hat{T}_0}{4\pi}\sqrt{\frac{(2n+1)(2m+1)}{4\pi}}
\int_{\mathbb{S}^2}\!\!\!^{\dos}\mathbf{e}\,\cos\theta
\mathscr{P}_n(\cos\theta)\mathscr{P}_m(\cos\theta)\nonumber
\end{eqnarray}
\begin{eqnarray}
&&\hspace{2.1cm}=\left\{\begin{array}{l}\displaystyle-\frac{(m+1)\hat{T}_0}{\sqrt{4\pi(2m+1)(2m+3)}}
\quad\mathrm{if}\,n=m+1\\
\displaystyle-\frac{m\hat{T}_0}{\sqrt{4\pi(2m+1)(2m-1)}}\quad\mathrm{if}\,n=m-1\\
0\quad\quad\mathrm{otherwise}\end{array}\right.\nonumber\\
&&\hspace{-0.5cm}\{p_{\hat{\gamma}_n},C[Y_m]\}\approx\hat{T}_0\sqrt{\frac{(2n+1)(2m+1)}{(4\pi)^3}}
\int_{\mathbb{S}^2}\!\!\!^{\dos}\mathbf{e}\,\mathscr{P}_n(\cos\theta)\big(\mathscr{P}_m(\cos\theta)+
\cos\theta \mathscr{P}_m^\prime(\cos\theta)\big)\nonumber\\
&&\hspace{1.9cm}=\left\{\begin{array}{l}\displaystyle0\quad\quad\mathrm{if}\,\,m=0\,\, \mathrm{or}\,\,m<n\nonumber\\
\displaystyle-\hat{T}_0\frac{n+1}{\sqrt{4\pi}}\quad\mathrm{if}\,\,m=n\\
\displaystyle\star\quad \quad\mathrm{otherwise}\end{array}\right.\nonumber\\
&&\hspace{-0.5cm}\{p_{\hat{\gamma}_n},C_\theta[Y_m]\}\approx
p_{\hat{\gamma}_0}\sqrt{\frac{(2n+1)(2m+1)}{(4\pi)^3}}
\int_{\mathbb{S}^2}\!\!\!^{\dos}\mathbf{e}\,\mathscr{P}_n(\cos\theta)\big(\mathscr{P}_m(\cos\theta)-\sin^2\theta
\mathscr{P}_m^\prime(\cos\theta)\big)\nonumber\\
&&\hspace{2.0cm}=\left\{\begin{array}{l}\displaystyle0\quad\quad\mathrm{if}\,\,m=0\,\,
\mathrm{or}\,\,m<n-1\nonumber\\
\displaystyle\ast\quad\quad\mathrm{otherwise}\end{array}\right.\nonumber
\end{eqnarray}
where $n\in\mathbb{N}$ and $m\in\mathbb{N}_0$. The symbol $\approx$
denotes equality on the hypersurface defined by the gauge fixing
conditions and the constraints, the so-called gauge-fixing surface
$\Gamma_{G}\subset\Gamma_c$. The $\star$ and $\ast$ symbols denote
terms (computable in closed form but with somewhat complicated
expressions) that are not needed in the following discussion. It is
convenient to display the previous results in a table form as below.
\begin{center}
\begin{tabular}{|c|c c|c c|c|}
\hline\hline
&$\hat{T}_1=0$&$p_{\hat{\gamma}_1}=0$&$\hat{T}_{2}=0$&$p_{\hat{\gamma}_2}=0$&$\hdots$\\\hline
$C[Y_0]$&$0$&$0$&$0$&$0$&$\hdots$\\
$C_\theta[Y_0]$&$\frac{\hat{T}_0}{2\sqrt{3\pi}}$&$0$&$0$&$0$&$\hdots$\\\hline
$C[Y_1]$&$-\frac{p_{\hat{\gamma}_0}}{2\sqrt{\pi}}$&$\frac{\hat{T}_0}{\sqrt{\pi}}$&$0$&$0$&$\hdots$\\
$C_\theta[Y_1]$&$0$&$\ast$&$\frac{\hat{T}_0}{\sqrt{15\pi}}$&$0$&$\hdots$\\\hline
$C[Y_2]$&$0$&$\star$&$-\frac{p_{\hat{\gamma}_0}}{2\sqrt{\pi}}$&$\frac{3\hat{T}_0}{\sqrt{4\pi}}   $&$\hdots$\\
$C_\theta[Y_2]$&$\frac{\hat{T}_0}{\sqrt{15\pi}}$&$\ast$&$0$&$\ast$&$\hdots$\\\hline
$\vdots$&$\vdots$&$\vdots$&$\vdots$&$\vdots$&$\ddots$\\
\hline\hline
\end{tabular}
\end{center}
One must also check if the polar constraints are gauge fixed by
conditions (\ref{gauge3S2}). To this end, we compute
\begin{eqnarray}
&&\{\hat{T}_n,\hat{T}e^{-\hat{\gamma}/2}\}\approx0\,,\nonumber\\
&&\{p_{\hat{\gamma}_n},\hat{T}e^{-\hat{\gamma}/2}\}\approx
\frac{1}{2}\hat{T}e^{-\hat{\gamma}/2}\sqrt{\frac{2n+1}{4\pi}}\mathscr{P}_n(\cos\theta)\,.\nonumber
\end{eqnarray}
The last Poisson bracket is different from zero at the poles
$(\theta=0,\pi)$ for all values of $n\in\mathbb{N}$. Therefore, the
only constraint that is not gauge-fixed by the conditions introduced
above, as long as $p_{\hat{\gamma}_0}\neq0$ and $\hat{T}_0\neq0$, is
$C[1]$,
\begin{equation}\label{Cleft}
\int_{\mathbb{S}^{2}}{^{\dos}}\mathbf{e}\left(-p_{\hat{\gamma}}p_{\hat{T}}+\hat{\gamma}^\prime\hat{T}\cos\theta-2\hat{T}
+\frac{1}{2}\sum_{i}\bigg(\frac{p_{\hat{\phi}_i}^2}{\hat{T}}+\hat{T}\hat{\phi}_i^{\prime2}\sin^2\theta\bigg)\right)\approx0\,.
\end{equation}
This is in contrast with the situation for the 3-torus case, where
one is left with two constraints instead of just one. The final
description of our system is then considerably simpler that in the
$\mathbb{T}^3$ case. This fact will obviously facilitate the
canonical quantization of these models, as well. We now pullback
every relevant geometric object to the submanifold $\Gamma_{G}$
defined by the gauge fixing conditions with the aim of eliminating
some of the variables in our model. Denoting by
$\iota:\Gamma_{G}\rightarrow\Gamma$ the immersion map, the pullback
of the (weakly) symplectic form (\ref{omegaS2XS1}) becomes
\begin{eqnarray}
\iota^{*}\omega=\mathrm{d}\hat{\gamma}_0\wedge \mathrm{d}
p_{\hat{\gamma}_0}+\mathrm{d}\hat{T}_0\wedge \mathrm{d}
p_{\hat{T}_0}+\sum_{i}\int_{\mathbb{S}^2}{^{\dos}}\mathbf{e}\,
\delta\phi_i\wedge \delta p_{\phi_i}\,.\label{symplecticS2}
\end{eqnarray}
The pullback of the constraint (\ref{Cleft}) is
\begin{eqnarray}
\label{remCS2}\mathcal{C}&:=&-p_{\hat{\gamma}_0}p_{\hat{T}_0}+\hat{T}_0
\Big(4\sqrt{\pi}\Big(\log\frac{\hat{T}_0}{\sqrt{4\pi}}-1\Big)-\hat{\gamma_0}\Big)
\\&+&\frac{1}{2}\sum_{i}\int_{\mathbb{S}^2}\!\!\!^{\dos}\mathbf{e}\left(\frac{\sqrt{4\pi}}{\hat{T}_0}p_{\hat{\phi}_i}^2+
\frac{\hat{T}_0}{\sqrt{4\pi}}\hat{\phi}_i^{\prime2}\sin^2\theta\right)\approx0\,.\nonumber
\end{eqnarray}
The gauge transformations generated by this constraint in the
variables $\hat{T}_0$ and $p_{\hat{\gamma}_0}$ are
\begin{equation}
\{\hat{T}_0,\mathcal{C}\}=-p_{\hat{\gamma}_0}\,,\quad\{p_{\hat{\gamma}_0},\mathcal{C}\}=\hat{T}_0\,,\nonumber
\end{equation}
so if we parameterize the gauge orbits with $s\in(0,\pi)$ we see
that on them we have $\hat{T}_0=p\sin s$ and
$p_{\hat{\gamma}_0}=-p\cos s$, $p\neq0$. This suggests that a
notable simplification of our models will occur if we introduce (a
series of) canonical transformations substituting $\hat{T}_{0}$ and
$p_{\hat{\gamma}_0}$ for new canonical variables. First, consider
$[12]$
\begin{eqnarray}\label{canonic1}
&&\displaystyle\hat{T}_0=P\sin T\,,\hspace{2.7cm}p_{\hat{T}_0}=\frac{p_T}{P}\cos T-Q\sin T\,,\nonumber\\
&&\displaystyle\hat{\gamma}_0=-Q\cos T-\frac{p_T}{P}\sin
T\,,\hspace{0.4cm}p_{\hat{\gamma}_0}=-P\cos T\,,
\end{eqnarray}
where $(Q,P)$ and $(T,p_{T})$ denote canonically conjugate pairs. It
is straightforward to check that this is indeed a canonical
transformation, i.e., (\ref{symplecticS2}) coincides with
$\mathrm{d}Q\wedge \mathrm{d}P+\mathrm{d}T\wedge \mathrm{d}
p_{T}+\sum_{i}\int_{\mathbb{S}^2}{^{\dos}}\mathbf{e}\,
\delta\phi_i\wedge \delta p_{\phi_i}$. It is possible to write the
remaining constraint $\mathcal{C}$ in a more pleasant form by
performing a further canonical transformation (here, again,
$(\tilde{Q},\tilde{P})$ and $(\varphi_i,p_{\varphi_i})$ are
canonical pairs)
\begin{eqnarray}\label{canonic2}
&\displaystyle\tilde{Q}:=PQ+\frac{1}{2}\sum_{i}\int_{\mathbb{S}^2}{^{\dos}}\mathbf{e}\,p_{\hat{\phi}_i}\hat{\phi}_i\,,
&\displaystyle\tilde{P}:=\log P\,,\nonumber\\
&\displaystyle\varphi_i=(4\pi)^{-1/4}\sqrt{P}\hat{\phi}_i\,,\hspace{1.7cm}&p_{\varphi_i}=(4\pi)^{1/4}\frac{p_{\hat{\phi}_i}}{\sqrt{P}}\,,
\end{eqnarray}
giving
\begin{equation}
\mathcal{C}=p_{T}+4\sqrt{\pi}e^{\tilde{P}}\Big(\log \frac{\sin
T}{\sqrt{4\pi}}+\tilde{P}-1\Big)\sin
T+\frac{1}{2}\sum_{i}\int_{\mathbb{S}^{2}}{^{\dos}}\mathbf{e}\left(\frac{p_{\varphi_{i}}^{2}}{\sin
T}+\varphi_{i}^{\prime2}\sin T\sin^{2}\theta\right)\approx 0\,.
\label{mathcalC}
\end{equation}
The 2-form (\ref{symplecticS2}) then becomes
\begin{equation}\label{omegaf}
\iota^{*}\omega=\mathrm{d}\tilde{Q}\wedge\mathrm{d}\tilde{P}+\sum_{i}\int_{\mathbb{S}^{2}}
{^{\dos}}\mathbf{e}\,\delta\varphi_{i}\wedge \delta
p_{\varphi_{i}}+\mathrm{d}T\wedge\mathrm{d}p_{T}\,.
\end{equation}
Expressions (\ref{mathcalC}) and (\ref{omegaf}), in particular, the
linearity of the first one in the momentum $p_{T}$, allow us to
interpret the 4-tuple
$((0,\pi)\times\Gamma_R,\mathrm{d}t,\omega_R,H_R)$ as a
nonautonomous Hamiltonian system with $T=t$ as the time parameter
$[13]$. The resulting phase space $\Gamma_{R}$ is coordinatized by
the canonical pairs
$(\tilde{Q},\tilde{P};\varphi_{i},p_{\varphi_i})$ and is endowed
with the (weakly) symplectic form
\begin{equation}\label{omegaR}
\omega_{R}:=\mathrm{d}\tilde{Q}\wedge\mathrm{d}\tilde{P}+\sum_{i}\int_{\mathbb{S}^{2}}
{^{\dos}}\mathbf{e}\,\delta\varphi_{i}\wedge \delta
p_{\varphi_{i}}\,.
\end{equation}
Here, the canonical pair $(\tilde{Q},\tilde{P})$ describes a global
degree of freedom. The dynamics is given by the time-dependent
Hamiltonian $H_{R}(t):\Gamma_{R}\rightarrow\mathbb{R}$
\begin{equation}
H_{R}(t)=4\sqrt{\pi}e^{\tilde{P}}\Big(\log \frac{\sin
t}{\sqrt{4\pi}}+\tilde{P}-1\Big)\sin
t+\frac{1}{2}\sum_{i}\int_{\mathbb{S}^{2}}{^{\dos}}\mathbf{e}\left(\frac{p_{\varphi_{i}}^{2}}{\sin
t}+\varphi_{i}^{\prime2}\sin t\sin^{2}\theta\right).\label{HR}
\end{equation}
Note that from the point of view of the phase-space description of
the dynamics developed here, we are able to understand in very
simple terms the appearance of both initial and final singularities
in the spacetime metrics of these models. The singularities that
must be present as a consequence of the Hawking-Penrose theorems
$[14]$ can be understood as coming from the behavior of the
Hamiltonian, which is singular whenever $\sin t=0$. This means that
if we pick the initial time $t_0\in(0,\pi)$ in order to write the
Cauchy data we meet a past singularity at $t=0$ and a future
singularity at $t=\pi$. The evolution vector field corresponding to
(\ref{HR}) is
\begin{eqnarray}
E_{H_R}&=&\frac{\partial}{\partial
t}+4\sqrt{\pi}e^{\tilde{P}}(\log\frac{\sin
t}{\sqrt{4\pi}}+\tilde{P})\sin
t\frac{\partial}{\partial\tilde{Q}}\nonumber\\
&&\quad+\sum_{i}\int_{\mathbb{S}^{2}}{^{(2)}}\mathbf{e}\left(\frac{p_{\varphi_{i}}}{\sin
t}\frac{\delta}{\delta
\varphi_{i}}+(\sin^{2}\theta\varphi_i^{\prime})^{\prime}\sin
t\frac{\delta}{\delta p_{\varphi_{i}}}\right).
\end{eqnarray}
It is possible to recover the original 4-dimensional spacetime from
this 3-dimensional formulation. The fact that the gauge fixing
conditions (\ref{gauge3S2}) defining the deparameterization must be
preserved under the dynamics generated by the Hamiltonian
$H[\hat{N},\hat{N}^{\theta}]$ (i.e.,
$\{\hat{T}_{n},H[\hat{N},\hat{N}^{\theta}]\}\approx0$ and
$\{p_{\hat{\gamma}_{n}},H[\hat{N},\hat{N}^{\theta}]\}\approx0$, for
all $n\in\mathbb{N}$) forces $\hat{N}$ to be $\theta$-independent
and $\hat{N}^{\theta}$ to be zero. A suitable redefinition of the
time parameter allows us to express the metric $g_{ab}$ found after
the deparameterization as
\begin{equation}\label{metricdepar1}
g_{ab}=e^\gamma\Big(-(\mathrm{d}t)_a(\mathrm{d}t)_b+(\mathrm{d}\theta)_a(\mathrm{d}\theta)_b\Big)+
\frac{P^2}{4\pi}\sin^2 t
\sin^2\theta(\mathrm{d}\sigma)_a(\mathrm{d}\sigma)_b\,,
\end{equation}
defined on $(0,\pi)\times \mathbb{S}^2$, with singular behavior at
$t=0,\pi$. Once we integrate the Hamiltonian equations corresponding
to (\ref{HR}), undo the canonical transformation defined above, and
solve the constraints in order to obtain the $\gamma$ function, we
uniquely determine the 3-metric (\ref{metricdepar1}) and, thereby,
the original 4-metric.
\\
\linebreak \indent Finally, we point out the possibility of
reinterpreting the dynamics of these models as simple massless
scalar field theories in conformally stationary backgrounds. This
will allow us to use well-known techniques of quantum field theory
in curved backgrounds in order to quantize the systems. Let us start
by giving a simple way to solve equations (\ref{ecs}). Given a
specific solution $(\mathring{g}_{ab},\mathring{\phi}_{i})$,
whenever condition $\mathcal{L}_{\sigma}\phi_{i}=0$ is satisfied the
following equivalence
\begin{eqnarray*}
g^{ab}\nabla_a\nabla_b\phi_i=0 \,\,\Leftrightarrow\,\,
\mathring{g}^{ab} \mathring{\nabla}_a\mathring{\nabla}_b\phi_i=0
\end{eqnarray*}
holds. On can solve the last equation in some convenient background
$\mathring{g}_{ab}$ and then use equation
$R_{ab}=\frac{1}{2}\sum_{i}(\mathrm{d}\phi_{i})_{a}(\mathrm{d}\phi_{i})_{b}$
just to give integrability conditions allowing us to recover
$g_{ab}$. In the 3-handle case, the metric found after
deparameterization is given by (\ref{metricdepar1}); a possible (non
unique) choice for $(\mathring{g}_{ab},\mathring{\phi}_{i})$ is in
this case
\begin{eqnarray*}
\mathring{g}_{ab}&=&\sin^2t\Big(-(\mathrm{d}t)_a(\mathrm{d}t)_a+(\mathrm{d}\theta)_a(\mathrm{d}\theta)_b+\sin^2\theta(\mathrm{d}\sigma)_a(\mathrm{d}\sigma)_b\Big)\,,\\
\mathring{\phi}_1&=&\log\sin(t/2)-\log\cos(t/2)\,,\quad
\mathring{\phi}_i=0\,,\,\,\,i\neq1\,.
\end{eqnarray*}
It is important to notice that even though the metric
$\mathring{g}_{ab}$ is not stationary, it is conformal to the
Einstein static metric on $(0,\pi)\times\mathbb{S}^2$. The scalar
field dynamics generated by the nonautonomous Hamiltonian (\ref{HR})
corresponds exactly to the one defined by the Klein–-Gordon
equations on the background given by $\mathring{g}_{ab}$.

\section{$\mathbb{S}^3$ Gowdy models coupled to massless scalars}{\label{S3}}

\indent Let us now consider the case where the spatial slices have
the topology of a 3-sphere $\mathbb{S}^3$, described as
$\mathbb{S}^3=\{(z_1,z_2)\in\mathbb{C}^2:|z_1|^2+|z_2|^2=1\}$. A
useful parametrization of $\mathbb{S}^3$, in terms of the so-called
Hopf coordinates, is $z_1=e^{i\sigma}\sin(\theta/2)$,
$z_2=e^{i\xi}\cos(\theta/2)$, with $\theta\in[0,\pi]$,
$\xi,\sigma\in\mathbb{R}(\mathrm{mod}\,2\pi)$. We define the
following action of $G^{\dos}$ on $\mathbb{S}^{3}$
\begin{eqnarray}
(g_1,g_2)\cdot(z_1,z_2)=(e^{ix_1},e^{ix_2})\cdot(z_1,z_2)&=&(e^{ix_1}z_1,e^{ix_2}z_2)\nonumber\\
&=&(e^{i(x_1+\sigma)}\sin(\theta/2),e^{i(x_2+\xi)}\cos(\theta/2))\,.\nonumber
\label{S3action_1}
\end{eqnarray}
The action of the two $U(1)$ subgroup factors is
\begin{equation*}
(e^{ix},1)\cdot(z_1,z_2)=(e^{ix}z_1,z_2)\,,\quad
(1,e^{ix})\cdot(z_1,z_2)=(z_1,e^{ix}z_2)\,.
\end{equation*}
The corresponding tangent vectors at each point of $\mathbb{S}^3$,
obtained by differentiating the previous expressions with respect to
$x$ at $x=0$, are now
\begin{equation*}
(iz_1,0)\,,\quad (0,iz_2).\nonumber
\end{equation*}
These are commuting vector fields. As we can see, they vanish
respectively at $z_1=0$ and $z_2=0$, i.e., at the circles
$(0,e^{i\xi})$ and $(e^{i\sigma},0)$. We proceed now to construct a
spacetime
$({^{\cuatro}}\mathcal{M}\simeq\mathbb{R}\times\mathbb{S}^{3},{^{\cuatro}}g_{ab})$
in the same way as in the 3-handle models. Here, however, we face
the fact that the Killing vectors fields
$\xi^a=(\partial/\partial\xi)^{a}$ and
$\sigma^a=(\partial/\partial\sigma)^{a}$ vanish alternatively in two
different circles when trying to perform a Geroch reduction.
Consider in particular, the Killing field $\xi^{a}$ which vanishes
at (the one-dimensional submanifold of $\mathbb{S}^3$) $\theta=\pi$
and is nonzero at $\theta=0$. Let ${^{\cuatro}}\tilde{\mathcal{M}}$
be the space ${^{\cuatro}}\mathcal{M}$ with the submanifold where
$\xi^{a}$ vanishes (diffeomorphic to
$\mathbb{R}\times\mathbb{S}^{1}$) removed. This substraction does
not affect the (1+3)-dimensional Einstein-Hilbert action for gravity
coupled to matter, since we have removed a zero-measure set. Of
course, one must take into account the fact that the fields in the
new integration region cannot be completely arbitrary but should be
subject to some restrictions (regularity conditions) reflecting the
fact that they should extend to the full ${^{\cuatro}}\mathcal{M}$
in a smooth way. By performing a Geroch reduction on
${^{\cuatro}}\tilde{\mathcal{M}}/U(1)$ with respect to $\xi^{a}$, we
obtain an action of the form (\ref{act}) where the 2-dimensional
spatial sections are now diffeomorphic to the open disc $D(0;\pi)$.

\begin{figure}[t] \centering
\includegraphics[width=8cm]{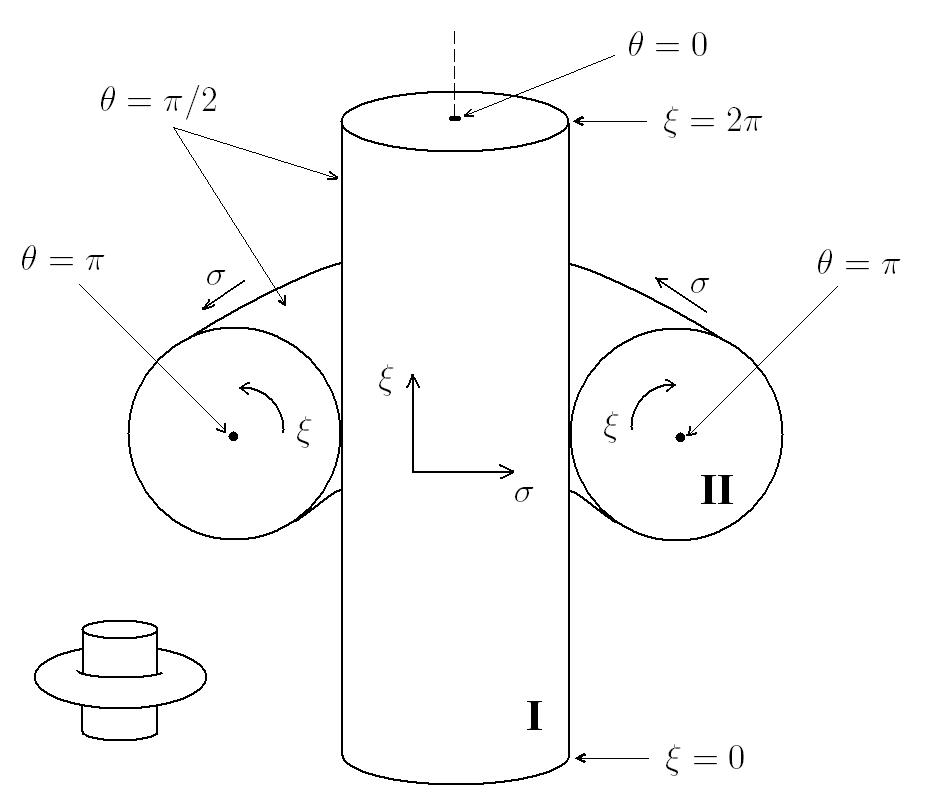}
\caption{Cylindrical coordinates-patches on the 3-sphere
$\mathbb{S}^{3}$, which has been sliced along the surface
$\theta=\pi/2$. One of the resulting solid toroids (number I) has
been further sliced at $\xi=0$ and rendered as a solid cylinder. The
half of toroid II has been cut away to display the behavior of the
coordinates. The complete figure is showed in the lower left. For
any fixed value of $\theta\in(0,\pi)$, the coordinates
$(\sigma,\xi)$ parameterize a 2-torus. In the degenerate cases
corresponding to $\theta=0$ and $\theta=\pi$ these coordinates
describe circles.}
\end{figure}

As in the 3-handle case, we are going to use
$(t^{a},\theta^{a},\sigma^{a})$ as coordinate vector fields. We will
write now $\theta^a=f\hat{\theta}^a$ and
$t^a=(Nn^a+N^\theta\hat{\theta}^a)f$. Here, the scalars $f>0$,
$N>0$, and $N^\theta$ are supposed to be smooth fields on
$\mathbb{R}\times D(0;\pi)$ subject to some regularity conditions
that we must specify. Note that we write $f$ instead of $e^\gamma/2$
foreseing the vanishing of this function at the disc boundary.
Again, $(N,N^\theta,\gamma,\phi_{i})$ are constant on the orbits of
the remaining Killing field $\sigma^a$ and, hence, they only depend
on the coordinates $(t,\theta)$. The commutation relations verified
by $(t^{a},\theta^{a},\sigma^{a})$ yield
\begin{eqnarray}
&&N\mathcal{L}_\theta f+f\mathcal{L}_\theta
N+Nf^2n^an^b\nabla_a\hat{\theta}_b=0\,,\label{proy_1_3sph}\\
&&N^\theta\mathcal{L}_\theta f+f\mathcal{L}_\theta
N^\theta-\mathcal{L}_t f+
Nf^2\hat{\theta}^a\hat{\theta}^b\nabla_a n_b=0\,,\label{proy_2_3sph}\\
&&\hat{\theta}^a\sigma^b\nabla_an_b=0\,.\label{proy_3_3sph}
\end{eqnarray}
Using the coordinates system $(t,\theta,\sigma,\xi)$ we can write
the original 4-metric ${^{\cuatro}}g_{ab}$ as
\begin{eqnarray}
\label{4metricS3}{^{\cuatro}}g_{ab}&=&\frac{f^2}{\lambda_\xi}\big((N^{\theta
2}-N^{2})(\mathrm{d}t)_a(\mathrm{d}t)_b+
2N^{\theta}(\mathrm{d}t)_{(a}(\mathrm{d}\theta)_{b)}+
(\mathrm{d}\theta)_a(\mathrm{d}\theta)_b\big)\\
&+&
\frac{\tau^{2}}{\lambda_\xi}(\mathrm{d}\sigma)_a(\mathrm{d}\sigma)_b+
\lambda_{\xi}(\mathrm{d}\xi)_a(\mathrm{d}\xi)_b\,.\nonumber
\end{eqnarray}
Next, we impose the regularity conditions to be satisfied by this
metric. At $\theta=0$, where $\sigma^{a}$ vanishes, the regularity
conditions should be of the same type as the ones that we have
already used in the $\mathbb{S}^1\times\mathbb{S}^2$ case. It
suffices to use the expression (\ref{Mmunu}) and identify
$t\leftrightarrow t$, $z\leftrightarrow\xi$,
$r\leftrightarrow2\sin(\theta/2)$, and
$\varphi\leftrightarrow\sigma$. Here, however, we also have to
impose regularity conditions when we approach the boundary of the
the filled torus that we obtained by removing the circle where the
Killing $\xi^{a}$ vanishes. In this case, we must simply identify
$t\leftrightarrow t$, $z\leftrightarrow\sigma$,
$r\leftrightarrow2\cos(\theta/2)$, and $\varphi\leftrightarrow\xi$.
Note that we use the functions $\sin(\theta/2)$ and $\cos(\theta/2)$
because they alternatively vanish on the circles where the Killings
themselves become zero; in addition, they have the dependence of a
regular scalar function in terms of the \emph{radial} coordinates
$\theta$ or $\pi-\theta$ on the circles where the Killing fields do
not vanish. According to this, we find
\begin{eqnarray}
&&\frac{f^2}{\lambda_{\xi}}(N^{\theta 2}-N^2)=A(t,\cos\theta)\,,\label{reg1_S3}\\
&&\frac{f^2}{\lambda_{\xi}}N^\theta=B(t,\cos\theta)\sin\theta\,, \label{reg2_S3}\\
&&\lambda_{\xi}=4\cos^2(\theta/2)\big(F(t,\cos\theta)-G(t,\cos\theta)\cos^2(\theta/2)\big)\,,\label{reg3_S3}\\
&&\frac{f^2}{\lambda_{\xi}}=D(t,\cos\theta)+E(t,\cos\theta)\sin^2(\theta/2)\nonumber\\
&&\hspace{0.6cm}=F(t,\cos\theta)+G(t,\cos\theta)\cos^2(\theta/2)\,,\label{reg4_S3}\\
&&\frac{\tau^2}{\lambda_{\xi}}=4\sin^2(\theta/2)\big(D(t,\cos\theta)-E(t,\cos\theta)\sin^2(\theta/2)
\big)\,,\label{reg5_S3}
\end{eqnarray}
where $A$, $B$, $D$, $E$, $F$, and $G$ are analytic in their
arguments. The cosine dependence of these functions is dictated by
regularity at certain submanifolds diffeomorphic to
$\mathbb{R}\times\mathbb{S}^{1}$. This will prove to be very
important because we will be able to describe the system in terms of
these fields, and having $\cos\theta$ as their argument they can be
interpreted as functions on $\mathbb{S}^{2}$ as in the 3-handle
case. Note that they are not independent because they are
constrained to satisfy (\ref{reg4_S3}). In addition,
$D(t,\cos\theta)\pm E(t,\cos\theta)\sin^{2}\theta>0$ and
$F(t,\cos\theta)\pm G(t,\cos\theta)\sin^{2}\theta>0$, so that $D>0$
and $F>0$; finally,
$B^{2}(t,\cos\theta)-A(t,\cos\theta)\big(D(t,\cos\theta)+
E(t,\cos\theta)\sin^{2}\theta\big)>0$. The conditions that the
fields themselves must satisfy (dropping the $t$-dependence) are
\begin{eqnarray}
&&\lambda_{\xi}=e^{\phi_{1}}=e^{\hat{\phi}_1(\cos\theta)}\cos^2(\theta/2)\,,\label{condfields2_S3}\\
&&\phi_2=\hat{\phi}_2(\cos\theta)\,,\label{condfields1_S3}\\
&&f=\cos(\theta/2)e^{\hat{\gamma}(\cos\theta)/2}\,,\label{condfields3_S3}\\
&&N^\theta=\hat{N}^\theta(\cos\theta)\sin\theta\,,\label{condfields4_S3}\\
&&N=\hat{N}(\cos\theta)\,,\label{condfields5_S3}\\
&&\tau=\hat{T}(\cos\theta)\sin\theta\,,\label{condfields6_S3}\\
&&\hat{T}^2e^{-\hat{\gamma}}=\frac{D(\cos\theta)-E(\cos\theta)\sin^2(\theta/2)}
{D(\cos\theta)+E(\cos\theta)\sin^2(\theta/2)}\,,\label{condfields7_S3}\\
&&e^{2\hat{\phi}_1-\hat{\gamma}}=4\frac{F(\cos\theta)-G(\cos\theta)\cos^2(\theta/2)}
{F(\cos\theta)+G(\cos\theta)\cos^2(\theta/2)}\,,\label{condfields8_S3}
\end{eqnarray}
where we have used $ \sin\theta=2\sin(\theta/2)\cos(\theta/2)$.
Here, as in the $\mathbb{S}^1\times\mathbb{S}^2$ case, we have that
$\hat{\phi}_{i},\hat{\gamma},\hat{N}^{\theta},\hat{N},\hat{T}:[-1,1]\rightarrow\mathbb{R}$
($\hat{N}>0$). They must be $C^\infty$ in $(-1,1)$ with bounded
derivative. Conditions (\ref{condfields7_S3}) and
(\ref{condfields8_S3}) imply the \emph{polar constraints} for the
$\mathbb{S}^{3}$ models
$$\hat{T}(+1)e^{-\hat{\gamma}(+1)/2}=1\,\quad \mathrm{and}\quad
e^{2\hat{\phi}_1(-1)-\hat{\gamma}(-1)}=4\,.$$ Note that in this case
the resulting conditions involve different pairs of objets at each
pole $\theta=0$ or $\theta=\pi$. Our starting point is now the
action (we take again units such that $16\pi G_{3}=1$)
\begin{eqnarray}\label{act3esf}
{^{\scriptstyle{(3)}}}S(N,N^{\theta},f,\tau,\phi_{i})&=&\int_{(t_0,t_1)}\mathrm{d}t\int_{
D(0;\pi)}{^{\scriptstyle{(2)}}}\mathbf{e}\,\Bigg\{
\underbrace{\frac{2N}{f\sin\theta}\left((\mathcal{L}_{\theta}f)\mathcal{L}_{\theta}\tau-f\mathcal{L}_{\theta}^{2}\tau\right)}_{(a)}\nonumber\\
&&-\underbrace{\frac{2}{Nf\sin\theta}
\left(\dot{f}-f\mathcal{L}_{\theta}N^{\theta}-N^{\theta}\mathcal{L}_{\theta}f
\right)\left(\dot{\tau}-N^\theta\mathcal{L}_{\theta}\tau\right)}_{(b)}\nonumber
\\
&&+\underbrace{\frac{\tau}{2N\sin\theta}\sum_{i\neq1}\hspace*{-0.05cm}\left[\dot{\phi}^2_i-2N^\theta\dot{\phi}_i\mathcal{L}_{\theta}\phi_{i}\hspace*{-0.05cm}+\hspace*{-0.05cm}
\left((N^{\theta})^2\hspace*{-0.05cm}-\hspace*{-0.05cm}N^{2}\right)(\mathcal{L}_{\theta}\phi_{i})^{2}\right]}_{(c)}\nonumber\\
&&+\underbrace{\frac{\tau}{2N\lambda_{\xi}^{2}\sin\theta}\hspace*{-0.05cm}\left[\dot{\lambda}^2_{\xi}-2N^\theta\dot{\lambda}_{\xi}\mathcal{L}_{\theta}\lambda_{\xi}\hspace*{-0.05cm}+\hspace*{-0.05cm}
\left((N^{\theta})^2\hspace*{-0.05cm}-\hspace*{-0.05cm}N^{2}\right)(\mathcal{L}_{\theta}\lambda_{\xi})^{2}\right]}_{(d)}\hspace*{-0.05cm}\Bigg\}\,,\nonumber
\end{eqnarray}
where the Lagrangian can be easily deduced from (\ref{act3b}) by
substituting $\phi_{1}\rightarrow\log\lambda_{\xi}$ and
$\gamma\rightarrow2\log f$. Here, as in the case of the 3-handle, we
choose the fiducial volume element $^{\dos}\mathbf{e}$ to be
compatible with the auxiliary round metric on the 2-sphere
$\mathbb{S}^2$, i.e.,
${^{\dos}}\mathbf{e}=\sin\theta\mathrm{d}\theta\wedge\mathrm{d}\sigma$,
with
${^{\dos}}\mathbf{e}_{ab}\theta^{a}\sigma^{b}=Nf^2\tau/|g|^{1/2}=\sin\theta$.
In terms of the fields
$(\hat{N},\hat{N}^{\theta},\hat{\gamma},\hat{T},\hat{\phi}_{i})$,
and using again the prime derivative defined in
(\ref{primederivative}), we have
\begin{eqnarray*}
(a)&=&
\hat{N}\Big[\Big(2-\frac{\tan(\theta/2)}{\tan\theta}\Big)\hat{T}
+\hat{T}^{\prime}+(5\hat{T}^{\prime}-\hat{\gamma}^{\prime}\hat{T})\cos\theta
+(\hat{\gamma}^{\prime}\hat{T}^{\prime}-2\hat{T}^{\prime\prime})\sin^{2}\theta\Big]\,,\\
(b)&=&\frac{1}{\hat{N}}\Big[\dot{\hat{T}}-\hat{N}^{\theta}\hat{T}\cos\theta+\hat{N}^{\theta}\hat{T}^{\prime}\sin^{2}\theta\Big]
\Big[\dot{\hat{\gamma}}+(2\hat{N}^{\theta\prime}+\hat{N}^{\theta}\hat{\gamma}^{\prime})\sin^{2}\theta
+(1-3\cos\theta)\hat{N}^{\theta}\Big]\,,\\
(c)&=&\frac{\hat{T}}{2\hat{N}}\sum_{i\neq1}\Big[\dot{\hat{\phi}}_{i}^{2}
+2\hat{N}^{\theta}\dot{\hat{\phi}}_{i}\hat{\phi}_{i}^{\prime}\sin^{2}\theta+\big((\hat{N}^{\theta})^{2}\sin^{2}\theta-\hat{N}^{2}\big)
\hat{\phi}_{i}^{\prime2}\sin^{2}\theta\Big]\,,\\
(d)&=&\frac{\hat{T}}{2\hat{N}}\Big[\dot{\hat{\phi}}_{1}^{2}
+2\hat{N}^{\theta}\dot{\hat{\phi}}_{1}\hat{\phi}_{1}^{\prime}\sin^{2}\theta+\big((\hat{N}^{\theta})^{2}\sin^{2}\theta-\hat{N}^{2}\big)
\hat{\phi}_{1}^{\prime2}\sin^{2}\theta\\
&&+2(1-\cos\theta)\big(\hat{N}^{\theta}\dot{\hat{\phi}}_{1}+\big((\hat{N}^{\theta})^{2}\sin^{2}\theta-\hat{N}^{2}\big)\hat{\phi}_{1}^{\prime}\big)
+(1-\cos\theta)^{2}\hat{N}^{\theta2}-\tan^{2}(\theta/2)\hat{N}^{2}\Big]\,.
\end{eqnarray*}
We can change the spatial integration region in the action from
$D(0;\pi)$ to $\mathbb{S}^2$ because the Lagrangian can be written
as a smooth function on the 2-sphere in terms of the
\textit{hat}-fields, that are smoothly extendable to $\mathbb{S}^2$.
We arrive at this result after several nontrivial cancellations of
terms that would diverge at the poles. Note, in particular, that the
fist term in (a) and the last term in (d) involving $\tan(\theta/2)$
yield
$(2-\tan(\theta/2)/\tan\theta-(1/2)\tan^{2}(\theta/2))\hat{T}=(3/2)\hat{T}$
when they are added. In this way,
\begin{eqnarray}
&&\hspace{-.5cm}{^{\tres}}S(\hat{N},\hat{N}^\theta,\hat{\gamma},\hat{T},\hat{\phi}_i)=\int_{t_0}^{t_1}\!\!\!\!\!
\mathrm{d}t\int_{\mathbb{S}^2}\!\!\!^{\dos}\mathbf{e}\bigg(
\hat{N}[(\hat{\gamma}^\prime\hat{T}^\prime-2\hat{T}^{\prime\prime})\sin^2\theta+
(5\hat{T}^{\prime}-\hat{\gamma}^\prime\hat{T})\cos\theta+\hat{T}^\prime+\frac{3}{2}\hat{T}]\nonumber\\
&&+\frac{1}{\hat{N}}[\hat{N}^\theta
\hat{T}\cos\theta-\dot{\hat{T}}-\hat{N}^\theta
\hat{T}^\prime\sin^2\theta][\dot{\hat{\gamma}}
+(2\hat{N}^{\theta\prime}+\hat{N}^\theta\hat{\gamma}^\prime)\sin^2\theta+(1-3\cos\theta)\hat{N}^\theta]\nonumber\\
&&+\frac{\hat{T}}{2\hat{N}}\sum_{i}\bigg[\dot{\hat{\phi}}^2_i
+2\hat{N}^\theta\dot{\hat{\phi}}_i\,\hat{\phi}_i^\prime\sin^2\theta+
(\hat{N}^{\theta2}\sin^2\theta-\hat{N}^2)\hat{\phi}_i^{\prime2}\sin^2\theta\bigg]\nonumber\\
&&+\frac{\hat{T}}{2\hat{N}}[2(1-\cos\theta)(\hat{N}^\theta
\dot{\hat{\phi}}_1+(\hat{N}^{\theta2}\sin^2\theta-\hat{N}^2)\hat{\phi}_1^\prime)
+(1-\cos\theta)^2\hat{N}^{\theta2}] \bigg)\,.\label{act_esf}
\end{eqnarray}
The Hamiltonian of the system can be readily obtained by performing
the Legendre transformation
\begin{eqnarray*}
&&p_{\hat{N}}=0=p_{\hat{N}^{\theta}}\,,\\
&&p_{\hat{\gamma}}=\frac{1}{\hat{N}}\Big(\hat{N}^{\theta}\hat{T}\cos\theta-\dot{\hat{T}}-\hat{N}^{\theta}\hat{T}^{\prime}\sin^{2}\theta\Big)\,,\\
&&p_{\hat{T}}=-\frac{1}{\hat{N}}\Big(\dot{\hat{\gamma}}+\big(2\hat{N}^{\theta\prime}+\hat{N}^{\theta}\hat{\gamma}^{\prime}\big)\sin^{2}\theta
+(1-3\cos\theta)\hat{N}^{\theta}\Big)\,,\\
&&p_{\hat{\phi}_{1}}=\frac{\hat{T}}{\hat{N}}\Big(\dot{\hat{\phi}}_{1}+\hat{N}^{\theta}\hat{\phi}_{1}^{\prime}+(1-\cos\theta)\hat{N}^{\theta}\Big)\,,\quad
p_{\hat{\phi}_{i}}=\frac{\hat{T}}{\hat{N}}\Big(\dot{\hat{\phi}}_{i}+\hat{N}^{\theta}\hat{\phi}_{i}^{\prime}\Big)\,,\,\,i\neq1\,.
\end{eqnarray*}
It is important to highlight that gravitational and matter modes
(encoded by $\hat{\phi}_{1}$ and $\hat{\phi}_{i}$, $i\neq1$,
respectively) cease to play a symmetric role in this particular
description, at variance with the 3-handle case. This issue will be
further discussed at the end of the section. The Hamiltonian is
given by
$$H=\int_{\mathbb{S}^{2}}{^{\dos}}\mathbf{e}\left(\hat{N}C+\hat{N}^{\theta}C_{\theta}\right),$$
with
\begin{eqnarray}
C&:=&-p_{\hat{\gamma}}p_{\hat{T}}+\big(2\hat{T}^{\prime\prime}-\hat{\gamma}^\prime\hat{T}^\prime\big)\sin^2\theta
+(\hat{\gamma}^\prime\hat{T}-5\hat{T}^\prime)\cos\theta-\frac{3}{2}\hat{T}-\hat{T}^\prime\nonumber\\
&&+\frac{1}{2}\sum_{i}\bigg(\frac{p_{\hat{\phi}_i}^2}{\hat{T}}
+\hat{T}\hat{\phi}_i^{\prime2}\sin^2\theta\bigg)
+(1-\cos\theta)\hat{T}\hat{\phi}_1^\prime\,,\nonumber\\
\label{esc_const_S3} C_\theta&=&
\Big(2p^\prime_{\hat{\gamma}}-\hat{\gamma}^\prime p_{\hat{\gamma}}-
\hat{T}^\prime p_{\hat{T}}-\sum_{i}\hat{\phi}_i^\prime
p_{\hat{\phi}_i}\Big)\sin^2\theta +\big(\hat{T}
p_{\hat{T}}-p_{\hat{\gamma}}+p_{\hat{\phi_1}}\big)\cos\theta-p_{\hat{\gamma}}-p_{\hat{\phi}_1}\,.
\nonumber
\end{eqnarray}
The two previous expressions, together with the conditions at the
poles $\hat{T}(+1)e^{-\hat{\gamma}(+1)/2}=1$ and
$e^{2\hat{\phi}_1(-1)-\hat{\gamma}(-1)}=4$, define the constraints
of the system. As before, the polar constraints are necessary
conditions to guarantee the differentiability of the (weighted)
constraints $C[\hat{N}_g]$ and $C_\theta[\hat{N}_g^\theta]$.
Concretely, it is straightforward to check that the surface term
which appears when calculating the exterior derivative $\delta
C[\hat{N}_{g}]$, namely,
$-(1+\cos\theta)\hat{N}_{g}\delta\hat{T}+\hat{N}_{g}\hat{T}\cos\theta\delta\hat{\gamma}
+(1-\cos\theta)\hat{N}_{g}\hat{T}\delta\hat{\phi}_{1}\big]_{\theta=0}^{\pi}$,
vanishes by virtue of these constraints. The gauge transformations
generated by $C[\hat{N}_g]$ and $C_\theta[\hat{N}_g^\theta]$ are
\begin{eqnarray*}
&&\{\hat{\gamma},C[\hat{N}_g]\}=-\hat{N}_g p_{\hat{T}}\,,
\nonumber\\
&&\{\hat{T},C[\hat{N}_g]\}=-\hat{N}_g p_{\hat{\gamma}}\,,\nonumber\\
&&\{\hat{\phi_i},C[\hat{N}_g]\}=\hat{N}_g\frac{p_{\hat{\phi}_i}}{\hat{T}}\,,\nonumber\\
&&\{p_{\hat{\gamma}},C[\hat{N}_g]\}=\hat{N}_{g}^{\prime}(\hat{T}\cos\theta-\hat{T}^\prime\sin^2\theta)
+\hat{N}_g(3\hat{T}^\prime\cos\theta+\hat{T}-\hat{T}^{\prime\prime}\sin^2\theta),\nonumber\\
&&\{p_{\hat{T}},C[\hat{N}_g]\}=
\hat{N}_g\big[\frac{1}{2}-\hat{\phi}_1^\prime+(\hat{\gamma}^\prime+\hat{\phi}_1^\prime)\cos\theta
-\hat{\gamma}^{\prime\prime}\sin^2\theta\big]+
\hat{N}_g^\prime(3\cos\theta-1-\hat{\gamma}^\prime\sin^2\theta)\nonumber\\
&&\hspace{2.5cm}-
2\hat{N}_g^{\prime\prime}\sin^2\theta+\frac{\hat{N}_g}{2}\sum_{i}\bigg(\frac{p_{\hat{\phi}_i}^2}{\hat{T}^2}
-\sin^2\theta\hat{\phi}_i^{\prime2}\bigg)\,,\nonumber\\
&&\{p_{\hat{\phi}_1},C[\hat{N}_g]\}=[\hat{N}_g\hat{T}(\hat{\phi}^\prime_2\sin^2\theta+1-\cos\theta)]^\prime\,,\nonumber\\
&&\{p_{\hat{\phi}_2},C[\hat{N}_g]\}=
(\hat{N}_g\hat{T}\hat{\phi}^\prime_2\sin^2\theta)^\prime\,,\nonumber
\end{eqnarray*}
and
\begin{eqnarray*}
&&\{\hat{\gamma},C_\theta[\hat{N}^\theta_g]\}=-2\hat{N}_g^{\theta\prime}\sin^2\theta+\hat{N}_g^\theta(3\cos\theta-\hat{\gamma}^\prime\sin^2\theta-1)\,,\nonumber\\
&&\{\hat{T},C_\theta[\hat{N}^\theta_g]\}=\hat{N}_g^\theta(\hat{T}\cos\theta-\hat{T}^\prime\sin^2\theta)\,,\nonumber\\
&&\{\hat{\phi}_1,C_\theta[\hat{N}^\theta_g]\}=\hat{N}_g^\theta(\cos\theta-1-\hat{\phi}_2^\prime\sin^2\theta)\,,\nonumber\\
&&\{\hat{\phi}_2,C_\theta[\hat{N}^\theta_g]\}=-\hat{N}_g^\theta\hat{\phi}_2^\prime\sin^2\theta\,,\nonumber\\
&&\{p_{\hat{\gamma}},C_\theta[\hat{N}^\theta_g]\}=-(\hat{N}_g^\theta p_{\hat{\gamma}}\sin^2\theta)^\prime\,,\nonumber\\
&&\{p_{\hat{T}},C_\theta[\hat{N}^\theta_g]\}=\hat{N}_g^\theta(p_{\hat{T}}\cos\theta-p_{\hat{T}}^\prime\sin^2\theta)
-\hat{N}_g^{\theta\prime}p_{\hat{T}}\sin^2\theta\,,
\nonumber\\
&&\{p_{\hat{\phi}_i},C_\theta[\hat{N}^\theta_g]\}=-(\hat{N}_g^\theta
p_{\hat{\phi_i}}\sin^2\theta)^\prime\,.\nonumber
\end{eqnarray*}
The Poisson brackets of these constraints give exactly the same
result that we obtained for the $\mathbb{S}^1\times\mathbb{S}^2$
topology and, hence, define a fist class constrained surface
$\Gamma_c\subset\Gamma$. Here, $(\Gamma,\omega)$ denotes the
canonical phase space of the system, coordinatized by the canonical
pairs
$(\hat{\gamma},p_{\hat{\gamma}};\hat{T},p_{\hat{T}};\hat{\phi}_{i},p_{\hat{\phi}_i})$,
endowed with the standard (weakly) symplectic form
(\ref{omegaS2XS1}). We must check now the stability of the polar
constraints. We do this by computing
\begin{eqnarray}
&&\hspace*{-0.6cm}\{\hat{T}e^{-\hat{\gamma}/2},C[\hat{N}_g]\}=\frac{1}{2}\hat{N}_ge^{-\hat{\gamma}/2}(\hat{T}p_{\hat{T}}-2p_{\hat{\gamma}})\,,\label{pb1}\\
&&\hspace*{-0.6cm}\{\hat{T}e^{-\hat{\gamma}/2},C_\theta[\hat{N}_g^\theta]\}=e^{-\hat{\gamma}/2}
\bigg(\frac{1}{2}\hat{N}_g^\theta\hat{T}(1-\cos\theta)+(\hat{N}_g^{\theta\prime}\hat{T}
-\hat{N}_g^\theta\hat{T}^\prime+\frac{1}{2}\hat{N}_g^\theta\hat{T}\hat{\gamma}^\prime)\sin^2\theta\bigg),\quad\quad\label{pb2}\\
&&\hspace*{-0.6cm}\{e^{2\hat{\phi}_1-\hat{\gamma}},C[\hat{N}_g]\}=\frac{\hat{N}^g}{\hat{T}}
e^{2\hat{\phi}_1-\hat{\gamma}}(2p_{\hat{\phi}_1}+\hat{T}p_{\hat{T}})\,,\label{pb3}\\
&&\hspace*{-0.6cm}\{e^{2\hat{\phi}_1-\hat{\gamma}},C_\theta[\hat{N}_g^\theta]\}=e^{2\hat{\phi}_1-\hat{\gamma}}
\Big(-\hat{N}_g^\theta(1+\cos\theta)+
(2\hat{N}_g^{\theta\prime}-2\hat{N}_g^\theta\hat{\phi}_2^\prime
+\hat{N}_g^\theta\hat{\gamma}^\prime)\sin^2\theta\Big)\,.\label{pb4}
\end{eqnarray}
The constraint $C_{\theta}=0$ at the poles $\theta=0,\pi$ gives
$\hat{T}(+1)p_{\hat{T}}(+1)-2p_{\hat{\gamma}}(+1)=0$, and
$\hat{T}(-1)p_{\hat{T}}(-1)+2p_{\hat{\phi}_1}(-1)=0$, respectively.
These relations guarantee that the Poisson bracket (\ref{pb1})
vanishes at $\theta=0$ and (\ref{pb3}) vanishes at $\theta=\pi$. The
vanishing of (\ref{pb2}) at $\theta=0$ is due to the presence of the
factors $1-\cos\theta$ and $\sin^2\theta$ and, finally, (\ref{pb4})
is zero at $\theta=\pi$ due to the factors $1+\cos\theta$ and
$\sin^2\theta$. As in the 3-handle case, we conclude that there are
no secondary constraints coming from the stability of these polar
constraints.

\subsection{Deparameterization}

\indent The deparameterization process in this case follows closely
the one developed for the $\mathbb{S}^1\times\mathbb{S}^2$ topology.
Particularly, the same gauge fixing conditions (\ref{gauge1S2}) work
in this case too. It suffices to check if the polar constraints are
gauge fixed. This only requires the computation of the Poisson
bracket
\begin{eqnarray*}
&&\{p_{\hat{\gamma}n},e^{2\hat{\phi}_1-\hat{\gamma}}\}=
e^{2\hat{\phi}_1-\hat{\gamma}}\sqrt{\frac{2n+1}{4\pi}}\mathscr{P}_n(\cos\theta)\,,
\end{eqnarray*}
which is different from zero at the poles. As we see, the situation
now is completely analogous to the 3-handle case. The only
constraint that is not gauge-fixed by the deparameterization
conditions is $C[1]$. The pull-back of the symplectic form to the
phase space hypersurface defined by the gauge fixing conditions is
given again by (\ref{symplecticS2}). We are left only with the
constraint
\begin{eqnarray}
\label{constS3}\mathcal{C} &:=&
-p_{\hat{\gamma}_{0}}p_{\hat{T}_{0}}+\hat{T}_{0}\left(\sqrt{4\pi}
\log\frac{\hat{T}_{0}}{\sqrt{4\pi}}-\hat{\gamma}_{0}-\sqrt{\pi}(2\log2+3)
+\hat{\phi}_{1_{0}}\right)\\
&+&\frac{1}{2}\sum_{i}\int_{\mathbb{S}^{2}}
{^{\scriptstyle{\dos}}}\mathbf{e}\left(\frac{\sqrt{4\pi}p_{\hat{\phi}_{i}}^{2}}{\hat{T}_{0}}
+\frac{\hat{T}_{0}}{\sqrt{4\pi}}\hat{\phi}_i^{\prime2}\right)\approx0\,.\nonumber
\end{eqnarray}
The gauge transformations generated by this constraint on the
variables $\hat{T}_0$ and $p_{\hat{\gamma}_0}$ are the same as for
the three-handle and, hence, we can use the canonical
transformations (\ref{canonic1}) and (\ref{canonic2}) introduced at
the end of the previous section to rewrite (\ref{constS3}) as
\begin{eqnarray}
\label{finconstS3}p_T&+&(4\pi)^{1/4}e^{\tilde{P}/2}{\varphi}_{1_0}\sin
T+2\sqrt{\pi}e^{\tilde{P}}\Big(\log \frac{\sin
T}{\sqrt{4\pi}}+\tilde{P}-\log2-\frac{3}{2}\Big)\sin
T\\
&+&\frac{1}{2}\sum_{i}\int_{\mathbb{S}^2}\!\!\!^{\dos}\mathbf{e}
\bigg(\frac{p_{{\varphi}_i}^2}{\sin T}+{\varphi}_i^{\prime2}\sin
T\sin^2\theta\bigg)\approx0.\nonumber
\end{eqnarray}
Again, it is possible to interpret the dynamics as being described
by a nonautonomous Hamiltonian system
$((0,\pi)\times\Gamma_R,\mathrm{d}t,\omega_R,H_R)$, where
$\Gamma_{R}$ denotes the reduced phase space coordinatized by the
canonical pairs $(\tilde{Q},\tilde{P};\varphi_{i},p_{\varphi_i})$,
endowed with the standard (weakly) symplectic form (\ref{omegaR}).
The dynamics is given by the time-dependent Hamiltonian
$H_{R}(t):\Gamma_{R}\rightarrow\mathbb{R}$
\begin{eqnarray*}
H_{R}(t)&=&(4\pi)^{1/4}e^{\tilde{P}/2}\varphi_{1_0}\sin
t+2\sqrt{\pi}e^{\tilde{P}}\Big(\log \frac{\sin
t}{\sqrt{4\pi}}+\tilde{P}-\log2-\frac{3}{2}\Big)\sin
t\\
&+&\frac{1}{2}\sum_{i}\int_{\mathbb{S}^2} {^{\dos}}\mathbf{e}\,
\bigg(\frac{p_{\varphi_i}^2}{\sin t}+\varphi_i^{\prime2}\sin
t\sin^2\theta\bigg)\,.
\end{eqnarray*}
Both initial and final singularities show up in the same way as for
the $\mathbb{S}^1\times\mathbb{S}^2$ topology. The corresponding
evolution vector field is given by
\begin{eqnarray*}
E_{H_R}&=&\frac{\partial}{\partial
t}+\left[(4\pi)^{1/4}e^{\tilde{P}/2}\varphi_{1_0}\sin
t+2\sqrt{\pi}e^{\tilde{P}}\sin t\left(\log\frac{\sin
t}{\sqrt{4\pi}}+\tilde{P}-\log2-\frac{1}{2}\right)\right]
\frac{\partial}{\partial\tilde{Q}}\\
&-&(4\pi)^{1/4}e^{\tilde{P}/2} \sin t\frac{\partial}{\partial
p_{\varphi_{1_0}}}+\sum_{i}\int_{\mathbb{S}^{2}}{^{(2)}}\mathbf{e}\left(\frac{p_{\varphi_{i}}}{\sin
t}\frac{\delta}{\delta
\varphi_{i}}+(\sin^{2}\theta\varphi_i^{\prime})^{\prime}\sin
t\frac{\delta}{\delta p_{\varphi_{i}}}\right).
\end{eqnarray*}
It is interesting at this point to compare the dynamics of this
model and the 3-handle one. First of all, we see that the global
mode has a different behavior now, in particular it couples to
$\varphi_{1_0}$ through the term $e^{\tilde{P}/2}\varphi_{1_0}\sin
t$. As we see, the gravitational and matter modes apparently cease
to play a symmetric role in this particular description, at variance
with the other topologies. However, we will see next that it is
possible to restore the symmetry between the gravitational and
matter scalars.
\\
\indent We proceed as at the end of the section devoted to the
$\mathbb{S}^{1}\times\mathbb{S}^{2}$ topology by introducing a
convenient auxiliary set $(\mathring{g}_{ab},\mathring{\phi}_{i})$
in terms of which we can solve the original Einstein-Klein-Gordon
equations. After the deparameterization procedure, and imposing the
consistency of the gauge fixing conditions under the dynamics
generated by the Hamiltonian $H[\hat{N},\hat{N}^{\theta}]$, the
3-metric $g_{ab}$ of the 3-sphere model can be written
$$
 g_{ab}=\cos^2(\theta/2)e^\gamma\bigg(-(\mathrm{d}t)_a(\mathrm{d}t)_b
+(\mathrm{d}\theta)_a(\mathrm{d}\theta)_b\bigg)+\frac{P^2}{4\pi}\sin^2t\sin^2\theta
(\mathrm{d}\sigma)_a(\mathrm{d}\sigma)_b\,,
$$
defined on $(0,\pi)\times D(0;\pi)$. In this case, a possible choice
of $(\mathring{g}_{ab},\mathring{\phi}_{i})$ is
\begin{eqnarray*}
\mathring
g_{ab}&=&\cos^2(\theta/2)e^{\mathring{\gamma}}\bigg(-(\mathrm{d}t)_a(\mathrm{d}t)_b
+(\mathrm{d}\theta)_a(\mathrm{d}\theta)_b\bigg)+\sin^2t\sin^2\theta (\mathrm{d}\sigma)_a(\mathrm{d}\sigma)_b\,,\\
\mathring{\phi}_1&=&\cos\theta \cos t\log(\tan(t/2))+\cos\theta+\log(\cos^2(\theta/2))+\log(2\sin t)\,,\\
\mathring{\phi}_{i}&=&0\,,\,\,\,i\neq0\,,
\end{eqnarray*}
where
\begin{eqnarray*}
\mathring{\gamma}&:=& \frac{\sin^2\theta}{4}\Big(\sin^2 t\log^2(\tan
t/2)-2\cos t\log(\tan t/2)-1\Big)+\log\left(\sin^2 t\right)\\
&-&\cos t\log(\tan(t/2))+\cos\theta \cos t\log(\tan(t/2))+\cos\theta
-1.
\end{eqnarray*}
In fact, the concrete functional form of $\mathring{\gamma}$ is
irrelevant here since whenever $\mathcal{L}_\sigma \phi_i=0$ we
have the following equivalence in
$(0,\pi)\times(\mathbb{S}^2-\{\theta=\pi\})$:
$$
\mathring g^{ab}\mathring \nabla_a\mathring \nabla_b
\phi_i=0\Leftrightarrow \breve{g}^{ab} \breve{\nabla}_a
\breve{\nabla}_b \phi_i=0\,,
$$
with
$$
\breve{g}_{ab}=\sin^2t\Big(-(\mathrm{d}t)_a(\mathrm{d}t)_b
+(\mathrm{d}\theta)_a(\mathrm{d}\theta)_b+\sin^2\theta(\mathrm{d}\sigma)_a(\mathrm{d}\sigma)_b\Big)\,.
$$
Here, the metric $\breve{g}_{ab}$ coincides with that one found for
the 3-handle case, conformal to the Einstein static metric on
$(0,\pi)\times \mathbb{S}^2$, but restricted now to the manifold
$(0,\pi)\times D(0;\pi)$ obtained by removing the pole $\theta=\pi$
from the 2-sphere. Note that the $\phi_1$ field cannot be extended
to the boundary of the disc, parameterized as $\theta=\pi$, because
(\ref{condfields2_S3}) forces it to behave as
$\log(\cos^2(\theta/2))$ when $\theta\rightarrow\pi$. However, if we
split $\phi_1$ as
$\phi_1=\phi_1^{\mathrm{sing}}+\phi_1^{\mathrm{reg}}$, with
$\phi_1^{\mathrm{sing}}:=\log(\cos^2(\theta/2))+\log(2\sin t)$
satisfying
$$
\breve{g}^{ab} \breve{\nabla}_a \breve{\nabla}_b
\phi^{\mathrm{sing}}_1=0\,,
$$
we guarantee that the gravitational degrees of freedom encoded by
$\phi^{\mathrm{reg}}_1$ still satisfy $\breve{g}^{ab}
\breve{\nabla}_a \breve{\nabla}_b \phi_1^{\mathrm{reg}}=0$ (just the
same equation as the matter fields $\phi_{i}$, $i\neq1$) and can be
extended to $(0,\pi)\times \mathbb{S}^2$. Both regularized
gravitational and matter fields, respectively
$\phi_1^{\mathrm{reg}}$ and $\phi_{i}$, $i\neq1$, are then well
behaved on $(0,\pi)\times\mathbb{S}^2$ and play a symmetric role
just as in the description of the 3-handle topology. Within the
Lagrangian formulation of the 3-sphere model, this can be attained
by introducing an extra $\hat{T}$ field on the right-hand side of
the regularity condition (\ref{condfields2_S3}).
\\
\linebreak \linebreak \noindent \textbf{References}
\begin{itemize}
\item [1] M. Nakahara, \emph{Geometry, Topology and Physics}, Second
Edition. Graduate Student Series in Physics, Taylor \& Francis
(2003).
\item [2] P. S. Mostert, Ann. of Math. \textbf{65}, 447 (1957); \textbf{66}, 589 (1957).
\item [3] P. T. Chru\'sciel, Ann. Phys. \textbf{202}, 100 (1990).
\item [4] S. W. Hawking and G. T. Horowitz, Class. Quant. Grav. \textbf{13}, 1487 (1996). S. W. Hawking and C. J. Hunter, Class. Quant. Grav. \textbf{13}, 2735 (1996).
\item [5] O. Rinne and J. M. Stewart, Class. Quant. Grav. 22, 1143 (2005).
\item [6] T. Thiemann, \emph{Modern Canonical Quantum General Relativity} (Cambridge Monographs on Mathematical
Physics), Cambridge University Press (2007).
\item [7] J. D. Romano and C. G. Torre, Phys. Rev. D \textbf{53}, 5634 (1996).
\item [8] A. Ashtekar, \emph{Lectures on Non-perturbative Canonical
Gravity}, Advanced series in Astrophysics and Cosmology, Vol. 6,
World Scientific (1991).
\item [9] C. W. Misner, Phys. Rev. D \textbf{8}, 3271 (1973).
\item [10] B. K. Berger, Ann. Phys. \textbf{83}, 458 (1974).
\item [11] M. Pierri, Int. J. Mod. Phys. D \textbf{11}, 135 (2002).
\item [12] J. Cortez and G. A. Mena Marug\'an, Phys. Rev. D \textbf{72}, 064020 (2005).
\item [13] D. Chinea, M. de Le\'on, and J.C. Marrero, ``Locally conformal cosymplectic manifolds and time-dependent Hamiltonian systems'', Comment. Math. Univ. Carolinae, 32, 2 (1991), 383-387.
\item [14] S. W. Hawking and R. Penrose, Proc. Roy. Soc. Lond. A \textbf{267}, 529-48 (1970).
\end{itemize}

\newpage
\thispagestyle{plain} \mbox{}


\chapter{Fock Space Quantization}\label{ChapterII}
\begin{flushright}
\small{\vspace*{-0.9cm} \textbf{J. F. Barbero G., D. G. Vergel, and E. J. S. Villase\~{n}or}\\ \textbf{Classical and Quantum Gravity \textbf{25}, 085002 (2008)}\\\vspace*{0.2cm} \textbf{D. G. Vergel and E. J. S. Villase\~{n}or}\\
\textbf{Classical and Quantum Gravity 25, 145008 (2008)}}
\end{flushright}


\pagestyle{fancy} \fancyhf{}

\fancyhead[LO]{\textsc{Chapter 2. Fock Space Quantization}}
\fancyhead[RO,LE]{\thepage}
\renewcommand{\headrulewidth}{0.6pt}


\vspace*{0.5cm} \indent In this chapter, we will proceed to exactly
quantize the classical theory developed in \emph{Chapter 1} by
promoting the canonical variables
$(\tilde{Q},\tilde{P};\varphi_{i},p_{\varphi_{i}})$ characterizing
the global and local degrees of freedom to quantum operators
$(\hat{\tilde{Q}},\hat{\tilde{P}};\hat{\varphi}_{i},\hat{p}_{\varphi_{i}})$.
Our starting point is given by the interpretation of the compact
$\mathbb{S}^{1}\times\mathbb{S}^{2}$ and $\mathbb{S}^{3}$ Gowdy
models coupled to matter as scalar field theories in curved
backgrounds conformally equivalent to the $(1+2)$-dimensional
Einstein metric on $(0,\pi)\times\mathbb{S}^2$. Here, the
corresponding conformal factor will be a simple function of $t$,
specifically, $\sin t$. This description will be used to gain useful
insights into the problem of the unitary implementability of the
quantum time evolution for these models.
\\
\indent Let
$\mathscr{H}=\mathscr{H}_{0}\otimes\mathscr{H}_{c}=\mathscr{H}_{0}\otimes(\otimes_{i}\mathscr{F}_{i})$
be the Hilbert space of the system. The absence of extra constraints
for the topologies under consideration was proved in the previous
chapter. From here follows the unnecessary distinction between
kinematical and physical Hilbert spaces.\footnote{Nevertheless, in
spite of the apparent simplicity of the phase space description
after deparameterization given the absolute decoupling of
gravitational and matter degrees of freedom, the full
$(1+3)$-dimensional metric that solves the original
Einstein-Klein-Gordon equations depends on both types of modes in a
nontrivial way.} This is in contrast with the 3-torus case, where an
extra $U(1)$ symmetry generated by a residual global constraint
still remains after deparameterization. The global modes
$(\tilde{Q},\tilde{P})$ can be quantized in a straightforward way in
terms of standard position and momentum operators with dense domains
in $\mathscr{H}_{0}\cong L^{2}(\mathbb{R},\mathrm{d}q)$, such
that\footnote{In what follows, we will use units such that
$\hbar=1$.} $\hat{\tilde{Q}}\psi=q\psi$,
$\hat{\tilde{P}}\psi=-i\partial_{q}\psi$, for $\psi=\psi(q)\in
L^{2}(\mathbb{R})$. The Hilbert spaces for gravitational and matter
modes, on the other hand, adopt the structure of symmetric Fock
spaces $\mathscr{F}_{i}$ built from appropriate one-particle Hilbert
spaces. As they are all isomorphic, and all massless scalar fields
$\varphi_{i}$ satisfy the same Euler-Lagrange equation, the same
construction is valid for all of them. For this reason, we will omit
the $i$-index in the following. The local degrees of freedom
$(\varphi,p_{\varphi})$ are then promoted to operator-valued
distributions on $\mathbb{S}^{2}$ for each value of the time
parameter $t$. They act as the identity over $\mathscr{H}_{0}$.
Similarly, the $\hat{\tilde{Q}}$ and $\hat{\tilde{P}}$ operators act
as the identity over $\mathscr{F}$ .

\section{Canonical and covariant phase spaces}{\label{PhaseSpace}}

\indent The dynamics of the local degrees of freedom of both
$\mathbb{S}^{1}\times\mathbb{S}^{2}$ and $\mathbb{S}^{3}$ Gowdy
models can be described after deparameterization by the same
Euler-Lagrange equation in a fixed background metric conformal to
the (1+2)-dimensional Einstein static metric on
$(0,\pi)\times\mathbb{S}^{2}$,
\begin{eqnarray}
\mathring{g}_{ab}=\sin^2t\left(-(\mathrm{d}t)_a(\mathrm{d}t)_b+\gamma_{ab}\right),\label{g0}
\end{eqnarray}
where $\gamma_{ab}$ denotes the round unit metric on the 2-sphere
$\mathbb{S}^2$; in spherical coordinates $(\theta,\sigma)\in
(0,\pi)\times (0,2\pi)$ on $\mathbb{S}^2$,
$\gamma_{ab}=(\mathrm{d}\theta)_a(\mathrm{d}\theta)_b+\sin^2\theta(\mathrm{d}\sigma)_a(\mathrm{d}\sigma)_b$.
In addition, we must impose the invariance under the diffeomorphisms
generated by the Killing vector field
$\sigma^a=(\partial/\partial\sigma)^{a}$. The equation of motion can
be derived, by imposing the additional symmetry condition
$\mathcal{L}_\sigma\varphi=0$ on the solutions, from the action
\begin{eqnarray}
S(\varphi)&=&-\frac{1}{2}\int_{[t_0,t_1]\times \mathbb{S}^2}
|\mathring{g}|^{1/2}
\mathring{g}^{ab}(\mathrm{d}\varphi)_a(\mathrm{d}\varphi)_b\label{accion}
\\
&=&\frac{1}{2}\int_{t_0}^{t_1}
\mathrm{d}t\int_{\mathbb{S}^2}|\gamma|^{1/2}\sin
t\,\Big(\dot{\varphi}^2+\varphi\,\Delta_{\mathbb{S}^2}\varphi\Big)\,,\nonumber
\end{eqnarray}
where $\Delta_{\mathbb{S}^2}$ denotes the Laplace-Beltrami operator
on the round 2-sphere $\mathbb{S}^2$. The space of smooth and
symmetric real solutions to the corresponding massless Klein-Gordon
equation of motion has the structure of an infinite-dimensional
$\mathbb{R}$-vector space
\begin{eqnarray}
\mathcal{S}&:=&\{\varphi\in
C^\infty((0,\pi)\times\mathbb{S}^2;\mathbb{R})\,|\,\mathring{g}^{ab}\mathring\nabla_a\mathring\nabla_b\varphi=0;\,
\mathcal{L}_\sigma\varphi=0\}\label{solspace}\\
&=&\{\varphi\in
C^\infty((0,\pi)\times\mathbb{S}^2;\mathbb{R})\,|\,\ddot{\varphi}+\cot
t \dot{\varphi}-\Delta_{\mathbb{S}^2}\varphi=0;\,
\mathcal{L}_\sigma\varphi=0\}\,.\nonumber
\end{eqnarray}
The variational principle (\ref{accion}) gives rise to a natural
(weakly) symplectic structure $\Omega$ on $\mathcal{S}$ defined by
\begin{equation}
\Omega(\varphi_1,\varphi_2):=\sin t\int_{\mathbb{S}^2}
|\gamma|^{1/2}\imath^*_t\Big(\varphi_2\dot{\varphi}_1-\varphi_1\dot{\varphi}_2\Big)\,.
\label{sympS3}
\end{equation}
Here, $\imath_t:\mathbb{S}^2\rightarrow (0,\pi)\times \mathbb{S}^2$
denotes the inclusion given by $\imath_{t}(s)=(t,s)\in (0,\pi)\times
\mathbb{S}^2$. It is straightforward to show that $\Omega$ does not
depend upon the choice of the value of time $t$ used to define the
embedding
$\iota_{t}(\mathbb{S}^{2})\subset\mathbb{R}\times\mathbb{S}^{2}$. We
will refer to the infinite-dimensional linear symplectic space
$\Gamma:=(\mathcal{S},\Omega)$ as the \textit{covariant phase space}
of the system $[1]$. We will denote the usual \textit{canonical
phase space} as $\Upsilon :=(\mathbf{P},\omega)$. Here, $\mathbf{P}$
is the space of smooth and symmetric Cauchy data
$\mathbf{P}:=\{(Q,P)\in C^\infty(\mathbb{S}^2;\mathbb{R})\times
C^\infty(\mathbb{S}^2;\mathbb{R})\,|\, \mathcal{L}_\sigma
Q=\mathcal{L}_\sigma P=0\}$, endowed with the standard symplectic
structure
\begin{eqnarray}\label{cpsomega}
\omega((Q_1,P_1),(Q_2,P_2)):=\int_{\mathbb{S}^2} |\gamma|^{1/2}
(Q_2P_1-Q_1P_2)\,.
\end{eqnarray}
Given any value of $t$, it is possible to construct a
symplectomorphism between the spaces $\Gamma$ and $\Upsilon$. The
bijection $\mathfrak{I}_t:\Upsilon\rightarrow\Gamma$, that maps
every Cauchy data $(Q,P)$ to the unique solution $\varphi\in
\mathcal{S}$ such that
\begin{equation}\label{isomorf1}
\varphi(t,s)=Q(s)\quad \mathrm{and} \quad (\sin
t)\dot{\varphi}(t,s)=P(s)
\end{equation}
is, irrespective of the value of $t$, a linear symplectomorphism,
i.e., $\omega=\mathfrak{I}_t^*\Omega$.
\\
\linebreak \indent Any vector $\varphi\in\Gamma$ is a smooth
function on $\mathbb{R}\times\mathbb{S}^{2}$. Therefore, for each
value of $t\in(0,\pi)$ we have $\iota_{t}^{*}\varphi\in
C^{\infty}(\mathbb{S}^{2};\mathbb{R})$. It is well known that any
smooth \emph{symmetric} function on $\mathbb{S}^{2}$ can be written
in the form
\begin{equation}\label{varphi}
(\iota_{t}^{*}\varphi)(s)=\varphi(t,s)=\sum_{\ell=0}^\infty
A_\ell(t)Y_{\ell0}(s)\,,
\end{equation}
where $Y_{\ell 0}$ denote the spherical harmonics (\ref{Yn}) that in
standard spherical coordinates have the form
$$
Y_{\ell
0}(s)=\left(\frac{2\ell+1}{4\pi}\right)^{1/2}\mathscr{P}_\ell(\cos\theta(s))
$$
in terms of Legendre polynomials $\mathscr{P}_\ell$, satisfying the
equations
$$
\Delta_{\mathbb{S}^2}Y_{\ell 0}=-\ell(\ell+1)Y_{\ell
0}\,,\quad\mathcal{L}_\sigma Y_{\ell 0}=0\,,
$$
and verifying the $L^{2}(\mathbb{S}^{2})$-orthogonality
conditions\footnote{The bar denotes complex conjugation and
$\delta(\ell_1,\ell_2)$ is the Kronecker delta.}
$$\int_{\mathbb{S}^{2}}|\gamma|^{1/2}\bar{Y}_{\ell_{1}0}Y_{\ell_{2}0}=\delta(\ell_1,\ell_2)\,.$$
The coefficients $A_{\ell}(t)$ appearing in (\ref{varphi}) are
defined in terms of $\varphi$ through
$$A_{\ell}(t)=\int_{\mathbb{S}^{2}}|\gamma|^{1/2}\bar{Y}_{\ell0}\,\iota_{t}^{*}\varphi\,.$$
Given the reality of the field $\varphi$, it is clear that
$\bar{A}_{\ell}(t)=A_{\ell}(t)$. From the fact that, for any fixed
value of $t$ and for all $n\in\mathbb{N}_{0}$,
$\partial_{t}^{n}\varphi(t,\cdot)$ is a smooth function on
$\mathbb{S}^{2}$, we have
$$\lim_{\ell\rightarrow+\infty}\frac{1}{\ell^{p}}\frac{\mathrm{d}^{n}A_{\ell}(t)}{\mathrm{d}t^{n}}=0\,,\quad\forall\,p,n\in\mathbb{N}_{0}\,,\,\,\,\forall\,t\in(0,\pi)\,.$$
Then, it is clear that any $\varphi\in\Gamma$ can be expanded in the
form
\begin{equation}\label{expansionS3}
\varphi(t,s)=\sum_{\ell=0}^{\infty}\Big(a_{\ell}y_{\ell}(t)+\overline{a_{\ell}y_{\ell}(t)}\Big)Y_{\ell0}(s)\,.
\end{equation}
The massless Klein-Gordon equation defining $\mathcal{S}$ leads to
the following equation for the complex functions $y_\ell(t)$
\begin{equation}
\ddot{y}_\ell+(\cot t)\dot{y}_\ell+\ell(\ell+1)y_\ell=0\,.
\label{diffeqS3}
\end{equation}
We will always assume that, for each $\ell$, the real and imaginary
parts of $y_\ell$, denoted $u_\ell$ and $v_\ell$ respectively, are
two real linearly independent solutions of (\ref{diffeqS3}). The
complex coefficients $a_{\ell}$ must satisfy the fall-off conditions
$$\lim_{\ell\rightarrow+\infty}\frac{1}{\ell^{p}}\left(a_{\ell}\frac{\mathrm{d}y_{\ell}}{\mathrm{d}t^{n}}(t)
+\bar{a}_{\ell}\frac{\mathrm{d}\bar{y}_{\ell}}{\mathrm{d}t^{n}}(t)\right)=0\,,\quad\forall\,p,n\in\mathbb{N}_{0}\,,\,\,\,\forall\,
t\,\in(0,\pi)\,.$$ From the point of view of the classical theory,
these conditions are necessary to guarantee the smoothness of the
solution to the field equations. However, we do not need to know
them in detail to discuss the quantization of the models. In fact,
they will be relaxed to the milder condition
$\sum_{\ell}|a_{\ell}|^{2}<+\infty$ when we introduce the
one-particle Hilbert space. We will not make at this point any
specific choice for the complex functions $y_{\ell}$, but we will
fix their normalization in the following way. Let us substitute
first (\ref{expansionS3}) in the symplectic structure
(\ref{sympS3}). We find that
$$
\Omega(\varphi_1,\varphi_2)=\sin t\sum_{\ell=0}^\infty
\big(\bar{a}^{(1)}_{\ell}a^{(2)}_{\ell}-\bar{a}^{(2)}_{\ell}a^{(1)}_{\ell}\big)
\big(y_\ell(t)\dot{\bar{y}}_\ell(t)-\dot{y}_\ell(t)\bar{y}_\ell(t)\big)\,.
$$
By explicitly decomposing $y_\ell(t)=u_\ell(t)+iv_\ell(t)$ and
writing
$$
y_\ell(t)\dot{\bar{y}}_\ell(t)-\dot{y}_\ell(t)\bar{y}_\ell(t)=2i
\det\left[\begin{array}{cc}
\dot{u}_\ell(t)&u_\ell(t)\\
\dot{v}_\ell(t)&v_\ell(t)
\end{array}
\right]=:2iW(t;u_\ell,v_\ell)\,.
$$
we have that, by virtue of the differential equation
(\ref{diffeqS3}), the Wronskian $W$ satisfies
$$
\dot{W}+(\cot t) W=0\Rightarrow
W(t;u_\ell,v_\ell)=\frac{c_\ell}{\sin t}\,,\quad c_\ell\in
\mathbb{R}\,,
$$
and hence the symplectic structure takes the simple expression
\begin{equation}
\Omega(\varphi_1,\varphi_2)=2i\sum_{\ell=0}^\infty
c_\ell\big(\bar{a}^{(1)}_{\ell}a^{(2)}_{\ell}-\bar{a}^{(2)}_{\ell}a^{(1)}_{\ell}\big)\,.
\label{symp_simp_S3}
\end{equation}
Note that the time-independence of the symplectic structure is
explicit now. In the following, we will choose the pair of functions
$(u_\ell,v_\ell)$ to be normalized in such a way that $c_\ell=1/2$,
$\forall\,\ell\in\mathbb{N}_0$, i.e.,
\begin{equation}\label{norm}
W(t;u_\ell,v_\ell)=\frac{1}{2\sin t}\,,\quad\forall\,
(u_\ell,v_\ell)\,,\,\,\,\ell\in\mathbb{N}_{0}\,.
\end{equation}
It could appear that this condition is rather arbitrary but, as we
will see, it is expedient to make this choice in order to ensure
that the modes
$(\varphi_\ell:=y_{\ell}Y_{\ell0})_{\ell\in\mathbb{N}_0}$ define an
orthogonal basis of the one-particle Hilbert space used to construct
the Fock space for the quantum theory. We also obtain in this way a
very convenient expression for the symplectic structure $\Omega$
that will be our starting point for the quantization of the models.
A concrete election satisfying (\ref{norm}) is given by
\begin{eqnarray}
u_{0\ell}(t)=\frac{1}{\sqrt{2}}\mathscr{P}_\ell(\cos t),\quad
v_{0\ell}(t)=\frac{1}{\sqrt{2}}\mathscr{Q}_\ell(\cos t)\,,
\,\,\,\ell\in\mathbb{N}_{0}\,,\label{u0v0}
\end{eqnarray}
with $\mathscr{P}_\ell$ and $\mathscr{Q}_\ell$ denoting the first
and second class Legendre functions, respectively.

\subsection{Classical dynamics}

The classical time evolution from the embedding
$\imath_{t_0}(\mathbb{S}^{2})\subset(0,\pi)\times\mathbb{S}^2$ to
$\imath_{t_1}(\mathbb{S}^{2})\subset(0,\pi)\times\mathbb{S}^2$ is
implemented on the canonical phase space $\Upsilon$ by the
symplectic transformation
$\tau_{(t_1,t_0)}:\Upsilon\rightarrow\Upsilon$ defined as
\begin{equation}\label{tau}
\tau_{(t_1,t_0)}:=\mathfrak{J}_{t_1}^{-1}\circ \mathfrak{J}_{t_0}
\end{equation}
in terms of the symplectic maps
$\mathfrak{J}_t:\Upsilon\rightarrow\Gamma$ introduced in
(\ref{isomorf1}) and their inverses. The maps $\mathfrak{J}_t$ can
be easily computed in terms of the Fourier coefficients $a_{\ell}$
of $\varphi$ (\ref{expansionS3}) as
\begin{equation}
a_\ell(t)=- i\sin t
\dot{\bar{y}}_\ell(t)\int_{\mathbb{S}^2}|\gamma|^{1/2} Y_{\ell
0}Q+i\bar{y}_\ell(t)\int_{\mathbb{S}^2}|\gamma|^{1/2} Y_{\ell
0}P\,.\label{I_direc}
\end{equation}
The operator $\tau_{(t_1,t_0)}$ acts as follows: (i) first, it takes
initial Cauchy data on $\imath_{t_0}(\mathbb{S}^{2})$, (ii) evolves
them to the corresponding solution in $\mathcal{S}$, and (iii) finds
the Cauchy data induced by this solution on
$\imath_{t_1}(\mathbb{S}^{2})$. On the other hand, time evolution
can also be viewed as a symplectic transformation on the covariant
phase space, $\mathcal{T}_{(t_1,t_0)}:\Gamma\rightarrow\Gamma$,
defined by
\begin{equation}
\mathcal{T}_{(t_1,t_0)}:=\mathfrak{J}_{t_1}\circ\tau_{(t_1,t_0)}\circ\mathfrak{J}_{t_1}^{-1}
=\mathfrak{J}_{t_0}\circ\mathfrak{J}_{t_1}^{-1}\,, \label{eee}
\end{equation}
that (i) takes a solution of $\mathcal{S}$, (ii) finds the induced
Cauchy data on $\imath_{t_1}(\mathbb{S}^{2})$, and (iii) takes those
data as initial data on $\imath_{t_0}(\mathbb{S}^{2})$, finding the
unique solution of $\mathcal{S}$ which corresponds to these initial
conditions. In our case, combining (\ref{isomorf1}) and
(\ref{I_direc}), it is straightforward to check that the action of
the operator $\mathcal{T}_{(t_1,t_0)}$ is given by
\begin{eqnarray}
(\mathcal{T}_{(t_1,t_0)}\varphi)(t,s)&=&\sum_{\ell=0}^\infty
\Big(\mathfrak{a}_\ell(t_1,t_0)
y_\ell(t)+\overline{\mathfrak{a}_\ell(t_1,t_0)
y_\ell(t)}\Big)Y_{\ell0}(s)\,,\label{T(t0t1)}
\end{eqnarray}
where
\begin{eqnarray}
\mathfrak{a}_\ell(t_1,t_0) &:=&i\big(\sin t_1
\bar{y}_\ell(t_0)\dot{y}_\ell(t_1)-\sin
t_0y_\ell(t_1)\dot{\bar{y}}_\ell(t_0)\big)a_\ell
\label{timeevolution}
\\
&+&i\big(\sin t_1\bar{y}_\ell(t_0)\dot{\bar{y}}_\ell(t_1)-\sin
t_0\bar{y}_\ell(t_1)\dot{\bar{y}}_\ell(t_0)\big)\bar{a}_\ell
\,.\nonumber
\end{eqnarray}
In the next sections we will try to find out if this classical
evolution can be unitarily implemented in a Fock quantization of the
system.

\section{Fock quantization}{\label{FockQuantization}}

In the passage to the quantum theory, we have to introduce a Hilbert
space for our system. This will have the structure of a symmetric
Fock space $\mathscr{F}$ built from some appropriate one-particle
Hilbert space. We will review in this section the Fock quantization
techniques based on the covariant phase space description of the
model, following the quantization steps discussed in section 2.3 of
reference $[2]$. It is well known that for a system of a finite
number of uncoupled quantum harmonic oscillators this procedure
provides a quantum theory unitarily equivalent to the usual tensor
product of one-particle Hilbert spaces. However, for the case of a
system of infinitely many uncoupled quantum harmonic oscillators,
the tensor product of infinite number of one-particle Hilbert spaces
gives rise to nonseparable Hilbert spaces, as well as reducible
representations of the canonical commutation relations. The Fock
quantization process analyzed below provides a better approach to
deal with the infinitely many degrees of freedom present in these
models, for it avoids the aforementioned difficulties. As expected
for scalar fields in nonstationary curved background spacetimes, the
Fock representation obtained in this way is highly non-unique; this
is a problem that will be discussed in detail.
\\
\indent In order to define the one-particle Hilbert space used to
build the Fock space $\mathscr{F}$, let
$\mathcal{S}_{\mathbb{C}}:=\mathbb{C}\otimes \mathcal{S}$ denote the
$\mathbb{C}$-vector space obtained by the complexification of the
solution space $\mathcal{S}$ (\ref{solspace}). The elements of
$\mathcal{S}_{\mathbb{C}}$ are ordered pairs of objects
$(\varphi_1,\varphi_2)\in\mathcal{S}\times\mathcal{S}$ that we will
write in the form\footnote{Here, $i\in\mathbb{C}$ denotes the
imaginary unit.} $\Phi:=\varphi_1+i\varphi_2$ with the natural
definition for their sum. Multiplication by complex scalars
$\mathbb{C}\ni\lambda=\lambda_1+i\lambda_2$,
$\lambda_1,\lambda_2\in\mathbb{R}$, is defined as
$$
\lambda\Phi:=(\lambda_1\varphi_1-\lambda_2\varphi_2)+i(\lambda_2\varphi_1+\lambda_1\varphi_2)\,.
$$
We also introduce the conjugation
$\bar{\phantom{z}}:\mathcal{S}_{\mathbb{C}}\rightarrow
\mathcal{S}_{\mathbb{C}}:(\varphi_1+i\varphi_2)\mapsto(\varphi_1-i\varphi_2)$.
Every vector $\Phi\in\mathcal{S}_{\mathbb{C}}$ can be expanded with
the help of the basis $(\varphi_\ell:=y_\ell Y_{\ell 0},
\bar{\varphi}_\ell:=\overline{y_\ell Y_{\ell
0}})_{\ell\in\mathbb{N}_0}$ introduced above as
$$\Phi=\sum_{\ell=0}^\infty\Big(a_\ell y_\ell
Y_{\ell 0}+b_\ell\overline{y_\ell Y_{\ell 0}}\,\Big)\,,$$ with
$a_\ell,b_\ell\in\mathbb{C}$. We extend the symplectic structure
(\ref{sympS3}) to $\mathcal{S}_{\mathbb{C}}$ by complex linearity in
each variable,
$$
\Omega_{\mathbb{C}}(\Phi_1,\Phi_2):=i\sum_{\ell=0}^\infty\big(b^{(1)}_{\ell}a^{(2)}_{\ell}-b^{(2)}_{\ell}a^{(1)}_{\ell}\big)\,.
$$
For each pair $\Phi_1,\Phi_2\in\mathcal{S}_{\mathbb{C}}$, consider
now the sesquilinear map
$\langle\cdot|\cdot\rangle:\mathcal{S}_{\mathbb{C}}\times
\mathcal{S}_{\mathbb{C}}\rightarrow\mathbb{C}$ defined by
\begin{equation}
\langle\Phi_1|\Phi_2\rangle:=-i\Omega_{\mathbb{C}}(\bar{\Phi}_1,\Phi_2)\,.
\label{escprod}
\end{equation}
It is antilinear in the first argument and linear in the second,
satisfying all the properties of an inner product on
$\mathcal{S}_{\mathbb{C}}$ except that it fails to be positive
definite. There are, however, linear subspaces of
$\mathcal{S}_{\mathbb{C}}$ where $\langle\cdot|\cdot\rangle$ is
positive definite (and, hence, defines an inner product). Consider,
in particular, the Lagrangian subspace
\begin{equation}
\mathcal{P}:=\Big\{\Phi\in\mathcal{S}_{\mathbb{C}}\,\big|\,\Phi=\sum_{\ell=0}^\infty
a_\ell\varphi_\ell\Big\}\,. \label{P}
\end{equation}
Here, the restriction $\langle\cdot|\cdot\rangle|_\mathcal{P}$
defines an inner product given by
\begin{equation}\label{inprod}
\langle \Phi_1|\Phi_2\rangle=\sum_{\ell=0}^\infty
\bar{a}^{(1)}_{\ell}a^{(2)}_{\ell}\,,\quad
\Phi_{1},\Phi_{2}\in\mathcal{P}\,.
\end{equation}
The separable and infinite-dimensional \emph{one-particle Hilbert
space} $\mathscr{H}_\mathcal{P}\cong\ell^{2}(\mathbb{C})$ is then
obtained by Cauchy completion of
$(\mathcal{P},\langle\cdot|\cdot\rangle|_\mathcal{P})$ with respect
to the norm defined by the inner product
$\langle\cdot|\cdot\rangle|_{\mathcal{P}}$,
$$\mathscr{H}_{\mathcal{P}}:=\overline{(\mathcal{P},\langle\cdot|\cdot\rangle|_{\mathcal{P}})}^{\langle\cdot|\cdot\rangle|_{\mathcal{P}}}=
\Big\{\Phi=\sum_{\ell=0}^{\infty}a_{\ell}\varphi_{\ell}\,\,\big|\,\,
a_{\ell}\in \mathbb{C}\,,\,
\sum_{\ell=0}^{\infty}|a_{\ell}|^2<+\infty\Big\}\,.
$$
Note that the set $(\varphi_\ell=y_\ell Y_{\ell 0})_{\ell\in
\mathbb{N}_0}$ becomes an orthonormal basis of
$\mathscr{H}_\mathcal{P}$, satisfying $\langle
\varphi_{\ell_1}\,|\,\varphi_{\ell_2
}\rangle|_{\mathcal{P}}=\delta(\ell_1,\ell_2)$. The Hilbert space of
the quantum theory is finally given by the symmetric Fock space
$$
\mathscr{F}_{+}(\mathscr{H}_\mathcal{P})=\displaystyle
\bigoplus_{n=0}^{\infty}\mathscr{P}_{+}^{(n)}\left(\mathscr{H}_{\mathcal{P}}^{\otimes
n}\right),
$$
where $\mathscr{H}_{\mathcal{P}}^{\otimes0}:=\mathbb{C}$ and
$\mathscr{P}_{+}^{(n)}\left(\mathscr{H}_{\mathcal{P}}^{\otimes
n}\right)$ denotes the $n$-fold symmetric tensor product of
$\mathscr{H}_{\mathcal{P}}$ (see \emph{appendix
\ref{AppendixFock}}). Associated with the orthonormal modes
$\varphi_\ell\in\mathscr{H}_\mathcal{P}$, we define the
corresponding annihilation $\hat{a}_\ell:=A(\varphi_\ell)$ and
creation operators $\hat{a}_\ell^\dagger:=C(\varphi_\ell)$, with
nonvanishing commutation relations given by
$[\hat{a}_{\ell_1},\hat{a}_{\ell_2}^\dagger]=\delta(\ell_1,\ell_2)$.
As usual, we will denote as $|0\rangle_{\mathcal{P}}$ the Fock
vacuum $1\oplus 0\oplus 0\oplus \cdots \in
\mathscr{F}_{+}(\mathscr{H}_\mathcal{P})$ whose only nonzero
component is $1\in\mathbb{C}$. The vacuum is in the domain of all
finite products of creation and annihilation operators, and the
vectors
$$
|{^1}\! n_{\ell_1}\, ^2\!n_{\ell_2}\,\cdots \,^k\! n_{\ell_k}\rangle
:=\frac{1}{\sqrt{ ^1n!\,^2n!\,\cdots\,^kn!}}
(\hat{a}_{\ell_1}^{\dagger})^{ ^1 \!
n}(\hat{a}_{\ell_1}^{\dagger})^{ ^2\! n} \cdots
(\hat{a}_{\ell_k}^{\dagger})^{ ^k \! n} |0\rangle_{\mathcal{P}}\in
\mathscr{F}_{+}(\mathscr{H}_\mathcal{P}) \,,
$$
where $k\in\mathbb{N}_{0}$, $(\,\!^1\!n,\, ^2\!n,\dots,\, ^k\!n)\in
\mathbb{N}^k$, and $\ell_i\neq \ell_j$ for $i\neq j$, provide a
basis of $\mathscr{F}_{+}(\mathscr{H}_\mathcal{P})$. The basis
vectors are normalized according to
\begin{eqnarray*}
\langle \, ^1\! n_{\ell_1}\,\cdots \,^k\! n_{\ell_k} \,|\,^1\!
m_{\ell'_1}\,\cdots \,^r\! m_{\ell'_r} \rangle &=& \delta(k, r)
\sum_{\sigma\in\Pi_k} \delta( ^1\! n, ^{\pi(1)}\! m) \cdots  \delta(
^k\!n, ^{\pi(k)}\! m)\\
&&\quad\quad\quad\quad\,\,\,\, \times\delta(\ell_1,\ell'_{\pi(1)})
\cdots \delta( \ell_k,\ell'_{\pi(k)})\,,
\end{eqnarray*}
where $\Pi_k$ denotes the set of permutations $\sigma$ of the $k$
symbols $\{1,2,\dots,k\}$. The creation and annihilation operators
satisfy
$$
\hat{a}^\dagger_\ell
|n_\ell\rangle=\sqrt{n+1}\,|(n+1)_\ell\rangle\,,\quad \hat{a}_\ell
|n_\ell\rangle=\sqrt{n}\,|(n-1)_\ell\rangle\,.
$$
Using the notation introduced above, the modes $\varphi_\ell$  of
the one particle Hilbert space $\mathscr{H}_\mathcal{P}$ can now be
considered as one-particle states that we will denote as
$|1_\ell\rangle:=a^\dagger_\ell|0\rangle_\mathcal{P}\in
\mathscr{F}_{+}(\mathscr{H}_\mathcal{P})$.
\\
\linebreak \indent Note that every choice (\ref{P}) of the
Lagrangian subspace $\mathcal{P}$ corresponds to the specification
of a complex structure $J_{\mathcal{P}}$ on the space of solutions
$\mathcal{S}$. Indeed, owing to the fact that
$\mathcal{P}\cap\bar{\mathcal{P}}=\{0\}$, where $\bar{\mathcal{P}}$
is the complex conjugate space of $\mathcal{P}$, it follows that
$\mathcal{S}_{\mathbb{C}}=\mathcal{P}\oplus\bar{\mathcal{P}}$ and,
hence, any vector $\varphi\in\mathcal{S}$ can be uniquely decomposed
as $\varphi=\Phi+\bar{\Phi}$, with $\Phi\in \mathcal{P}$,
$\bar{\Phi}\in\bar{\mathcal{P}}$. Then, given $\mathcal{P}$ and
$\bar{\mathcal{P}}$, we can define the complex structure
$J_{\mathcal{P}}:\mathcal{S}\rightarrow\mathcal{S}$ by
$J_{\mathcal{P}}\varphi:=i(\Phi-\bar{\Phi})$. This map is a linear
canonical transformation on $\Gamma=(\mathcal{S},\Omega)$ --i.e.,
$J_{\mathcal{P}}$ on $\mathcal{S}$ is compatible with $\Omega$--,
with $J_{\mathcal{P}}^{2}=J_{\mathcal{P}}\circ
J_{\mathcal{P}}=-\mathrm{Id}_{\mathcal{S}}$.
\\
\indent Denote by $\mathcal{S}_{J_{\mathcal{P}}}$ the complex vector
space $\mathcal{S}$ where, given any $\varphi\in\mathcal{S}$, the
product by complex scalars $\mathbb{C}\ni z=x+iy$,
$x,y\in\mathbb{R}$, is defined by the rule
$z\cdot\varphi:=x\varphi+yJ_{\mathcal{P}}\varphi$. We have that the
formula
\begin{equation}\label{mu}
\mu_{J_{\mathcal{P}}}(\varphi_1,\varphi_2):=\frac{1}{2}\Omega(J_{\mathcal{P}}\varphi_1,\varphi_2)
\end{equation}
defines a positive definite bilinear symmetric form on
$\mathcal{S}$. We then conclude that the sesquilinear map
\begin{equation}\label{innerprod}
\langle
\varphi_1|\varphi_2\rangle_{J_{\mathcal{P}}}:=\mu_{J_{\mathcal{P}}}(\varphi_1,\varphi_2)-\frac{i}{2}\Omega(\varphi_1,\varphi_2)
\end{equation}
is an inner product on $\mathcal{S}_{J_{\mathcal{P}}}$ $[3]$. In
this context, the one-particle Hilbert space
$\mathscr{H}_{\mathcal{P}}$ is given by the Cauchy completion of the
Euclidean space
$(S_{J_{\mathcal{P}}},\langle\cdot|\cdot\rangle_{J_{\mathcal{P}}})$.
It is straightforward to check that the Cauchy completions of
$(\mathcal{P},\langle\cdot|\cdot\rangle)$ and
$(S_{J_{\mathcal{P}}},\langle\cdot|\cdot\rangle_{J_{\mathcal{P}}})$
are isomorphic: Indeed, the $\mathbb{C}$-linear map
$\kappa:S_{J_\mathcal{P}}\rightarrow \mathcal{P}$ such that
$\kappa(\varphi)=\Phi$, $\varphi\in\mathcal{S}_{J_\mathcal{P}}$,
$\Phi\in\mathcal{P}$, defines a unitary transformation of
$\mathcal{S}_{J_\mathcal{P}}$ onto $\mathcal{P}$, i.e.,
$\langle\varphi_1|\varphi_2\rangle_{J_\mathcal{P}}=\langle\kappa(\varphi_1)|\kappa(\varphi_2)\rangle=\langle
\Phi_1|\Phi_2\rangle$,
$\forall\,\varphi_1,\varphi_2\in\mathcal{S}_{J_{\mathcal{P}}}$.
Finally, we will obtain a relation that is relevant for the
algebraic formulation of the quantum theory. By the Schwartz
inequality, for all
$\varphi_1,\varphi_2\in\mathcal{S}_{J_{\mathcal{P}}}$ we have
$\|\varphi_1\|_{J_{\mathcal{P}}}^{2}\|\varphi_2\|_{J_{\mathcal{P}}}^{2}\ge|\langle\varphi_1|\varphi_2\rangle_{J_{\mathcal{P}}}|^{2}
\ge|\mathrm{Im}[\langle\varphi_1|\varphi_2\rangle_{J_{\mathcal{P}}}]|^{2}$,
where $\|\cdot\|_{J_{\mathcal{P}}}$ denotes the norm associated with
the inner product (\ref{innerprod}), so that $\mu_{J_{\mathcal{P}}}$
satisfies
\begin{equation}\label{muJp}
\mu_{J_{\mathcal{P}}}(\varphi_1,\varphi_1)\,\mu_{J_{\mathcal{P}}}(\varphi_2,\varphi_2)\ge\frac{1}{4}\big(\Omega(\varphi_1,\varphi_2)\big)^{2}\,.
\end{equation}

\subsection{Complex structures}{\label{Complex forms}}

\indent In practice, the definition of the complex structure
$J_{\mathcal{P}}$ is complete once a choice of complex functions
$(y_{\ell})_{\ell\in\mathbb{N}_0}$ satisfying (\ref{diffeqS3}) and
(\ref{norm}) is given. In that case, we can construct an orthonormal
basis $(\varphi_{\ell}=y_{\ell}Y_{\ell0})_{\ell\in\mathbb{N}_0}$ for
the one-particle Hilbert space $\mathscr{H}_{\mathcal{P}}$ and
define $J_{\mathcal{P}}$ by imposing that the complex structure be
diagonalized in $\mathcal{S}_{\mathbb{C}}$,
\begin{equation}\label{Jpdiagonal}
J_{\mathcal{P}}\varphi_{\ell}=i\varphi_{\ell}\,,\quad
J_{\mathcal{P}}\bar{\varphi}_{\ell}=-i\bar{\varphi}_{\ell}\,.
\end{equation}
Different choices for $(y_{\ell})_{\ell\in\mathbb{N}_0}$ give rise
in general to different complex structures. With the aim of
characterizing the freedom in the election, consider the family
$(y_{0\ell})_{\ell\in\mathbb{N}_0}$ defined by (\ref{u0v0})
satisfying the normalization condition (\ref{norm}). Denote by
$J_{0}$ the corresponding complex structure. For any other
normalized election of a family of linearly independent functions
$(y_\ell=u_\ell+i v_\ell)_{\ell\in\mathbb{N}_0}$ we can write (in
terms of the $u_{0\ell}$ and $v_{0\ell}$)
\begin{equation}
y_\ell(t)=u_\ell (t)+iv_\ell (t)=\alpha_\ell u_{0\ell}(t)+\beta_\ell
v_{0\ell}(t)+i\big(\gamma_\ell u_{0\ell}(t)+\delta_\ell
v_{0\ell}(t)\big)\,.\label{alphabeta}
\end{equation}
The normalization that we are choosing (\ref{norm}) yields the
following condition for the real coefficients $\alpha_\ell$,
$\beta_\ell$, $\gamma_\ell$, and $\delta_\ell$,
\begin{equation}
\alpha_\ell\delta_\ell-\beta_\ell\gamma_\ell=1\,,\quad
\forall\,\ell\in\mathbb{N}_0\,,\label{cond}
\end{equation}
i.e.,
$$
\left[
\begin{array}{cc}
\alpha_\ell&\beta_\ell\\
\gamma_\ell&\delta_\ell
\end{array}
\right]\in SL(2;\mathbb{R})\,,\quad
\forall\,\ell\in\mathbb{N}_{0}\,.
$$
It is well-known that $SL(2,\mathbb{R})$ is bijective (as a set) to
$\mathbb{S}^1\times\mathbb{R}^2$, in the sense that any element of
$SL(2,\mathbb{R})$ can be uniquely decomposed as the product of a
rotation and an upper triangular matrix with unit determinant,
\begin{equation}
SL(2,\mathbb{R})\ni\left[
\begin{array}{cc}
\alpha_\ell&\beta_\ell\\
\gamma_\ell&\delta_\ell
\end{array}
\right]= \left[
\begin{array}{cc}
\cos\theta_\ell&-\sin\theta_\ell\\
\sin\theta_\ell&\cos\theta_\ell
\end{array}
\right]\cdot\left[
\begin{array}{cc}
\rho_\ell&\nu_\ell\\
0&\rho_\ell^{-1}
\end{array}
\right] \label{polar}
\end{equation}
for a unique choice of $\rho_\ell>0$, $\nu_\ell\in\mathbb{R}$,
$\theta_\ell\in[0,2\pi)$. Different choices of the triplet
$(\rho_\ell,\nu_\ell,\theta_\ell)$ correspond, in principle, to
different complex structures on $\mathcal{S}$, defined through
(\ref{Jpdiagonal}) with
\begin{eqnarray*}
y_\ell(t)&=&\rho_\ell\cos\theta_\ell u_{0\ell}(t)+(\nu_\ell\cos\theta_\ell-\rho^{-1}_\ell\sin\theta_\ell)v_{0\ell}(t)\\
&+&i\Big(\rho_\ell\sin\theta_\ell
u_{0\ell}(t)+(\nu_\ell\sin\theta_\ell+\rho^{-1}_\ell\cos\theta_\ell)v_{0\ell}(t)\Big)\,.\nonumber
\end{eqnarray*}
However, this is not always the case. For instance, if we obtain
$y_\ell$ from $y_{0\ell}$ by the rotation appearing in the
decomposition (\ref{polar}),
$$
y_\ell=u_\ell+iv_\ell=\cos\theta_\ell u_{0\ell}-\sin\theta_\ell
v_{0\ell}+i(\sin\theta_\ell u_{0\ell}+\cos \theta_\ell v_{0\ell}
)=e^{i\theta_\ell} y_{0\ell}\,,
$$
the set $(\varphi_{\ell}=y_\ell Y_{\ell 0})_{\ell\in\mathbb{N}_0}$
defines a complex structure $J$ through $J\varphi_{\ell
}:=i\varphi_{\ell}$ and
$J\bar{\varphi}_{\ell}:=-i\bar{\varphi}_{\ell}$. Now, it is
straightforward to see that $J\varphi_{\ell}=i\varphi_{\ell}
\Leftrightarrow J
e^{i\theta_\ell}\varphi_{0\ell}=ie^{i\theta_\ell}\varphi_{0\ell }$,
and $\mathbb{C}$-linearity implies
$J\varphi_{0\ell}=i\varphi_{0\ell}$, i.e., $J=J_0$. Therefore, two
different choices of the form $(\rho_\ell,\nu_\ell,\theta_\ell)$ and
$(\rho_\ell,\nu_\ell,\tilde{\theta}_\ell)$, with
$\theta_\ell\neq\tilde{\theta}_{\ell}$, give rise to the same
complex structure. Then, in the following we will omit the angular
part of $(\rho_\ell,\nu_\ell,\theta_\ell)$ by choosing
$\theta_\ell=0$ in (\ref{polar}). The complex structures defined
trough $(\rho_\ell,\nu_\ell)$ and, hence, the corresponding
Lagrangian subspaces $\mathcal{P}$, will generally yield irreducible
\emph{unitarily nonequivalent} Fock representations. This a well
known property of any QFT in a generic curved spacetime, and can be
considered as a serious drawback to the formulation of the theory.
Obviously, this is not the case for a system of finite number of
degrees of freedom, where the Stone-von Neumann's theorem can be
applied $[4]$: For any Lagrangian subspace $\mathcal{P}$ one obtains
a quantum theory unitarily equivalent to the standard tensor product
construction. Also, for the case of a massless scalar field evolving
in a fixed \emph{stationary} spacetime, there exists a preferred
choice of Lagrangian subspace by virtue of the time translation
symmetry $[2]$. In our case, in absence of this symmetry (or any
extra constraint obtained after deparameterization, that would
generate residual symmetries useful to select a preferred
representation of the canonical commutation relations, as in the
3-torus case), no natural, preferred election of $\mathcal{P}$ is
available. In other words, due to the time-dependence of the
Hamiltonian, the solutions of $\mathcal{S}$ do not oscillate
harmonically and, thus, it is not possible to uniquely define
subspaces of positive and negative frequency solutions.

\subsubsection{$SO(3)$-invariant complex structures}{\label{Invariant
complex forms}}

\indent Our purpose now is to characterize those complex structures
on the real solution space $\mathcal{S}^{KG}$ of the field equation
$\mathring{g}^{ab}\mathring{\nabla}_a\mathring{\nabla}_b\varphi=0$,
invariant under the symmetries of $\mathbb{S}^2$ --the spatial
manifold in our $(1+2)$-dimensional description-- \textit{without
imposing} the symmetry condition $\mathcal{L}_\sigma\varphi=0$. As
we will show, once this is done it is straightforward to restrict
them to the solution space $\mathcal{S}$. In particular, we will
prove that all complex structures $J_\mathcal{P}$ as defined in
previous sections are $SO(3)$-invariant; similarly, any
$SO(3)$-invariant complex structure has an associated Lagrangian
subspace $\mathcal{P}$ characterized by definite pairs
$(\rho_\ell,\nu_\ell)$. With this aim in mind, let us consider the
complexified solution space
$\mathcal{S}^{KG}_{\mathbb{C}}=\mathcal{P}^{KG}_0\oplus\bar{\mathcal{P}}^{KG}_0$
where \begin{eqnarray*}
\mathcal{P}^{KG}_1&:=&\mathcal{P}^{KG}_0=\mathrm{Span}\{y_{0\ell}
Y_{\ell m}\,|\, \ell\in \mathbb{N}_0,\,
m\in\{-\ell,\dots,\ell\}\}\,,\\
\mathcal{P}^{KG}_2&:=&\bar{\mathcal{P}}^{KG}_0=\mathrm{Span}\{\bar{y}_{0\ell}
Y_{\ell m} \,|\, \ell\in \mathbb{N}_0,\,
m\in\{-\ell,\dots,\ell\}\}\,.
\end{eqnarray*}
Here, $Y_{\ell m}$ are the usual spherical harmonics on
$\mathbb{S}^2$. There are two antilinear maps connecting the spaces
$\mathcal{P}_{1}^{KG}$ and $\mathcal{P}_{2}^{KG}$ that we denote (in
a slight notational abuse) with the same symbol
$\bar{\phantom{z}}:\mathcal{P}_{1}^{KG}\rightarrow\mathcal{P}_{2}^{KG}:\psi_1
\mapsto \bar{\psi}_1$ and
$\bar{\phantom{z}}:\mathcal{P}_2\rightarrow
\mathcal{P}_{1}^{KG}:\psi_2\mapsto \bar{\psi}_2$. Each one of these
maps is the inverse of the other and their composition is the
identity for every element of $\mathcal{P}_{1}^{KG}$ or
$\mathcal{P}_{2}^{KG}$ (i.e., $\bar{\bar{\psi}}=\psi$). With their
help, we can write the conjugation
$\bar{\phantom{z}}:\mathcal{S}_{\mathbb{C}}\rightarrow\mathcal{S}_{\mathbb{C}}$
according to
$$
\Psi= \left[\begin{array}{c}\psi_1\\\psi_2\end{array}\right]\mapsto
\bar{\Psi}:=\left[\begin{array}{c}\bar{\psi}_2\\\bar{\psi}_1\end{array}\right],
$$
with $\psi_1\in{\mathcal{P}}_{1}^{KG}$ and
$\psi_2\in{\mathcal{P}}_{2}^{KG}$. The elements in the original
\emph{real} solution space $\mathcal{S}^{KG}$ can be easily
characterized by using the previous conjugation as those of the form
$$
\Phi=\left[\begin{array}{c}\varphi\\\bar{\varphi}\end{array}\right]
$$
or, alternatively, as the real linear subspace of
$\mathcal{S}_{\mathbb{C}}^{KG}$ given by
$\mathcal{S}^{KG}=\{\Phi\in\mathcal{S}_{\mathbb{C}}^{KG}\,|\,
\Phi=\bar{\Phi}\}$. The elements $\varphi_a\in \mathcal{P}^{KG}_a$,
$a=1,2$, are complex functions $\varphi_a(t,s)$ defined on
$(0,\pi)\times \mathbb{S}^2$. There is a natural representation
$D_a$  of $SO(3)$ in $\mathcal{P}^{KG}_a$ defined by
$(D_a(g)\varphi)(t,s)=\varphi(t,g^{-1}\cdot s)$, where $g^{-1}\cdot
s$ denotes the action of the rotation $g^{-1}\in SO(3)$ on the point
$s\in \mathbb{S}^2$.  Then, the natural representation of $SO(3)$ in
$\mathcal{S}^{KG}_{\mathbb{C}}=\mathcal{P}^{KG}_1\oplus\mathcal{P}^{KG}_2$
can be written in matrix form as
$$
D(g)=\left[\begin{array}{cc} D_1(g)&0\\0&D_2(g)
\end{array}\right],\quad g\in SO(3)\,,
$$
in terms of the representations $(D_a,\mathcal{P}^{KG}_a)$. Consider
now a $\mathbb{C}$-linear map
$J:\mathcal{S}_{\mathbb{C}}^{KG}\rightarrow\mathcal{S}_{\mathbb{C}}^{KG}$,
in matrix notation
$$
J=\left[\begin{array}{cc}J_{11}&J_{12}\\J_{21}&J_{22}\end{array}\right],
$$
where the maps $J_{ab}:\mathcal{P}^{KG}_b\rightarrow
\mathcal{P}^{KG}_a$ are $\mathbb{C}$-linear for $a,b\in\{1,2\}$. The
invariance of $J$ under the action of the $SO(3)$ group implies
$$
D(g)J=J D(g)\Leftrightarrow \left[\begin{array}{cc}
J_{11}D_1(g)&J_{12}D_2(g)\\J_{21}D_1(g)&J_{22}D_2(g)
\end{array}\right]=\left[\begin{array}{cc}
D_1(g)J_{11}&D_1(g)J_{12}\\D_2(g)J_{21}&D_2(g)J_{22}
\end{array}\right],\,\,\, \forall\, g\in SO(3)\,.
$$
We require that the restriction of $J$ to $\mathcal{S}^{KG}$ is
$\mathbb{R}$-linear, i.e., $J\Phi=\overline{J\Phi}$ for every
$\Phi\in\mathcal{S}^{KG}$, which implies
$J_{11}\varphi=\overline{J_{22}\bar{\varphi}}$ and
$J_{21}\varphi=\overline{J_{12}\bar{\varphi}}$, this is,
\begin{equation}
J_{22}=\bar{J}_{11}\,,\quad J_{12}=\bar{J}_{21}\,,\label{JbarJ}
\end{equation}
where we have used the notation
$\bar{A}\varphi:=\overline{A\bar{\varphi}}$ to denote the
$\mathbb{C}$-linear map $\bar{A}:P_b^{KG}\rightarrow P_a^{KG}$
($a\neq b$) obtained from the $\mathbb{C}$-linear map
$A:P_a^{KG}\rightarrow P_b^{KG}$. Finally, we impose that $J$ be a
complex structure, $J^2=-\mathrm{Id}_{\mathcal{S}_{\mathbb{C}}}$.
This requires
\begin{equation}\label{Jcompstruct}
J^2_{11}+\bar{J}_{21}J_{21}=-\mathrm{Id}_1\,,\quad
J_{21}J_{11}+\bar{J}_{11}J_{21}=0\,.
\end{equation}
It is convenient now to expand the vector spaces
$\mathcal{P}^{KG}_a$ as
\begin{eqnarray*}
\mathcal{P}^{KG}_a=\bigoplus_{\ell=0}^{\infty}
\mathcal{P}^\ell_a\,,\quad a=1,2\,,
\end{eqnarray*}
with
\begin{eqnarray*}
\mathcal{P}^\ell_1&:=&\mathrm{Span}\{y_{0\ell}\}\otimes\mathrm{Span}\{Y_{\ell
m}\,|\, m\in\{-\ell,\dots,\ell\}\}\,, \\
\mathcal{P}_2^\ell&:=&\mathrm{Span}\{\bar{y}_{0\ell}\}\otimes\mathrm{Span}\{Y_{\ell
m}\,|\, m\in\{-\ell,\dots,\ell\}\}\,.
\end{eqnarray*}
This is useful because the operators $D_a(g)$ can be written as
$D_a=\bigoplus_{\ell=0}^\infty D_a^\ell$, where each pair
$(\mathcal{P}_a^\ell,D_a^\ell)$ is an irreducible representation.
Denoting as $\Pi_a^\ell$ the projectors on the linear spaces
$\mathcal{P}_a^\ell$, we define
$$
J_{ab}^{\ell_1\ell_2}:=\Pi_a^{\ell_1}J_{ab}\Pi_b^{\ell_2}:\mathcal{P}_b^{\ell_2}\rightarrow
\mathcal{P}_a^{\ell_1}\,.
$$
In order to proceed with the characterization of the complex
structures, we establish the following lemma.
\begin{lema}[Schur]
Let $D_1(g)$ and $D_2(g)$ be two finite dimensional, irreducible
representations of the group $G$ in the complex finite-dimensional
linear spaces $V_1$ and $V_2$. Let us suppose that a linear operator
$L:V_1\rightarrow V_2$ \emph{commutes} with these representations,
i.e., $D_2(g)L=LD_1(g)$, $\forall\,g\in G$. Then, either $L$ is zero
or it is invertible. In the latter case, both representations are
equivalent and $L$ is uniquely determined modulo a multiplicative
constant.
\end{lema}
\noindent This lemma directly implies that
$J_{ab}^{\ell_1\ell_2}=0$, whenever $\ell_1\neq \ell_2$, and
$J_{aa}^{\ell\ell}=\jmath_{aa}^\ell I_{aa}^\ell$, where
$\jmath_{aa}^\ell\in\mathbb{C}$ and $I_{aa}^\ell$ denotes the
identity on $\mathcal{P}_a^\ell$, with
$\jmath_{22}^\ell=\bar{\jmath}^\ell_{11}$ as a consequence of
(\ref{JbarJ}). Also,
$$
J_{12}^{\ell\ell}(\bar{y}_{0\ell}\otimes v)=\jmath_{12}^\ell
y_{0\ell}\otimes v,\quad J_{21}^{\ell\ell}(y_{0\ell}\otimes
v)=\jmath_{21}^\ell \bar{y}_{0\ell}\otimes v,\quad
\jmath_{12}^\ell,\,\jmath_{21}^\ell\in\mathbb{C}\,,$$ with
$\jmath_{12}^\ell=\bar{\jmath}_{21}^\ell$ again as a consequence of
(\ref{JbarJ}). In conclusion, the general form of the mapping $J$ is
given by
$$
J=\bigoplus_{\ell=0}^{\infty}\left[
\begin{array}{cc}
\jmath_{11}^\ell I_{11}^\ell&\jmath_{12}^\ell I_{12}^\ell\\
\bar{\jmath}_{12}^\ell I_{21}^\ell&\bar{\jmath}_{11}^\ell
I_{22}^\ell
\end{array}
\right],
$$
where $I_{aa}^\ell$ denotes the identity operator in
$\mathcal{P}_a^\ell$ and the linear operators
$I_{ab}^\ell:\mathcal{P}_b^\ell\rightarrow\mathcal{P}_a^\ell$ act
according to $I_{12}^\ell(\bar{y}_{0\ell}\otimes v)=y_{0\ell}\otimes
v$ and $I_{21}^\ell(y_{0\ell}\otimes v)=\bar{y}_{0\ell}\otimes v$.
Conditions (\ref{Jcompstruct}) defining $J$ as a complex structure
finally yield the following restrictions on $\jmath_{11}^\ell$ and
$\jmath_{12}^\ell$,
\begin{equation}
|\jmath_{11 }^\ell|\,^2-|\jmath_{12}^\ell|\,^2=1,\quad \jmath_{11
}^\ell\in i \mathbb{R}\smallsetminus\{0\}\,,\quad
\jmath_{12}^\ell\in \mathbb{C}\,. \label{conds}
\end{equation}
The previous considerations apply to solutions of the Klein-Gordon
equation without imposing the additional axial symmetry. This can be
trivially taken into account at this point by realizing that it
suffices to restrict ourselves to the one-dimensional subspaces (for
each value of $\ell$) spanned by the spherical harmonics
$Y_{\ell0}$. Note that on each subspace
$\mathcal{P}^\ell_1\oplus\mathcal{P}^\ell_2$ the complex structure
is completely fixed by a pair of complex parameters $(\jmath_{11
}^\ell,\jmath_{12}^\ell)$ subject to the conditions (\ref{conds});
the remaining freedom is then parameterized by two real numbers.
This is what we found before by explicitly considering the solution
space and the choice of the families of functions $u_\ell$ and
$v_\ell$.\footnote{ Here, the choice $\jmath_{11}^\ell\in
i\mathbb{R}_+$ is equivalent to the normalization for the Wronskian
of $u_\ell$ and $v_\ell$ introduced in equation (\ref{norm}) and
guarantees that the sesquilinear form (\ref{escprod}) restricted to
the subspace corresponding to the $i$ eigenvalue of $J$ (that we
have denotes as $\mathcal{P}$ in previous sections) defines an inner
product. Changing the sign in the Wronskian corresponds to taking
$\jmath_{11}^\ell\in i\mathbb{R}_-$.} According to (\ref{conds}), it
suffices to take $\jmath_{11}^{\ell}=i$ and $\jmath_{12}^{\ell}=0$,
$\forall\,\ell\in\mathbb{N}_{0}$, to conclude that all complex
structures $J_{\mathcal{P}}$ naturally defined by these families of
functions are, in fact, $SO(3)$ invariant. Similarly, in accordance
with the previous arguments it is also clear that any
$SO(3)$-invariant complex structure, characterized by pairs
$(\jmath_{11}^{\ell},\jmath_{12}^{\ell})$ verifying (\ref{conds}),
has an associated Lagrangian subspace $\mathcal{P}$ defined by a set
$(y_{\ell}=\rho_{\ell}u_{0\ell}+(\nu_\ell+i\rho_{\ell}^{-1})v_{0\ell})_{\ell\in\mathbb{N}_0}$.
The formulas that relate the parameters $\rho_\ell$ and $\nu_\ell$
to the definition of the invariant complex structure discussed in
this section are calculated as follows. Once a fiducial basis
$\varphi_{0\ell}=y_{0\ell}Y_{\ell0}$ is chosen (\ref{u0v0}), any
other complex structure defined by a different basis --satisfying
the normalization condition (\ref{norm})-- can be written in terms
of $\varphi_{0\ell}$, by using (\ref{alphabeta}) and (\ref{polar}),
as
\begin{eqnarray}
J\left[\begin{array}{c}\varphi_{0\ell}\\\bar{\varphi}_{0\ell}\end{array}\right]=\left[\begin{array}{cc} \jmath_{11}^\ell I_{11}^\ell & \jmath_{12}^\ell \, I_{12}^\ell\\
 \bar{\jmath}_{12}^\ell \, I_{21}^\ell& \bar{\jmath}_{11}^\ell\,
 I_{22}^\ell\end{array}\right]\cdot\left[\begin{array}{c}\varphi_{0\ell}\\\bar{\varphi}_{0\ell}\end{array}\right],
\label{JJ0}
\end{eqnarray}
where
\begin{eqnarray}
\hspace*{-0.5cm}\jmath_{11}^\ell&=&\frac{i}{2}(\alpha^2_\ell+\beta^2_\ell+\gamma^2_\ell+\delta^2_\ell)
=\frac{i}{2}(\nu_\ell^2+\rho^{-2}_\ell+\rho^2_\ell)\,,\label{j1}\\
\hspace*{-0.5cm}\jmath_{12}^\ell&=&
-(\alpha_\ell\beta_\ell+\gamma_\ell\delta_\ell)+\frac{i}{2}(\beta^2_\ell+\delta^2_\ell-\alpha^2_\ell-\gamma^2_\ell)
=-\rho_\ell\nu_\ell+\frac{i}{2}(\nu_\ell^2+\rho^{-2}_\ell-\rho^2_\ell)\,.
\label{j2}
\end{eqnarray}
Note that, as expected, the complex structures defined by (\ref{j1})
and (\ref{j2}) do not depend on the parameters
$\theta_\ell\in[0,2\pi)$ appearing in (\ref{polar}) but only on the
pairs $(\rho_\ell,\nu_\ell)\in (0,+\infty)\times \mathbb{R}$.

\subsection{Canonical commutation relations}

\indent The canonical field operators  associated with a given time
$t\in(0,\pi)$ are defined as operator-valuated distributions on
$\mathbb{S}^{2}$,
\begin{eqnarray}
\hat{Q}(t,s)&=&\sum_{\ell=0}^{\infty}\left(y_{\ell}(t)\hat{a}_{\ell}+\bar{y}_{\ell}(t)\hat{a}_{\ell}^{\dag}\right)Y_{\ell0}(s)\,,\label{opQ}\\
\hat{P}(t,s)&=&\sin t
\sum_{\ell=0}^{\infty}\left(\dot{y}_{\ell}(t)\hat{a}_{\ell}+\dot{\bar{y}}_{\ell}(t)\hat{a}_{\ell}^{\dag}\right)Y_{\ell0}(s)\,.\label{opP}
\end{eqnarray}
In practice, these expressions can be obtained by formally promoting
the Fourier coefficients in (\ref{expansionS3}) to the creation and
annihilation operators --$\hat{a}_{\ell}^{\dag}$ and
$\hat{a}_{\ell}$, respectively-- associated with the basic vectors
$\varphi_{\ell}\in\mathscr{H}_\mathcal{P}$. The infinite sums
appearing in (\ref{opQ}) and (\ref{opP}) do not converge, and the
theory actually does not admit observables co\-rresponding to the
values of the field and its momentum at a given spacetime point
$(t,s)$. Rather, they must be interpreted in a distributional sense.
Only the spacetime average of these expressions weighted by smooth
functions on the 2-sphere are mathematically well-defined.
Nevertheless, the above expressions can be used to perform formal
calculations provided that only linear operations of them be
involved. Given any pair of smooth axially-symmetric real-valued
functions on the 2-sphere, $g_1,g_2\in
C^{\infty}(\mathbb{S}^{2};\mathbb{R})$ with $\mathcal{L}_\sigma
g_k=0$, $k=1,2$, the above distributions define canonical field
operators $(\hat{Q}_{t}[g_1],\hat{P}_{t}[g_2])$ just by multiplying
the formal expressions (\ref{opQ}) and (\ref{opP}) by $g_1$ and
$g_2$, respectively, and integrating then over the round 2-sphere
$\mathbb{S}^{2}$. In this way, we obtain
\begin{equation}\label{Q&PFock}
\hat{Q}_{t}[g_1]=\sum_{\ell=0}^{\infty}g_{\ell}^{(1)}\left(y_{\ell}(t)\hat{a}_{\ell}+\bar{y}_{\ell}(t)\hat{a}_{\ell}^{\dag}\right)\,,\quad
\hat{P}_{t}[g_2]=\sin t
\sum_{\ell=0}^{\infty}g_{\ell}^{(2)}\left(\dot{y}_{\ell}(t)\hat{a}_{\ell}+\dot{\bar{y}}_{\ell}(t)\hat{a}_{\ell}^{\dag}\right)\,,
\end{equation}
where $g_{k}=\sum_{\ell=0}^{\infty}g_{\ell}^{(k)}Y_{\ell0}$,
$g_{\ell}^{(k)}\in\mathbb{R}$, $k=1,2$. It is straightforward to
check that, by construction, the states with a finite number of
particles define a common, invariant, dense domain of analytic
vectors for these configuration and momentum operators, so that
their essential self-adjointness is guaranteed and, hence, the
existence of unique self-adjoint extensions for these operators (see
Nelson's analytic vector theorem in $[5]$). Consider first the
$\hat{Q}_{t}[g_1]$ operator. If
$\psi^{(n)}\in\mathscr{P}_{+}^{(n)}(\mathscr{H}_{\mathcal{P}}^{\otimes
n})$, then $\psi^{(n)}\in\mathscr{D}_{\hat{Q}_{t}[g_1]^{m}}$, for
all $m$, and
$\hat{Q}_{t}[g_1]\psi^{(n)}\in\mathscr{P}_{+}^{(n-1)}(\mathscr{H}_{\mathcal{P}}^{\otimes(n-1)})
\oplus\mathscr{P}_{+}^{(n+1)}(\mathscr{H}_{\mathcal{P}}^{\otimes(n+1)})$.
Denote
$\hat{a}_{t}(g_1):=\sqrt{2}\sum_{\ell=0}^{\infty}g_{\ell}^{(1)}y_{\ell}(t)\hat{a}_{\ell}$,
so that
$\hat{Q}_{t}[g_1]=\big(\hat{a}_{t}(g_1)+\hat{a}_{t}^{\dag}(g_1)\big)/\sqrt{2}$.
Relations (\ref{CfAf}) of \emph{appendix \ref{AppendixFock}} imply
$\|\hat{Q}_{t}[g_1]^{m}\psi^{(n)}\|\le2^{m/2}(n+m)^{1/2}(n+m-1)^{1/2}\cdots(n+1)^{1/2}\|g_1\|^{m}\|\psi^{(n)}\|\,,$
where we have denoted
$\|g_1\|^{2}:=\sum_{\ell=0}^{\infty}|g_{\ell}^{(1)}y_{\ell}(t)|^{2}$
which converges given the square-summability of
$(g_{\ell}^{(1)})_{\ell\in\mathbb{N}_0}$ (recall that $g$ is smooth)
and the asymptotic behavior of $y_{\ell}(t)=O(\ell^{-1/2})$ (see
next section). Therefore,
$$\sum_{m=0}^{\infty}\frac{|z|^{m}}{m!}\|\hat{Q}_{t}[g_1]^{m}\psi^{(n)}\|
\le\sum_{m=0}^{\infty}\frac{(\sqrt{2}|z|)^{m}}{m!}\left(\frac{(n+m)!}{n!}\right)^{1/2}\|g_1\|^{m}\|\psi^{(n)}\|<+\infty\,,
$$
for all $z\in\mathbb{C}$. It directly follows that each vector
$\psi=(\psi^{(n)})_{n\ge0}$ in the dense domain of finite number of
particles is analytic for $\hat{Q}_{t}[g_1]$. We can proceed in the
same way to prove a similar result for the momentum operator
$\hat{P}_{t}[g_2]$. In the next section, we will elucidate if the
functional dependence in $t$ of $\hat{Q}(t,s)$ and $\hat{P}(t,s)$
can be obtained by the action of a unitary operator in the Fock
space $\mathscr{F}_{+}(\mathscr{H}_{\mathcal{P}})$.

\section{Unitarity of the quantum time evolution}{\label{UnitarityEvolution}}

We discuss in this section the unitarity of the quantum evolution
for the models under consideration. A particularly convenient
approach to this issue is given by the algebraic formulation of QFT,
since the notion of unitary implementability of linear symplectic
transformations on Hilbert spaces can be analyzed in a natural way
within this framework. As explained in \emph{appendix \ref{MSPT}},
the basic ingredients in the algebraic approach are (i) a unital
$C^{*}$-algebra $\mathfrak{A}$, with observables\footnote{Some
relevant physical observables, such as the stress-energy tensor of
the quantum field, will not be represented as elements of
$\mathfrak{A}$. In this sense, the $\mathfrak{A}$ algebra will
encompass a \emph{minimal} collection of physical observables
sufficiently large to enable the theory to be formulated.} defining
the subset of $*$-invariant elements, and (ii) positive normalized
linear functionals (states)
$\varpi:\mathfrak{A}\rightarrow\mathbb{C}$ with
$\varpi(\mathbf{1})=1$ and $\varpi(A^{*}A)\ge0$ for all
$A\in\mathfrak{A}$. The value of a state $\varpi$ on an observable
$A\in\mathfrak{A}$, $A=A^{*}$, is interpreted as the expectation
value of that observable in the physical state of the system
represented by $\varpi$, $\langle A\rangle_{\varpi}=\varpi(A)$.
\\
\linebreak \indent The construction of the appropriate
$C^{*}$-algebra for free (linear) fields is straightforward.
Consider the covariant phase space $\Gamma=(\mathcal{S},\Omega)$ of
smooth real classical solutions to the equation of motion given by
(\ref{solspace}) and (\ref{sympS3}); $\Gamma$ has the structure of
an infinite-dimensional symplectic vector space. As a consequence of
the linearity of $\mathcal{S}$, the set of elementary classical
observables $\mathcal{O}_{c}$ can be identified with the
$\mathbb{R}$-vector space generated by linear functionals on
$\Gamma$. Every vector $\varphi\in\Gamma$ has an associated
functional $F_{\varphi}:\Gamma\rightarrow\mathbb{R}$ such that, for
all $\psi\in\Gamma$,
$$F_{\varphi}(\psi):=\Omega(\varphi,\psi)=\sin t\int_{\mathbb{S}^{2}}|\gamma|^{1/2}\iota_{t}^{*}\Big(\psi
\dot{\varphi}-\varphi\dot{\psi}\Big)\,.$$ Therefore,
$\mathcal{O}_{c}=\mathrm{Span}\big\{\mathbb{I},F_{\varphi}\big\}_{\varphi\in\Gamma}$.
As expected $[6]$, this set satisfies the condition that any regular
function on $\Gamma$ can be obtained as a (suitable limit of) sum of
products of elements in $\mathcal{O}_{c}$, and also that it is
closed under Poisson brackets,
$\{F_{\varphi}(\cdot),F_{\psi}(\cdot)\}=F_{\varphi}(\psi)\mathbb{I}$.
The abstract quantum algebra of observables is then given by the
usual Weyl $C^{*}$-algebra $\mathscr{W}(\Gamma)$ on $\Gamma$
generated by the elements $W(\varphi)=\exp(iF_{\varphi}(\cdot))$,
$\varphi\in\Gamma$, satisfying for all
$\varphi_1,\varphi_2\in\Gamma$ the following relations
\begin{equation*}
\displaystyle W(\varphi_1)^{*}=W(-\varphi_1)\,,\quad
W(\varphi_1)W(\varphi_2)=\exp\left(-\frac{i}{2}\Omega(\varphi_1,\varphi_2)\right)W(\varphi_1+\varphi_2)\,,
\end{equation*}
which contain the information about the canonical commutation
relations. The GNS construction (see \emph{Theorem \ref{GNS}} in
\emph{appendix \ref{MSPT} }) establishes that, given any state
$\varpi_0$ on the algebra $\mathscr{W}(\Gamma)$, there exist a
Hilbert space $\mathfrak{H}_{\varpi_0}$, a representation
$\pi_{\varpi_0}:\mathscr{W}(\Gamma)\rightarrow\mathscr{B}(\mathfrak{H}_{\varpi_0})$
from the Weyl algebra to the collection of bounded linear operators
on $\mathfrak{H}_{\varpi_0}$, and a cyclic vector
$\Psi_{\varpi_0}\in\mathfrak{H}_{\varpi_0}$ such that
$\varpi_{0}(A)=\langle\Psi_{\varpi_0}\,|\,\pi_{\varpi_0}(A)\Psi_{\varpi_0}\rangle_{\mathfrak{H}_{\varpi_0}}$,
$\forall\,A\in\mathscr{W}(\Gamma)$. Moreover, the triplet
$(\mathfrak{H}_{\varpi_0},\pi_{\varpi_0},\Psi_{\varpi_0})$ with
these properties is unique up to unitary equivalence. The
cons\-truction of the Fock spaces
$\mathscr{F}_{+}(\mathscr{H}_{\mathcal{P}})$ depending on the
Lagrangian subspaces $\mathcal{P}$ discussed in previous sections
proves equivalent to the GNS construction for the (faithful) state
$\varpi_0:\mathscr{W}(\Gamma)\rightarrow\mathbb{C}$ defined as
\begin{equation}\label{omega0}
\varpi_0(W(\varphi)):=\exp\left(-\frac{1}{2}\mu_{J_\mathcal{P}}(\varphi,\varphi)\right),
\end{equation}
with $\mu_{J_\mathcal{P}}$ given by equation (\ref{mu}) and the
vacuum $\Psi_{\varpi_0}=|0\rangle_{\mathcal{P}}$ serving as the
cyclic vector. Indeed, the value of the state $\varpi_0$ acting on
the Weyl generators $W(\varphi)$ is interpreted as the expectation
value of the associated operator $\pi_{\varpi_0}(W(\varphi))$ on the
vacuum state $|0\rangle_{\mathcal{P}}$, i.e.,
$\varpi_{0}(W(\varphi))=\langle\Psi_{\varpi_0}\,|\,\pi_{\varpi_0}(W(\varphi))\Psi_{\varpi_0}\rangle$.
We have
$\pi_{\varpi_0}(W(\varphi))=\exp(i\hat{\Omega}(\varphi,\cdot))$,
where the fundamental observables are defined as
$\hat{\Omega}(\varphi,\cdot)=i\big(A(\kappa(\varphi))-C(\kappa(\varphi))\big)$.
Here, $C$ and $A$ are the creation and annihilation operators
associated with $\kappa(\varphi)$, with
$\kappa:\mathcal{S}_{\mathbb{C}}\rightarrow\mathcal{P}$ being the
$\mathbb{C}$-linear projector defined by the splitting
$\mathcal{S}_{\mathbb{C}}=\mathcal{P}\oplus\bar{\mathcal{P}}$. By
using the Baker-Campbell-Hausdorff (BCH) relation, and taking into
account that
$[A(\kappa(\varphi)),C(\kappa(\varphi))]=\langle\varphi|\varphi\rangle_{J_{\mathcal{P}}}=\mu_{J_{\mathcal{P}}}(\varphi,\varphi)$,
we finally get
\begin{eqnarray*}
\langle\Psi_{\varpi_0}\,|\,\pi_{\varpi_0}(W(\varphi))\Psi_{\varpi_0}\rangle\!\!&=&\!\!\langle\Psi_{\varpi_0}\,|\,\exp(C(\kappa(\varphi)))\exp(-A(\kappa(\varphi)))\Psi_{\varpi_0}\rangle\exp\Big(-\frac{1}{2}\mu_{J_{\mathcal{P}}}(\varphi,\varphi)\Big)\\
&=&\langle\Psi_{\varpi_0}\,|\,\Psi_{\varpi_0}\rangle\exp\Big(-\frac{1}{2}\mu_{J_{\mathcal{P}}}(\varphi,\varphi)\Big)\,.
\end{eqnarray*}
Since the vacuum state is normalized, we have
$\langle\Psi_{\varpi_0}|\Psi_{\varpi_0}\rangle=1$ and recover the
expression (\ref{omega0}). The state $\varpi_0$ is uniquely extended
to $\mathscr{W}(\Gamma)$ by linearity and continuity. It clearly
satisfies the positivity condition $\varpi_0(A^{*}A)\ge0$,
$A\in\mathscr{W}(\Gamma)$, for the basic elements $W(\varphi)$,
although it does not automatically verify this condition for
arbitrary complex linear combinations of these elements. A necessary
and sufficient condition for positivity is in fact given by relation
(\ref{muJp}).
\\
\linebreak \indent Every linear symplectic transformation
$\mathcal{T}\in\mathrm{SP}(\Gamma)$,
$\mathcal{T}^{*}\Omega(\varphi_1,\varphi_2)=\Omega(\mathcal{T}\varphi_1,\mathcal{T}\varphi_2)=\Omega(\varphi_1,\varphi_2)$,
defines a (unique) $*$-automorphism
$\alpha_{\mathcal{T}}\in\mathrm{Aut}(\mathscr{W}(\Gamma))$ such that
$\alpha_{\mathcal{T}}\cdot W(\varphi):=W(\mathcal{T}\varphi)$. Given
any point $ \varphi=\sum_{\ell=0}^\infty \left(\, a_\ell y_\ell
+\overline{ a_\ell y_\ell}\,\right)Y_{\ell0}\in \Gamma$, the action
of $\mathcal{T}$ can be written in the form
$$
\mathcal{T}\varphi=\sum_{\ell=0}^\infty \Big(
\mathfrak{a}_{\ell}(a,\bar{a})y_{\ell}+
\overline{\mathfrak{a}_{\ell}(a,\bar{a})y_{\ell}}\Big)Y_{\ell_{1}0}\,,
$$
with the complex coefficients
$$
\mathfrak{a}_{\ell}(a,\bar{a})=\sum_{\ell^\prime=0}^\infty
\Big(\alpha(\ell,\ell^\prime)
a_{\ell^\prime}+\beta(\ell,\ell^\prime)\bar{a}_{\ell^\prime}\Big)\,,\,\,\,\ell\in\mathbb{N}_{0}\,.
$$
Given a concrete Hilbert space representation
$(\mathfrak{H},\pi,\Psi)$ of the Weyl algebra $\mathscr{W}(\Gamma)$,
for instance the Fock-like one, a (continuous) linear symplectic
transformation $\mathcal{T}\in\mathrm{SP}(\Gamma)$ is said to be
\emph{unitarily implementable} on the Hilbert space $\mathfrak{H}$
if $\pi$ and $\pi\circ\alpha_{\mathcal{T}}$ (or, equivalently,
$\pi\circ\alpha_\mathcal{T}^{-1}$) are unitarily equivalent
representations, i.e., there exists a unitary operator
$\hat{U}_{\mathcal{T}}:\mathfrak{H}\rightarrow\mathfrak{H}$ such
that
$\hat{U}_{\mathcal{T}}^{-1}\,\pi(W(\varphi))\,\hat{U}_{\mathcal{T}}=\pi(\alpha_{\mathcal{T}}\cdot
W(\varphi))=\pi(W(\mathcal{T}\varphi))$, for all
$W(\varphi)\in\mathscr{W}(\Gamma)$. Concretely, for the Fock
representation space
$\mathfrak{H}=\mathscr{F}_{+}(\mathscr{H}_{\mathcal{P}})$, one has
\begin{equation}
\hat{U}_\mathcal{T}^{-1}\,\hat{a}_{\ell}\,\hat{U}_\mathcal{T}=
\sum_{\ell^\prime=0}^\infty \Big(\alpha(\ell,\ell^\prime)
\hat{a}_{\ell^\prime}+\beta(\ell,\ell^\prime)\hat{a}^\dag_{\ell^\prime}\Big)\,,\,\,\,\ell\in\mathbb{N}_{0}\,.
\end{equation}
It is well known $[7]$ that not every linear symplectic
transformation $\mathcal{T}$ defined on the infinite dimensional
symplectic linear space $\Gamma$ can be unitarily implemented in the
Fock space $\mathscr{F}_{+}(\mathscr{H}_{\mathcal{P}})$. This is so
because $\pi_{\varpi_0}$ and its transform
$\pi_{\varpi_{0}}\circ\alpha_{\mathcal{T}}$, whose action on basic
observables is given by
$$
\langle\Psi_{\varpi_{0}}\,|\,\pi_{\varpi_{0}}\circ\alpha_{\mathcal{T}}(W(\varphi))\Psi_{\varpi_{0}}\rangle=\exp\left(-\frac{1}{4}\Omega\big((\mathcal{T}^{-1}\circ
J_{\mathcal{P}}\circ\mathcal{T})\varphi,\varphi\big)\right),
$$
do not necessarily yield unitarily equivalent Fock space
representations. This will be the case if and only if
$J_{\mathcal{P}}-\mathcal{T}^{-1}\circ
J_{\mathcal{P}}\circ\mathcal{T}$ is a Hilbert-Schmidt operator on
the one-particle Hilbert space $\mathscr{H}_{\mathcal{P}}$ $[2,7]$.
This immediately translates into the condition
$$
\sum_{\ell=0}^\infty\sum_{\ell^\prime=0}^\infty|\beta(\ell,\ell^\prime)|^2<+\infty\,.
$$
In particular, we will focus on the symplectic transformations
$\mathcal{T}_{(t_1,t_0)}$ defined in (\ref{T(t0t1)}) representing
the time evolution from $\iota_{t_0}(\mathbb{S}^{2})$ to
$\iota_{t_1}(\mathbb{S}^{2})$. The corresponding condition for the
unitary implementability of this transformation can be written from
(\ref{T(t0t1)}) and (\ref{timeevolution}) as
\begin{equation}
\sum_{ \ell=0}^{\infty}|\beta_\ell(t_1,t_0\,|\,y_\ell)|^2<+\infty\,,
\label{(A)}
\end{equation}
for all $t_0,t_1\in(0,\pi)$, where
$$\beta_\ell(t_1,t_0\,|\,y_\ell):=i\big(\sin t_1
\bar{y}_\ell(t_0)\dot{\bar{y}}_\ell(t_1)-\sin t_0
\bar{y}_\ell(t_1)\dot{\bar{y}}_\ell(t_0)\big)\,.$$ Equivalently, we
have to check if
$$\sum_{\ell=0}^\infty\mathrm{Re}^{2}[\beta_{\ell}(t_1,t_0\,|\,y_{\ell})]<+\infty\,\,\,\,\mathrm{and}\,\,\,\,
\sum_{\ell=0}^\infty\mathrm{Im}^{2}[\beta_{\ell}(t_1,t_0\,|\,y_{\ell})]<+\infty\,.$$
If these conditions are indeed verified, there will exist a unitary
operator
$\hat{U}(t_1,t_0):\mathscr{F}_{+}(\mathscr{H}_{\mathcal{P}})\rightarrow\mathscr{F}_{+}(\mathscr{H}_{\mathcal{P}})$,
the so-called \emph{unitary evolution operator}, such that
$$
\hat{U}^{-1}(t_1,t_0)\hat{Q}(t_0,s)\hat{U}(t_1,t_0)=\hat{Q}(t_1,s)\,,\,\,\,\hat{U}^{-1}(t_1,t_0)\hat{P}(t_0,s)\hat{U}(t_1,t_0)=\hat{P}(t_1,s)\,,
$$
with the field and momentum operators defined as in equations
(\ref{opQ}) and (\ref{opP}). At this point, we have to study the
convergence of the previous series. Note, in particular, that their
square summability depends only on its ultraviolet behavior (the
zero mode corresponding to $\ell=0$ plays no role in this context).
Let us consider the real part of the $\beta_\ell$ coefficients. By
using the expression for $y_\ell$ in terms of the $\rho_{\ell}$ and
$\nu_\ell$ coefficients, $y_\ell=\rho_\ell
u_{0\ell}+(\nu_\ell+i\rho_\ell^{-1})v_{0\ell}$, it is possible to
identify the dependence of
$\mathrm{Re}[\beta_\ell(t_1,t_0\,|\,y_\ell)]$ on the choice of
complex structures. This is given by
\begin{eqnarray}
\mathrm{Re}[\beta_\ell(t_1,t_0\,|\,y_\ell)]&=&
A_\ell(t_1,t_0)+2\rho^{-1}_\ell\nu_\ell B_\ell(t_1,t_0)\,,
\end{eqnarray}
where
\begin{eqnarray*}
A_\ell(t_1,t_0)&:=& \sin
t_1\big(u_{0\ell}(t_0)\dot{v}_{0\ell}(t_1)+v_{0\ell}(t_0)\dot{u}_{0\ell}(t_1)\big)-\sin
t_0\big(u_{0\ell}(t_1)\dot{v}_{0\ell}(t_0)+\dot{u}_{0\ell}(t_0)v_{0\ell}(t_1)\big)\,,
\\
B_\ell(t_1,t_0)&:=&\sin t_1 v_{0\ell}(t_0)\dot{v}_{0\ell}(t_1)-\sin
t_0v_{0\ell}(t_1)\dot{v}_{0\ell}(t_0)\,.
\end{eqnarray*}
The explicit form of $A_\ell$ and $B_\ell$, derived in a
straightforward way from (\ref{u0v0}), is
\begin{eqnarray*}
A_\ell(t_1,t_0) &=&\frac{\ell+1}{2}\Big(
\mathscr{P}_{\ell+1}(\cos t_1)\mathscr{Q}_\ell(\cos t_0)-\mathscr{P}_{\ell+1}(\cos t_0)\mathscr{Q}_\ell(\cos t_1)\\
&&\hspace{19mm}+\mathscr{P}_\ell(\cos t_1)\big(\cos t_0-\cos t_1)\mathscr{Q}_\ell(\cos t_0)-\mathscr{Q}_{\ell+1}(\cos t_0)\big)\nonumber\\
&&\hspace{19mm}+\mathscr{P}_\ell(\cos t_0)\big((\cos t_0-\cos
t_1)\mathscr{Q}_\ell(\cos t_1)-\mathscr{Q}_{\ell+1}(\cos
t_1)\big)\Big)\,,\nonumber
\\
B_\ell(t_1,t_0)
&=&-\frac{\ell+1}{2}\Big(\mathscr{Q}_\ell(\cos t_1)\mathscr{Q}_{\ell+1}(\cos t_0)\nonumber\\
& & \hspace{15mm}-\big((\cos t_0-\cos t_1)\mathscr{Q}_\ell(\cos
t_1)+\mathscr{Q}_{\ell+1}(\cos t_1)\big)\mathscr{Q}_\ell(\cos
t_0)\Big)\,.\nonumber
\end{eqnarray*}
By using the following asymptotic expansions for the first and
second class Legendre functions ($\varepsilon<t<\pi-\varepsilon$,
$\varepsilon>0$) $[8]$
\begin{eqnarray}
\mathscr{P}_\ell(\cos t) & = & \sqrt{\frac{2}{\pi\ell\sin
t}}\cos\big((\ell+1/2)t-\pi/4\big)-\sqrt{\frac{1}{8\pi\ell^{3}\sin t}}\Big(\cos\big((\ell+1/2)t-\pi/4\big)\nonumber\\
&-&\frac{1}{2}\cot t\sin\big((\ell+1/2)-\pi/4\big)\Big)+O\big(\ell^{-5/2}\big)\,,\nonumber\\
\mathscr{Q}_\ell(\cos t) & = & \sqrt{\frac{\pi}{2\ell\sin
t}}\cos\big((\ell+1/2)t+\pi/4\big)-\sqrt{\frac{\pi}{32\,\ell^{3}\sin t}}\Big(\cos\big((\ell+1/2)t+\pi/4\big)\nonumber\\
&-&\frac{1}{2}\cot
t\sin\big((\ell+1/2)+\pi/4\big)\Big)+O\big(\ell^{-5/2}\big)\,,\label{asintoticoQP}
\end{eqnarray}
we find that
\begin{eqnarray*}
\mathrm{Re}[\beta_\ell(t_1,t_0\,|\,y_\ell)]& \sim & -\frac{1}{2}\frac{\sin t_1-\sin t_0}{\sqrt{\sin t_0\sin t_1}}\sin[(\ell+1/2)(t_0+t_1)]\nonumber\\
&&-\frac{\pi \nu_\ell\rho_\ell^{-1}}{2\sqrt{\sin t_0 \sin
t_1}}\Big(\sin t_1
\cos[(\ell+1/2)t_0+\pi/4]\sin[(\ell+1/2)t_1+\pi/4]\nonumber\\
&&\hspace{2.4cm}-\sin t_0
\cos[(\ell+1/2)t_1+\pi/4]\sin[(\ell+1/2)t_0+\pi/4]\Big)\,,\nonumber
\end{eqnarray*}
as $\ell\rightarrow+\infty$. The asymptotic behavior of
$\mathrm{Re}[\beta_\ell(t_1,t_0\,|\,y_\ell)]$ leads us to conclude
that, irrespective of the choice of $(\rho_\ell,\nu_\ell)$, it is
not square summable and, hence, the quantum time evolution cannot be
unitarily implemented for any choice of $SO(3)$-invariant complex
structure.

\subsection{Conformal field redefinitions}

\indent The negative conclusion concerning the impossibility of
unitarily implementing the time evolution can be overcome much in
the same way as in the three-torus 3-torus case, i.e., by
introducing a redefinition of the fields in terms of which the model
is formulated $[9]$. In our approach, this redefinition is
appropriately interpreted from a geometrical point of view, being
suggested by the functional form of the conformal factor $\sin t$
which appears in the auxiliary metric $\mathring{g}_{ab}$
(\ref{g0}). In the following, we will reintroduce the index $i$ that
labels the gravitational scalar ($i=0$) and the matter scalars
($i\neq0$) and consider the new fields
\begin{equation}\label{xi}
\xi_i:=\sqrt{\sin t}\varphi_i\,.
\end{equation}
The equations of motion are now
\begin{equation}
-\ddot{\xi}_i+\Delta_{\mathbb{S}^2}\xi_i=\frac{1}{4}(1+\csc^2t)\xi_i\,,\quad
\mathcal{L}_\sigma\xi_i=0\,. \label{(B)}
\end{equation}
These can be interpreted as the equations for scalar, axially
symmetric fields with time-dependent mass term
$\frac{1}{4}(1+\csc^2t)$, evolving in $(0,\pi)\times \mathbb{S}^2$
with the \textit{regular} --i.e. extensible to $\mathbb{R}\times
\mathbb{S}^2$-- fixed \emph{static} background metric
\begin{equation}\label{eta_ab}
\mathring{\eta}_{ab}=-(\mathrm{d}t)_a(\mathrm{d}t)_b+\gamma_{ab}\,.
\end{equation}
Note that the singular behavior introduced by the conformal factor
$\sin t$ in (\ref{g0}) is translated, in terms of the redefined
fields, into the behavior of the time-dependent potential term,
which is singular at $t=0$ and $t=\pi$. In spite of being singular
at these instants of time, we expect to attain unitary dynamics if
this potential is well behaved enough. In particular, it has the
\textit{correct} positive sign for all $t\in(0,\pi)$. Otherwise, the
modes would satisfy harmonic oscillator equations with a negative
time dependent square frequency; this would introduce a
non-oscillatory behavior of the modes that, at the end of the day,
may become again responsible for the failure of the unitarity
condition. The field redefinition (\ref{xi}) can be incorporated in
the model at the Lagrangian level by substituting
$\phi_i=\xi_i/\sqrt{\sin t}$ in the action (\ref{accion}) to get the
corresponding variational problem in terms of the new fields
\begin{eqnarray}
s(\xi_i)&=&-\frac{1}{2}\sum_{i}\int_{[t_0,t_1]\times \mathbb{S}^2} |\mathring{\eta}|^{1/2} \mathring{\eta}^{ab} \Big((\mathrm{d}\xi_i)_a(\mathrm{d}\xi_i)_b-(\mathrm{d}\log \sin t)_a(\mathrm{d}\xi_i)_b\xi_i\label{action_xi}\\
& & \hspace{5.5cm}+\frac{1}{4}(\mathrm{d}\log\sin
t)_a(\mathrm{d}\log\sin t)_b \xi_i^2\Big)\,.\nonumber
\end{eqnarray}
Next, we will follow the method used in the preceding sections for
the original $\varphi$ fields. Some details will be omitted owing to
their similarity with the previous derivations. The canonical phase
space for the $\xi$-field equations is given again by
$\Upsilon=(\mathbf{P},\omega)$, with $\mathbf{P}:=\{(Q,P)\in
C^\infty(\mathbb{S}^2;\mathbb{R})\times
C^\infty(\mathbb{S}^2;\mathbb{R})\,|\, \mathcal{L}_\sigma
Q=0=\mathcal{L}_\sigma P\}$ and $\omega$ given by (\ref{cpsomega}).
We define the space $\mathcal{S}_\xi$ of smooth and symmetric real
solutions to the Klein-Gordon equation (\ref{(B)}) and expand
$\xi\in\mathcal{S}_\xi$ as
\begin{equation}
\xi(t,s)=\sum_{\ell=0}^\infty\Big(b_\ell z_\ell(t)+\overline{b_\ell
z_\ell(t)}\Big)Y_{\ell0}(s)\,,\label{fourier_xi}
\end{equation}
where $z_\ell(t)$ are complex functions satisfying the differential
equations
\begin{equation}
\ddot{z}_\ell+\left(\frac{1}{4}(1+\csc^2t)+\ell(\ell+1)\right)
z_\ell=0\,. \label{(C)}
\end{equation}
The functions $z_\ell$ can be easily written in terms the functions
$y_\ell$ appearing in (\ref{diffeqS3}) and satisfying (\ref{norm}),
$$z_\ell(t)=\sqrt{\sin t} \,y_\ell(t)\,,$$
the Wronskian being now normalized as
\begin{eqnarray}
z_\ell\dot{\bar{z}}_\ell-\bar{z}_\ell\dot{z}_\ell=i\,.\label{norm_zs}
\end{eqnarray}
This allows us to write the symplectic structure in
$\mathcal{S}_\xi$, naturally derived from the varia\-tional
principle (\ref{action_xi}), as
\begin{eqnarray*}
\Omega_\xi(\xi_1,\xi_2)&=&\int_{\mathbb{S}^2}|\gamma|^{1/2}\imath^*_t
\Big(\xi_2\dot{\xi}_1- \xi_1\dot{\xi}_2\Big)=
i\sum_{\ell=0}^\infty(\bar{b}_{\ell}^{(1)}b_{\ell}^{(2)}-\bar{b}_{\ell}^{(2)}b_{\ell}^{(1)})\,,
\quad\xi_1,\xi_2\in\mathcal{S}_\xi\,.
\end{eqnarray*}

\noindent \textbf{Classical evolution}

\bigskip

\indent We now consider the classical functional time evolution
operator $\mathcal{T}_{(t_1,t_0)}:\Gamma_\xi\rightarrow\Gamma_\xi$
in the covariant phase space
$\Gamma_\xi=(\mathcal{S}_\xi,\Omega_\xi)$. As before, we will write
it in the form
\begin{eqnarray}
(\mathcal{T}_{(t_1,t_0)}\xi)(t,s):=\sum_{\ell=0}^\infty
\Big(\mathfrak{b}_\ell(t_1,t_0)
z_\ell(t)+\overline{\mathfrak{b}_\ell(t_1,t_0)
z_\ell(t)}\Big)Y_{\ell0}(s)\,. \label{eq1}
\end{eqnarray}
In this case, the map
$\mathcal{T}_{(t_1,t_0)}=\mathfrak{J}_{t_0}\circ
\mathfrak{J}_{t_1}^{-1}$ is constructed from
\begin{eqnarray}
\mathfrak{J}^{-1}_{t_1}:\Gamma_\xi\rightarrow \Upsilon\,,\quad
\xi\mapsto(Q,P)=\mathfrak{J}^{-1}_{t_1}(\xi)\,,
\end{eqnarray}
defined by
\begin{eqnarray}
Q(s)&:=&\xi(t_1,s)=\sum_{\ell=0}^\infty \Big( b_\ell z_\ell(t_1)+ \overline{b_\ell z_\ell(t_1)}\Big)Y_{\ell 0}(s)\label{Q_xi}\,,\\
P(s)&:=&\dot{\xi}(t_1,s)-\frac{1}{2}\cot t_1 \xi(t_1,s)\label{P_xi}\\
&=& \sum_{\ell=0}^\infty \Big(b_\ell
\Big(\dot{z}_\ell(t_1)-\frac{1}{2}\cot t_1
z_\ell(t_1)\Big)+\overline{b_\ell\Big(\dot{z}_\ell(t_1)-\frac{1}{2}\cot
t_1 z_\ell(t_1)\Big)}\Big)Y_{\ell0}(s)\,,\nonumber
\end{eqnarray}
and from
\begin{eqnarray}
\mathfrak{J}_{t_0}: \Upsilon\rightarrow \Gamma_\xi\,,\quad
(Q,P)\mapsto\xi=\mathfrak{J}_{t_0}(Q,P)\,,
\end{eqnarray}
defined, in terms of the Fourier coefficients $b_\ell$ of $\xi$
(\ref{fourier_xi}), by
$$
b_\ell(t_0)=-i\Big(\dot{\bar{z}}_\ell(t_0)-\frac{1}{2}\cot t_0
\bar{z}_\ell(t_0)\Big)\int_{\mathbb{S}^2}|\gamma|^{1/2}Y_{\ell0} Q
+i\bar{z}_\ell(t_0)\int_{\mathbb{S}^2}|\gamma|^{1/2}Y_{\ell0}P\,.
$$
From these expressions, we finally obtain
\begin{eqnarray}
\mathfrak{b}_\ell(t_1,t_0)&=&i\Big(\bar{z}_\ell(t_0)\Big(\dot{z}_\ell(t_1)-\frac{1}{2}\cot
t_1
z_\ell(t_1)\Big)-z_\ell(t_1)\Big(\dot{\bar{z}}_\ell(t_0)-\frac{1}{2}\cot
t_0\bar{z}_\ell(t_0)\Big)
\Big)\, b_\ell\phantom{espacio}\label{eq2}\\
&+&i\Big(\bar{z}_\ell(t_0)\Big(\dot{\bar{z}}_\ell(t_1)-\frac{1}{2}\cot
t_1
\bar{z}_\ell(t_1)\Big)-\bar{z}_\ell(t_1)\Big(\dot{\bar{z}}_\ell(t_0)-\frac{1}{2}\cot
t_0 \bar{z}_\ell(t_0)\Big)\Big)\, \bar{b}_\ell\,.\nonumber
\end{eqnarray}

\newpage

\noindent \textbf{Quantum evolution}

\bigskip

\indent The unitarity condition for the quantum evolution in the
corresponding Fock space quantization becomes
\begin{eqnarray}
\sum_{\ell=0}^\infty|\beta^\xi_\ell(t_1,t_0\,|\,z_\ell)|^2=\sum_{\ell=0}^\infty\Big(
\mathrm{Re}^2[\beta^\xi_\ell(t_1,t_0\,|\,z_\ell)]
+\mathrm{Im}^2[\beta^\xi_\ell(t_1,t_0\,|\,z_\ell)]\Big)<+\infty\,,
\label{(D)}
\end{eqnarray}
for all $t_0$,$t_1\in(0,\pi)$, where
\begin{eqnarray}
\beta^\xi_\ell(t_1,t_0\,|\,z_\ell):=i\bar{z}_\ell(t_0)
\Big(\dot{\bar{z}}_\ell(t_1)-\frac{1}{2}\cot
t_1\bar{z}_\ell(t_1)\Big)-i\bar{z}_\ell(t_1)
\Big(\dot{\bar{z}}_\ell(t_0)-\frac{1}{2}\cot
t_0\bar{z}_\ell(t_0)\Big)\,.\label{betaxi}
\end{eqnarray}
The general solution to equation (\ref{(C)}) under the normalization
condition (\ref{norm_zs}) can be written again in terms of
associated Legendre functions (\ref{u0v0}) in the form
\begin{eqnarray*}
z_\ell(t)&=&\rho_\ell \sqrt{\sin t}\, u_{0\ell}(t)+(\nu_\ell+i\rho^{-1}_\ell)\sqrt{\sin t}\, v_{0\ell}(t)\\
&=& \rho_\ell
\tilde{u}_{0\ell}(t)+(\nu_\ell+i\rho^{-1}_\ell)\tilde{v}_{0\ell}(t)\,,
\end{eqnarray*}
where, as above, $SO(3)$-invariant complex structures differ from
each other just in the pairs $(\rho_\ell,\nu_\ell)$, with
$\rho_\ell>0$ and $\nu_\ell\in \mathbb{R}$, and we have defined
\begin{equation}\label{tildeuyv}
\tilde{u}_{0\ell}:=\sqrt{\sin
t}\,u_{0\ell}\,\,\,\,\mathrm{and}\,\,\,\,\tilde{v}_{0\ell}:=\sqrt{\sin
t}\,v_{0\ell}\,.
\end{equation}
We have to discuss now the convergence condition expressed in
(\ref{(D)}). Let us first consider the real part
\begin{eqnarray*}
\mathrm{Re}[\beta^\xi_\ell(t_1,t_0\,|\,z_\ell)]&=&\tilde{A}_\ell(t_{1},t_{0})
+2\nu_{\ell}\rho_{\ell}^{-1}\tilde{B}_{\ell}(t_{1},t_{0})
\end{eqnarray*}
where
\begin{eqnarray*}
\tilde{A}_\ell(t_{1},t_{0})&:=&
\tilde{u}_{0\ell}(t_{0})\dot{\tilde{v}}_{0\ell}(t_{1})-\tilde{u}_{0\ell}(t_{1})\dot{\tilde{v}}_{0\ell}(t_{0})
+\dot{\tilde{u}}_{0\ell}(t_{1})\tilde{v}_{0\ell}(t_{0})-\dot{\tilde{u}}_{0\ell}(t_{0})\tilde{v}_{0\ell}(t_{1})\\
&-&\frac{1}{2}(\cot t_{1}-\cot t_{0})\left(\tilde{u}_{0\ell}(t_{1})\tilde{v}_{0\ell}(t_{0})+\tilde{u}_{0\ell}(t_{0})\tilde{v}_{0\ell}(t_{1})\right)\,,\\
\tilde{B}_{\ell}(t_{1},t_{0})&:=&
\tilde{v}_{0\ell}(t_{0})\dot{\tilde{v}}_{0\ell}(t_{1})-\tilde{v}_{0\ell}(t_{1})\dot{\tilde{v}}_{0\ell}(t_{0})
-\frac{1}{2}(\cot t_{1}-\cot
t_{0})\tilde{v}_{0\ell}(t_{0})\tilde{v}_{0\ell}(t_{1})\,.
\end{eqnarray*}
The asymptotic behaviors of $\tilde{A}_\ell$ and $\tilde{B}_\ell$ as
$\ell\rightarrow+\infty$ can be obtained from (\ref{u0v0}) and
(\ref{asintoticoQP}),
\begin{eqnarray}
\tilde{A}_\ell(t_{1},t_{0})& \sim & -\frac{1}{4\ell}(\cot t_{1}-\cot
t_{0})\cos\big((\ell+1/2)(t_{0}+t_{1})\big)\,, \label{asint_A}\\
\tilde{B}_\ell(t_{1},t_{0})& \sim &
-\frac{\pi}{4}\sin\big((\ell+1/2)(t_{1}-t_{0})\big)\,.\label{asint_B}
\end{eqnarray}
We then conclude that
$\mathrm{Re}[\beta_\ell^\xi(t_1,t_0\,|\,z_\ell)]$ is square summable
if and only if
\begin{equation}\label{sqsmm1}
\big(\nu_\ell\rho_\ell^{-1}\big)_{\ell\in\mathbb{N}_0}\in\ell^{2}(\mathbb{R})\,.
\end{equation}
For the imaginary part of the
$\beta_{\ell}^{\xi}(t_1,t_0\,|\,z_\ell)$ coefficients, we have
\begin{eqnarray*}
\mathrm{Im}[\beta^\xi_\ell(t_1,t_0\,|\,z_\ell)]&=&\rho_{\ell}\nu_{\ell}
\tilde{A}_{\ell}(t_{1},t_{0})+(\nu_{\ell}^{2}-\rho_{\ell}^{-2})\tilde{B}_{\ell}(t_{1},t_{0})
+\rho_{\ell}^{2}\tilde{C}_{\ell}(t_{1},t_{0})\,,
\end{eqnarray*}
where
\begin{eqnarray}
\tilde{C}_{\ell}(t_{1},t_{0})&:=&
\tilde{u}_{0\ell}(t_{0})\dot{\tilde{u}}_{0\ell}(t_{1})-\tilde{u}_{0\ell}(t_{1})\dot{\tilde{u}}_{0\ell}(t_{0})
-\frac{1}{2}(\cot t_{1}-\cot t_{0})\tilde{u}_{0\ell}(t_{0})\tilde{u}_{0\ell}(t_{1})\nonumber\\
& \sim & -\frac{1}{\pi}\sin\big((\ell+1/2)(t_{1}-t_{0})\big),\,\,\,
\mathrm{when}\,\,\, \ell\rightarrow+\infty. \label{asint_C}
\end{eqnarray}
The asymptotic behavior as $\ell\rightarrow+\infty$ of
$\mathrm{Im}[\beta^\xi_\ell(t_1,t_0\,|\,z_\ell)]$ can be obtained
now from (\ref{asint_A}), (\ref{asint_B}), and (\ref{asint_C}). The
imaginary part is then square summable if and only if
\begin{equation}\label{sqsmm2}
\big(\rho_{\ell}\nu_{\ell}/\ell\big)_{\ell\in\mathbb{N}_0}\in\ell^{2}(\mathbb{R})
\,\,\,\,\mathrm{and}\,\,\,\,\big(\nu_{\ell}^{2}+4\rho_{\ell}^{2}/\pi^2-\rho_{\ell}^{-2}\big)_{\ell\in\mathbb{N}_0}\in\ell^{2}(\mathbb{R})\,.
\end{equation}
Taking into account that conditions (\ref{sqsmm1}) and
(\ref{sqsmm2}) must be satisfied jointly, we conclude that
$\beta_\ell^\xi(t_1,t_0\,|\,z_\ell)$ is square summable if and only
if
\begin{equation}
\rho_\ell=\sqrt{\frac{\pi}{2}}+x_\ell>0\,,\,\,\,(x_\ell)_{\ell\in\mathbb{N}_0}\in\ell^{2}(\mathbb{R})\,,
\,\,\,\,\mathrm{and}\,\,\,\,(\nu_{\ell})_{\ell\in\mathbb{N}_0}\in\ell^{2}(\mathbb{R})\,.
\label{condiciones}
\end{equation}
We end this section by showing that the linear symplectic map
$\mathcal{T}_{(t_1,t_0)}$ is continuous in the norm
$||\cdot||=\sqrt{\langle\cdot|\cdot\rangle|_{\mathcal{P}}}$
associated with the inner product (\ref{inprod}) for all complex
structures characterized by pairs $(\rho_\ell,\nu_\ell)$ verifying
(\ref{condiciones}). That is, there exists some $K(t_1,t_0)>0$ such
that
$$\|\kappa(\mathcal{T}_{(t_1,t_0)}\xi)\|\leq K(t_1,t_0)\|\kappa(\xi)\|\,,$$
for all $\xi\in \mathcal{S}_\xi$, where
$\kappa:\mathcal{S}_{\xi_{\mathbb{C}}}\rightarrow\mathcal{P}_{\xi}$
is the $\mathbb{C}$-linear projector defined by the splitting
$\mathcal{S}_{\xi_\mathbb{C}}=\mathcal{P}_\xi\oplus\bar{\mathcal{P}}_\xi$.
By using (\ref{eq1}) and (\ref{eq2}), it is straightforward to show
that
\begin{eqnarray}
\|\kappa(\mathcal{T}_{(t_1,t_0)}\xi)\|^2=\sum_{\ell=0}^\infty
|\mathfrak{b}_\ell(t_1,t_0)|^2\leq \sum_{\ell=0}^\infty
\Big(|\alpha_\ell(t_1,t_0\,|\,z_\ell)|^2+|\beta^\xi_\ell(t_1,t_0\,|\,z_\ell)|^2\Big)
|b_\ell|^2\,,\label{cont}
\end{eqnarray}
where
\begin{eqnarray*}
\alpha^\xi_\ell(t_1,t_0\,|\,z_\ell)&:=&i\bar{z}_\ell(t_0)
\Big(\dot{z}_\ell(t_1)-\frac{1}{2}\cot t_1 z_\ell(t_1)\Big)-
iz_\ell(t_1) \Big(\dot{\bar{z}}_\ell(t_0)-\frac{1}{2}\cot t_0
\bar{z}_\ell(t_0) \Big)\,,
\end{eqnarray*}
and the $\beta^\xi_\ell(t_1,t_0\,|\,z_\ell)$ coefficients are given
by (\ref{betaxi}). We have shown above that the sequence
$(|\beta^\xi_\ell(t_1,t_0\,|\,z_\ell)|)_{\ell\in\mathbb{N}_0}$ is
bounded (actually square summable) so in the case when the sequence
$(|\alpha^\xi_\ell(t_1,t_0\,|\,z_\ell)|)_{\ell\in\mathbb{N}_0}$ be
also bounded, the continuity of $\mathcal{T}_{(t_1,t_0)}$ follows
directly from equation (\ref{cont}). It suffices to remember that
$\alpha^\xi_\ell(t_1,t_0\,|\,z_\ell)$ and
$\beta^\xi_\ell(t_1,t_0\,|\,z_\ell)$ are Bogoliubov coefficients
satisfying
$|\alpha^\xi_\ell(t_1,t_0\,|\,z_\ell)|^{2}-|\beta^\xi_\ell(t_1,t_0\,|\,z_\ell)|^{2}=1$
for all $\ell$ to conclude that
$|\alpha^\xi_\ell(t_1,t_0\,|\,z_\ell)|\sim1$ as
$\ell\rightarrow+\infty$. Therefore, it is clear that there exists a
$K^2(t_1,t_0)>0$ such that
$|\alpha^\xi_\ell(t_1,t_0\,|\,z_\ell)|^2+|\beta^\xi_\ell(t_1,t_0\,|\,z_\ell)|^2\leq
K^2(t_1,t_0)\,,\,\,\, \forall\,\ell\in\mathbb{N}_{0}\,.$ Then, using
(\ref{cont}), we get that
$\|\kappa(\mathcal{T}_{(t_1,t_0)}\xi)\|^2\leq
K^2(t_1,t_0)\|\kappa(\xi)\|^2$ and, hence, $\mathcal{T}_{(t_1,t_0)}$
is continuous. In conclusion, by imposing suitable conditions
(\ref{condiciones}) on the parameters $\rho_\ell$ and $\nu_\ell$, it
is possible to find $SO(3)$-complex structures (equivalently,
subspaces $\mathcal{P}$) such that the quantum dynamics can be
unitarily implemented in $\mathcal{F}_+(\mathscr{H}_{\mathcal{P}})$.

\section{Uniqueness of the Fock quantization}{\label{uniqueness}}

\indent In the case of the $\mathbb{T}^3$ Gowdy models, the presence
of an extra constraint remaining after deparameterization, and the
corresponding $U(1)$ symmetry generated by it, gives the possibility
of introducing a physically sensible criterion to select a preferred
complex structure, namely, invariance under this symmetry $[10]$.
This is not the case for the other compact topologies
$\mathbb{S}^{1}\times\mathbb{S}^{2}$ and $\mathbb{S}^{3}$ for which,
as showed in \emph{Chapter \ref{ChapterI}}, there are no extra
constraints after deparameterization. In these cases, we will use
the $SO(3)$ symmetry associated with the background metric to select
a preferred class of complex structures. Once we require that the
quantum dynamics is unitary, we will find that all of them are
unitarily equivalent.\footnote{See also $[11]$ for an independent
proof of this result.}
\\
\indent To this end, let us recall some properties of the
$SO(3)$-invariant complex structures considered in section
\ref{Invariant complex forms}. Any invariant complex structure $J$
is related to the complex structure $J_{0}$ (defined for the $\xi$
field by the set of functions
$(z_{0\ell}(t)=\tilde{u}_{0\ell}(t)+i\tilde{v}_{0\ell})_{\ell\in\mathbb{N}_0}$
corresponding to $\rho_{\ell}=1$ and $\nu_{\ell}=0$) through a
linear symplectic transformation $T_{J}$, so that $J=T_{J}\circ
J_{0}\circ T_{J}^{-1}$. Explicitly, making use of relation
(\ref{JJ0}), a direct calculation shows
\begin{equation}
T_{J}=\bigoplus_{\ell=0}^{\infty}\left[\begin{array}{cc}
(\tau_{1}^{\ell})_{J}I_{11}^{\ell}&(\tau_{2}^{\ell})_{J}I_{12}^{\ell}\\
(\bar{\tau}_{2}^{\ell})_{J}I_{21}^{\ell}&(\bar{\tau}_{1}^{\ell})_{J}I_{22}^{\ell}\end{array}\right],
\end{equation}
with
\begin{eqnarray*}
(\tau_{1}^{\ell})_{J}&:=&\sqrt{(1+|\jmath^\ell_{11}|)/2}\quad
\textrm{(up to
multiplicative phase)}\,,\\
(\tau_{2}^{\ell})_{J}&:=&\frac{i\jmath^\ell_{12}}{2(\tau_{1}^{\ell})_J}\,.
\end{eqnarray*}
Note that $J_0$ does not lead to a unitary implementation of the
dynamics. In this context, it is fixed just to compare different
complex structures. Let us then consider any two $SO(3)$-invariant
complex structures, $J$ and $J'$, for which the dynamics is unitary
--they are thus characterized by pairs $(\rho_\ell,\nu_\ell)$
satisfying (\ref{condiciones}). They will define unitarily
equivalent quantum theories if and only if the linear symplectic
transformation $T_{J,J'}:=T_{J}\circ T_{J'}^{-1}$ connecting them
through $J=T_{J,J'}\circ J'\circ T_{J,J'}^{-1}$ is unitarily
implementable. This is the case if and only if the sequence
$$
\big((\tau_{2}^{\ell})_{J}(\tau_{1}^{\ell})_{J'}
-(\tau_{1}^{\ell})_{J}(\tau_{2}^{\ell})_{J'}\big)_{\ell\in\mathbb{N}_0}
$$
is square summable. Taking into account the relations (\ref{j1}) and
(\ref{j2}), as well as the asymptotic behaviors (\ref{condiciones}),
the previous condition is verified, so the quantum theories defined
by $J$ and $J'$ are, indeed, unitarily equivalent. The simplicity of
this result typifies the usefulness of the employed formalism.

\section{Self-adjointness and domain of quantum
Hamiltonians}\label{Self-adjointness}

\indent We analyze here an interesting feature of the quantum
dynamics for these systems: The fact that, even though the evolution
is unitarily implemented, the time-dependent quantum Hamiltonian,
proved to be self-adjoint for each value of the time parameter, has
the striking property that Fock space vectors corresponding to a
finite number of particle-like excitations do not belong to its
domain. We will then discuss the possibility of modifying the
expression of the Hamiltonian at the classical level in order to
avoid these problems regarding the domain of its quantum
counterpart.
\\
\linebreak \indent The classical Hamiltonian governing the dynamics
on the canonical phase space $\Upsilon=(\mathbf{P},\omega)$ in the
$\xi$-description of the system is derived from the action
(\ref{action_xi}). It is given by the time-dependent
\emph{indefinite}\footnote{Note the appearance of a cross term
involving $Q$ and $P$.} quadratic form
\begin{equation}
H(Q,P;t)=\frac{1}{2}\int_{\mathbb{S}^2}|\gamma|^{1/2}\left(P^2+\cot
t \,PQ-Q\Delta_{\mathbb{S}^2}Q\right). \label{Hamiltoniano}
\end{equation}
Note that, due to the nonautonomous nature of the classical
Hamiltonian, the time evolution does not define a one-parameter
symplectic group on $\Upsilon$ and we cannot apply Stone's theorem
to justify the self-adjointness of the corresponding (one-parameter
family of) operators in the quantum theory. Nevertheless, it is
possible to show that the quantum Hamiltonian is self-adjoint for
each value of the time parameter $t$ by analyzing the unitary
implementability on $\mathscr{F}_{+}(\mathscr{H}_\mathcal{P})$ of
the one-parameter symplectic group generated by the
\emph{autonomous} Hamiltonian $H(\tau)$, once a value
$t=\tau\in(0,\pi)$ has been fixed. Here, we will follow the
efficient procedure employed in $[12]$ for the Gowdy
$\mathbb{T}^{3}$ model, subsequently generalized in $[13]$ to
discuss the self-adjointness of general quadratic operators in this
context. We start by considering the auxiliary system
$(\mathbf{P},\omega,H(\tau))$, where the dynamics is governed by the
classical autonomous Hamiltonian
\begin{equation}
H(\tau)=\frac{1}{2}\sum_{\ell=0}^{\infty}\big(K_{\ell}(\tau)b_{\ell}^{2}+\bar{K}_{\ell}(\tau)\bar{b}_{\ell}^{2}+2G_{\ell}(\tau)\bar{b}_{\ell}b_{\ell}\big)\,,
\end{equation}
with
\begin{eqnarray}\label{K&G}
K_{\ell}(\tau)&:=&\Big(\dot{z}_{\ell}(\tau)-\frac{1}{2}\cot
\tau\,z_{\ell}(\tau)\Big)^{2}+\ell(\ell+1)z_{\ell}^{2}(\tau)+\cot
\tau\,\Big(\dot{z}_{\ell}(\tau)-\frac{1}{2}\cot
\tau\,z_{\ell}(\tau)\Big)z_{\ell}(\tau)\,,\nonumber\\
G_{\ell}(\tau)&:=&\Big|\dot{z}_{\ell}(\tau)-\frac{1}{2}\cot
\tau\,z_{\ell}(\tau)\Big|^{2}+\ell(\ell+1)|z_{\ell}(\tau)|^{2}\nonumber\\
&+&\frac{1}{2}\cot\tau\,\left(\Big(\dot{z}_{\ell}(\tau)-\frac{1}{2}\cot
\tau\,z_{\ell}(\tau)\Big)\bar{z}_{\ell}(\tau)+\Big(\dot{\bar{z}}_{\ell}(\tau)-\frac{1}{2}\cot
\tau\,\bar{z}_{\ell}(\tau)\Big)z_{\ell}(\tau)\right).
\end{eqnarray}
The modes $b_{\ell},\bar{b}_{\ell}$ are defined through the
relations
$Q_{\ell}:=\int_{\mathbb{S}^{2}}|\gamma|^{1/2}QY_{\ell0}=z_{\ell}(\tau)b_{\ell}+\bar{z}_{\ell}(\tau)\bar{b}_{\ell}$
and
$P_{\ell}:=\int_{\mathbb{S}^{2}}|\gamma|^{1/2}PY_{\ell0}=\big(\dot{z}_{\ell}(\tau)-(1/2)\cot\tau\,z_{\ell}(\tau)\big)b_{\ell}+\big(\dot{\bar{z}}_{\ell}(\tau)-(1/2)\cot\tau\,\bar{z}_{\ell}(\tau)\big)\bar{b}_{\ell}$.
Their evolution in a fictitious time parameter $s\in\mathbb{R}$ is
given by the linear equations\footnote{Here, $\{\cdot,\cdot\}$
denotes the Poisson bracket defined from (\ref{cpsomega}), with
$\{b_{\ell},\bar{b}_{\ell^{\prime}}\}=-i\delta(\ell,\ell^{\prime})\mathbb{I}$.}
\begin{eqnarray}
\frac{\mathrm{d}b_{\ell}}{\mathrm{d}s}&=&\{b_{\ell},H(\tau)\}=-i\big(G_{\ell}(\tau)b_{\ell}+\bar{K}_{\ell}(\tau)\bar{b}_{\ell}\big)\,,\\
\frac{\mathrm{d}\bar{b}_{\ell}}{\mathrm{d}s}&=&\{\bar{b}_{\ell},H(\tau)\}=i\big(K_{\ell}(\tau)b_{\ell}+G_{\ell}(\tau)\bar{b}_{\ell}\big)\,.\nonumber
\end{eqnarray}
Using the normalization condition (\ref{norm_zs}), we easily obtain
the second-order differential equation
\begin{equation}\label{dderiva}
\frac{\mathrm{d}^{2}b_{\ell}}{\mathrm{d}s^{2}}=-\Big(\ell(\ell+1)-\frac{1}{4}\cot^{2}\tau\Big)b_{\ell}\,,
\end{equation}
whose solutions have a linear dependence on the initial conditions
$b_{\ell}(s_0)$ and $\bar{b}_{\ell}(s_0)$,
\begin{eqnarray}\label{transforma}
b_{\ell}(s)=\alpha_{\ell}(s,s_0)b_{\ell}(s_0)+\beta_{\ell}(s,s_0)\bar{b}_{\ell}(s_0)\,,\,\,\,\bar{b}_{\ell}(s)=\overline{b_{\ell}(s)}\,.
\end{eqnarray}
This symplectic transformation is unitarily implementable on
$\mathscr{F}_{+}(\mathscr{H}_\mathcal{P})$ for each
$s\in\mathbb{R}$, i.e., there exists a unitary operator
$\hat{u}(s,s_0):\mathscr{F}_{+}(\mathscr{H}_\mathcal{P})\rightarrow\mathscr{F}_{+}(\mathscr{H}_\mathcal{P})$
such that
$\hat{u}^{-1}(s,s_0)\hat{b}_{\ell}\hat{u}(s,s_0)=\alpha_{\ell}(s,s_0)\hat{b}_{\ell}+\beta_{\ell}(s,s_0)\hat{b}_{\ell}^{\dag}$
and
$\hat{u}^{-1}(s,s_0)\hat{b}_{\ell}^{\dag}\hat{u}(s,s_0)=\bar{\beta}_{\ell}(s,s_0)\hat{b}_{\ell}+\bar{\alpha}_{\ell}(s,s_0)\hat{b}_{\ell}^{\dag}$,
if and only if the Bogoliubov coefficients $\beta_{\ell}$ are square
summable $[7]$,
\begin{equation}\label{betacoeff}
\sum_{\ell=0}^{\infty}|\beta_{\ell}(s,s_0)|^{2}<+\infty\,.
\end{equation}
Here, $\hat{b}^{\dag}_{\ell}$ and $\hat{b}_{\ell}$ are the creation
and annihilation operators associated with the modes
$\xi_\ell=z_{\ell}Y_{\ell0}$, respectively. Note that, for each
value of $\tau\in(0,\pi)$, there exists $\ell_{0}\in\mathbb{N}_{0}$
such that
\begin{equation*}
\lambda_{\ell}^{2}:=\ell(\ell+1)-\frac{1}{4}\cot^{2}\tau>0\,,\,\,\,\forall\,\ell>\ell_{0}\,.
\end{equation*}
In this case,
\begin{eqnarray*}
\alpha_{\ell}(s,s_0)&=&\cos\big(\lambda_{\ell}(s-s_0)\big)-i\lambda_{\ell}^{-1}G_{\ell}(\tau)\sin\big(\lambda_{\ell}(s-s_0)\big)\,,\\
\beta_{\ell}(s,s_0)&=&-i\lambda_{\ell}^{-1}\bar{K}_{\ell}(\tau)\sin\big(\lambda_{\ell}(s-s_0)\big)\,.
\end{eqnarray*}
It suffices to consider the modes corresponding to $\ell>\ell_{0}$,
since the convergence of the series (\ref{betacoeff}) depends, in
practice, only on the high-frequency behavior of the $\beta_{\ell}$
coefficients. Taking into account the asymptotic expansions in
$\ell$
\begin{eqnarray}\label{asymp}
&&z_{\ell}(t)=\frac{1}{\sqrt{2\ell}}\exp\left(-i[(\ell+1/2)t-\pi/4]\right)+O(\ell^{-3/2})\,,\\
&&\dot{z}_{\ell}(t)-\frac{1}{2}\cot t\,z_{\ell}(t)=
-i\sqrt{\frac{\ell}{2}}\exp\left(-i[(\ell+1/2)t-\pi/4]\right)+O(\ell^{-1/2})\,,\nonumber
\end{eqnarray}
we have $K_{\ell}(\tau)=O(1)$, so that
$\sum_{\ell>\ell_0}\lambda_{\ell}^{-2}|K_{\ell}(\tau)|^{2}\sin^{2}\big(\lambda_{\ell}(s-s_0)\big)<+\infty$,
$\forall\,s\in\mathbb{R}$ and, hence, condition (\ref{betacoeff}) is
verified. Finally, the transformation (\ref{transforma}) is
implementable as a continuous, unitary, one-parameter group if it
verifies the strong continuity condition in the auxiliary parameter
$s$
\begin{equation}
\lim_{s\rightarrow
s_{0}}\sum_{\ell=0}^{\infty}|b_{\ell}(s)-b_{\ell}(s_0)|^{2}=0\,,\,\,\,s_{0}\in\mathbb{R}\,.
\end{equation}
Again, we can restrict ourselves to the modes $\ell>\ell_0$. It is
straightforward to check that this condition holds for the solution
(\ref{transforma}) with square summable initial data $b_{\ell}$ and
$\bar{b}_{\ell}$. Therefore, we have obtained a strongly continuous
and unitary one-parameter group whose generator is self-adjoint
according to Stone's theorem.
\\
\linebreak \indent The quantum Hamiltonian of the models under
consideration can be explicitly calculated as the strong limit
\begin{equation}\label{s-lim}
\mathrm{s-}\!\!\lim_{t_0\rightarrow
t}\frac{\hat{U}(t,t_0)-\hat{\mathbb{I}}}{t-t_0}f=-i\hat{H}(t)f\,,\,\,\,
f\in\mathscr{D}_{\hat{H}(t)}\,,
\end{equation}
where $\hat{U}(t,t_0)$ denotes the quantum evolution operator on
$\mathscr{F}_{+}(\mathscr{H}_{\mathcal{P}})$, that can be univocally
derived from the evolution of creation operators in the Heisenberg
picture
\begin{eqnarray*}
&&\hat{U}^{-1}(t,t_0)\,\hat{b}_\ell\,\hat{U}(t,t_0)=\alpha_\ell^\xi(t,t_0\,|\,z_\ell)\hat{b}_\ell+\beta_\ell^\xi(t,t_0\,|\,z_\ell)\hat{b}_\ell^\dag\,,\\
&&\hat{U}^{-1}(t,t_0)\,\hat{b}_\ell^\dag\,\hat{U}(t,t_0)=\bar{\beta}_\ell^\xi(t,t_0\,|\,z_\ell)\hat{b}_\ell+\bar{\alpha}_\ell^\xi(t,t_0\,|\,z_\ell)\hat{b}_\ell^\dag\,,
\end{eqnarray*}
and the evolution of the vacuum state
$|0\rangle_{\mathcal{P}}:=1\oplus
0\oplus0\oplus\cdots\in\mathscr{F}_{+}(\mathscr{H}_{\mathcal{P}})$,
that can be written in closed form as (see also $[14-16]$)
$$
\hat{U}(t,t_0)|0\rangle_{\mathcal{P}}=N(t,t_0)\exp\left(-\frac{1}{2}\sum_{\ell=0}^{\infty}\frac{\beta_{\ell}^\xi(t_0,t\,|\,z_\ell)}{\alpha_{\ell}^\xi(t_0,t\,|\,z_\ell)}\,\hat{b}_{\ell}^{\dag2}\right)|0\rangle_{\mathcal{P}}\,,
$$
where $N(t,t_0)$ is fixed (up to an irrelevant multiplicative phase)
by normalization
\begin{equation*}
|N(t,t_0)|=\prod_{\ell\in\mathbb{N}_0}\frac{1}{\sqrt{\big|\alpha_\ell^\xi(t_0,t\,|\,z_\ell)\big|}}\,.
\end{equation*}
Note in particular that, as expected in a nonautonomous system, the
vacuum state (and, hence, states with a finite number of particles)
is not stable under time evolution. The previous result ensures the
self-adjointness of the quantum Hamiltonian $\hat{H}(t)$ and the
existence of a dense domain
$\mathscr{D}_{\hat{H}(t)}\subset\mathscr{F}_{+}(\mathscr{H}_{\mathcal{P}})$,
for each value of the time parameter $t\in(0,\pi)$. Unfortunately,
the method employed here does not provide us with a characterization
of such domains or the concrete expression of the quantum
Hamiltonian. This can be done in a mathematically rigorous way by
studying the differentiability (\ref{s-lim}) of $\hat{U}(t,t_0)$.
Nevertheless, given the quadratic nature of the classical
Hamiltonian, it is expected that this limit coincides with the
operator directly promoted from the classical function up to normal
ordering,
\begin{eqnarray}
\hat{H}(t)&=&\frac{1}{2}\sum_{\ell=0}^\infty \Big(
K_\ell(t)\,\hat{b}_\ell^2 +\bar{K}_\ell(t)\,\hat{b}_\ell^{\dag2}
+2G_\ell(t)\,\hat{b}_\ell^\dag \hat{b}_\ell\Big)\,. \label{qham}
\end{eqnarray}
However, this operator does not have the Fock vacuum state
$|0\rangle_{\mathcal{P}}$ in its domain due to the non-square
summability of the $K_{\ell}$ sequence defined in (\ref{K&G}).
Indeed, the action of the quantum Hamiltonian on the vacuum
$|0\rangle_{\mathcal{P}}$ is
$$
\hat{H}(t)|0\rangle_{\mathcal{P}}=\frac{1}{\sqrt{2}}\sum_{\ell=0}^\infty
\bar{K}_\ell(t)|2_\ell\rangle\,,
$$
where $|2_\ell\rangle=2^{-1/2}\hat{b}_\ell^{\dag
2}\,|0\rangle_{\mathcal{P}}$. The state
$\hat{H}(t)|0\rangle_{\mathcal{P}}$ is normalizable if and only if
$\sum_{\ell=0}^\infty|K_\ell(t)|^2\\<+\infty$, a condition which is
not verified since that $K_\ell(t)=O(1)$. As a consequence, the
action of the operator is not defined either on the dense subspace
of states with a finite number of particles. This difficulty can be
overcome right from the start by realizing that the covariant phase
space $\Gamma_\xi$ defined by (\ref{action_xi}) can be equivalently
derived from the simpler action
\begin{eqnarray}
s_0(\xi)&=&-\frac{1}{2}\int_{[t_0,t_1]\times \mathbb{S}^2}
|\mathring{\eta}|^{1/2} \mathring{\eta}^{ab}
\Big((\mathrm{d}\xi)_a(\mathrm{d}\xi)_b+\frac{1}{4}(1+\csc^2t)
\,\xi^2\Big)\,.\label{action_xi0}
\end{eqnarray}
This variational principle gives now a time-dependent,
\textit{positive definite}, diagonal Hamiltonian of the form
\begin{equation}\label{NewClassicHamilt}
H_{0}(Q,P;t):=\frac{1}{2}\int_{\mathbb{S}^{2}}|\gamma|^{1/2}\left(P^{2}+Q\Big[\frac{1}{4}(1+\csc^{2}t)-\Delta_{\mathbb{S}^{2}}\Big]Q\right).
\end{equation}
The Hamiltonians (\ref{Hamiltoniano}) and (\ref{NewClassicHamilt})
obviously govern the same classical evolution. Note, however, that
they are connected by a time-dependent symplectic transformation
that, in principle, is not unitarily implementable. As a
consequence, one possibly obtains nonequivalent quantum theories
from them. The corresponding quantum Hamiltonian is given, after
normal ordering, by
\begin{equation}\label{QuantHamiltonian}
\hat{H}_{0}(t)=\frac{1}{2}\sum_{\ell=0}^{\infty}\left(K_{0{\ell}}(t)\hat{b}_{\ell}^{2}+\bar{K}_{0\ell}(t){\hat{b}_{\ell}}^{\dag
2}+2G_{0\ell}(t)\hat{b}_{\ell}^{\dag}\hat{b}_{\ell}\right),
\end{equation}
where
\begin{eqnarray}\label{K0&G0}
K_{0\ell}(t)&:=&\dot{z}_{\ell}^{2}(t)+\left(\frac{1}{4}\big(1+\csc^{2}t\big)+\ell(\ell+1)\right)z_{\ell}^{2}(t)\,,\\
G_{0\ell}(t)&:=&\big|\dot{z}_{\ell}(t)\big|^{2}+\left(\frac{1}{4}\big(1+\csc^{2}t\big)+\ell(\ell+1)\right)|z_{\ell}(t)|^{2}\,.\nonumber
\end{eqnarray}
There are no subtleties associated with the domain of this new
quantum Hamiltonian in the sense that now the Fock space vacuum
belongs to the its domain --in this case, $K_{0\ell}(t)$ defines a
square summable sequence for each value of $t$. Moreover, the
previous results concerning the unitary implementation of the time
evolution and the uniqueness of the Fock representation are also
valid in this case. Concretely, the biparametric family of complex
structures for which the dynamics is unitary is characterized again
by the pairs (\ref{condiciones}). In what follows, we will consider
the dynamics of the system to be described by
(\ref{NewClassicHamilt}).
\\
\linebreak \linebreak \noindent \textbf{References}
\begin{itemize}
\item [1] C. Crnkovi\'c and E. Witten, in \emph{300 Years of Gravitation}, edited by
S. Hawking and W. Israel, Cambridge University Press, Cambridge
(1987). C. Crnkovi\'c, Class. Quantum. Grav. \textbf{5} (1988)
1557-1575. J. Lee and R. Wald, J. Math. Phys. \textbf{31}, 725
(1990). A. Ashtekar, L. Bombelli, and O. Reula, in \emph{Mechanics,
Analysis and Geometry : 200 Years After Lagrange}, edited by M.
Francaviglia, North-Holland, New York (1991). G. Barnich, M.
Henneaux, and C. Schomblond, Phys. Rev. D \textbf{44}, R939 (1991).
\item [2] R. M. Wald, \emph{Quantum field theory in curved spacetime and black hole thermodynamics}, The
University of Chicago Press, USA (1994).
\item [3] N. M. J. Woodhouse, \emph{Geometric quantization}, Oxford University Press, USA; 2nd edition
(1997).
\item [4] E. Prugove\v{c}ki. Quantum mechanics in Hilbert space. Academic Press, New York and London
(1971).
\item [5] M. Reed and B. Simon, \emph{Methods of Modern Mathematical Physics}, Vol. I. Academic Press,
New York, (1980).
\item [6] A. Ashtekar, Commun. Math. Phys. \textbf{71}, 59-64 (1980).
\item [7] D. Shale, Trans. Am. Math. Soc. \textbf{103}, 149 (1962).
\item [8] G. N. Watson, ``Asymptotic expansions of hypergeometric functions'', Trans. Cambridge Philos. Soc. \textbf{22} (1918) 277-308.
\item [9] A. Corichi, J. Cortez, and G. A. Mena Marug\'an, Phys. Rev. D \textbf{73}, 041502 and 084020 (2006).
\item [10] A. Corichi, J. Cortez, G. A. Mena Marug\'an, and J. M. Velhinho, Class. Quant. Grav. \textbf{23},
6301 (2006).
\item [11] J. Cortez, G. A. Mena Marug\'an, and J. M. Velhinho, Class. Quant. Grav. \textbf{25}, 105005 (2008).
\item [12] C. G. Torre, Phys. Rev. D \textbf{66}, 084017 (2002).
\item [13] J. F. Barbero G., D. G. Vergel, and E. J. S. Villase\~{n}or, Phys. Rev. D \textbf{74}, 024003 (2006).
\item [14] L. Parker, Phys. Rev. \textbf{183}, 5 (1969).
\item [15] K. Pilch and N. P. Warner, Class. Quant. Grav. \textbf{4}, 1183 (1987).
\item [16] C. G. Torre and M. Varadarajan, Phys. Rev. D \textbf{58}, 064007 (1998).
\end{itemize}

\newpage
\thispagestyle{plain} \mbox{}


\chapter{Schr\"{o}dinger Quantization}
\begin{flushright}
\small{\vspace*{-0.9cm} \textbf{D. G. Vergel}\\ \textbf{Classical and Quantum Gravity \textbf{25}, 175016 (2008)}\\\vspace*{0.2cm} \textbf{J. F. Barbero G., D. G. Vergel, and E. J. S. Villase\~{n}or}\\
\textbf{Physical Review D \textbf{74}, 024003
(2006)}}\label{ChapterIII}
\end{flushright}


\pagestyle{fancy} \fancyhf{}

\fancyhead[LO]{\textsc{Chapter 3. Schr\"{o}dinger Quantization}}
\fancyhead[RO,LE]{\thepage}
\renewcommand{\headrulewidth}{0.6pt}


\vspace*{0.5cm} \indent We will consider the Schr\"{o}dinger
representation for the linearly polarized Gowdy
$\mathbb{S}^{1}\times\mathbb{S}^{2}$ and $\mathbb{S}^{3}$ models
coupled to massless scalar fields, where states act as functionals
on the quantum configuration space $\overline{\mathscr{C}}$ for a
fixed time $t_0$. Here, $\overline{\mathscr{C}}$ is an appropriate
distributional extension of the classical configuration space
$\mathscr{C}$, characterized in these cases by the space of tempered
distributions on the 2-sphere. The Hilbert space then takes the form
$\mathscr{H}_{s}(t_0)=L^{2}(\overline{\mathscr{C}},\mathrm{d}\mu_{t_0})$.
The identification of the Gaussian nature of the measure
$\mu_{t_0}$, the nonstandard representation of the momentum
operator, and the relation between Schr\"{o}dinger and Fock
representations were exhaustively analyzed in $[1]$ as a natural
extension to the functional description of the Fock quantization of
scalar fields in curved backgrounds $[2]$. In the QFT context, the
Schr\"{o}dinger representation has been historically pushed into the
background in favor of the usual Fock quantization due to the
difficulty in using it to address sensible questions regarding
physical scattering processes. However, it is certainly the most
natural representation in the context of canonical quantum gravity,
in view of the splitting of spacetime into spatial sections of
constant time. Furthermore, as was pointed out in $[3]$ for the
vacuum three-torus case, it provides a better understanding of the
properties of the quantized field, since it is possible to determine
the behavior of the typical field confi\-gurations through the study
of the measure support.

\section{Constructing the $L^2$ space}\label{SCHROD}

\indent Let us denote by $\mathscr{S}$ the Schwartz space of smooth
and symmetric test functions on the 2-sphere,
\begin{equation*}
\mathscr{S}:=\{f\in
C^{\infty}(\mathbb{S}^{2};\mathbb{R})\,\,|\,\,\mathcal{L}_{\sigma}f=0\}\,,
\end{equation*}
endowed with the standard nuclear topology. Note that every element
$f\in\mathscr{S}$ can be expanded as
\begin{equation}\label{f(s)}
f(s)=\sum_{\ell=0}^{\infty}f_{\ell}Y_{\ell
0}(s)\,,\,\,\,s\in\mathbb{S}^{2}\,,
\end{equation}
with $(f_\ell)_{\ell\in\mathbb{N}_0}$ being a sequence of rapidly
decreasing real coefficients, such that
$\lim_{\ell\rightarrow+\infty}\\\ell^{n}f_{\ell}=
0,\,\forall\,n\in\mathbb{N}_{0}$. As $\mathscr{S}$ is central to our
future considerations, it is very useful to have as complete a
characterization as possible. Concretely, we will revise the
equivalent description of the topological structure of $\mathscr{S}$
in terms of the locally convex space of rapidly decreasing sequences
in section \ref{METRIC}.\footnote{For more details, the reader can
consult $[4]$} The quantum configuration space used to define the
Schr\"{o}dinger representation is then the topological dual
$\mathscr{S}^{\prime}$ space of continuous linear functionals
(tempered distributions) on $\mathscr{S}$. Note that this space
includes the delta functions and their derivatives. Given a time of
embedding $t_0$, the Schr\"{o}dinger representation is introduced by
defining a suitable Hilbert space
$L^{2}(\mathscr{S}^{\prime},\mathrm{d}\mu_{t_0})$, for a certain
measure $\mu_{t_0}$, in which the configuration observables act as
\emph{multiplication} operators. Here, the measure $\mu_{t_0}$ is
implicitly assumed to be defined on the $\sigma$-algebra
$\sigma(\mathrm{Cyl}(\mathscr{S}^{\prime}))$ generated by the
cylinder sets. As we will see later, given the Gaussian nature of
the measure $\mu_{t_0}$, the momentum operators will differ from the
usual ones in terms of derivatives by a multiplicative term
depending on the configuration variables,
\begin{equation*}
\hat{P}_{\ell}(t_0)\Psi=-i\partial_{q_\ell}\Psi+\mathrm{multiplicative}\,\,\mathrm{term}\,.
\end{equation*}
\indent We saw in \emph{Chapter \ref{ChapterII}} that the phase
space of the models under consideration can be alternatively
described by solutions to the equation of motion in the covariant
scheme or in terms of Cauchy data in the canonical formalism. In the
present approach, it is especially convenient to construct the Weyl
$C^{*}$-algebra of quantum observables from the canonical phase
space scheme. The arguments used here will be, in any case,
analogous to those employed in \emph{section
\ref{UnitarityEvolution}} within the covariant formalism. We start
by constructing the set $\mathcal{O}_{c}$ of elementary classical
observables of the theory. Again, the election is particularly
simple given the linearity of the space
$\mathbf{P}=\mathscr{S}\times\mathscr{S}$. In this case,
$\mathcal{O}_{c}$ can be identified with the $\mathbb{R}$-vector
space generated by linear functionals on $\mathbf{P}$. Every pair
$\lambda:=(-g,f)\in\mathbf{P}$, $f,g\in\mathscr{S}$, has an
associated functional $L_{\lambda}:\mathbf{P}\rightarrow\mathbb{R}$
such that, for all $X=(Q,P)\in\mathbf{P}$,
\begin{equation}
L_{\lambda}(X):=\omega(\lambda,X)=\int_{\mathbb{S}^{2}}|\gamma|^{1/2}(fQ+gP)\,,
\end{equation}
with the symplectic structure $\omega$ defined as in equation
(\ref{cpsomega}). In this way,
$\mathcal{O}_{c}=\mathrm{Span}\{\mathbb{I},L_{\lambda}\}_{\lambda\in\mathbf{P}}$.
Again, this set satisfies the condition that any regular function on
$\mathbf{P}$ can be obtained as a (suitable limit of) sum of
products of elements in $\mathcal{O}_{c}$, and also that it is
closed under Poisson brackets,
$\{L_{\lambda}(\cdot),L_{\nu}(\cdot)\}=L_{\nu}(\lambda)\mathbb{I}$.
Specifically, the configuration and momentum observables are objects
of this type defined by the pairs $\lambda=(0,f)$ and
$\lambda=(-g,0)$, respectively
\begin{eqnarray}
Q(f)&:=&L_{(0,f)}(Q,P)=\int_{\mathbb{S}^{2}}|\gamma|^{1/2}fQ=\sum_{\ell=0}^{\infty}f_{\ell}Q_{\ell}\,,\\
P(g)&:=&L_{(-g,0)}(Q,P)=\int_{\mathbb{S}^{2}}|\gamma|^{1/2}gP=\sum_{\ell=0}^{\infty}g_{\ell}P_{\ell}\,,
\end{eqnarray}
where the symmetric test functions have been expanded as explained
at the beginning of this section --see the equation (\ref{f(s)}).
Here, with the aim of simplifying the notation, we have used the
same symbol to denote the canonical inclusion
$\mathscr{S}\hookrightarrow\mathscr{S}^{\prime}$ of $\mathscr{S}$
into $\mathscr{S}^{\prime}$. In this way,
$L_{(-g,f)}(Q,P)=Q(f)+P(g)$. Given the canonical phase space
$\Upsilon=(\mathbf{P},\omega)$, the corresponding Weyl
$C^{*}$-algebra $\mathscr{W}(\Upsilon)$ is generated by the elements
$W(\lambda)=\exp(iL_{\lambda}(\cdot))$, $\lambda\in\mathbf{P}$,
satisfying the conditions
\begin{equation}
\displaystyle W(\lambda_1)^{*}=W(-\lambda_1)\,,\quad
W(\lambda_1)W(\lambda_2)=\exp\big(-\frac{i}{2}\omega(\lambda_1,\lambda_2)\big)W(\lambda_1+\lambda_2)\,,
\end{equation}
for all $\lambda_1,\lambda_2\in\mathbf{P}$. Since the generators of
this algebra and the one defined in \emph{section
\ref{UnitarityEvolution}} satisfy the same Weyl relations, there
exists a unique $*$-isomorphism connecting them.
\\
\linebreak \indent From now on, we will implicitly assume the use of
a concrete $SO(3)$-invariant complex structure $J_{\mathcal{P}}$
satisfying the conditions (\ref{condiciones}), so that the dynamics
is unitarily implemented on the Fock space
$\mathscr{F}_{+}(\mathscr{H}_{\mathcal{P}})$. Let
$\mathfrak{I}_{t_0}:\mathbf{P}\rightarrow\mathcal{S}_{\xi}$,
$t_0\in(0,\pi)$, be the symplectomorphism introduced in
\emph{section \ref{PhaseSpace}} which defines for each pair of
Cauchy data $(Q,P)\in\mathbf{P}$ the unique solution
$\xi\in\mathcal{S}_{\xi}$ such that $\xi(t_0,s)=Q(s)$ and
$\dot{\xi}(t_0,s)=P(s)$ under the evolution given by the Hamiltonian
(\ref{NewClassicHamilt}). This is,
\begin{equation}\label{I0QP}
\xi(t,s)=(\mathfrak{I}_{t_0}(Q,P))(t,s)=\sum_{\ell=0}^{\infty}\left(b_{\ell}(t_0)z_{\ell}(t)+\overline{b_{\ell}(t_0)z_{\ell}(t)}\right)Y_{\ell
0}(s)\in\mathcal{S}_{\xi}\,,
\end{equation}
with
\begin{equation}\label{alt0}
b_{\ell}(t_0):=i\bar{z}_{\ell}(t_0)P_{\ell}-i\dot{\bar{z}}_{\ell}(t_0)Q_{\ell}\,.
\end{equation}
This map gives rise to a natural $\omega$-compatible complex
structure on the canonical phase space,
\begin{equation*}
J_{t_0}:=\mathfrak{I}_{t_0}^{-1}\circ
J_{\mathcal{P}}\circ\mathfrak{I}_{t_0}:\mathbf{P}\rightarrow\mathbf{P}\,,
\end{equation*}
such that
\begin{equation}\label{J0}
(Q,P)\in\mathbf{P}\mapsto
J_{t_0}(Q,P)=(A(t_0)Q+B(t_0)P,D(t_0)Q+C(t_0)P)\in\mathbf{P}\,,
\end{equation}
where
$A(t_0),B(t_0),C(t_0),D(t_0):\mathscr{S}\rightarrow\mathscr{S}$ are
linear operators satisfying, by virtue of the
$\omega$-compatibility, the relations $[5]$
\begin{eqnarray*}
\langle f,B(t_0)f^{\prime}\rangle=\langle
B(t_0)f,f^{\prime}\rangle\,,\,\,\,\,\langle
g,D(t_0)g^{\prime}\rangle=\langle
D(t_0)g,g^{\prime}\rangle\,,\,\,\,\,\langle
f,A(t_0)g\rangle=-\langle C(t_0)f,g\rangle\,,
\end{eqnarray*}
for all $f,g,f^{\prime},g^{\prime}\in\mathscr{S}$. Here, we have
denoted
$$\langle f,g\rangle:=\int_{\mathbb{S}^{2}}|\gamma|^{1/2}fg\,.$$
Also, given the condition $J_{t_0}^{2}=-\mathrm{Id}_{\mathbf{P}}$,
we have
\begin{eqnarray*}
&& A^{2}(t_0)+B(t_0)D(t_0)=-\mathbf{1}\,,\,\,\,\,\,A(t_0)B(t_0)+B(t_0)C(t_0)=0\,,\\
&&
C^{2}(t_0)+D(t_0)B(t_0)=-\mathbf{1}\,,\,\,\,\,\,D(t_0)A(t_0)+C(t_0)D(t_0)=0\,.
\end{eqnarray*}
Thereby, assuming $B(t_0)$ invertible, the $C(t_0)$ and $D(t_0)$
operators can be expressed in terms of the $A(t_0)$ and $B(t_0)$
operators through the relations
$$C(t_0)=-B^{-1}(t_0)A(t_0)B(t_0)\,\,\,\,\mathrm{and}\,\,\,\,
D(t_0)=-B^{-1}(t_0)(\mathbf{1}+A^{2}(t_0))\,,$$ respectively, in
such a way that the complex structure $J_{t_0}$ is fully
characterized by them. Using the normalization condition
(\ref{norm_zs}) and equation (\ref{I0QP}) it is straightforward to
obtain\footnote{Note that the zero mode $\ell=0$ has been included
into the spherical harmonic expansion of the test functions. The
$B(t_0)$ operator is well defined even for this mode, ultimately as
a consequence of equation (\ref{(C)}) verified by the $z_{\ell}$
functions, where the squared frequency is positive definite
$\forall\,t\in(0,\pi)$ when $\ell=0$.}
\begin{eqnarray}\label{A&B}
\big(A(t_0)Q\big)(s)&=&\sum_{\ell=0}^{\infty}\big(\dot{\bar{z}}_{\ell}(t_0)z_{\ell}(t_0)+\dot{z}_{\ell}(t_0)\bar{z}_{\ell}(t_0)\big)Q_{\ell}Y_{\ell 0}(s)\,,\\
\big(B(t_0)P\big)(s)&=&-2\sum_{\ell=0}^{\infty}|z_{\ell}(t_0)|^{2}P_{\ell}Y_{\ell
0}(s)\,.\nonumber
\end{eqnarray}
It is worth noting that, given the rapidly decreasing nature of the
sequences $(Q_{\ell})_{\ell\in\mathbb{N}_0}$ and
$(P_{\ell})_{\ell\in\mathbb{N}_0}$, as well as the asymptotic
behavior of the $z_{\ell}$ functions decaying like (\ref{asymp}),
the $A(t_0)$ and $B(t_0)$ operators are well defined on
$\mathscr{S}$. In addition, $B(t_0)$ has an inverse operator
$B^{-1}(t_0):\mathscr{S}\rightarrow\mathscr{S}$ given by
\begin{equation}\label{invB}
\big(B^{-1}(t_0)P\big)(s)=-\frac{1}{2}\sum_{\ell=0}^{\infty}|z_{\ell}(t_0)|^{-2}P_{\ell}Y_{\ell
0}(s)\,.
\end{equation}
Summarizing, a fixed complex structure
$J_{\mathcal{P}}:\mathcal{S}_{\xi}\rightarrow\mathcal{S}_{\xi}$ on
the covariant phase space determines a one-parameter family of
complex structures $J_{t}:\mathbf{P}\rightarrow\mathbf{P}$,
$t\in(0,\pi)$, on the canonical phase space. Once a time of
embedding $t_0$ is fixed, the corresponding complex structure
$J_{t_0}$ is fully characterized by the pairs (\ref{A&B}). The
Schr\"{o}dinger quantization associated with time $t_0$ consists
then in a representation of the canonical commutation relations in
terms of self-adjoint operators in a space of complex-valued
functionals $\Psi:\mathscr{S}^{\prime}\rightarrow\mathbb{C}$
belonging to a certain Hilbert space
$\mathscr{H}_{s}(t_0)=L^{2}(\mathscr{S}^{\prime},\mathrm{d}\mu_{t_0})$.
The functionals representing the pure states of the system are,
thus, square integrable with respect to the measure $\mu_{t_0}$. Due
to the infinite dimensionality of the quantum configuration space
$\overline{\mathscr{C}}=\mathscr{S}^{\prime}$, it is not possible to
define a Lebesgue-type translation invariant measure $\mu_{t_0}$
--such a measure does not exists-- but rather a probability one,
i.e., a measure satisfying
$\int_{\mathscr{S}^{\prime}}\mathrm{d}\mu_{t_0}=1$. This
representation is constructed in such a way that it is associated
with the state
$\varpi_{t_0}:\mathscr{W}(\Upsilon)\rightarrow\mathbb{C}$ on the
Weyl algebra $\mathscr{W}(\Upsilon)$ whose action on the elementary
observables is given by
\begin{equation}\label{statesigma}
\varpi_{t_0}(W(\lambda))=\exp\left(-\frac{1}{4}\omega(J_{t_0}(\lambda),\lambda)\right),\,\,\,\lambda\in\mathbf{P}\,.
\end{equation}
We will check in section \ref{UNITARY} that the Schr\"{o}dinger
representations corresponding to different values of the time
parameter are unitarily equivalent owing to the unitary
implementability of the dynamics. We require that the configuration
observables are represented as \emph{multiplication} operators, so
that for $\lambda=(0,f)\in\mathbf{P}$,
\begin{equation}\label{Qoperator}
\displaystyle \pi_{s}(t_0)\cdot
W(\lambda)|_{\lambda=(0,f)}=\exp(i\hat{Q}_{t_0}[f])\,,\quad
\left(\hat{Q}_{t_0}[f]\Psi\right)[\tilde{Q}]=\tilde{Q}(f)\Psi[\tilde{Q}]\,,
\end{equation}
where $\tilde{Q}\in\mathscr{S}^{\prime}$ denotes a generic
distribution of $\mathscr{S}^{\prime}$ and $\tilde{Q}(f)$ gives the
usual pairing between $\mathscr{S}$ and $\mathscr{S}^{\prime}$,
$\Psi\in\mathscr{D}_{\hat{Q}_{t_0}[f]}\subset\mathscr{H}_{s}(t_{0})$
(the self-adjointness of the configuration and momentum operators
will be discussed in subsection \ref{CCRSchr}), and
$\pi_{s}(t_0):\mathscr{W}(\Upsilon)\rightarrow
\mathscr{B}(\mathscr{H}_{s}(t_0))$ is the map from the Weyl algebra
to the collection of bounded linear operators on the Hilbert space
$\mathscr{H}_{s}(t_0)$. The expectation value (\ref{statesigma})
evaluated at $\lambda=(0,f)$ yields
\begin{equation}\label{B1}
\varpi_{t_0}(W(\lambda))|_{\lambda=(0,f)}=\exp\left(\frac{1}{4}\langle
f,B(t_0)f\rangle\right).
\end{equation}
The left hand side of this equation corresponds to the vacuum
expectation value of the $\pi_{s}(t_0)\cdot W(\varphi)$ operator, so
it must coincide with the integral
\begin{equation}\label{B2}
\int_{\mathscr{S}^{\prime}}\bar{\Psi}^{(t_0)}_{0}\big(\exp(i\hat{Q}_{t_0}[f])\Psi^{(t_0)}_{0}\big)\,\mathrm{d}\mu_{t_0}[\tilde{Q}]=
\int_{\mathscr{S}^{\prime}}e^{i\tilde{Q}(f)}\mathrm{d}\mu_{t_0}[\tilde{Q}]\,,
\end{equation}
where $\Psi^{(t_0)}_{0}\in\mathscr{H}_{s}(t_0)$ is the normalized
vacuum state. Comparing (\ref{B1}) and (\ref{B2}), we finally get
\begin{equation}\label{generatfunct}
\int_{\mathscr{S}^{\prime}}e^{i\tilde{Q}(f)}\,\mathrm{d}\mu_{t_0}[\tilde{Q}]=\exp\left(\frac{1}{4}\langle
f,B(t_0)f\rangle\right).
\end{equation}
In order to interpret this result, let us introduce the covariance
operator $\check{\mathcal{C}}_{t_0}:\mathbf{P}\rightarrow\mathbb{R}$
defined as $\check{\mathcal{C}}_{t_0}(f,g):=\langle
f,-B(t_0)g/2\rangle$, $f,g\in\mathscr{S}$. Since
$|z_{\ell}(t_0)|^{2}$ is bounded and positive definite
$\forall\,t\in(0,\pi)$ and $\forall\,\ell\in\mathbb{N}_{0}$, it
follows that $\check{\mathcal{C}}_{t_0}$ is a nondegenerate,
positive, definite, and continuous bilinear form on the topological
vector space $\mathscr{S}$. Now, the Bochner-Minlos theorem, which
plays a key role in the characterization of measures on functional
spaces, states $[6]$:
\begin{theorem}[Bochner-Minlos]\label{B-MTheorem}
Let $\check{\mathcal{C}}$ be a positive continuous nondegenerate
bili\-near form on $\mathscr{S}\times\mathscr{S}$, written
symbolically $\check{\mathcal{C}}(f,g)=\langle
f,\mathcal{C}g\rangle$. Then, there exists a unique Gaussian
integration measure $\mathrm{d}\mu_{\mathcal{C}}$ on
$\mathscr{S}^{\prime}$ with covariance $\mathcal{C}$ and mean zero.
The corresponding generating function (or Fourier transform) is
given explicitly as
$\int_{\mathscr{S}^\prime}\exp(i\phi(f))\mathrm{d}\mu_{\mathcal{C}}[\phi]=\exp(-\langle
f,\mathcal{C}f\rangle/2)$.
\end{theorem}
\noindent Thus, according to this result, we have that $\mu_{t_0}$
is the unique Gaussian integration measure (with covariance
$\mathcal{C}(t_0):=-B(t_0)/2$) defined by the covariance operator
$\check{\mathcal{C}}_{t_0}$. The generating function is given in
this case by equation (\ref{generatfunct}).

\subsection{Properties of the measure}\label{METRIC}

\indent In order to easily visualize the nature of the measure
$\mu_{t_0}$, note that upon restriction on any number of coordinate
directions in $\mathscr{S}^{\prime}$, say
$\tilde{Q}_{\ell}=\tilde{Q}(Y_{\ell 0})$, $\ell=0,1,\ldots,n$, we
obtain
\begin{equation}\label{restrict}
\displaystyle
\mathrm{d}\mu_{t_0}|_{(\tilde{Q}_{\ell})_{\ell=0}^{n}}=\prod_{\ell=0}^{n}\,\frac{1}{\sqrt{2\pi}}\,|z_{\ell}(t_0)|^{-1}
\exp\left(-\frac{1}{2}|z_{\ell}(t_0)|^{-2}\tilde{Q}_{\ell}^{2}\right)\mathrm{d}\tilde{Q}_{\ell}\,,
\end{equation}
in terms of the Lebesgue measures $\mathrm{d}\tilde{Q}_{\ell}$
$[6]$. Now, we will prove that the support of the
$\mu_{t_0}$-measure is smaller than $\mathscr{S}^{\prime}$.
Concretely, it is given by the topological dual of the subspace of
symmetric functions in the Sobolev space
$H^{\epsilon}(\mathbb{S}^{2})$ on the 2-sphere, for any
$\epsilon>0$. With this aim, we will use some well-established
consequences of the Bochner-Minlos theorem, closely relying on the
analysis developed in $[7]$. We first point out that the space of
test functions $\mathscr{S}$ is topologically isomorphic to
$\varsigma=\bigcap_{r\in\mathbb{Q}}\varsigma_{r}$, where
\begin{equation*}
\varsigma_{r}:=\Big\{f=\{f_{\ell}\}_{\ell=0}^{\infty}\,\,\big|\,\,\|f\|_{r}^{2}:=\sum_{\ell=0}^{\infty}(\ell+1/2)^{2r}f_{\ell}^{2}<+\infty\Big\}\,,
\end{equation*}
endowed with the Fr\'{e}chet topology induced by the norms
$(\|\cdot\|_{r})_{r\in\mathbb{Q}}$. As a corollary to the
Bochner-Minlos theorem, one has that if the covariance
$\check{\mathcal{C}}_{t_0}$ is continuous in the norm associated
with some $\varsigma_{r}$, then the associated Gaussian measure
$\mu_{t_0}$ has support on any set of the form $[7]$
\begin{equation}\label{setr}
\Big\{f\,\,\big|\,\,\sum_{\ell=0}^{\infty}(\ell+1/2)^{-2r-1-2\epsilon}f_{\ell}^{2}<+\infty\,,\,\,\epsilon>0\Big\}\subset\displaystyle\bigcup_{r\in\mathbb{Q}}\varsigma_{r}=\varsigma^{\prime}\,,
\end{equation}
where $\varsigma^{\prime}$ is the topological dual of
$\varsigma$.\footnote{Here, $g\in\varsigma^{\prime}$ is associated
with the linear functional
$L_{g}(f):=\sum_{\ell=0}^{\infty}f_{\ell}g_{\ell}$,
$f\in\varsigma$.} In particular, given the asymptotic behavior of
the $z_{\ell}$ functions (\ref{asymp}), it is straightforward to
check the continuity in the norm corresponding to $r=-1/2$, i.e.,
\begin{equation*}
\langle f,\mathcal{C}(t_0)f\rangle\le
N(t_0)\sum_{\ell=0}^{\infty}(\ell+1/2)^{-1}f_{\ell}^{2}
\end{equation*}
for a certain constant $N(t_0)\in\mathbb{R}_{+}$. According to this,
the measure $\mu_{t_0}$ is concentrated on the set (\ref{setr}) for
$r=-1/2$, which can be identified with the topological dual
$\mathfrak{h}_{\epsilon}^{\prime}$ of the subspace of symmetric
functions in the Sobolev space\footnote{This is,
$H^{\epsilon}(\mathbb{S}^{2})=\big\{f\in
L^{2}(\mathbb{S}^{2})\,\,|\,\,\|f\|_{\epsilon}^{2}=\sum_{\ell}(\ell+1/2)^{2\epsilon}\sum_{m=-\ell}^{\ell}f_{\ell
m}^{2}<+\infty\big\}$, where the spherical Fourier transform
$f\mapsto f_{\ell m}$ is defined as $f_{\ell
m}:=\int_{\mathbb{S}^{2}}|\gamma|^{1/2}f(s)\overline{Y}_{\ell
m}(s)$.} $H^{\epsilon}(\mathbb{S}^{2})$, for any $\epsilon>0$,
\begin{equation*}
\mathfrak{h}_{\epsilon}:=\Big\{f\in
H^{\epsilon}(\mathbb{S}^{2})\,\,\big|\,\,\mathcal{L}_{\sigma}f=0
\,,\,\,\|f\|_{\epsilon}^{2}:=\sum_{\ell=0}^{\infty}(\ell+1/2)^{2\epsilon}f_{\ell}^{2}<+\infty\Big\}\,,\,\,\,\epsilon>0\,,
\end{equation*}
where $f_\ell$ are the Fourier coefficients of the function $f$. We
then conclude that the typical field configurations are not as
singular as the delta functions or their derivatives. Note, however,
that the subset
$\mathfrak{b}\subset\mathfrak{h}_{\epsilon}^{\prime}$ of symmetric
$L^{2}(\mathbb{S}^{2})$ functions has also measure zero. Indeed,
consider the characteristic function $\chi_{\mathfrak{b}}$ of the
measurable set $\mathfrak{b}$, defined by
\begin{equation}
\chi_{\mathfrak{b}}[\tilde{Q}]:=\lim_{\alpha\rightarrow
+0}\,\exp\left(-\alpha\sum_{\ell=0}^{\infty}\tilde{Q}_{\ell}^{2}\right),
\end{equation}
so that $\chi_{\mathfrak{b}}[\tilde{Q}]=1$, for $\tilde{Q}\in
\mathfrak{b}$, vanishing anywhere else. Making use of the
restriction (\ref{restrict}) and applying the Lebesgue monotone
convergence theorem it is straightforward to obtain
\begin{equation}
\mu_{t_0}(\mathfrak{b})=\int_{\mathscr{S}^{\prime}}\chi_{\mathfrak{b}}[\tilde{Q}]\,\mathrm{d}\mu_{t_0}[\tilde{Q}]=\lim_{\alpha\rightarrow
+0}\,\lim_{n\rightarrow+\infty}\,\prod_{\ell=0}^{n}\frac{1}{\sqrt{1+2\alpha|z_{\ell}(t_0)|^{2}}}\,.
\end{equation}
The limit of the product vanishes as $n\rightarrow+\infty$ due to
the nonconvergence of the series
$\sum_{\ell=0}^{\infty}\log(1+2\alpha|z_{\ell}(t_0)|^{2})$, and
hence $\mu_{t_0}(\mathfrak{b})=0$. Since
$\mathscr{S}\hookrightarrow\mathfrak{b}$, we have that, as usual for
a field theory, the  $\mu_{t_0}$-measure is not supported on the
classical configuration space $\mathscr{S}$. This is precisely the
reason why a suitable distributional extension of $\mathscr{S}$ must
be chosen as measure space in order to construct the $L^{2}$ space
for the Schr\"{o}dinger representation.
\\
\linebreak \indent Finally, note that the Bochner-Minlos theorem
(see \emph{Theorem \ref{B-MTheorem}}) gives the support of the
measure as a linear subspace of the original measure space. In order
to find a finer (nonlinear) characterization of this support, one
can apply the methods developed in $[8]$ for countable products of
Gaussian measures. According to that reference, given a sequence
$\{\Delta_{k}\}$, $\Delta_{k}>1$, $k=1,2,\ldots$, the
$\mu_{t_0}$-measure of the set
$$
Z_{t_0}(\{\Delta_{k}\}):=\big\{\tilde{Q}\in\mathscr{S}^{\prime}\,\,|\,\,\exists\,
N\in\mathbb{N}\,\,\,\mathrm{s.t.}\,\,\,|\tilde{Q}_{\ell}|<|z_{\ell}(t_0)|\sqrt{2\log\Delta_{\ell}}\,,\,\,\,\mathrm{for}\,\,\ell\ge
N\big\}
$$
is one (resp. zero) if $\sum_{k}1/(\Delta_{k}\sqrt{\log\Delta_{k}})$
converges (resp. diverges). Concretely, this condition is satisfied
for $\Delta^{(\alpha)}_{\ell}:=(1+\ell)^{\alpha}$, with $\alpha>1$,
whereas it is not verified for the same sequence with $\alpha=1$.
Thus, $\mu_{t_0}(Z_{t_0}(\{\Delta^{(\alpha)}_{\ell}\}))=1$ for
$\alpha>1$, and the measure vanishes on the set corresponding to
$\alpha=1$.

\section{Canonical commutation relations}\label{CCRSchr}

\indent Next, we will obtain the representation of the basic
momentum observables following the discussion developed in $[1]$,
adapted here to our definitions and conventions. We will realize
that both the representation of the momentum operator and the choice
of the $\mu_{t_0}$ measure are interrelated, in the sense that the
information on the complex structure $J_{\mathcal{P}}$ used to
construct the Fock representation is encoded in both of them. We
start by computing the expectation value
\begin{eqnarray*}
\varpi_{t_0}(W(\lambda))|_{\lambda=(-g,f)}&=&\big\langle\Psi_{0}^{(t_0)}\big|\exp\left(i\big(\hat{Q}_{t_0}[f]+\hat{P}_{t_0}[g]\big)\right)\Psi_{0}^{(t_0)}\big\rangle_{\mathscr{H}_{s}(t_0)}\nonumber\\
&=&\exp\left(\frac{i}{2}\langle f,g
\rangle\right)\big\langle\Psi_{0}^{(t_0)}\big|\exp\left(i\hat{Q}_{t_0}[f]\right)\exp\left(i\hat{P}_{t_0}[g]\right)\Psi_{0}^{(t_0)}\big\rangle_{\mathscr{H}_{s}(t_0)}\,,
\end{eqnarray*}
where we have used the BCH relation to separate operators. We take
the general expression
$$\left(\hat{P}_{t_0}[g]\Psi\right)[\tilde{Q}]=-i(D_{\tilde{Q}}\Psi)[g]-i\tilde{Q}\left(M(t_0)g\right)\Psi[\tilde{Q}]\,,$$
consisting of the expected directional derivative of the functional
$\Psi\in\mathscr{D}_{\hat{P}_{t_0}[g]}\subset\mathscr{H}_{s}(t_{0})$
in the direction defined by $\tilde{Q}\in\mathscr{S}^{\prime}$ (this
will acquire a definite sense in terms of the modes
$\tilde{Q}_{\ell}$) plus an extra linear multiplicative term. Now,
we must determine the expression of the new operator
$M(t_0):\mathscr{S}\rightarrow\mathscr{S}$. Taking into account that
for Gaussian measures the vacuum state is a constant functional, and
using again the BCH formula, it is straightforward to obtain
\begin{eqnarray*}
\varpi_{t_0}(W(\lambda))|_{\lambda=(-g,f)}&=&\exp\left(\frac{i}{2}\langle
f,g\rangle-\frac{1}{2}\langle g,M(t_0)g\rangle\right)
\int_{\mathscr{S}^{\prime}}e^{i\tilde{Q}(f-iM(t_0)g)}\mathrm{d}\mu_{t_0}[\tilde{Q}]\\
&=&\exp\left(\frac{i}{2}\langle f,g\rangle-\frac{1}{2}\langle
g,M(t_0)g\rangle\right)\exp\Big(\frac{1}{4}\big(\langle f,B(t_0)f\rangle\\
&&-\langle M(t_0)g,B(t_0)M(t_0)g\rangle-2i\langle
f,B(t_0)M(t_0)g\rangle\big)\Big)\,,
\end{eqnarray*}
where we have used the generating function definition
(\ref{generatfunct}) and also the relation $\langle
M(t_0)g,B(t_0)f\rangle=\langle f,B(t_0)M(t_0)g\rangle$. The left
hand side of the above equation can be easily calculated by using
the expressions (\ref{J0}) and (\ref{statesigma}), as well as the
relation $\langle f,A(t_0)g\rangle=-\langle g,C(t_0)f\rangle$,
$$
\varpi_{t_0}(W(\lambda))|_{\lambda=(-g,f)}=\exp\left(\frac{1}{4}\big(\langle
f,B(t_0)f\rangle-\langle g,D(t_0)g\rangle-2\langle
f,A(t_0)g\rangle\big)\right).
$$
A simple comparison between the two expressions obtained for
$\varpi_{t_0}(W(\lambda))$, $\lambda=(-g,f)\in\mathbf{P}$, leads us
to conclude that, for all $f,g\in\mathscr{S}$,
\begin{eqnarray}
\langle f,A(t_0)g\rangle & = & -i\langle f,g\rangle+i\langle
M(t_0)g,B(t_0)f\rangle\,,\label{Comp1}\\
\langle g,D(t_0)g\rangle & = & -2\langle g,M(t_0)g\rangle+\langle
M(t_0)g,B(t_0)M(t_0)g\rangle\,.\label{Comp2}
\end{eqnarray}
From (\ref{Comp1}) we finally obtain
$$M(t_0)=B^{-1}(t_0)\big(\mathbf{1}-iA(t_0)\big)\,,$$
whereas (\ref{Comp2}) simply becomes a consistency relation.
Therefore, the representation of the momentum observable is given by
\begin{eqnarray}
&&\pi_{s}(t_0)\cdot
W(\lambda)|_{\lambda=(-g,0)}=\exp(i\hat{P}_{t_0}[g])\,,\nonumber\\
&&\left(\hat{P}_{t_0}[g]\Psi\right)[\tilde{Q}]=-i(D_{\tilde{Q}}\Psi)[g]-i\tilde{Q}\left(B^{-1}(t_0)(\mathbf{1}-i
A(t_0))g\right)\Psi[\tilde{Q}]\,.\label{P0operator}
\end{eqnarray}
Note that the unusual multiplicative term depends both on the
measure $\mu_{t_0}$ --uniquely characterized by the operator
$B(t_0)$-- and the operator $A(t_0)$. It guarantees that the
momentum operator is symmetric with respect to the inner product
$\langle\cdot|\cdot\rangle_{\mathscr{H}_{s}(t_0)}$. Indeed, just by
using the Gaussian integration by parts formula
$$\int_{\mathscr{S}^{\prime}}(D_{\tilde{Q}}\Psi)[f]\,\mathrm{d}\mu_{t_0}[\tilde{Q}]=\int_{\mathscr{S}^{\prime}}\tilde{Q}(\mathcal{C}^{-1}(t_0)f)\Psi[\tilde{Q}]\,\mathrm{d}\mu_{t_0}[\tilde{Q}]\,,$$
that can be easily deduced from (\ref{restrict}), we obtain
\begin{eqnarray*}
\big\langle\Phi\,\big|\,\hat{P}_{t_0}[g]\Psi\big\rangle_{\mathscr{H}_{s}(t_0)}&=&i\big\langle
(D_{\tilde{Q}}\Phi)[g]\,\big|\,\Psi\big\rangle_{\mathscr{H}_{s}(t_0)}+i\big\langle\Phi\,\big|\,\tilde{Q}\big(B^{-1}(t_0)(\mathbf{1}+iA(t_0))g\big)\Psi\big\rangle_{\mathscr{H}_{s}(t_0)}\\
&=&i\big\langle(D_{\tilde{Q}}\Phi)[g]+\tilde{Q}\big(B^{-1}(t_0)(\mathbf{1}-iA(t_0))g\big)\Phi\,\big|\,\Psi\big\rangle_{\mathscr{H}_{s}(t_0)}\\
&=&\big\langle\hat{P}_{t_0}[g]\Phi\,\big|\,\Psi\big\rangle_{\mathscr{H}_{s}(t_0)}\,,\,\,\,\forall\,\Phi,\Psi\in\mathscr{D}_{\hat{P}_{t_0}[g]}\,.
\end{eqnarray*}
Let us now denote $\hat{Q}_{\ell}(t_{0}):=\hat{Q}_{t_0}[Y_{\ell 0}]$
and $\hat{P}_{\ell}(t_0):=\hat{P}_{t_0}[Y_{\ell 0}]$, where the
$\hat{Q}_{t_0}[f]$ operator has been defined in (\ref{Qoperator}).
By considering the normalization condition (\ref{norm_zs}) and
equation (\ref{invB}), we get
\begin{equation*}
\big(B^{-1}(t_0)(\mathbf{1}-iA(t_0))Y_{\ell
0}\big)(s)=i\frac{\dot{\bar{z}}_{\ell}(t_0)}{\bar{z}_{\ell}(t_0)}Y_{\ell
0}(s)\,,
\end{equation*}
and, hence, we finally obtain
\begin{equation}\label{QlyPl}
\hat{Q}_{\ell}(t_0)\Psi=\tilde{Q}_{\ell}\Psi\,,\,\,\,\,\hat{P}_{\ell}(t_0)\Psi=-i\frac{\partial\Psi}{\partial
\tilde{Q}_\ell}+\frac{\dot{\bar{z}}_{\ell}(t_0)}{\bar{z}_{\ell}(t_0)}\tilde{Q}_{\ell}\Psi\,,
\end{equation}
where $\Psi$ is a functional of the components $\tilde{Q}_{\ell}$.
The canonical commutation relations
$[\hat{Q}_{\ell}(t_0),\hat{P}_{\ell^{\prime}}(t_0)]=i\delta(\ell,\ell^{\prime})\hat{\mathbb{I}}$
and
$[\hat{Q}_{\ell}(t_0),\hat{Q}_{\ell^{\prime}}(t_0)]=0=[\hat{P}_{\ell}(t_0),\hat{P}_{\ell^{\prime}}(t_0)]$
are obviously satisfied on the appropriate domains.

\bigskip

\indent It is possible to relate the Fock and Schr\"{o}dinger
representations through the action of the annihilation and creation
operators on wave functionals $[1]$. Making use of equations
(\ref{alt0}) and (\ref{QlyPl}), we get
\begin{equation}\label{aadag}
\hat{b}_{\ell}(t_0)=\bar{z}_{\ell}(t_0)\frac{\partial}{\partial
\tilde{Q}_{\ell}}\,,\quad\hat{b}_{\ell}^{\dag}(t_0)=-z_{\ell}(t_0)\frac{\partial}{\partial
\tilde{Q}_{\ell}}+\frac{1}{\bar{z}_{\ell}(t_0)}\tilde{Q}_{\ell}\,.
\end{equation}
In particular, the vacuum state is given by the unit constant
functional (up to multiplicative phase)
\begin{equation*}
\Psi^{(t_0)}_{0}[\tilde{Q}]=1\,,\,\,\,\forall\,\tilde{Q}\in\mathscr{S}^{\prime}\,.
\end{equation*}
There exists a map
$\hat{V}_{t_0}:\mathscr{F}_+(\mathscr{H}_{\mathcal{P}})\rightarrow\mathscr{H}_{s}(t_0)$
that unitarily connects the creation and annihilation operators of
both representations $[6]$. Given the annihilation and creation
operators associated with the modes $z_{\ell}Y_{\ell0}$,
$\hat{b}_{\ell}$ and $\hat{b}_{\ell}^{\dag}$ respectively, the
expressions (\ref{aadag}) correspond to
$\hat{V}_{t_0}\,\hat{b}_{\ell}\,\hat{V}^{-1}_{t_0}$ and
$\hat{V}_{t_0}\,\hat{b}_{\ell}^{\dag}\,\hat{V}^{-1}_{t_0}$,
respectively. These relations, and the action
$\Psi^{(t_0)}_{0}=\hat{V}_{t_0}|0\rangle_{\mathcal{P}}$ on the Fock
vacuum state
$|0\rangle_{\mathcal{P}}\in\mathscr{F}_+(\mathscr{H}_\mathcal{P})$,
univocally characterize the unitary transformation $\hat{V}_{t_0}$.
The general procedure that we have followed guarantees the
self-adjointness of the configu\-ration and momentum operators in
the Schr\"{o}dinger representation. Indeed, the self-adjoint
operators $\hat{Q}_{t_0}[f]$ and $\hat{P}_{t_0}[g]$ with dense
domain in the Fock space
$\mathscr{F}_{+}(\mathscr{H}_{\mathcal{P}})$ corresponding to the
fixed value $t=t_0$ (see equation (\ref{Q&PFock})) are unitarily
related to (\ref{Qoperator}) and (\ref{P0operator}) through the
unitary transformation $\hat{V}_{t_0}$.

\bigskip

\indent Finally, the probabilistic interpretation of the models is
given by the usual Born correspondence rules $[9]$: Given
$f\in\mathscr{S}$, the theoretical probability that a measurement
carried out in the pure state $\Psi$ at a certain time to determine
the value of $\tilde{Q}(f)$ will yield a result contained in the
Borel set $\Delta\in\mathrm{Bor}(\mathbb{R})$ for some
$\tilde{Q}\in\mathscr{S}^{\prime}$ is given by
\begin{equation}\label{probability}
\mathrm{P}_{\Psi}^{\hat{Q}_{t_0}[f]}(\Delta)=\|\Psi\|_{\mathscr{H}_{s}(t_0)}^{-2}\big\langle\Psi\,\big|\,E^{\hat{Q}_{t_0}[f]}(\Delta)\Psi\big\rangle_{\mathscr{H}_{s}(t_0)}
=\|\Psi\|_{\mathscr{H}_{s}(t_0)}^{-2}\int_{V_{f,\,\Delta}}\big|\Psi[\tilde{Q}]\big|^{2}\,\mathrm{d}\mu_{t_0}[\tilde{Q}]\,,
\end{equation}
where $E^{\hat{Q}_{t_0}[f]}(\Delta)$ is the spectral measure
univocally associated with $\hat{Q}_{t_0}[f]$, defined by
$\big(E^{\hat{Q}_{t_0}[f]}(\Delta)\Psi\big)[\tilde{Q}]=\chi_{V_{f,\,\Delta}}[\tilde{Q}]\,\Psi[\tilde{Q}]$,
with $\chi_{V_{f,\,\Delta}}$ being the characteristic function of
the measu\-rable set
$V_{f,\,\Delta}:=\{\tilde{Q}\in\mathscr{S}^{\prime}\,|\,\tilde{Q}(f)\in\Delta\}\in\sigma(\mathrm{Cyl}(\mathscr{S}^{\prime}))$.
Here, $\|\cdot\|_{\mathscr{H}_{s}(t_0)}$ denotes the norm associated
with the inner product
$\langle\cdot|\cdot\rangle_{\mathscr{H}_{s}(t_0)}$. According to
this, the measure $\mu_{t_0}$ admits the following physical
interpretation: It defines the probability measure
(\ref{probability}) for the vacuum state $\Psi^{(t_0)}_{0}$.

\section{Unitary equivalence of Schr\"{o}dinger representations}\label{UNITARY}

\indent Let us consider the symplectomorphism (\ref{tau})
$\tau_{(t_1,t_0)}:=\mathfrak{I}_{t_1}^{-1}\circ\mathfrak{I}_{t_0}:\mathbf{P}\rightarrow\mathbf{P}$,
$t_1>t_0$, which implements the classical time evolution from the
embedding $\iota_{t_0}(\mathbb{S}^{2})$ to
$\iota_{t_1}(\mathbb{S}^{2})$ on the canonical phase space. It
induces a one-parameter family of states on the Weyl algebra:
Starting from the initial state $\varpi_{t_0}$ defined in equation
(\ref{statesigma}), the dynamical evolution in the algebraic
formulation of the theory is given by
$\varpi_{t_1}=\varpi_{t_0}\circ\alpha_{(t_1,t_0)}^{-1}$ in the
Schr\"{o}dinger picture, with
$\alpha_{(t_1,t_0)}:\mathscr{W}(\Upsilon)\rightarrow\mathscr{W}(\Upsilon)$
being the $*$-automorphism univocally associated with the symplectic
transformation $\tau_{(t_1,t_0)}$ defined by
$\alpha_{(t_1,t_0)}\cdot W(\lambda):=W(\tau_{(t_1,t_0)}(\lambda))$.
The evolved state $\varpi_{t_1}$ acts on the elementary observables
as
$$\varpi_{t_1}(W(\lambda))=\exp\big(-\omega(J_{t_1}(\lambda),\lambda)/4\big)\,,$$
where the complex structure
\begin{equation*}
J_{t_{1}}:=\tau_{(t_1,t_0)}\circ
J_{t_0}\circ\tau^{-1}_{(t_1,t_0)}=\mathfrak{I}_{t_1}^{-1}\circ
J_{\mathcal{P}}\circ\mathfrak{I}_{t_1}:\mathbf{P}\rightarrow\mathbf{P}
\end{equation*}
defines the Schr\"{o}dinger representation\footnote{Here, we will
make a notational abuse and simply denote the triplet
$\big(\mathscr{H}_{s}(t),\pi_{s}(t),\Psi^{(t)}_{0}\big)$ as
$\mathscr{H}_{s}(t)$.} $\mathscr{H}_{s}(t_1)$ corresponding to the
time value $t_1$. The condition of unitary equivalence of
Schr\"{o}dinger representations corresponding to different values
$t_0<t_1$ of the time parameter clearly amounts to demanding the
unitary implementability of the symplectic transformation
$\tau_{(t_1,t_0)}$ in the $\mathscr{H}_{s}(t_0)$ representation. In
that case, $J_{t_1}-J_{t_0}$ is a Hilbert-Schmidt operator in the
one-particle Hilbert space constructed from $J_{t_0}$ (or
equivalently $J_{t_1}$), and there exists a unitary transformation
$\hat{V}_{(t_1,t_0)}:\mathscr{H}_{s}(t_0)\rightarrow\mathscr{H}_{s}(t_1)$
mapping the configuration and momentum operators from one
representation into the other, in such a way that
\begin{eqnarray}\label{Vt0t1}
\hat{V}_{(t_1,t_0)}\,\hat{b}_{\ell}(t_0)\,\hat{V}^{-1}_{(t_1,t_0)}&=&\alpha_{\ell}(t_1,t_0\,|\,z_\ell)\hat{b}_{\ell}(t_1)+\beta_{\ell}(t_1,t_0\,|\,z_\ell)\hat{b}^{\dag}_{\ell}(t_1)\,,\\
\hat{V}_{(t_1,t_0)}\,\hat{b}^{\dag}_{\ell}(t_0)\,\hat{V}^{-1}_{(t_1,t_0)}&=&\bar{\beta}_{\ell}(t_1,t_0\,|\,z_\ell)\hat{b}_{\ell}(t_1)+\bar{\alpha}_{\ell}(t_1,t_0\,|\,z_\ell)\hat{b}^{\dag}_{\ell}(t_1)\,,\nonumber
\end{eqnarray}
where
\begin{equation*}\label{alpha&beta}
\alpha_{\ell}(t_1,t_0\,|\,z_\ell):=i\Big(\bar{z}_{\ell}(t_0)\dot{z}_{\ell}(t_1)-z_{\ell}(t_1)\dot{\bar{z}}_{\ell}(t_0)\Big),\,\,\,\,
\beta_{\ell}(t_1,t_0\,|\,z_\ell):=i\Big(\bar{z}_{\ell}(t_0)\dot{\bar{z}}_{\ell}(t_1)-\bar{z}_{\ell}(t_1)\dot{\bar{z}}_{\ell}(t_0)\Big).
\end{equation*}
This is precisely ensured by the square summability of the
$\beta_{\ell}$ coefficients appearing in the Bogoliubov
transformation (\ref{Vt0t1}), exactly the same condition that
guarantees the unitary implementation of the quantum time evolution
in the Fock space $\mathscr{F}_{+}(\mathscr{H}_\mathcal{P})$. We
then conclude that the unitarity of the quantum dynamics in the Fock
representation guarantees the equivalence of the Schr\"{o}dinger
representations corresponding to different times $t_0,t_1$. The map
$\hat{V}_{(t_1,t_0)}=\hat{V}_{t_1}\,\hat{U}(t_1,t_0)\,\hat{V}_{t_0}^{-1}$
relating them is completely characterized by the relations
(\ref{Vt0t1}) and the action on the vacuum state
$\Psi^{(t_0)}_{0}\in\mathscr{H}_{s}(t_0)$, given by
\begin{equation}\label{VPsi0}
\Big(\hat{V}_{(t_1,t_0)}\Psi^{(t_0)}_{0}\Big)[\tilde{Q}]=\prod_{\ell=0}^{\infty}\frac{|z_{\ell}(t_1)|^{1/2}}{|z_{\ell}(t_0)|^{1/2}}
\exp\left(-\frac{1}{2}\frac{\beta_{\ell}(t_1,t_0\,|\,z_\ell)}{\bar{z}_{\ell}(t_0)\bar{z}_{\ell}(t_1)}
\tilde{Q}_{\ell}^{2}\right)\in\mathscr{H}_{s}(t_1)\,,
\end{equation}
where we have used the fact that
$\hat{b}_{\ell}(t_0)\Psi^{(t_0)}_{0}=0$,
$\forall\,\ell\in\mathbb{N}_{0}$, and the expressions
(\ref{norm_zs}), (\ref{aadag}), and (\ref{Vt0t1}) to obtain the
differential equations verified by this state; namely, $$\partial
\hat{V}_{(t_1,t_0)}\Psi^{(t_0)}_{0}/\partial\tilde{Q}_{\ell}=-\big(\beta_{\ell}(t_1,t_0\,|\,z_\ell)/\bar{z}_{\ell}(t_0)\bar{z}_{\ell}(t_1)\big)\tilde{Q}_{\ell}
\hat{V}_{(t_1,t_0)}\Psi^{(t_0)}_{0}\,,\,\,\,
\ell\in\mathbb{N}_{0}\,.$$ Equation (\ref{VPsi0}) must be
interpreted as the limit in the $\mathscr{H}_{s}(t_1)$-norm of the
Cauchy sequence of normalized vectors $f_{n}\in\mathscr{H}_{s}(t_1)$
obtained by extending the product (\ref{VPsi0}) to a finite integer
$n\in\mathbb{N}$.

\bigskip

\indent The mutual absolute continuity of any two Gaussian measures
associated with different times $t_0,t_1\in(0,\pi)$ is also
verified, i.e., they have the same zero measure sets. This property
requires that the operator $\mathcal{C}(t_1)-\mathcal{C}(t_0)$ is
Hilbert-Schmidt $[8,10,11]$, which is satisfied in our case. Indeed,
it is straightforward to check that the sequence
$$\big(|z_{\ell}(t_1)|^{2}-|z_{\ell}(t_0)|^2\big)_{\ell\in\mathbb{N}_0}$$
is square summable. In fact, it is possible to show that the
equivalence of measures is a necessary condition for the unitary
equivalence between Schr\"{o}dinger representations, and that any
possible unitary equivalence between them is of the form
$\Psi\mapsto\big(\mathrm{d}\mu_{t_1}/\mathrm{d}\mu_{t_0}\big)^{1/2}\exp(iF)\Psi$,
with $\mathrm{d}\mu_{t_1}/\mathrm{d}\mu_{t_0}$ denoting the
Radon-Nikodym derivative of $\mu_{t_1}$ with respect to $\mu_{t_0}$
and $F$ being a real functional $[12]$.
\\
\indent On the contrary, for the original scalar field
$\varphi=\xi/\sqrt{\sin t}$, for which the time evolution is not
unitary, we get the nonequivalence of the representations obtained
for different times, and also the impossibility of such continuity.
In this case, the mutual singularity of measures can be expected, as
proved for the vacuum Gowdy $\mathbb{T}^{3}$ model in $[3]$. This
typifies the advantage of using the re-scaled fields making the
quantum dynamics unitary, for in this case it is possible to obtain
a unique (up to unitary equivalence) Schr\"{o}dinger representation
for these models and, as a direct consequence, the mutual absolute
continuity of the measures corresponding to different times. Neither
of these properties can be attained for the original variables. In
this last situation, even if the failure of the unitarity of time
evolution and the mutual singularity of measures are not serious
obstacles for a suitable probabilistic interpretation of the models
$[3,13]$, we must face the lack of uniqueness of the representation.

\bigskip

\indent Note that the map
$\hat{V}_{t_0}:\mathscr{F}_+(\mathscr{H}_{\mathcal{P}})\rightarrow\mathscr{H}_{s}(t_0)$
introduced in subsection \ref{CCRSchr} does not connect the
configuration and momentum operators of the Fock representation,
$\hat{Q}_{\ell}(t)=z_{\ell}(t)\hat{b}_{\ell}+\bar{z}_{\ell}(t)\hat{b}_{\ell}^{\dag}$
and
$\hat{P}_{\ell}(t)=\dot{z}_{\ell}(t)\hat{b}_{\ell}+\dot{\bar{z}}_{\ell}(t)\hat{b}_{\ell}^{\dag}$,
respectively, with those of the Schr\"{o}dinger one (except for
$t=t_0$). However, owing to the unitary implementability of the
dynamics, there exists also a unitary transformation
$\hat{W}_{t_0}(t):\mathscr{F}_+(\mathscr{H}_\mathcal{P})\rightarrow\mathscr{H}_{s}(t_0)$,
such that
\begin{eqnarray*}
\hat{W}_{t_0}(t)\,
\hat{b}_{\ell}\,\hat{W}_{t_0}^{-1}(t)&=&\alpha_{\ell}(t,t_0\,|\,z_\ell)\hat{b}_{\ell}(t_0)+\beta_{\ell}(t,t_0\,|\,z_\ell)\hat{b}_{\ell}^{\dag}(t_0)\,,\\
\hat{W}_{t_0}(t)\,
\hat{b}^{\dag}_{\ell}\,\hat{W}_{t_0}^{-1}(t)&=&\bar{\beta}_{\ell}(t,t_0\,|\,z_\ell)\hat{b}_{\ell}(t_0)+\bar{\alpha}_{\ell}(t,t_0\,|\,z_\ell)\hat{b}_{\ell}^{\dag}(t_0)\,,
\end{eqnarray*}
relating these operators. In terms of the unitary evolution operator
on $\mathscr{F}_{+}(\mathscr{H}_\mathcal{P})$, we have
$\hat{W}_{t_0}(t)=\hat{V}_{t_0}\,\hat{U}^{-1}(t,t_0)$. Finally,
given the quantum Hamiltonian (\ref{QuantHamiltonian}) in the Fock
representation, with dense domain
$\mathscr{D}_{\hat{H}_{0}(t)}\subset\mathscr{F}_{+}(\mathscr{H}_{\mathcal{P}})$
spanned by the states with a finite number of particles, the
corresponding operator in the $\mathscr{H}_{s}(t_0)$ representation
is given by
$\hat{W}_{t_0}(t)\,\hat{H}_{0}(t)\,\hat{W}_{t_0}^{-1}(t)$,
\begin{equation*}\label{SchquantHamilt}
\frac{1}{2}\sum_{\ell=0}^{\infty}\Bigg(-\frac{\partial^{2}}{\partial\tilde{Q}_{\ell}^{2}}
-2i\frac{\dot{\bar{z}}_{\ell}(t_0)}{\bar{z}_{\ell}(t_0)}\tilde{Q}_{\ell}\frac{\partial}{\partial\tilde{Q}_{\ell}}
+\left(\frac{\dot{\bar{z}}^{2}_{\ell}(t_0)}{\bar{z}_{\ell}^{2}(t_0)}+\frac{1}{4}\big(1+\csc^{2}t\big)+\ell(\ell+1)\right)\big(\tilde{Q}_{\ell}^{2}-|z_{\ell}(t_0)|^{2}\big)\Bigg)\,,
\end{equation*}
modulo an irrelevant real term proportional to the identity. Note,
by contrast, that the complex independent term appearing in the
previous expression is necessary to ensure that the operator is
self-adjoint. This Hamiltonian is defined in the dense subspace
$\hat{W}_{t_0}(t)\mathscr{D}_{\hat{H}_{0}(t)}=\big\{\hat{W}_{t_0}(t)f\,|\,f\in\mathscr{D}_{\hat{H}_{0}(t)}\big\}\subset\mathscr{H}_{s}(t_0)$
generated by the cyclic vector
$\hat{W}_{t_0}(t)|0\rangle_{\mathcal{P}}\in\mathscr{H}_{s}(t_0)$.

\section{Unitary evolution operator}\label{UnitaryEvolutOper}

\indent In this final section we provide a general procedure to
obtain the unitary evolution operator
$\hat{U}_{t_0}(t,t^{\prime}):\mathscr{H}_s(t_0)\rightarrow\mathscr{H}_s(t_0)$
written explicitly in closed form in terms of the field and momentum
operators. The strategy that we follow is to generalize the results
already known for a single harmonic oscillator with time-dependent
frequency to a system consisting of an infinite number of uncoupled
harmonic oscillators $[14,15]$. The reader is referred to
\emph{appendix \ref{AppendixTDHO}} for a detailed study of these
topics. Analogously to the one-dimensional case, when the dynamics
is unitarily implementable we define the time evolution propagator
through the relation
\begin{equation*}
\big(\hat{U}_{t_0}(t,t^\prime)\Psi\big)[\tilde{Q}]=\int_{\mathscr{S}^\prime}K_{t_0}\big(\tilde{Q},t;\tilde{Q}^\prime,t^\prime\big)
\Psi[\tilde{Q}^\prime]\,\mathrm{d}\mu_{t_0}[\tilde{Q}^\prime]\,.
\end{equation*}
where a straightforward calculation formally provides\footnote{The
reader may wish to compare this expression with equations (\ref{K1})
and (\ref{Kalphabeta}) in \emph{appendix \ref{AppendixTDHO}}.}
\begin{eqnarray*}
K_{t_0}\big(\tilde{Q},t;\tilde{Q}^\prime,t^\prime\big)
&=&\prod_{\ell=0}^\infty\sqrt{2\pi}|z_\ell(t_0)|\exp\left(\frac{i}{2}\left(\frac{\dot{z}_\ell(t_0)}{z_\ell(t_0)}\tilde{Q}_\ell^{\prime2}
-\frac{\dot{\bar{z}}_\ell(t_0)}{\bar{z}_\ell(t_0)}
\tilde{Q}_{\ell}^2\right)\right)\\
&\times&
K_\ell\big(\tilde{Q}_{\ell},t;\tilde{Q}_{\ell}^\prime,t^\prime\big)\exp\left(-i\int_{t^\prime}^{t}\mathrm{d}\tau\,\vartheta_\ell(\tau)\right),
\end{eqnarray*}
with $K_\ell$ denoting the well-known Feynman propagator
(\ref{K_qq0a}) associated with the one-dimensional oscillator of
squared frequency $\kappa_\ell(t):=\ell(\ell+1)+(1+\csc^2t)/4$,
written in terms of the $c_\ell(t,t^\prime)$ and
$s_\ell(t,t^\prime)$ solutions to the equation of motion (\ref{(C)})
(see equation (\ref{c&srelations}) in \emph{appendix
\ref{AppendixTDHO}}). These functions are the unique solutions to
(\ref{(C)}) such that $c_\ell(t^\prime,t^\prime)=1$, $\partial_t
c_\ell(t^\prime,t^\prime)=0$, $s_\ell(t^\prime,t^\prime)=0$, and
$\partial_t s_\ell(t^\prime,t^\prime)=1$. They are given in terms of
the associated Legendre functions by expressiones
(\ref{c&sremaintopol}) in \emph{appendix \ref{AppendixTDHO}},
substituting $\omega=\sqrt{\ell(\ell+1)}$. Finally, the term
$\vartheta_\ell$ comes from taking normal order in the quantum
Hamiltonian; explicitly,
$$\vartheta_\ell(t)=-\frac{1}{2}\big(|\dot{z}_\ell(t_0)|^2+\kappa_\ell(t)|z_\ell(t_0)|^2\big)\sim-\frac{\ell}{2}\,\,\,\,\,\mathrm{as}\,\,\,\,\,\ell\rightarrow+\infty\,.$$
\indent At first sight, the application of the techniques employed
in \emph{appendix \ref{AppendixTDHO}} for the single oscillator
should allow us to factorize the evolution operator in the form
\begin{equation}\label{UFactorized}
\hat{U}_{t_0}(t,t^\prime)=\hat{T}_{t_0,\bm{\rho}}^{-1}(t)\,\hat{R}_{t_0,\bm{\rho}}(t,t^\prime)\,\hat{T}_{t_0,\bm{\rho}}(t^\prime)\,,
\end{equation}
where, given an \emph{arbitrary} sequence
$\bm{\rho}(t)=(\rho_\ell(t))_{\ell\in\mathbb{N}_0}$ of solutions to
the auxiliary Ermakov-Pinney equations $[16,17]$
$$\ddot{\rho}_\ell+\kappa_\ell(t)\rho_\ell=1/\rho_\ell^3\,,$$the
$\hat{T}_{\bm{\rho}}(t)$ and $\hat{R}_{\bm{\rho}}(t,t^\prime)$
operators are univocally characterized up to phases by their action
on annihilation and creation operators,
\begin{eqnarray*}
\hat{T}_{t_0,\bm{\rho}}^{-1}(t)\,\hat{b}_\ell(t_0)\,\hat{T}_{t_0,\bm{\rho}}(t)&=&i\left(\dot{z}_\ell(t_0)\bar{z}_\ell(t_0)\rho_\ell(t)-\frac{z_\ell(t_0)\dot{\bar{z}}_\ell(t_0)}{\rho_\ell(t)}
-|z_\ell(t_0)|^2\dot{\rho}_\ell(t)\right)\hat{b}_\ell(t_0)\\
&+&i\left(\bar{z}_\ell(t_0)\dot{\bar{z}}_\ell(t_0)\Big(\rho_\ell(t)-\frac{1}{\rho_\ell(t)}\Big)-\bar{z}_\ell^2(t_0)
\dot{\rho}_\ell(t)\right)\hat{b}_\ell^\dag(t_0)\,,\\
\hat{R}_{t_0,\bm{\rho}}^{-1}(t,t^\prime)\,\hat{b}_\ell(t_0)\,\hat{R}_{t_0,\bm{\rho}}(t,t^\prime)&=&
\Big(\cos\gamma_\ell(t,t^\prime)-i(|z_\ell(t_0)|^2+|\dot{z}_\ell(t_0)|^2)\sin\gamma_\ell(t,t^\prime)\Big)\hat{b}_\ell(t_0)\\
&-&i(\bar{z}_\ell^2(t_0)+\dot{\bar{z}}_\ell^2(t_0))\sin\gamma_\ell(t,t^\prime)\hat{b}_\ell^{\dag}(t_0)\,,
\end{eqnarray*}
and similarly for $\hat{b}_\ell^{\dag}(t_0)$. Here, we have denoted
$$\gamma_\ell(t,t^\prime):=\int_{t^\prime}^{t}\frac{\mathrm{d}\tau}{\rho_\ell^{2}(\tau)}\,.$$
Nevertheless, even in the case of $\hat{U}_{t_0}(t,t^\prime)$ being
well-defined as unitary operator, the factorization
(\ref{UFactorized}) is ill-defined. Indeed, the necessary and
sufficient condition for $\hat{T}_{t_0,\bm{\rho}}(t)$ to be unitary
for each value of $t$ is given by
\begin{eqnarray}
\sum_{\ell=0}^\infty\big|z_\ell(t_0)\dot{z}_\ell(t_0)\big(\rho_\ell(t)-1/\rho_\ell(t)\big)-z_\ell^2(t_0)
\dot{\rho}_\ell(t)\big|^2<+\infty\,,\,\,\,\,\forall\,t\in(0,\pi)\,.\label{uniTt0}
\end{eqnarray}
Similarly, it is straightforward to show that
$\hat{R}_{t_0,\bm{\rho}}(t,t_0)$ is unitarily implementable if and
only if
\begin{eqnarray}
\sum_{\ell=0}^\infty\Big|(z_\ell^2(t_0)+\dot{z}_\ell^2(t_0))
\sin\gamma_\ell(t,t^\prime)\Big|^2<+\infty\,,\,\,\,\,\forall\,t,t_0\in(0,\pi)\,.\label{uniRt0}
\end{eqnarray}
\noindent The asymptotic expansions (\ref{asymp}) lead us to
conclude that conditions (\ref{uniTt0}) and (\ref{uniRt0}) are not
verified and, hence, neither $\hat{T}_{t_0,\bm{\rho}}(t)$ nor
$\hat{R}_{t_0,\bm{\rho}}(t,t^\prime)$ are unitary for those systems.
In the case of $\hat{R}_{t_0,\bm{\rho}}(t,t^\prime)$, this
conclusion follows readily, irrespective of $\bm{\rho}(t)$. For
$\hat{T}_{t_0,\bm{\rho}}(t)$, a necessary condition for
(\ref{uniTt0}) to be satisfied is given by
$$\sum_{\ell=0}^\infty|\rho_\ell(t)-1/\rho_\ell(t)|^2<+\infty\,\,\Leftrightarrow\,\,
\lim_{\ell\rightarrow+\infty}\rho_\ell(t)=1\,,\,\,\,\,\forall\,t\in(0,\pi)\,,$$
where we have taken into account the fact that the real sequence
$\bm{\rho}(t)$ is positive and bounded for all $t$. According to
equation (\ref{s_rho}) in \emph{appendix \ref{AppendixTDHO}}, this
implies $s_\ell(t,t_0)\sim\sin C(t,t_0)$ as
$\ell\rightarrow+\infty$, where $C(t,t_0)$ is a \emph{nonzero}
function whose form we do not need to specify. This is in conflict
with the asymptotic behavior of $s_\ell(t,t_0)$ for the systems
under study, given by $s_\ell(t,t_0)\sim0$ as
$\ell\rightarrow+\infty$ for all $t,t_0\in(0,\pi)$. In the context
of the search of semiclassical states for the Gowdy models, the
nonunitarity of the $\hat{T}_{t_0,\bm{\rho}}(t)$ operator makes it
difficult to apply the techniques developed in \emph{subsection
\ref{SemiclStates}} of \emph{appendix \ref{AppendixTDHO}} for a
single time-dependent harmonic oscillator. This point will be
discussed in depth in the conclusions of the thesis, where we will
take advantage of the unitary implementability of the dynamics in
order to define a family of coherent states for these systems.
Obviously, this does not prevent us from defining other well-defined
factorizations for $\hat{U}_{t_0}(t,t^\prime)$ different from
(\ref{UFactorized}). A particularly convenient choice is given by
$$\hat{U}_{t_0}(t,t^\prime)=\hat{\mathcal{D}}_{t_0,\bm{\rho}}(t,t^\prime)\,\hat{\mathcal{R}}_{t_0,\bm{\rho}}(t,t^\prime)\,\hat{\mathcal{S}}_{t_0,\bm{\rho}}(t,t^\prime)\,,$$
with
\begin{eqnarray*}
&&\hat{\mathcal{D}}_{t_0,\bm{\rho}}(t,t^\prime):=\hat{D}_{t_0,\bm{\rho}}^{-1}(t)\,\hat{D}_{t_0\bm{\rho}}(t^\prime)\,,\\
&&\hat{\mathcal{S}}_{t_0,\bm{\rho}}(t,t^\prime):=\hat{D}_{t_0,\bm{\rho}}^{-1}(t^\prime)\,
\hat{S}_{t_0,\bm{\rho}}^{-1}(t)\,\hat{T}_{t_0,\bm{\rho}}(t^\prime)\,,\\
&&\hat{\mathcal{R}}_{t_0,\bm{\rho}}(t,t^\prime):=\hat{T}_{t_0,\bm{\rho}}^{-1}(t^\prime)\,
\hat{R}_{t_0,\bm{\rho}}(t,t^\prime)\,\hat{T}_{t_0,\bm{\rho}}(t^\prime)\,,
\end{eqnarray*}
where $\hat{D}_{t_0,\bm{\rho}}(t)$ and $\hat{S}_{t_0,\bm{\rho}}(t)$
are displacement and squeeze operators of the type defined in
\emph{subsection \ref{ConstructingU}} of \emph{appendix
\ref{AppendixTDHO}}, in such a way that
\begin{eqnarray*}
\hat{\mathcal{D}}^{-1}_{t_0,\bm{\rho}}(t,t^\prime)\,\hat{b}_\ell(t_0)\,\hat{\mathcal{D}}_{t_0,\bm{\rho}}(t,t^\prime)&=&\left(1+i|z_\ell(t_0)|^2\left(\frac{\dot{\rho}_\ell(t)}{\rho_\ell(t)}
-\frac{\dot{\rho}_\ell(t^\prime)}{\rho_\ell(t^\prime)}\right)\right)\hat{b}_\ell(t_0)\nonumber
\\&+&i\bar{z}_\ell^2(t_0)\left(\frac{\dot{\rho}_\ell(t)}{\rho_\ell(t)}-\frac{\dot{\rho}_\ell(t^\prime)}{\rho_\ell(t^\prime)}\right)\hat{b}_\ell^\dag(t_0)\,,\label{TransfD}
\\
\hat{\mathcal{S}}^{-1}_{t_0,\bm{\rho}}(t,t^\prime)\,\hat{b}_\ell(t_0)\,\hat{\mathcal{S}}_{t_0,\bm{\rho}}(t,t^\prime)
&=&i\Bigg(\dot{z}_\ell(t_0)\bar{z}_\ell(t_0)\frac{\rho_\ell(t^\prime)}{\rho_\ell(t)}-z_\ell(t_0)\dot{\bar{z}}_\ell(t_0)\frac{\rho_\ell(t)}{\rho_\ell(t^\prime)}\nonumber
\\&+&|z_\ell(t_0)|^2\frac{\dot{\rho}_\ell(t^\prime)}{\rho_\ell(t^\prime)}\left(\frac{\rho_\ell(t)}{\rho_\ell(t^\prime)}-\frac{\rho_\ell(t^\prime)}{\rho_\ell(t)}\right)\Bigg)\hat{b}_\ell(t_0)
\nonumber\\
&+&i\bar{z}_\ell(t_0)\left(
\bar{z}_\ell(t_0)\frac{\dot{\rho}_\ell(t^\prime)}{\rho_\ell(t^\prime)}-\dot{\bar{z}}_\ell(t_0)\right)\left(\frac{\rho_\ell(t)}{\rho_\ell(t^\prime)}-\frac{\rho_\ell(t^\prime)}{\rho_\ell(t)}\right)\hat{b}^\dag_\ell(t_0)\,,\label{TransfS}
\end{eqnarray*}
\begin{eqnarray*}
\hat{\mathcal{R}}^{-1}_{t_0\bm{\rho}}(t,t^\prime)\,\hat{b}_\ell(t_0)\,\hat{\mathcal{R}}_{t_0,\bm{\rho}}(t,t^\prime)&=&\Bigg(\cos\gamma_\ell(t,t^\prime)
+i\Bigg((z_\ell(t_0)\dot{\bar{z}}_\ell(t_0)+\dot{z}_\ell(t_0)\bar{z}_\ell(t_0))\dot{\rho}_\ell(t^\prime)\rho_\ell(t^\prime)\nonumber
\\&-&|z_\ell(t_0)|^2\left(\dot{\rho}^2_\ell(t^\prime)+\frac{1}{\rho_\ell^2(t^\prime)}\right)
-|\dot{z}_\ell(t_0)|^2\rho_\ell^2(t^\prime)\Bigg)\\
&\times&\sin\gamma_\ell(t,t^\prime)\Bigg)\,\hat{b}_\ell(t_0)+i\Bigg(2\bar{z}_\ell(t_0)\dot{\bar{z}}_\ell(t_0)\dot{\rho}_\ell(t^\prime)\rho_\ell(t^\prime)\nonumber\\
&-&\dot{\bar{z}}_\ell^2(t_0)
\rho^2_\ell(t^\prime)-\bar{z}_\ell^2(t_0)\left(\dot{\rho}_\ell^2(t^\prime)+\frac{1}{\rho_\ell^2(t^\prime)}\right)
\Bigg)\sin\gamma_\ell(t,t^\prime)\,\hat{b}_\ell^{\dag}(t_0)\,,\label{TransfR}
\end{eqnarray*}
and similarly for $\hat{b}_\ell^{\dag}(t_0)$. Here, the solutions
$\rho_\ell$ to the Ermakov-Pinney equations are conveniently
selected as
\begin{equation}\label{rhoS1xS2&S3}
\rho_\ell(t)=\sqrt{\frac{\sin t}{2}\left(\pi \mathscr{P}^2_\ell(\cos
t)+\frac{4}{\pi}\mathscr{Q}^2_\ell(\cos
t)\right)}\,,\,\,\,\ell\in\mathbb{N}_0\,,
\end{equation}
with the asymptotic expansions
\begin{equation*}
\rho_\ell(t)=1/\sqrt{|\ell|}+O\big(|\ell|^{-3/2}\big)\,,\quad
\dot{\rho}_\ell(t)= C(t)/|\ell|^{5/2}+O\big(|\ell|^{-7/2}\big)\,,
\end{equation*}
as $\ell\rightarrow+\infty$. Here, $C(t)$ is a function of time
whose form we do not need to specify. Recall that the unitary
evolution operator does not depend on the concrete choice of
$\bm{\rho}$. The election (\ref{rhoS1xS2&S3}) is motivated by the
fact that the usual $\rho_\ell$ solutions to the Ermakov-Pinney
equations for Minkowskian free scalar fields evolving in a spacetime
$\mathbb{R}\times\mathbb{T}^3$ with closed spatial sections are,
precisely, $\rho_\ell(t)=1/\sqrt{|\ell|}$,
$\ell\in\mathbb{Z}\setminus\{0\}$. In this way, the functions
$z_\ell$ and $\dot{z}_\ell$ (\ref{asymp}), as well as $\rho_\ell$,
approach those corresponding to the free Minkowskian system at high
frequencies, for which the evolution is well defined and unitary. It
is straightforward to check the unitary implementability of the
above transformations in the Hilbert space $\mathscr{H}_{s}(t_0)$.
\\
\linebreak \linebreak \noindent \textbf{References}
\begin{itemize}
\item [1] A. Corichi, J. Cortez, and H. Quevedo, Phys. Rev. D \textbf{66}, 085025 (2002); Class. Quant. Grav.
\textbf{20}, L83 (2003); Annals Phys. \textbf{313}, 446478 (2004).
\item [2] R. M. Wald, \emph{Quantum Field Theory in Curved Spacetime and Black Hole
Thermodynamics}, The University of Chicago Press, USA (1994).
\item [3] C. G. Torre, Class. Quant. Grav. \textbf{24}, 1 (2007).
\item [4] D. A. Dubin and M. A. Hennings, \emph{Quantum Mechanics, Algebras and Distributions}, Longman
Scientific \& Technical (1990).
\item [5] A. Ashtekar and A. Magnon, Proc. Roy. Soc. Lond. A\textbf{346}, 375 (1975).
\item [6] J. Glimm and A. Jaffe, \emph{Quantum Physics. A Functional Integral Point of View} (Second
Edition), Springer (1984).
\item [7] B. Simon, \emph{Functional Integration and Quantum Physics}, Academic Press, New York, (1974).
\item [8] J. M. Mour\~{a}o, T. Thiemann, and J. M. Velhinho, J. Math. Phys. \textbf{40}, 2337 (1999).
\item [9] E. Prugove\v{c}ki, \emph{Quantum Mechanics in Hilbert Space} (Second Edition), Academic Press, New
York, (1981).
\item [10] B. Simon, \emph{The $P(\phi)_2$ Euclidean (Quantum) Field Theory}, Princeton University Press, Princeton,
NJ, (1974).
\item [11] Y. Yamasaki, \emph{Measures on Infinite Dimensional Spaces}, World Scientific, Singapure, (1985).
\item [12] A. Corichi, J. Cortez, G. A. Mena Marug\'an, and J. M. Velhinho, Phys. Rev. D \textbf{76}, 124031 (2007).
\item [13] C. G. Torre, Phys. Rev. D \textbf{66}, 084017 (2002).
\item [14] H. R. Lewis Jr. and W. B. Riesenfeld, J. Math. Phys. \textbf{10}, 1458 (1969).
\item [15] M. Fern\'andez Guasti and H. Moya-Cessa, J. Phys. A \textbf{36}, 2069 (2003).
\item [16] V. P. Ermakov, Univ. Izv. Kiev. \textbf{20}, 1 (1880); see Appl. Anal.
Discrete Math. \textbf{2}, 123-145 (2008) for an English translation
of Ermakov's original paper.
\item [17] E. Pinney, Proc. Am. Math. Soc. \textbf{1}, 681 (1950).
\end{itemize}

\newpage
\thispagestyle{plain} \mbox{}


\chapter*{Conclusions}
\addcontentsline{toc}{chapter}{Conclusions}


\pagestyle{fancy} \fancyhf{}

\fancyhead[LO]{\textsc{Conclusions}} \fancyhead[RO,LE]{\thepage}
\renewcommand{\headrulewidth}{0.6pt}
\setlength{\headheight}{1.5\headheight}

\begin{figure}[t] \centering
\includegraphics[width=15.5cm]{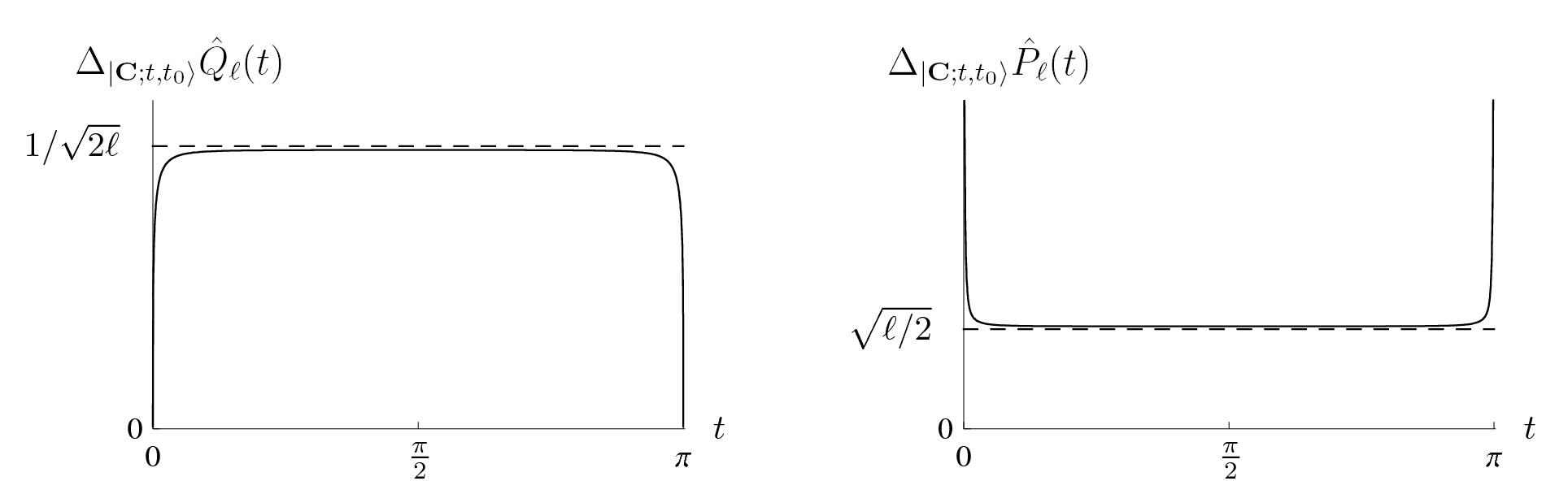}
\caption{Asymptotic behavior of the variances of the field and
momentum operators $\hat{Q}_{\ell}(t)$ and $\hat{P}_{\ell}(t)$ in
the states $|\mathbf{C};t,t_0\rangle$, $t,t_0\in(0,\pi)$, at high
frequencies. These graphics can be considered as the limit of
\emph{figure \ref{GrGowdyS1S2&S3}} in \emph{appendix
\ref{AppendixTDHO}} when
$\ell\rightarrow+\infty$.}\label{FigureAsymp}
\end{figure}

\indent Throughout this thesis, we have studied the linearly
polarized $\mathbb{S}^{1}\times\mathbb{S}^{2}$ and $\mathbb{S}^{3}$
Gowdy models coupled to massless scalar fields in a rigorous and
self-contained way, paying special attention to the mathematical
aspects of their classical formulations and exact quantizations.
Concretely, we have carefully applied modern (symplectic)
differential-geometric techniques to the description of these
dynamical systems. The presence of both initial and final
singularities, as well as the inhomogeneity and anisotropy of space,
justify the great interest in these models. Firstly, in
\emph{Chapter \ref{ChapterI}}, we have analyzed their Hamiltonian
formalism as a necessary first step towards their quantization by
gauge fixing and deparameterization.\footnote{In addition, our study
would allow us to follow other roads to quantization such as the
viewpoint pioneered by M. Varadarajan in $[1]$.} After performing a
Geroch symmetry reduction and an appropriate conformal
transformation, these models can be interpreted as (1+2)-dimensional
gravity systems coupled to a set of massless scalar fields with
axial symmetry. As analyzed in this chapter, the description of
these models requires a careful discussion of the regularity
conditions that the metric must satisfy on the symmetry axis. These
conditions give rise to the so-called \emph{polar constraints},
which are shown to be first class and necessary to guarantee the
differentiability of the other constraints present in the models.
\\
\indent An important issue to analyze on this type of cosmological
models is the so-called \emph{problem of time} in general
relativity. Since there are not preferred foliations of the
spacetime, one has to consider all of them jointly with the aim of
satisfying the principle of general covariance. As a consequence,
the well-known Hamiltonian constraint is directly obtained within
the canonical ADM formalism of the theory. In the case of a closed
universe, the time evolution is purely gauge and the Hamiltonian of
the system is restricted to vanish on the physical phase space.
Thus, in order to recover the dynamics, one has to apply some
\emph{ad hoc} procedure such as deparameterization, based on a
partial gauge fixation of the system. Obviously, different
deparameterizations give rise in general to nonequivalent quantum
theories. In our case, by imposing gauge fixing conditions similar
to those employed in the literature for the familiar 3-torus case,
and after a suitable series of canonical transformations suggested
by very simple gauge transformations verified by some natural
variables, one arrives at a reduced phase-space description where
the dynamics of the systems is governed by nonautonomous quadratic
Hamiltonians depending on a time parameter $t\in(0,\pi)$. The
function $\sin t$ in their denominators explicitly shows that both
initial and final singularities are present in these models. This is
in contrast with the 3-torus case, where only an initial (or final)
singularity appears. The usefulness of the deparameterization
employed in this text consists in encoding the local degrees of
freedom of the systems in massless scalar fields evolving in the
\emph{same} fixed background spacetime, conformal to the Einstein
metric on $(0,\pi)\times\mathbb{S}^{2}$, so that one can apply
standard techniques of QFT in order to exactly quantize the models.
In this context, the time singularities of the metric are described
by the time-conformal factor, given again by the function $\sin t$,
so that the metric becomes singular whenever it cancels.
\\
\indent An interesting feature of both the 3-handle and the 3-sphere
models is the fact that after deparameterization there are no
constraints left, so that they can be completely described by the
time-dependent Hamiltonian. This is again in contrast with the
situation for the 3-torus topology where, in addition to the
dynamics generated by the nonautonomous Hamiltonian, there is an
extra $U(1)$ symmetry generated by a residual global constraint that
must be appropriately taken into account.
\\
\linebreak \indent In \emph{Chapter \ref{ChapterII}}, we have
studied the Fock quantization of the models. Concretely, we have
focused our attention on the problem of unitarily implementing the
quantum time evolution. It is expedient to tackle this roblem within
the algebraic formalism of QFT. As a first result, we have proved
the impossibility to get unitary dynamics when the systems are
written in terms of their original variables, irrespectively of the
$SO(3)$-invariant complex structures adapted to the round
$\mathbb{S}^2$ background used to construct the symmetric Fock
spaces. This result generalizes the conclusion reached in $[2]$ for
the 3-torus case to the topologies under consideration in this
thesis. The lack of unitary dynamics could certainly lead us to
conclude that the models quantized in this way are not physically
acceptable. However, one may adopt the point of view proposed in
$[3]$ and $[4]$ according to which these nonunitary Heisenberg
formulations, though pathological in other respects, remain
physically viable thanks to the fact that some relevant quantum
observables can be described as self-adjoint operators and, hence,
their probability interpretations are safe. Indeed, this is the case
for the field and momentum operators, and also for the quantum
Hamiltonian which is self-adjoint for each value of the time
parameter.\footnote{Recall that the dynamics does not generate a
one-parameter symplectic group on the phase space for these models.
Due to this fact, Stone's theorem does not apply and the existence
of a self-adjoint Hamiltonian is not in conflict with the absence of
unitary time evolution.} Nevertheless, it is possible to overcome
the failure of dynamics to be unitarily implemented by performing a
suitable time-dependent redefinition of the fields at the Lagrangian
level involving \emph{precisely} the conformal factor $\sin t$
mentioned above. Furthermore, by demanding the unitarity of the
dynamics and invariance under the $SO(3)$ symmetry associated with
the background metric, the existence of a unique (up to unitary
equivalence) Fock representation can be easily proved for these
systems.
\\
\indent It is important to point out that this method, successfully
applied to the Gowdy models in order to obtain unitary dynamics, is
not of general validity and cannot be generalized to other highly
symmetric spacetime backgrounds. Consider, for instance, the case of
a massless scalar field evolving in a fixed de Sitter background. As
shown in reference $[5]$, it is impossible to find a time-dependent
conformal redefinition of the field leading to a unitary
implementation of the dynamics on any Fock space built from an
$SO(4)$-symmetric complex structure. The ultimate reason that the
method employed in the Gowdy models does not work in this case is
the fact that the time-dependent potential terms appearing in the
equations of motion after the field redefinitions are not as well
behaved as the ones that show up in the treatment of the Gowdy
models. A common situation in this case is that the field ends up
verifying a Klein-Gordon equation with a \emph{tachyonic}
time-dependent mass term.
\\
\linebreak \indent Finally, \emph{Chapter \ref{ChapterIII}} has been
devoted to the construction of the Schr\"{o}dinger representation
for the 3-handle and 3-sphere Gowdy models, completing in this way
the quantization of these systems previously performed within the
Fock scheme. Here, the Hilbert space takes the form of a $L^{2}$
probability space of tempered distributions on the 2-sphere, endowed
with a time-dependent Gaussian measure, whose support is analyzed by
applying the Bochner-Minlos theorem. In particular, we have shown
that the interrelation between measure theory and representation of
quantum operators involves the appearance of unusual linear
multiplicative terms in the momentum operators.
\\
\indent It is important in this context to highlight the advantage
of using the re-scaled fields that make the quantum dynamics
unitary. In this case, the Schr\"{o}dinger representations
corresponding to different values of the time parameter are
unitarily equivalent. This guarantees at the same time the mutual
absolute continuity of the corresponding measures.
\\
\indent As far as the support of the measure or the unitary
implementability of the dynamics are concerned, the discussions and
results obtained for these models are analogous to those found for
the vacuum 3-torus model in $[6]$ and $[7]$. It could be argued that
this similarity is somehow expected owing to the fact that the
critical features of these systems are determined by their
ultraviolet behaviors, and these should not be sensitive to the
topology of the spacetimes. This argument can be found, for
instance, in $[8]$ concerning the simplest generalization of
Minkowski space quantum field theory to the
$\mathbb{R}\times\mathbb{T}^{3}$ spacetime with closed spatial
sections. This compactification can modify the long-wavelength
behavior of the system, but not the ultraviolet one, so that both
spacetimes suffer from the same ultraviolet divergence properties.
Such statement is clearly intuitive, but it is not obvious to what
extent it is also true for quantum field theories in non-locally
isometric spacetimes, like those corresponding to the Gowdy models.
In this respect, the similarity of the results is probably due to
the similar structure of the differential equations verified by the
mode functions.
\\
\indent The final issue discussed in this chapter is the
construction of the unitary evolution operator, written explicitly
in closed form in terms of the field and momentum operators.
Although the resulting operator has been calculated in the
restricted context of the Schr\"{o}dinger quantization, it is clear
that this study remains valid for other faithful representations of
the Weyl algebra, offering the possibility to explore different
choices for the quantization of the systems such as the polymer one
$[9]$.
\\
\linebreak \indent We will conclude by commenting some open problems
and questions that will be tackled elsewhere. To this end, let us
first probe the existence of semiclassical states for the models
under consideration. As shown in \emph{appendix \ref{AppendixTDHO}},
the explicit expression of the quantum unitary evolution for the
single harmonic oscillator as a product of unitary operators turns
out to be very useful to construct semiclassical states for some
relevant one-dimensional dynamical systems (including Gowdy-type
oscillators). However, there are obstructions that arise when
dealing with systems of infinite oscillators --particularly, the
nonunitarity of the $T_{\bm{\rho}}(t)$ operator (see \emph{section
\ref{UnitaryEvolutOper}} in \emph{Chapter \ref{ChapterIII}})--,
making the application of the techniques developed in \emph{appendix
\ref{AppendixTDHO}} particularly difficult. In order to avoid these
difficulties, we will probe an alternative procedure to construct
semiclassical states that takes advantage of the unitary
implementability of the quantum time evolution. We start by
constructing the analogs of the minimal wave packets of the
one-dimensional harmonic oscillator (see for example $[10]$). In
what follows, we will assume the use of a complex structure
$J_{\mathcal{P}}$ such that the dynamics is unitarily implementable
on the associated symmetric Fock space
$\mathscr{F}_{+}(\mathscr{H}_{\mathcal{P}})$ (see \emph{Chapter
\ref{ChapterII}}). Given a square summable sequence
$\mathbf{C}:=(C_{\ell})_{\ell\in\mathbb{N}_0}\in\ell^2(\mathbb{C})$
belonging to the one-particle Hilbert space, consider the state
\begin{equation*}\label{Cstates}
|\mathbf{C}\rangle=\mathrm{e}^{-\|\mathbf{C}\|^{2}/2}\exp\left(\sum_{\ell=0}^{\infty}C_{\ell}\hat{b}_{\ell}^{\dag}\right)|0\rangle_{\mathcal{P}}\,,
\end{equation*}
where $|0\rangle_{\mathcal{P}}$ is the vacuum state, corresponding
in this context to $\mathbf{C}=\mathbf{0}$, and
$\|\mathbf{C}\|^2=\sum_{\ell=0}^{\infty}|C_{\ell}|^{2}$. Vectors
defined in this way appear as coherent superpositions of states with
arbitrary number of particles. Let $\hat{U}(t,t_0)$ be the unitary
evolution operator in the Fock representation, with $t_0\in(0,\pi)$
being a fixed initial value of the time parameter. We can now
introduce the annihilation and creation operators in the Heisenberg
picture corresponding to evolution \emph{backwards} in time,
\begin{eqnarray*}
\hat{\mathfrak{b}}_{\ell}(t_0,t)&:=&\hat{U}(t,t_0)\,\hat{b}_{\ell}\,\hat{U}^{-1}(t,t_0)=\bar{\alpha}_{\ell}(t,t_0|z_\ell)\hat{b}_{\ell}-
\beta_{\ell}(t,t_0|z_{\ell})\hat{b}_{\ell}^{\dag}\,,\\
\hat{\mathfrak{b}}^{\dag}_{\ell}(t_0,t)&:=&\hat{U}(t,t_0)\,\hat{b}_{\ell}^{\dag}\,\hat{U}^{-1}(t,t_0)=-\bar{\beta}_{\ell}(t,t_0|z_\ell)\hat{b}_{\ell}+
\alpha_{\ell}(t,t_0|z_{\ell})\hat{b}_{\ell}^{\dag}\,,
\end{eqnarray*}
satisfying the Heisenberg algebra for all $t,t_0\in(0,\pi)$. Here,
$\alpha_\ell$ and $\beta_\ell$ are the Bogoliubov coefficients
appearing in equation (\ref{Vt0t1}) of \emph{Chapter
\ref{ChapterIII}}. We then evolve the $|\mathbf{C}\rangle$ states in
the Schr\"{o}dinger picture to obtain
\begin{eqnarray*}
|\mathbf{C};t,t_0\rangle:=\hat{U}(t,t_0)|\mathbf{C}\rangle&=&e^{-\|\mathbf{C}\|^{2}/2}\hat{U}(t,t_0)\exp\left(\sum_{\ell=0}^{\infty}C_{\ell}\hat{b}_{\ell}^{\dag}\right)|0\rangle_{\mathcal{P}}\\
&=&e^{-\|\mathbf{C}\|^{2}/2}\exp\left(\sum_{\ell=0}^{\infty}C_{\ell}\hat{\mathfrak{b}}_{\ell}^{\dag}(t_0,t)\right)|0;t,t_0\rangle_{\mathcal{P}}\,,
\end{eqnarray*}
with
$\hat{\mathfrak{b}}_{\ell}(t_0,t)|\mathbf{C};t,t_0\rangle=C_\ell|\mathbf{C};t,t_0\rangle\,$,
$\forall\,\ell\in\mathbb{N}_0$, and
$|0;t,t_0\rangle_{\mathcal{P}}:=\hat{U}(t,t_0)|0\rangle_{\mathcal{P}}$.
By definition, the one-parameter family of states obtained in this
way verifies the Schr\"{o}dinger equation with initial condition
$|\mathbf{C};t_0,t_0\rangle=|\mathbf{C}\rangle$, and is closed under
time evolution as well,
$\hat{U}(t_2,t_1)|\mathbf{C};t_1,t_0\rangle=|\mathbf{C};t_2,t_0\rangle$.
We can now calculate the uncertainties for the field and momentum
operators in the $|\mathbf{C};t,t_0\rangle$ states. We easily obtain
(see \emph{figure \ref{FigureAsymp}})

\begin{eqnarray*}
\Delta_{|\mathbf{C};t,t_0\rangle}\hat{Q}_{\ell}(t)&=&\big|z_\ell(t)\alpha_\ell(t,t_0\,|\,z_\ell)+\bar{z}_\ell(t)\bar{\beta}_\ell(t,t_0\,|\,z_\ell)\big|\sim\frac{1}{\sqrt{2\ell}}\,\,\,\,\,\mathrm{when}\,\,\,\,\,\ell\rightarrow+\infty\,,\\
\Delta_{|\mathbf{C};t,t_0\rangle}\hat{P}_{\ell}(t)&=&\big|\dot{z}_\ell(t)\alpha_\ell(t,t_0\,|\,z_\ell)+\dot{\bar{z}}_\ell(t)\bar{\beta}_\ell(t,t_0\,|\,z_\ell)\big|\sim\sqrt{\frac{\ell}{2}}\,\,\,\,\,\,\,\mathrm{when}\,\,\,\,\,\ell\rightarrow+\infty\,,
\end{eqnarray*}
where, for fixed values of $t_0$, these asymptotic behaviors
converge uniformly in $t$ for time intervals away from the classical
singularities at $t=0$ and $t=\pi$. We then conclude that the
$|\mathbf{C};t,t_0\rangle$ vectors are states of minimum uncertainty
far enough from the singularities. They can be used to probe the
existence of large quantum gravity effects in several ways. For
instance, one may construct suitable regularized operators to
represent the (3- or 4-dimensional) metric of these models by using
arguments similar to those employed in the linearly polarized
Einstein-Rosen waves $[11,12]$ and the Schmidt model $[13]$ (see the
\emph{Table \ref{TableI}} in the \emph{Introduction} to this
thesis). Calculating the expectation values of these operators in
the coherent states, one may deduce the additional conditions (if
any) that the sequences $\mathbf{C}\in\ell^2(\mathbb{C})$ should
satisfy in order to admit an approximate classical behavior. In this
respect, it would be important to analyze if the metric quantum
fluctuation are relevant for all states.
\\
\indent In addition, one may proceed as in $[14]$ for the 3-torus
case by appropriately promoting the quadratic invariant
${^{\cuatro}}R_{abcd}{^{\cuatro}}R^{abcd}$ into a quantum mechanical
operator. According to that reference, one should be able to
unambiguously fix the operator order by requiring that the
expectation values of this quantity in the coherent states exactly
reproduce the classical results far from the singularities. In
analogy with the results of $[14]$, even if the expectation values
in other states (such as linear combinations of coherent states)
give nonclassical results, it is expected that the classical
singularities persist in all cases. This physical consideration is
supported by the purely quantum behavior of the uncertainties of the
field and momentum operators in the coherent states at the classical
spacetime singularities.
\\
\linebreak \indent As a natural extension of the work developed in
this thesis, it would be interesting to couple gravity to different
types of matter, for instance to electromagnetic fields $[15]$, and
find out if it is still possible to exactly solve the resulting
systems. In this case, the dynamics is expected to be entirely
described by the transverse part of the gravitational field and the
components of the electromagnetic vector potential coupled in a
nonlinear way. This nonlinearity should cause an evolution
significantly different from that of the models in vacuum or coupled
to scalar fields. It may be useful in this context to follow a
classification of solutions similar to the one given in $[16]$ for
the 3-torus case. A fact that will play a relevant role here is the
possibility of describing again these reduced models in the
different Gowdy spatial topologies as field theories in certain
conformally stationary curved backgrounds.
\\
\indent Another important issue to study at the classical level is
the explicit characterization of all observables of these models.
The objective here would be to obtain all functions on the phase
space which have (weakly) vanishing Poisson brackets with the
Hamiltonian, momentum, and polar constraints, following the general
procedure outlined in $[17]$ for the vacuum 3-torus case.
\\
\indent In addition, in order to complete the quantization of the
$\mathbb{S}^{1}\times\mathbb{S}^{2}$ and $\mathbb{S}^{3}$ Gowdy
models, one could perform a discussion similar to the one developed
in $[18]$ for the vacuum 3-torus topology in order to prove that the
redefinition of the fields involving the conformal factor $\sin t$
is, in fact, the only reasonable one (up to multiplicative
constants) providing unitary dynamics under the $SO(3)$-invariance
condition.
\\
\indent Finally, one can go beyond the Gowdy cosmologies and cover
more general dynamical systems by considering generic nonautonomous
quadratic Hamiltonians. These can be analyzed from the perspective
of some recent works on this subject $[19]$ in which Lie systems in
quantum mechanics are studied from a geometrical point of view,
developing methods to obtain the time evolution operators associated
with time-dependent Schr\"{o}dinger equations of Lie-type. These
techniques may be successfully applied to infinite-dimensional
quadratic Hamiltonian systems by following a functional description
similar to the one performed in \emph{Chapter \ref{ChapterIII}}. In
particular, the different resulting factorizations for the time
evolution operators may be especially useful to define alternative
families of semiclassical states for these systems.
\\
\linebreak \indent The study of the Gowdy models has made a notable
contribution to the current development of advanced theoretical
cosmology. In particular, we want to remark the usefulness of the
$\mathbb{S}^{1}\times\mathbb{S}^{2}$ and $\mathbb{S}^{3}$ Gowdy
models as testing grounds for quantum gravity theories such as loop
quantum gravity. Even if the classical Hamiltonian formulation is
more complicated for these topologies than for the 3-torus case, one
finally obtains nonautonomous quadratic Hamiltonian systems without
extra constraints. This fact provides a notable simplification of
the quantization process given the unnecessary distinction between
kinematical and physical Hilbert spaces, which is precisely one of
the difficulties encountered in the treatment of the 3-torus model.
In conclusion, we expect that the reader has convinced himself of
the importance of the Gowdy models and other symmetry reductions as
useful systems to gain valuable insights into the mathematical
aspects of general relativity and the current formulation of quantum
field theory in curved spacetimes, as well as to probe the behavior
of gravity in its quantum regime.
\\
\linebreak \linebreak \noindent \textbf{References}
\begin{itemize}
\item [1] M. Varadarajan, Phys. Rev. D \textbf{75}, 044018 (2007).
\item [2] A. Corichi, J. Cortez, and G. A. Mena Marug\'an, Phys. Rev. D \textbf{73}, 084020 (2006).
\item [3] C. G. Torre, Phys. Rev. D \textbf{66}, 084017 (2002).
\item [4] T. Jacobson, in \emph{Conceptual Problems of Quantum Gravity}, edited by A. Ashtekar and
J. Stachel, Birkh\"{a}user, Boston (1991).
\item [5] D. G. Vergel and E. J. S. Villase\~{n}or, Class. Quant. Grav. \textbf{25}, 145008 (2008).
\item [6] C. G. Torre, Class. Quant. Grav. \textbf{24}, 1 (2007).
\item [7] A. Corichi, J. Cortez, G. A. Mena Marug\'an, and J. M. Velhinho, Phys. Rev. D \textbf{76}, 124031 (2007).
\item [8] N. D. Birrell and P. C. W. Davies, \emph{Quantum Fields in Curved Space}, Cambridge University Press, (1982).
\item [9] A. Ashtekar and J. Lewandowski, Class. Quant. Grav. \textbf{18}, L117-28 (2001). A Ashtekar, J. Lewandowski, and H.
Sahlmann, Class. Quant. Grav. \textbf{20}, L11-1 (2003).
\item [10] C. Itzykson and J.-B. Zuber, \emph{Quantum Field Theory}, Dover Publications (2005).
\item [11] A. Ashtekar, Phys. Rev. Lett. \textbf{77}, 4864-4867 (1996).
\item [12] M. E. Angulo and G. A. Mena Marug\'an, Int. J. Mod.Phys. D \textbf{9}, 669-686 (2000).
\item [13] C. Beetle, Adv. Theor. Math. Phys. \textbf{2}, 471-495 (1998).
\item [14] V. Husain, Class. Quant. Grav. \textbf{4}, 1587 (1987).
\item [15] M. Carmeli, Ch. Charach, and S. Malin, Phys. Rep. \textbf{76}, 79-156 (1981).
\item [16] Ch. Charach, Phys. Rev. D \textbf{19}, 3516-3523 (1979).
\item [17] C. G. Torre, Class. Quant. Grav. \textbf{23}, 1543-1556 (2006).
\item [18] J. Cortez, G. A. Mena Marug\'an, and J. M. Velhinho, Phys. Rev. D \textbf{75}, 084027 (2007).
\item [19] J. F. Cari\~{n}ena, J. Grabowski, and G. Marmo, Rept. Math. Phys. \textbf{60}, 237-258
(2007). J. F. Cari\~{n}ena, J. de Lucas, and M. F. Ra\~{n}ada, J.
Phys. A: Math. Theor. \textbf{41}, 304029 (2008). J. F.
Cari\~{n}ena, J. de Lucas, and A. Ramos, accepted for publication in
the Int. J. Theor. Phys., preprint at arXiv:0811.4386.
\end{itemize}


\appendix

\chapter{Symmetry Reduction in General
Relativity}\label{AppendixGeroch}


\pagestyle{fancy} \fancyhf{}

\fancyhead[LO]{\textsc{Appendix A. Symmetry Reduction in General Relativity}} \fancyhead[RO,LE]{\thepage}
\renewcommand{\headrulewidth}{0.6pt}


\indent In this appendix, the method of symmetry reduction developed
by Geroch in $[1]$ is generalized for a nonvacuum 4-dimensional
spacetime with a spacelike hypersurface orthogonal Killing vector
field.\footnote{In reference $[1]$, the reader will find a more
extensive analysis of the vacuum case than the one performed here,
allowing the Killing vector field to be either spacelike or
timelike, and not necessarily hypersurface orthogonal.} Concretely,
we couple gravity to a massless scalar field. At the end of this
study, we introduce a conformal transformation that provides a
notable simplification of the equations of motion. Specifically,
they become equivalent to the 3-dimensional Einstein-Klein-Gordon
equations corresponding to two uncoupled symmetric massless scalar
fields, one of them related to the logarithm of the norm of the
Killing vector field and the other being simply proportional to the
original field. The reader will find \emph{Chapter 10} and
\emph{Appendices C} and \emph{D} of reference $[2]$ particularly
useful in this context.

\section{Geroch reduction technique}

\indent Let
$\left({^{\scriptstyle{(4)}}}\mathcal{M},{^{\scriptstyle{(4)}}}g_{ab}\right)$
be a spacetime with an everywhere spacelike and hypersurface
orthogonal Killing vector field $\xi^{a}$. Let $M$ denote the
collection of all integral curves of $\xi^{a}$, called the
\emph{space of orbits} of $\xi^{a}$ --an element of $M$ is,
therefore, a curve in ${^{\scriptstyle{(4)}}}\mathcal{M}$ everywhere
tangential to $\xi^{a}$. We will assume in the following that $M$ is
a smooth manifold and can be identified with one of the
3-dimensional hypersurfaces embedded in
${^{\scriptstyle{(4)}}}\mathcal{M}$ which is everywhere orthogonal
to the $\xi^{a}$ trajectories, so that each orbit intersects $M$ in
exactly one point. The metric that ${^{\scriptstyle{(4)}}}g_{ab}$
induces on $M$ is
\begin{equation*}
{^{\scriptstyle{(3)}}}g_{ab}={^{\scriptstyle{(4)}}}g_{ab}-\lambda^{-1}\xi_{a}\xi_{b}\,,
\end{equation*}
where $\xi_{a}:={^{\scriptstyle{(4)}}}g_{ab}\xi^{b}$,
$\lambda:={^{\scriptstyle{(4)}}}g_{ab}\xi^{a}\xi^{b}>0$.
\\
\linebreak \indent Let ${^{\scriptstyle{(3)}}}R_{ab}$,
${^{\scriptstyle{(3)}}}\nabla_{a}$ and
${^{\scriptstyle{(3)}}}\square:={^{\scriptstyle{(3)}}}g^{ab}{^{\scriptstyle{(3)}}}\nabla_{a}{^{\scriptstyle{(3)}}}\nabla_{b}$
be, respectively, the Ricci tensor, the Levi-Civita connection and
the d'Alembert operator associated with
${^{\scriptstyle{(3)}}}g_{ab}$. Given a tensor field on the manifold
$M$, say ${T^{a_1\cdots a_m}}_{b_1\cdots b_n}$, the metric
connection ${^{\scriptstyle{(3)}}}\nabla_{a}$ acts according to the
formula $[2]$
\begin{displaymath}
{^{\scriptstyle{(3)}}}\nabla_{e}{T^{a_1\cdots a_m}}_{b_1\cdots
b_n}={^{\scriptstyle{(3)}}}{g^{a_1}}_{a_1^\prime}\cdots{^{\scriptstyle{(3)}}}{g^{a_m}}_{a_m^\prime}
{^{\scriptstyle{(3)}}}{g_{b_1}}^{b_1^\prime}\cdots{^{\scriptstyle{(3)}}}{g_{b_n}}^{b_n^\prime}{^{\scriptstyle{(3)}}}{g_{e}}^{f}\,{^{\scriptstyle{(4)}}}\nabla_{f}{T^{a_1^\prime\cdots
a_m^\prime}}_{b_1^\prime\cdots b_n^\prime}\,,
\end{displaymath}
where ${^{\scriptstyle{(4)}}}\nabla_{a}$ is the Levi-Civita
connection associated with the original $4$-dimensional metric
${^{\scriptstyle{(4)}}}g_{ab}$. Note that the indices of any tensor
field on $M$ can be raised or lowered with either the metric
${^{\scriptstyle{(3)}}}g_{ab}$ or ${^{\scriptstyle{(4)}}}g_{ab}$.
\\
\linebreak \indent We proceed now to calculate the Riemann tensor on
$M$. Let $k^{a}$ be an arbitrary vector field defined on $M$. Then,
\begin{eqnarray}\label{DDk}
{^{\scriptstyle{(3)}}}\nabla_{a}{^{\scriptstyle{(3)}}}\nabla_{b}k_{c}
& = &
{^{\scriptstyle{(3)}}}{g_{a}}^{p}{^{\scriptstyle{(3)}}}{g_{b}}^{q}{^{\scriptstyle{(3)}}}{g_{c}}^{r}{^{\scriptstyle{(4)}}}\nabla_{p}\left({^{\scriptstyle{(3)}}}{g_{q}}^{s}{^{\scriptstyle{(3)}}}{g_{r}}^{t}{^{\scriptstyle{(4)}}}\nabla_{s}k_{t}\right) \nonumber\\
& = &
{^{\scriptstyle{(3)}}}{g_{a}}^{p}{^{\scriptstyle{(3)}}}{g_{b}}^{s}{^{\scriptstyle{(3)}}}{g_{c}}^{t}{^{\scriptstyle{(4)}}}\nabla_{p}{^{\scriptstyle{(4)}}}\nabla_{s}k_{t} \nonumber\\
&-&
\lambda^{-1}{^{\scriptstyle{(3)}}}{g_{a}}^{p}{^{\scriptstyle{(3)}}}{g_{b}}^{q}{^{\scriptstyle{(3)}}}{g_{c}}^{r}\left({^{\scriptstyle{(4)}}}\nabla_{p}\xi_{q}\right)\xi^{s}{^{\scriptstyle{(4)}}}\nabla_{s}k_{r} \nonumber\\
&-&
\lambda^{-1}{^{\scriptstyle{(3)}}}{g_{a}}^{p}{^{\scriptstyle{(3)}}}{g_{b}}^{q}{^{\scriptstyle{(3)}}}{g_{c}}^{r}\left({^{\scriptstyle{(4)}}}\nabla_{p}\xi_{r}\right)\xi^{t}{^{\scriptstyle{(4)}}}\nabla_{q}k_{t}\,,
\end{eqnarray}
where $k_{a}:={^{\scriptstyle{(3)}}}g_{ab}k^{b}$. Next, we
antisymmetrize over indices $a$ and $b$ and eliminate the
derivatives of $k_{c}$ on the right hand side of (\ref{DDk}) by
using the vanishing of the Lie derivative $\mathcal{L}_{\xi}k_{r}=0$
for the second term, and the orthogonality condition
$\xi^{t}k_{t}=0$ for the third one. In this way, we get
\begin{eqnarray*}
{^{\scriptstyle{(3)}}}\nabla_{[a}{^{\scriptstyle{(3)}}}\nabla_{b]}k_{c}
& = &
{^{\scriptstyle{(3)}}}{g_{a}}^{p}{^{\scriptstyle{(3)}}}{g_{b}}^{q}{^{\scriptstyle{(3)}}}{g_{c}}^{r}{^{\scriptstyle{(4)}}}\nabla_{[p}{^{\scriptstyle{(4)}}}\nabla_{q]}k_{r}
\nonumber\\
&+&
\lambda^{-1}{^{\scriptstyle{(3)}}}{g_{a}}^{p}{^{\scriptstyle{(3)}}}{g_{b}}^{q}{^{\scriptstyle{(3)}}}{g_{c}}^{r}\left({^{\scriptstyle{(4)}}}\nabla_{p}\xi_{q}\right)\left({^{\scriptstyle{(4)}}}\nabla_{r}\xi_{s}\right)k^{s}
\nonumber\\
&+&
\lambda^{-1}{^{\scriptstyle{(3)}}}{g_{[a}}^{p}{^{\scriptstyle{(3)}}}{g_{b]}}^{q}{^{\scriptstyle{(3)}}}{g_{c}}^{r}\left({^{\scriptstyle{(4)}}}\nabla_{p}\xi_{r}\right)\left({^{\scriptstyle{(4)}}}\nabla_{q}\xi_{t}\right)k^{t}\,.
\end{eqnarray*}
Since $k_{c}$ is arbitrary, the Riemann tensor
${^{\scriptstyle{(3)}}}R_{abcd}$ of the manifold $M$ is related to
${^{\scriptstyle{(4)}}}R_{abcd}$ of
${^{\scriptstyle{(4)}}}\mathcal{M}$ by the analogue of the first
Gauss-Codazzi relation\footnote{Here, we take the convention
$2{^{(n)}}\nabla_{[a}{^{(n)}}\nabla_{b]}k_{c}=:{{^{(n)}}R_{abc}}^{d}k_{d},\,\,{^{(n)}}R_{ab}:={^{(n)}}{R_{acb}}^{c}\,\,\,(n=3,4)$.
With the aim of obtaining the equation (\ref{Riemann}), we make use
of the identities
${^{\scriptstyle{(4)}}}R_{abcd}=-{^{\scriptstyle{(4)}}}R_{bacd}=-{^{\scriptstyle{(4)}}}R_{abdc}$,
as well as of the Killing equation
${^{\scriptstyle{(4)}}}\nabla_{(a}\xi_{b)}=0$.}
\begin{eqnarray}\label{Riemann}
{^{\scriptstyle{(3)}}}R_{abcd} & = &
{^{\scriptstyle{(3)}}}{g_{[a}}}^{p}{^{\scriptstyle{(3)}}}{g_{b]}}^{q}{^{\scriptstyle{(3)}}}{g_{[c}}^{r}{^{\scriptstyle{(3)}}}{{g_{d]}}^{s}\Big({^{\scriptstyle{(4)}}}R_{pqrs}+2\lambda^{-1}\left({^{\scriptstyle{(4)}}}\nabla_{p}\xi_{q}\right)\left({^{\scriptstyle{(4)}}}\nabla_{r}\xi_{s}\right)
\nonumber \\
& &
+2\lambda^{-1}\left({^{\scriptstyle{(4)}}}\nabla_{p}\xi_{r}\right)\left({^{\scriptstyle{(4)}}}\nabla_{q}\xi_{s}\right)\Big)\,.
\end{eqnarray}
By virtue of the hypersurface orthogonality of $\xi^{a}$, we
have\footnote{Take the expression
${^{\scriptstyle{(4)}}}\nabla_{a}\xi_{b}={\delta_{a}}^{[p}{\delta_{b}}^{q]}\left({^{\scriptstyle{(4)}}}\nabla_{p}\xi_{q}\right)=-\frac{1}{4\lambda}{^{\scriptstyle{(4)}}}\epsilon^{e_{1}e_{2}pq}{^{\scriptstyle{(4)}}}\epsilon_{e_{1}e_{2}ab}\xi_{r}\xi^{r}\left({^{\scriptstyle{(4)}}}\nabla_{p}\xi_{q}\right)$,
with ${^{\scriptstyle{(4)}}}\epsilon$ being the volume form
naturally associated with ${^{\scriptstyle{(4)}}}g_{ab}$ satisfying
${^{\scriptstyle{(4)}}}\nabla_{a}{^{\scriptstyle{(4)}}}\epsilon_{bcde}=0$
and
${^{\scriptstyle{(4)}}}\epsilon_{abcd}{^{\scriptstyle{(4)}}}\epsilon^{abcd}=4!$,
and make use of the relations
${^{\scriptstyle{(4)}}}\epsilon^{[e_{1}e_{2}pq}\xi^{r]}=0$ and
${^{\scriptstyle{(4)}}}\epsilon_{abcd}\xi^{b}{^{\scriptstyle{(4)}}}\nabla^{c}\xi^{d}=0$,
the last one being a direct consequence of the hypersurface
orthogonality of $\xi^{a}$.}
\begin{equation}\label{nablaxi}
{^{\scriptstyle{(4)}}}\nabla_{a}\xi_{b}=\lambda^{-1}\xi_{[b}{^{\scriptstyle{(3)}}}\nabla_{a]}\lambda\,.
\end{equation}
We also require the formula for the second derivative of the Killing
vector field $\xi^{a}$,
\begin{equation}\label{nablanablaxi}
{^{\scriptstyle{(4)}}}\nabla_{a}{^{\scriptstyle{(4)}}}\nabla_{b}\xi_{c}={^{\scriptstyle{(4)}}}{R_{cba}}^{d}\xi_{d}\,.
\end{equation}
Making use of equations (\ref{nablaxi}) and (\ref{nablanablaxi}), we
obtain
\begin{equation*}
{^{\scriptstyle{(3)}}}\square\lambda=\frac{1}{2}\lambda^{-1}{^{\scriptstyle{(3)}}}g^{ab}\left({^{\scriptstyle{(3)}}}\nabla_{a}\lambda\right)\left({^{\scriptstyle{(3)}}}\nabla_{b}\lambda\right)-2{^{\scriptstyle{(4)}}}R_{ab}\xi^{a}\xi^{b}\,.
\end{equation*}
Contracting (\ref{Riemann}) once, using (\ref{nablaxi}) and
(\ref{nablanablaxi}), and taking again the hypersurface
orthogonality of $\xi^{a}$ into account, we get
\begin{equation*}
{^{\scriptstyle{(3)}}}R_{ab}=\frac{1}{2}\lambda^{-1}{^{\scriptstyle{(3)}}}\nabla_{a}{^{\scriptstyle{(3)}}}\nabla_{b}\lambda-\frac{1}{4}\lambda^{-2}\left({^{\scriptstyle{(3)}}}\nabla_{a}\lambda\right)\left({^{\scriptstyle{(3)}}}\nabla_{b}\lambda\right)+{^{\scriptstyle{(3)}}}{g_{a}}^{c}{^{\scriptstyle{(3)}}}{g_{b}}^{d}{^{\scriptstyle{(4)}}}R_{cd}\,.
\end{equation*}
We can then enunciate the following theorem.
\begin{theorem}\label{Ger1}
Consider a system consisting in 4-dimensional gravity minimally
coupled to a massless scalar field $\phi$, so that
\begin{equation}
{^{\scriptstyle{(4)}}}R_{ab}=8\pi
G_{N}\left({^{\scriptstyle{(4)}}}\nabla_{a}\phi\right){^{\scriptstyle{(4)}}}\nabla_{b}\phi\,,\quad
{^{\scriptstyle{(4)}}}\square\phi:={^{\scriptstyle{(4)}}}g^{ab}{^{\scriptstyle{(4)}}}\nabla_{a}{^{\scriptstyle{(4)}}}\nabla_{b}\phi=0\,,\label{Einst}
\end{equation}
where $G_{N}$ denotes the Newton constant. The symmetry of the
system implies $\mathcal{L}_{\xi}\phi=0$. In terms of 3-dimensional
quantities, the basic equations of motion are given by
\begin{eqnarray}
{^{\scriptstyle{(3)}}}R_{ab} & = &
\frac{1}{2}\lambda^{-1}{^{\scriptstyle{(3)}}}\nabla_{a}{^{\scriptstyle{(3)}}}\nabla_{b}\lambda-\frac{1}{4}\lambda^{-2}\left({^{\scriptstyle{(3)}}}\nabla_{a}\lambda\right)\left({^{\scriptstyle{(3)}}}\nabla_{b}\lambda\right)\nonumber\\
&+& 8\pi G_{N}\left({^{\scriptstyle{(3)}}}\nabla_{a}\phi\right){^{\scriptstyle{(3)}}}\nabla_{b}\phi\,,\label{(3)Rab}\\
{^{\scriptstyle{(3)}}}\square\phi & = & -\frac{1}{2}\lambda^{-1}{^{\scriptstyle{(3)}}}g^{ab}\left({^{\scriptstyle{(3)}}}\nabla_{a}\phi\right){^{\scriptstyle{(3)}}}\nabla_{a}\lambda\,,\label{phi}\\
{^{\scriptstyle{(3)}}}\square\lambda & = &
\frac{1}{2}\lambda^{-1}{^{\scriptstyle{(3)}}}g^{ab}\left({^{\scriptstyle{(3)}}}\nabla_{a}\lambda\right)\left({^{\scriptstyle{(3)}}}\nabla_{b}\lambda\right).\label{lambda}
\end{eqnarray}
\end{theorem}
\begin{proof}
Indeed, we may rewrite the last relation in (\ref{Einst}) in the
form
\begin{eqnarray*}
{^{\scriptstyle{(4)}}}\square\phi & = &
{^{\scriptstyle{(3)}}}\square\phi+\lambda^{-1}\xi^{a}\xi^{b}{^{\scriptstyle{(4)}}}\nabla_{a}{^{\scriptstyle{(4)}}}\nabla_{b}\phi\\
&=&{^{\scriptstyle{(3)}}}\square\phi-\lambda^{-1}\xi^{a}({^{\scriptstyle{(4)}}}\nabla_{a}\xi^{b}){^{\scriptstyle{(4)}}}\nabla_{b}\phi\nonumber\\
& = &{^{\scriptstyle{(3)}}}\square\phi+
\frac{1}{2}\lambda^{-1}{^{\scriptstyle{(3)}}}g^{ab}\left({^{\scriptstyle{(3)}}}\nabla_{a}\phi\right){^{\scriptstyle{(3)}}}\nabla_{a}\lambda\,,
\end{eqnarray*}
where we have used the relations
${^{\scriptstyle{(4)}}}\nabla_{a}\left(\xi^{b}{^{\scriptstyle{(4)}}}\nabla_{b}\phi\right)=0$
and (\ref{nablaxi}), the first of them being a consequence of the
vanishing of the Lie derivative $\mathcal{L}_{\xi}\phi=0$.
\end{proof}

\section{Conformal transformation}

\indent Since $\xi^{a}$ is an everywhere spacelike vector field, its
norm $\lambda={^{\scriptstyle{(4)}}}g_{ab}\xi^{a}\xi^{b}$ is a
strictly positive function on ${^{\scriptstyle{(4)}}}\mathcal{M}$
and, hence, the metric
\begin{equation*}
g_{ab}:=\lambda{^{\scriptstyle{(3)}}}g_{ab}
\end{equation*}
arises from ${^{\scriptstyle{(3)}}}g_{ab}$ via a well-defined
conformal transformation. Denote by $R_{ab}$, $\nabla_{a}$ and
$\square$ the Riemann tensor, the Levi-Civita connection and the
d'Alembert operator associated with the new 3-metric $g_{ab}$,
respectively. Recall, in particular, that the action of the
derivative operators ${^{\scriptstyle{(3)}}}\nabla_{a}$ and
$\nabla_{a}$ coincide over scalars. For any scalar function $f$, we
have $[2]$
\begin{equation}\label{nablanablaf}
\nabla_{a}\nabla_{b}f={^{\scriptstyle{(3)}}}\nabla_{a}{^{\scriptstyle{(3)}}}\nabla_{b}f-{C^{c}}_{ab}{^{\scriptstyle{(3)}}}\nabla_{c}f\,,
\end{equation}
where
\begin{equation}\label{Ccab}
{C^{c}}_{ab}:={\delta^{c}}_{(a}{^{\scriptstyle{(3)}}}\nabla_{b)}\log\lambda-\frac{1}{2}{^{\scriptstyle{(3)}}}g_{ab}{^{\scriptstyle{(3)}}}g^{cd}{^{\scriptstyle{(3)}}}\nabla_{d}\log\lambda\,.
\end{equation}
Similarly, we have the relation between the Ricci tensors
\begin{eqnarray}\label{relRiemann}
{^{\scriptstyle{(3)}}}R_{ab} & = & R_{ab}+\frac{1}{2}\nabla_{a}\nabla_{b}\log\lambda + \frac{1}{2}g_{ab}g^{cd}\nabla_{c}\nabla_{d}\log\lambda\nonumber\\
&+&
\frac{1}{4}\left(\nabla_{a}\log\lambda\right)\left(\nabla_{b}\log\lambda\right)-\frac{1}{4}g_{ab}g^{cd}\left(\nabla_{c}\log\lambda\right)\left(\nabla_{d}\log\lambda\right).
\end{eqnarray}
Then, it is straightforward to prove the following theorem.
\begin{theorem}\label{GerochTheorem}
The 4-dimensional system considered in the \emph{Theorem \ref{Ger1}}
can be thought of as 3-dimensional general relativity coupled to two
uncoupled symmetric massless scalar fields $\phi_{1}:=\log\lambda$
and $\phi_{2}:=\sqrt{16\pi G_{N}}\phi$. This reduced model satisfies
the Einstein-Klein-Gordon equations
\begin{eqnarray}
R_{ab}&=&\frac{1}{2}\left(\nabla_{a}\log\lambda\right)\left(\nabla_{b}\log\lambda\right)+8\pi
G_{N}\left(\nabla_{a}\phi\right)\nabla_{b}\phi\,,\label{Rabconf}\\
\square\phi&=&0\,,\quad \square\log\lambda=0\,.\label{EKGlambda}
\end{eqnarray}
\end{theorem}
\begin{proof}
Substituting $f=\phi$ and $f=\log\lambda$ in (\ref{nablanablaf}),
and using (\ref{phi}) and (\ref{lambda}), it is straightforward to
obtain the equations (\ref{EKGlambda}). Finally, writing
${^{\scriptstyle{(3)}}}\nabla_{a}{^{\scriptstyle{(3)}}}\nabla_{b}\lambda$
in terms of quantities associated with the 3-metric $g_{ab}$ as in
(\ref{nablanablaf}) and (\ref{Ccab}), and making use of equations
(\ref{(3)Rab}), (\ref{relRiemann}), and (\ref{EKGlambda}), we
finally obtain (\ref{Rabconf}).
\end{proof}

\noindent \textbf{References}
\begin{itemize}
\item [1] R. Geroch, J. Math. Phys. \textbf{12}, 918-924 (1971); \textbf{13}, 394-404 (1972).
\item [2] R. M. Wald, \emph{General Relativity}, The University of Chicago Press, Chicago and London (1984).
\end{itemize}


\chapter{Symplectic Geometry Applied to Analytical
Mechanics}\label{AppendixSymplectic}


\pagestyle{fancy} \fancyhf{}

\fancyhead[LO]{\textsc{Appendix B. Symplectic Geometry Applied to
Analytical Mechanics}} \fancyhead[RO,LE]{\thepage}
\renewcommand{\headrulewidth}{0.6pt}


\indent The purpose of this appendix is to introduce the basics of
symplectic geometry, profusely applied throughout the thesis, and
show their usefulness for the description of Lagrangian and
Hamiltonian classical systems --assumed in what follows to have a
finite number of degrees of freedom with the aim of simplifying
their analysis. The interested reader will find extensive studies on
this subject in references $[1]$ and $[2]$.

\section{Symplectic manifolds}

\indent Let $V$ be a real $m$-dimensional vector space ($m<+\infty$)
and $\omega\in\Lambda^{2}V$ a 2-form on $V$. Consider the linear
mapping $\chi_{\omega}:V\rightarrow V^{*}$ (with $V^{*}$ the dual
space of $V$) defined as
\begin{equation}\label{chi}
\chi_{\omega}(u):=i_{u}\omega\,,\,\,\,(i_{u}\omega)(v)=\omega(u,v)\,,\,\,\,u,v\in
V\,,
\end{equation}
with $i_{u}\omega$ being the inner product of the vector $u$ by
$\omega$. Let $\mathrm{Im}\chi_{\omega}$ and
$\mathrm{ker}\chi_{\omega}:=\{u\in V\,|\,i_{u}\omega=0\}$ be the
image and the kernel of $\chi_{\omega}$, respectively. We define the
\emph{rank of $\omega$}, denoted $\mathrm{rank}\,\omega$, as the
dimension of $\mathrm{Im}\chi_{\omega}$. This is an even number
smaller or equal to $\mathrm{dim}V$. The dimension of the kernel
$\mathrm{ker}\chi_{\omega}$ is called the \emph{corank of $\omega$},
being denoted $\mathrm{corank}\omega$. If
$\mathrm{corank}\,\omega=0$, then
$\mathrm{dim}V=\mathrm{rank}\,\omega$, and $\omega$ is said to be
\emph{nondegenerate, regular} or \emph{of maximal rank}.
\\
\indent We will say that every nondegenerate 2-form $\omega$ on a
real $m$-dimensional vector space $V$ defines a \emph{symplectic
structure} on it. The form $\omega$ is called \emph{symplectic} in
this case, and the pair $(V,\omega)$ a \emph{symplectic vector
space}. The dimension of $V$ is then an even number, i.e., $m=2n$
for some integer $n$. It is possible to show that, given a real
vector space $V$ of even dimension $2n$ and a 2-form $\omega$ on
$V$, $\omega$ is nondegenerate iff the linear mapping (\ref{chi}) is
an isomorphism, or equivalently, iff
$\omega^{n}:=\omega\wedge\ldots\wedge\omega$ ($n$ times) defines a
volume form on $V$.
\\
\linebreak \indent Let $M^{m}$ be a smooth ($C^{\infty}$)
$m$-dimensional manifold, and $\omega$ a 2-form on $M^{m}$. The rank
(corank) of $\omega$ at a point $x\in M^{m}$ is defined as the rank
(corank) of the form $\omega(x)\in\Lambda^{2}(T_{x}M^{m})$. We will
say that $\omega$ is nondegenerate or of maximal rank if for every
point $x\in M^{m}$, $\omega(x)$ is nondegenerate; in this case, the
2-form $\omega$ defines an \emph{almost symplectic form} on $M^{m}$.
Then, $M^{m}$ has even dimension $m=2n$ and the pair
$(M^{2n},\omega)$ is called an \emph{almost symplectic manifold}.
Note that since $\omega^{n}$ is a volume form on $M^{2n}$ every
almost symplectic manifold is orientable. Furthermore, the linear
mapping $\chi_{\omega}:\mathfrak{X}(M^{2n})\rightarrow
\Lambda^{1}(M^{2n})$ defined by
\begin{equation}\label{isomorf}
\chi_{\omega}(X)=i_{X}\omega\,,\,\,\,X\in \mathfrak{X}(M^{2n})\,,
\end{equation}
is an isomorphism. An almost symplectic form $\omega$ on a manifold
$M^{2n}$ is said to be \emph{symplectic} if it is closed, i.e.,
$\mathrm{d}\omega=0$. In this case, the pair $(M^{2n},\omega)$ is
called a \emph{symplectic manifold}.\footnote{For systems with
infinite degrees of freedom, one must be careful with functional
analysis $[3]$. Consider a Banach space $\mathcal{E}$ and let
$\omega:\mathcal{E}\times\mathcal{E}\rightarrow\mathbb{R}$ be a
continuous linear mapping on it. Define the (also continuous and
linear) mapping
$\chi_{\omega}:\mathcal{E}\rightarrow\mathcal{E}^{*}$, where
$\mathcal{E}^{*}$ denotes the topological dual of $\mathcal{E}$,
such that $\mathcal{E}\ni e\mapsto\chi_{\omega}(e)\,:\,
\chi_{\omega}(e)\cdot f=\omega(e,f)$, $f\in\mathcal{E}$; $\omega$ is
said to be \emph{weakly nondegenerate} if $\chi_{\omega}$ is an
injection, i.e.,
$\omega(e,f)=0,\,\forall\,f\in\mathcal{E}\Rightarrow e=0$; $\omega$
is called \emph{strongly nondegenerate} if $\chi_{\omega}$ is an
isomorphism. Note that, if $\mathcal{E}$ is a finite-dimensional
space, the distinction between weak and strong nondegeneracy is
unnecessary. This is not the case for infinite-dimensional spaces.
Let $\mathcal{P}$ be a manifold modeled on a Banach space
$\mathcal{E}$; a 2-form $\omega$ on $\mathcal{P}$ is called
\emph{symplectic} if: (i) $\omega$ is exact, i.e.,
$\mathrm{d}\omega=0$; (ii) for each point $x\in\mathcal{P}$,
$\omega_{x}:T_{x}\mathcal{P}\times
T_{x}\mathcal{P}\rightarrow\mathbb{R}$ is a strongly nondegenerate
map. If $\omega_{x}$ is weakly nondegenerate, it is said to be
\emph{weakly symplectic}. Although in this appendix we have
restricted ourselves to the study of finite-dimensional systems, the
main results presented here can be properly generalized for weakly
symplectic manifolds.}
\\
\linebreak \indent Let $(M^{2n},\omega)$ and
$(\tilde{M}^{2n},\tilde{\omega})$ be two symplectic manifolds of
same dimension. A \emph{symplectic transformation} is a mapping
$\displaystyle\phi\in C^{\infty}(M^{2n};\tilde{M}^{2n})$ such that
$\phi^{*}\,\tilde{\omega}=\omega$, i.e.,
\begin{equation*}
\tilde{\omega}(\mathrm{d}\phi(x)X_{1},\mathrm{d}\phi(x)X_{2})=\omega(X_{1},X_{2})\,,
\end{equation*}
for all $x\in M^{2n}$, $X_{1},X_{2}\in T_{x}M^{2n}$. The map $\phi$
is a local diffeomorphism; if it is also a global diffeomorphism,
then it is called a \emph{symplectomorphism}. In particular, when
$M^{2n}=\tilde{M}^{2n}$, a symplectic transformation $\phi$
preserves the symplectic form, $\phi^{*}\omega=\omega$, and is
called a (global) \emph{canonical transformation}.
\\
\linebreak \indent Let $(M^{2n},\omega)$ be a symplectic manifold. A
vector field $X$ on $M^{2n}$ is called a \emph{symplectic vector
field} or \emph{infinitesimal canonical transformation} if its flow
consists of symplectic transformations. In this case, the following
statements are equivalent: (i) $X$ is a symplectic vector field;
(ii) $\mathcal{L}_{X}\omega=0$; (iii) $i_{X}\omega=\mathrm{d}f$
(locally) for some function $f$, i.e., $\mathrm{d}(i_{X}\omega)=0$.
The equivalence between (i) and (ii) is straightforward to show,
given the definition of the Lie derivative and the fact that the
flow of $X$, $\varphi_{t}$, is a symplectic transformation:
$\mathcal{L}_{X}\omega:=\mathrm{d}/\mathrm{d}t|_{t=0}(\varphi_{t}^{*}\omega)=\lim_{t\rightarrow
0}(\varphi^{*}_{t}\omega-\omega)/t=0$. The equivalence of (ii) and
(iii) follows from the H. Cartan formula,
$\mathcal{L}_{X}=i_{X}\mathrm{d}+\mathrm{d}i_{X}$, so that
$\mathcal{L}_{X}\omega=\mathrm{d}i_{X}\omega$, and the Poincar\'e
lemma. We can now prove the so-called \emph{Liouville theorem}:
\begin{theorem}[Liouville]
The flow $\varphi_{t}$ of an infinitesimal canonical transformation
preserves the volume form $\omega^{n}$, $\forall\,t\in
I\subseteq\mathbb{R}$.
\begin{proof}
Indeed, since $\varphi_{t}^{*}\omega=\omega$ by definition, we have
$\varphi_{t}^{*}\omega^{n}=\varphi_{t}^{*}(\omega\wedge\ldots\wedge\omega)=(\varphi_{t}^{*}\omega)\wedge\ldots\wedge(\varphi^{*}_{t}\omega)
=(\varphi_{t}^{*}\omega)^{n}=\omega^{n}$.
\end{proof}
\end{theorem}
\indent Finally, we enunciate the \emph{Darboux theorem}, as a
consequence of which any two symplectic manifolds of the same
dimension are locally symplectomorphic:
\begin{theorem}[Darboux]
An almost symplectic manifold $(M^{2n},\omega)$ is symplectic (i.e.,
$\mathrm{d}\omega=0$) iff for each point $x\in M^{2n}$ there exists
a coordinate neighborhood $U$ with local coordinates
$(x^{1},\ldots,x^{2n})$ such that the symplectic form can be written
\begin{equation*}
\omega=\sum_{i=1}^{n}\mathrm{d}x^{i}\wedge\mathrm{d}x^{n+i}
\end{equation*}
in $U$. These coordinates are called \emph{symplectic} or
\emph{canonical coordinates} on $M^{2n}$, and will be denoted from
now on as $x^{i}=q^{i}$, $x^{n+i}=p_{i}$, $1\le i \le n$.
\end{theorem}

\section{Hamiltonian systems}

\subsection{Autonomous Hamiltonian systems}

\indent Let $(M^{2n},\omega)$ be a symplectic manifold. Denote by
$\mathcal{F}(M^{2n})$ the ring of smooth functions
$C^{\infty}(M^{2n};\mathbb{R})$, endowed with the structure of an
infinite-dimensional $\mathbb{R}$-vector space with respect to the
basic vector operations of sum and products by real scalars. Given a
function $H\in\mathcal{F}(M^{2n})$, its exterior derivative
$\mathrm{d}H$ is a 1-form on $M^{2n}$ and, hence, there is by
virtue of the isomorphism (\ref{isomorf}) a unique vector field
$X_{H}:=\chi_{\omega}^{-1}(\mathrm{d}H)\in\mathfrak{X}(M^{2n})$ such
that
\begin{equation}\label{Hamvec}
i_{X_{H}}\omega=\mathrm{d}H
\end{equation}
called the \emph{Hamiltonian vector field associated with the
Hamiltonian function $H$}. The triplet $(M^{2n},\omega,H)$ is said
to characterize an \emph{autonomous Hamiltonian system}. In view of equation
(\ref{Hamvec}), it is clear that every Hamiltonian vector field on
$(M^{2n},\omega)$ defines an infinitesimal canonical
transformation.\footnote{In general, however, an infinitesimal
canonical transformation $X$ on $M^{2n}$ does not define a
Hamiltonian vector field, since an equation of the form
(\ref{Hamvec}) is not necessarily satisfied globally. Nevertheless,
there will be a neighborhood $U$ for each point $x\in M^{2n}$ and a
function $H$ in $U$ such that $X=X_{H}$ in $U$. Owing to this fact,
any infinitesimal canonical transformation is said to be
\emph{locally Hamiltonian}.} This is the more general geometric
framework for the des\-cription of time-independent classical
Hamiltonian systems; in this context, $M^{2n}$ is identified with the
\emph{phase space} of the system and each point $x\in M^{2n}$ represents a
possible (pure) state. Once a point $x_0$ of this space
is fixed as an initial state, the dynamical trajectory of the system
is simply given by the (unique) integral curve of the Hamiltonian
vector field $X_{H}$ crossing that point.
\\
\linebreak \indent Take canonical coordinates $(q^{i},p_{i})$ in
$(M^{2n},\omega)$ and consider the isomorphism
$\chi_{\omega}:X\in\mathfrak{X}(M^{2n})\mapsto
\chi_{\omega}(X)=i_{X}\omega\in\Lambda^{1}M^{2n}$. It is
straightforward to check that $\chi_{\omega}(\partial/\partial
q^{i})=\mathrm{d}p_{i}$, $\chi_{\omega}(\partial/\partial
p_{i})=-\mathrm{d}q^{i}$, so that
$\chi_{\omega}^{-1}(\mathrm{d}q^{i})=-\partial/\partial p_{i}$,
$\chi_{\omega}^{-1}(\mathrm{d}p_{i})=\partial/\partial q^{i}$. From
the previous equations, we deduce that given a vector field $X$ on
$M^{2n}$ with local expression $X=X^{i}\partial/\partial
q^{i}+\tilde{X}^{i}\partial/\partial p_{i}$ (in the following
formulae we will implicitly assume summation over $i$), then
$\chi_{\omega}(X)=-\tilde{X}^{i}\mathrm{d}q^{i}+X^{i}\mathrm{d}p_{i}$.
Similarly, given a 1-form $\alpha$ on $M^{2n}$ locally given by
$\alpha=\alpha_{i}\mathrm{d}q^{i}+\tilde{\alpha}_{i}\mathrm{d}p_{i}$,
then $\chi_{\omega}^{-1}(\alpha)=\tilde{\alpha}_{i}\partial/\partial
q^{i}-\alpha_{i}\partial/\partial p_{i}$. Since
$\mathrm{d}H=(\partial H/\partial q^{i})\mathrm{d}q^{i}+(\partial
H/\partial p_{i})\mathrm{d}p_{i}$, we obtain
\begin{equation}\label{XH}
X_{H}=\chi_{\omega}^{-1}(\mathrm{d}H)=
\sum_{i=1}^{n}\left(\frac{\partial H}{\partial
p_{i}}\frac{\partial}{\partial q^{i}}-\frac{\partial H}{\partial
q^{i}}\frac{\partial}{\partial p_{i}}\right).
\end{equation}
The time evolution (in the Heisenberg picture) of any observable
$f\in\mathcal{F}(M^{2n})$ is then given by
$\dot{f}:=\mathcal{L}_{X_{H}}f$. Note, in particular, that for an
autonomous Hamiltonian system the Hamiltonian $H$ is a \emph{first
integral}, i.e., it remains constant along every curve solution of
the system (indeed, $\mathcal{L}_{X_{H}}H=0$).
\\
\linebreak \indent Let $\sigma:I=(-\epsilon,\epsilon)\rightarrow
M^{2n}$, $\epsilon>0$, be an integral curve of $X_{H}$, i.e.,
$X_{H}|_{\sigma(t)}=\dot{\sigma}(t)$, $t\in I$; in local
coordinates, $\sigma(t)=(q^{i}(t),p_{i}(t))$,
$\dot{\sigma}(t)=(\mathrm{d}q^{i}/\mathrm{d}t)\partial/\partial
q^{i}+(\mathrm{d}p_{i}/\mathrm{d}t)\partial/\partial p_{i}$ (sum
over $i$). Making use of (\ref{XH}), we finally get the so-called
(canonical) \emph{Hamilton equations}
\begin{equation}\label{Hameq}
\frac{\mathrm{d}q^{i}}{\mathrm{d}t}=\frac{\partial H}{\partial
p_{i}}\,,\,\,\,\frac{\mathrm{d}p_{i}}{\mathrm{d}t}=-\frac{\partial
H}{\partial q^{i}}\,,\,\,\,1\le i\le n\,.
\end{equation}

\subsection{Poisson bracket}

\indent Let $(M^{2n},\omega)$ be a symplectic manifold. The
\emph{Poisson bracket} of two functions $f,g\in\mathcal{F}(M^{2n})$
is defined as
\begin{equation}\label{CPoisson}
\{f,g\}:=\omega(X_{f},X_{g})=(i_{X_{f}}\omega)(X_{g})=i_{X_{g}}i_{X_{f}}\omega\,,
\end{equation}
where $X_{f}$ and $X_{g}$ are the Hamiltonian vector fields
associated with the functions $f$ and $g$, respectively. Since
$(i_{X_{f}}\omega)(Y)=(\mathrm{d}f)Y$, then $\omega(X_{f},Y)=Y(f)$.
In particular, choosing $Y=X_{g}$, we get
$\mathcal{L}_{X_{g}}f=X_{g}(f)=\omega(X_{f},X_{g})=\{f,g\}$.
\\
\linebreak \indent The Poisson bracket satisfies, for all
$f,g,h\in\mathcal{F}(M^{2n})$: (i) $\{f,g\}=-\{g,f\}$; (ii)
$\{f,gh\}=\{f,g\}h+g\{f,h\}$; (iii)
$\{f,\{g,h\}\}+\{g,\{h,f\}\}+\{h,\{f,g\}\}=0$ (Jacobi identity), as
a result of the fact that $\omega$ is exact; (iv)
$\{af,g\}=a\{f,g\},\,\forall\,a\in\mathbb{R}$; (v)
$\{f+g,h\}=\{f,h\}+\{g,h\}$. This operation turns the real vector
space $\mathcal{F}(M^{2n})$ into a Lie algebra with the Poisson
bracket as the product. By virtue of the Jacobi identity (iii), it
is immediate to check that
$X_{\{f,g\}}=-[X_{f},X_{g}],\,\,f,g\in\mathcal{F}(M^{2n})$, where we
have defined the Lie bracket $[X,Y](f):=X(Y(f))-Y(X(f))$,
$X,Y\in\mathfrak{X}(M^{2n})$, i.e., the map $f\mapsto X_{f}$
associating to $f$ its corresponding Hamiltonian vector field
$X_{f}$ takes Poisson brackets of pairs of functions to Lie
commutators of vector fields.
\\
\linebreak \indent Taking canonical coordinates $(q^{i},p_{i})$ in
$M^{2n}$, and making use of (\ref{XH}), the Poisson bracket
(\ref{CPoisson}) is given by
\begin{equation}\label{cPoissonQyP}
\{f,g\}=\mathcal{L}_{X_{g}}f=\sum_{i=1}^{n}\left(\frac{\partial
f}{\partial q^{i}}\frac{\partial g}{\partial p_{i}}-\frac{\partial
f}{\partial p_{i}}\frac{\partial g}{\partial q^{i}}\right).
\end{equation}
In particular, we obtain the Poisson brackets of the canonical coordinates,
\begin{equation*}
\{q^{i},q^{j}\}=\{p_{i},p_{j}\}=0\,,\quad
\{q^{i},p_{j}\}=\delta^{i}_{j}\,,\quad 1\le i,j\le n\,.
\end{equation*}
Furthermore, Hamilton equations (\ref{Hameq}) can be rewritten as
\begin{equation}\label{HqnP}
\frac{\mathrm{d}q^{i}}{\mathrm{d}t}=\{q^{i},H\}\,,\,\,\,\frac{\mathrm{d}p_{i}}{\mathrm{d}t}=\{p_{i},H\}\,,\,\,\,1\le
i\le n\,.
\end{equation}
It is easy to show that the Poisson bracket is invariant under the
action of a canonical transformation $\phi:M^{2n}\rightarrow
M^{2n}$, i.e.,
$\phi^{*}\{f,g\}=\{f,g\}\circ\phi=\{f\circ\phi,g\circ\phi\}=\{\phi^{*}f,\phi^{*}g\}$.
As a consequence, canonical transformations preserve the form of the
Hamilton equations (\ref{HqnP}). Indeed, if
$\phi:(q^{i},p_{i})\rightarrow(\tilde{q}^{i},\tilde{p}_{i})$, where
$(q^{i},p_{i})$ and $(\tilde{q}^{i},\tilde{p}_{i})$ are canonical
coordinates, we have
\begin{eqnarray*}
\phi^{*}\{q^{i},H\}&=&\{q^{i}\circ\phi,H\circ\phi\}=\{\tilde{q}^{i},\tilde{H}\}=\frac{\mathrm{d}\tilde{q}^{i}}{\mathrm{d}t}\,,\\
\phi^{*}\{p_{i},H\}&=&\{p_{i}\circ\phi,H\circ\phi\}=\{\tilde{p}_{i},\tilde{H}\}=\frac{\mathrm{d}\tilde{p}_{i}}{\mathrm{d}t}\,,\,\,\,1\le
i\le n\,,
\end{eqnarray*}
with $\tilde{H}:=\phi^{*}H=H\circ\phi$.

\subsection{Cosymplectic manifolds}

\indent Let $V$ be a $(2n+1)$-dimensional real vector space,
$\eta\in\Lambda^{1}(V)$ a 1-form, and $\omega\in\Lambda^{2}(V)$ a
2-form on $V$, respectively. The triplet $(V,\eta,\omega)$ is called
a \emph{cosymplectic vector space} if $\eta\wedge\omega^{n}\neq 0$.
Consider the linear map $\chi_{\eta,\omega}:V\rightarrow V^{*}$,
where $V^{*}$ denotes the dual space of $V$,
\begin{equation*}
\chi_{\eta,\omega}(v):=i_{v}\omega+(\eta(v))\eta\,,\,\,\,\,\forall\,v\in
V\,.
\end{equation*}
This map defines a linear isomorphism iff: (i) $(V,\eta,\omega)$ is
a cosymplectic vector space in the case when $\mathrm{dim}V$ is an odd
number, or (ii) $(V,\omega)$ is a symplectic vector space in the case when
$V$ is even dimensional. Therefore, as for a cosymplectic vector space
$(V,\eta,\omega)$, there exists a unique $\mathcal{R}\in V$, called
the \emph{Reeb vector} of the cosymplectic vector space, such that
$\eta(\mathcal{R})=1$ and $i_{\mathcal{R}}\omega=0$, i.e.,
$\mathcal{R}=\chi_{\eta,\omega}^{-1}(\eta)$.
\\
\linebreak \indent Let $M^{2n+1}$ be a smooth (2n+1)-dimensional
manifold. $M^{2n+1}$ is said to be an \emph{almost cosymplectic
manifold} if there exist $\eta\in\Lambda^{1}(M^{2n+1})$ and
$\omega\in\Lambda^{2}(M^{2n+1})$ such that, for all $x\in M^{2n+1}$,
the triplet $(T_{x}(M^{2n+1}),\eta_{x},\omega_{x})$ is a
cosymplectic vector space. If the $p$-forms $\eta$ and $\omega$ are
also closed, $M^{2n+1}$ is called \emph{cosymplectic}.
\\
\linebreak \indent Let $(M^{2n+1},\eta,\omega)$ be an almost
cosymplectic manifold, $\mathcal{R}$ its Reeb vector field, and
$\chi_{\eta,\omega}:\mathfrak{X}(M^{2n+1})\rightarrow
\Lambda^{1}(M^{2n+1})$ its corresponding isomorphism. Denote by
$\mathcal{F}(M^{2n+1})$ the ring of differentiable real functions on
$M^{2n+1}$. By virtue of the isomorphism $\chi_{\eta,\omega}$, every
function $f\in \mathcal{F}(M^{2n+1})$ has a unique associated vector
field $X_{f}\in\mathfrak{X}(M^{2n+1})$, called the \emph{Hamiltonian
vector field with energy function $f$}, defined by
\begin{equation*}
X_{f}:=\chi_{\eta,\omega}^{-1}(\mathrm{d}f-\mathcal{R}(f)\eta)
\Leftrightarrow
i_{X_{f}}\eta=0\,,\,\,\,i_{X_{f}}\omega=\mathrm{d}f-\mathcal{R}(f)\eta\,.
\end{equation*}
Clearly, this construction generalizes the one corresponding to
Hamiltonian vector fields on symplectic manifolds. The
\emph{evolution vector field} associated with
$f\in\mathcal{F}(M^{2n+1})$ is given by
\begin{equation*}
E_{f}:=\mathcal{R}+X_{f}\,.
\end{equation*}
\noindent Let $(M^{2n+1},\eta,\omega)$ be a cosymplectic manifold.
It is then possible to define a \emph{Poisson bracket} on
$\mathcal{F}(M^{2n+1})$ by $\{f,g\}:=\omega(X_{f},X_{g})=
\mathcal{L}_{X_{g}}f$, with $f,g\in\mathcal{F}(M^{2n+1})$. In this
way, note that $E_{f}(g)=\mathcal{R}(g)+\{g,f\}$,
$g\in\mathcal{F}(M^{2n+1})$.

\subsection{Nonautonomous Hamiltonian systems}

\indent Let $(M^{2n},\omega)$ be a $2n$-dimensional symplectic
manifold. Consider the product manifold $\mathbb{R}\times M^{2n}$
and denote by $\pi:\mathbb{R}\times M^{2n}\rightarrow M^{2n}$ the
canonical projection $\pi(t,x)=x$ on the second factor, where
$t\in\mathbb{R}$ and $x\in M^{2n}$, with $t$ being a global
coordinate on $\mathbb{R}$. Define $\tilde{\omega}:=\pi^{*}\omega$
and $\eta:=\mathrm{d}t$; then, the triplet $(\mathbb{R}\times
M^{2n},\mathrm{d}t,\tilde{\omega})$ is a cosymplectic manifold.
\\
\indent Take a function $H\in\mathcal{F}(\mathbb{R}\times M^{2n})$
with $X_{H}\in\mathfrak{X}(\mathbb{R}\times M^{2n})$ being its
associated Hamiltonian vector field, univocally characterized by the
relations
\begin{equation*}
i_{X_{H}}\tilde{\omega}=\mathrm{d}H-\frac{\partial H}{\partial
t}\mathrm{d}t\,,\,\,\,\,i_{X_{H}}\mathrm{d}t=0\,,
\end{equation*}
where we have taken into account that the Reeb vector field is given
in this case by $\partial/\partial t$. The 4-tuple
$(\mathbb{R}\times M^{2n},\mathrm{d}t,\tilde{\omega},H)$ is said to
define a \emph{nonautonomous Hamiltonian system}. It provides the
proper geometric description for the generalized phase space of a
time-dependent classical Hamiltonian system. Let $(q^{i},p_{i})$ be
canonical coordinates in $M^{2n}$; $(t,q^{i},p_{i})$ are then the
induced coordinates in $\mathbb{R}\times M^{2n}$. Since
$\omega=\sum_{i=1}^{n}\mathrm{d}q^{i}\wedge\mathrm{d}p_{i}$,
$\tilde{\omega}$ takes the same expression,
$\tilde{\omega}=\sum_{i=1}^{n}\mathrm{d}q^{i}\wedge\mathrm{d}p_{i}$.
The corresponding evolution vector field (see the \emph{figure B.1})
is given by
\begin{equation}\label{EH}
E_{H}=\frac{\partial}{\partial t}+X_{H}=\frac{\partial}{\partial
t}+\sum_{i=1}^{n}\left(\frac{\partial H}{\partial
p_{i}}\frac{\partial}{\partial q^{i}}-\frac{\partial H}{\partial
q^{i}}\frac{\partial}{\partial p_{i}}\right).
\end{equation}
\begin{figure}[t] \centering
\includegraphics[width=4cm]{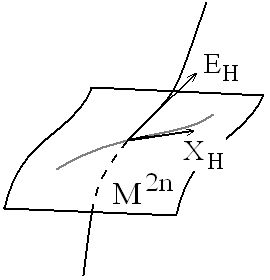}
\caption{Evolution vector field $E_{H}$ corresponding to the
Hamiltonian function $H$.}
\end{figure}
\noindent It is important to highlight that, by introducing the
(closed) 2-form on $\mathbb{R}\times M^{2n}$
\begin{equation}\label{omegaH}
\omega_{H}:=\tilde{\omega}+\mathrm{d}H\wedge\mathrm{d}t\,,
\end{equation}
$E_{H}$ is the unique vector field on $\mathbb{R}\times
M^{2n}$ satisfying
\begin{equation*}
i_{E_{H}}\omega_{H}=0\,,\,\,\,i_{E_{H}}\mathrm{d}t=1\,.
\end{equation*}
\noindent Note that the 2-form (\ref{omegaH}) has the feature that
the time parameter $t$ plays a role analogous to the one played by
the generalized coordinates $q^{i}$, but with minus the Hamiltonian
as its associated canonical
conjugate momentum (see the next section for a definition of these concepts).
\\
\linebreak \indent The integral curves of $E_{H}$, in local
coordinates $\sigma(t)=(a(t),q^{i}(t),p_{i}(t))$,
$\dot{\sigma}(t)=E_{H}|_{\sigma(t)}$, satisfy the equations
$\mathrm{d}a/\mathrm{d}t=1$, i.e., $a(t)=t+c$, $c\in\mathbb{R}$, and
\begin{equation*}
\frac{\mathrm{d}q^{i}}{\mathrm{d}t}=\{q^{i},H\}=\frac{\partial
H}{\partial
p_{i}}\,,\,\,\,\,\frac{\mathrm{d}p_{i}}{\mathrm{d}t}=\{p_{i},H\}=-\frac{\partial
H}{\partial q^{i}}\,,\,\,\,1\le i\le n\,.
\end{equation*}
The above relations are called the \emph{Hamilton equations} for the
nonautonomous Hamiltonian $H$. Here, $\mathcal{L}_{E_{H}}H=\partial
H/\partial t\neq0$ and, hence, the energy is not a constant quantity
for this type of (dissipative) systems.

\section{Autonomous Lagrangian systems}

\indent Let $\mathscr{C}$ be a $n$-dimensional manifold and
$(T\mathscr{C},\tau_{\mathscr{C}},\mathscr{C})$ its tangent bundle,
with $\tau_{\mathscr{C}}:T\mathscr{C}\rightarrow \mathscr{C}$ being
the canonical projection; this is the so-called \emph{phase space of
velocities} associated with the \emph{configuration space}
$\mathscr{C}$. Taking local coordinates $q^{i}$ in $\mathscr{C}$,
let $(q^{i},v^{i})$, $1\le i\le n$, be the induced coordinates in
$T\mathscr{C}$ (i.e., any vector $X$ can be expressed
$X=\sum_{i=1}^{n}v^{i}(\partial/\partial q^{i})$). Consider an
(autonomous) \emph{Lagrangian function}
$L\in\mathcal{F}(T\mathscr{C})$. We introduce the following
associated (closed) 2-form on $T\mathscr{C}$,
\begin{equation*}
\omega_{L}:=-\mathrm{d}\mathrm{d}_{J}L\,,
\end{equation*}
where $J$ denotes the \emph{canonical almost tangent structure} on
$T\mathscr{C}$, locally given by $J=(\partial/\partial
v^{i})\otimes(\mathrm{d}q^{i})$, and $\mathrm{d}_{J}$ is the
\emph{vertical differentiation} on $T\mathscr{C}$, in local
coordinates $d_{J}L=\sum_{i=1}^{n}(\partial L/\partial
v^{i})\mathrm{d}q^{i}$, with
$\mathrm{d}_{J}(\mathrm{d}q^{i})=\mathrm{d}_{J}(\mathrm{d}v^{i})=0$.
Thus, locally,
\begin{equation}\label{omegaL}
\omega_{L}=\sum_{i,j=1}^{n}\left(\frac{\partial^{2}L}{\partial
q^{i}\partial
v^{j}}\mathrm{d}q^{i}\wedge\mathrm{d}q^{j}+\frac{\partial^{2}L}{\partial
v^{i}\partial v^{j}}\mathrm{d}q^{i}\wedge\mathrm{d}v^{j}\right).
\end{equation}
The form $\omega_{L}$ is a symplectic form on $T\mathscr{C}$ iff the
Lagrangian function $L$ is \emph{regular} (or nondegenerate), i.e.,
iff the Hessian matrix
\begin{equation*}
\left[\frac{\partial^{2} L}{\partial v^{i}\partial
v^{j}}\right]_{i,j=1}^{n}
\end{equation*}
is invertible (of maximal rank) for any coordinate
system\footnote{Indeed, under this assumption
$\omega_{L}^{n}=c\,\mathrm{det}(\partial^{2}L/\partial v^{i}\partial
v^{j})\mathrm{d}q^{1}\wedge\ldots\wedge\mathrm{d}q^{n}\wedge\mathrm{d}v^{1}\wedge\ldots\wedge\mathrm{d}v^{n}$,
$c\in\mathbb{R}_{+}$, becomes a volume form for $T\mathscr{C}$.}
$(q^{i},v^{i})$. Otherwise, the function $L$ is said to be
\emph{singular}.
\\
\linebreak \indent We define the \emph{energy function} associated
with the Lagrangian function $L$ by
\begin{equation}\label{EL}
E_{L}:=C(L)-L\,,
\end{equation}
where $C$ denotes the \emph{Liouville vector field} on
$T\mathscr{C}$; in terms of the induced coordinates, it is given by
$C=\sum_{i=1}^{n}v^{i}(\partial/\partial v^{i})$. Consider now the
equation
\begin{equation}\label{EdEL}
i_{X}\omega_{L}=\mathrm{d}E_{L}\,.
\end{equation}
If $L$ is a regular Lagrangian function, then equation (\ref{EdEL})
admits a unique solution $X=\xi_{L}$ (for $\omega_{L}$ is
symplectic) called the \emph{Euler-Lagrange vector field}.
Furthermore, this field is a \emph{second order differential
equation}, i.e., $J\xi_{L}=C$, so that it can be written
\begin{equation}\label{xiL}
\xi_{L}=\sum_{i=1}^{n}\left(v^{i}\frac{\partial}{\partial
q^{i}}+\xi^{i}\frac{\partial}{\partial v^{i}}\right).
\end{equation}
Let $\dot{c}(t)=(q^{i}(t),\dot{q}^{i}(t))$ be an integral curve of
$\xi_{L}$, with $c(t)=(q(t))$ its projection in $\mathscr{C}$ (the
dot denotes time derivative). According to
(\ref{omegaL})-(\ref{xiL}), $c(t)$ verifies the \emph{Euler-Lagrange
equations}
\begin{equation}\label{Euler-Lagrange}
\frac{\partial L}{\partial
q^{i}}-\frac{\mathrm{d}}{\mathrm{d}t}\left(\frac{\partial
L}{\partial \dot{q}^{i}}\right)=0\,,\,\,\,1\le i\le n\,.
\end{equation}

\indent Let $L\in\mathcal{F}(T\mathscr{C})$ be a Lagrangian
function. Fix two points $q_{1}$, $q_{2}$ of $\mathscr{C}$ and some
interval $[a,b]\subset\mathbb{R}$. We define the \emph{path space}
from $q_{1}$ to $q_{2}$ by
\begin{equation}\label{espcaminos}
\Omega(q_{1},q_{2},[a,b]):=\big\{c:[a,b]\rightarrow\mathscr{C}\,\,|\,\,
c\,\,\mathrm{is}\,\,\mathrm{a}\,\,\mathrm{C^{\infty}}\,\,\mathrm{curve},\,c(a)=q_{1},\,c(b)=q_{2}\big\}\,.
\end{equation}
It can be shown that (\ref{espcaminos}) defines a smooth
infinite-dimensional manifold. Consider the map
$\mathrm{J}:\Omega(q_{1},q_{2},[a,b])\rightarrow\mathbb{R}$ given by
\begin{equation*}
\mathrm{J}(c)=\int_{a}^{b}\mathrm{d}t\,L(\dot{c}(t))\,,\,\,c\in\Omega(q_{1},q_{2},[a,b])\,,
\end{equation*}
where $\dot{c}(t)=(q(t),\dot{q}(t))$ is the natural prolongation to
$T\mathscr{C}$ of a curve $c(t)=(q(t))$ in $\mathscr{C}$. In this
context, $q^{i}$ and $\dot{q}^{i}$ are called \emph{generalized
coordinates} and \emph{velocities}, respectively. Then, we can state
the well-known \emph{variational principle} of Hamilton $[2]$, which
becomes the main result of Analytical Mechanics is terms of calculus
of variations:
\begin{theorem}[Hamilton's variational principle]
Let $L\in\mathcal{F}(T\mathscr{C})$ be a Lagrangian function and
$c_{0}\in\Omega(q_{1},q_{2},[a,b])$ a smooth curve joining
$q_{1}=c_{0}(a)$ to $q_{2}=c_{0}(b)$; $c_{0}$ satisfies the
Euler-Lagrange equations (\ref{Euler-Lagrange}) iff it is a
\emph{critical point} of the function
$\mathrm{J}:\Omega(q_{1},q_{2},[a,b])\rightarrow\mathbb{R}$, i.e.,
$\mathrm{d}\mathrm{J}(c_{0})=0$. For regular Lagrangian functions,
this condition amounts to demanding that $\dot{c}_{0}$ be an
integral curve of the field $\xi_{L}$.
\end{theorem}
\indent The Euler-Lagrange equations (\ref{Euler-Lagrange}) may be
written
\begin{equation*}
V_{i}(q,\dot{q})-\sum_{j=1}^{n}W_{ij}(q,\dot{q})\ddot{q}^{j}=0\,,\,\,\,1\le
i\le n\,,
\end{equation*}
where we have defined
\begin{equation*}
V_{i}:=\frac{\partial L}{\partial
q^{i}}-\sum_{j=1}^{n}\frac{\partial^{2}L}{\partial
\dot{q}^{i}\partial q^{j}}\dot{q}^{j}\,,\,\,\,\,
W_{ij}:=\frac{\partial^{2} L}{\partial
\dot{q}^{i}\partial\dot{q}^{j}}\,,\,\,\,1\le i,j\le n\,.
\end{equation*}
Note that the \emph{generalized accelerations} $(\ddot{q}^{i})$ at a
given time are uniquely determined as functions of the generalized
coordinates and velocities $(q^{i},\dot{q}^{i})$, provided that $L$
is regular. This is not the case for singular Lagrangian systems.

\subsection{Legendre transformation}

\indent Let $(T^{*}\mathscr{C},\pi_{\mathscr{C}},\mathscr{C})$ be
the cotangent bundle of the configuration space $\mathscr{C}$, with
$\pi_{\mathscr{C}}$ being its canonical projection; this is the
so-called \emph{phase space} (of momenta) associated with
$\mathscr{C}$. Let $(q^{i},p_{i})$, $1\le i\le n$, be induced
coordinates on $T^{*}\mathscr{C}$ (i.e., every 1-form $\alpha$ takes
the local expression $\alpha=\sum_{i=1}^{n}p_{i}\mathrm{d}q^{i}$).
The manifold $T^{*}\mathscr{C}$ is endowed with a natural symplectic
form,
\begin{equation}\label{Darboux}
\omega_{\mathscr{C}}:=\sum_{i=1}^{n}\mathrm{d}q^{i}\wedge\mathrm{d}p_{i}\,,
\end{equation}
so that the induced coordinates on $T^{*}\mathscr{C}$ are canonical.
As a concrete example, the phase space of a simple pendulum is given
by the cotangent bundle of $\mathbb{S}^{1}$,
$\mathbb{S}^{1}\times\mathbb{R}$, with pairs $(\theta,p_{\theta})$,
$\theta\in\mathbb{S}^{1}$, $p_{\theta}\in\mathbb{R}$; the symplectic
form is simply given in this case by
$\omega=\mathrm{d}\theta\wedge\mathrm{d}p_\theta$.
\\
\indent We will analyze now the relation between the Lagrangian
formulation on $T\mathscr{C}$ and a suitable Hamiltonian formulation
on $T^{*}\mathscr{C}$. Consider a regular Lagrangian function
$L\in\mathcal{F}(T\mathscr{C})$ and let $(q^{i},v^{i})$ and
$(q^{i},p_{i})$ be induced coordinates on $T\mathscr{C}$ and
$T^{*}\mathscr{C}$, respectively. For each tangent vector $v\in
T_{x}\mathscr{C}$, $x\in \mathscr{C}$, let us introduce the natural
identifications (isomorphisms)
$\varphi_{v}:T_{x}\mathscr{C}\rightarrow T_{v}(T_{x}\mathscr{C})$,
$\varphi^{*}_{v}:T_{x}^{*}\mathscr{C}\rightarrow
T_{v}^{*}(T_{x}\mathscr{C})$, given by
\begin{equation}\label{natur}
\varphi_{v}((\partial/\partial q^{i})_{x})=(\partial/\partial
q^{i})_{v}\,,\,\,\,\,\varphi_{v}^{*}((\mathrm{d}q^{i})_{x})=(\mathrm{d}q^{i})_{v}\,.
\end{equation}
Denote by $L_{x}:T_{x}\mathscr{C}\rightarrow\mathbb{R}$ the
restriction of $L$ to $T_{x}\mathscr{C}$. The \emph{Legendre
transformation} (or \emph{fiber derivative}) determined by the
Lagrangian function $L$ is the mapping
$Leg_{L}:T\mathscr{C}\rightarrow T^{*}\mathscr{C}$ such that
\begin{equation}\label{TransfLegendre}
Leg_{L}(v):=(\varphi_{v}^{*})^{-1}(\mathrm{d}L_{x}(v))\,.
\end{equation}
Thus, from (\ref{natur}), we have
\begin{equation*}
Leg_{L}(q^{i},v^{i})=(q^{i},p_{i})=(q^{i},\partial L/\partial v^{i})\,.
\end{equation*}
It follows that $\omega_{L}=(Leg_{L})^{*}\omega_{\mathscr{C}}$, with
the 2-forms $\omega_{L}$ and $\omega_{\mathscr{C}}$ given by
(\ref{omegaL}) and (\ref{Darboux}), respectively. The Legendre
transformation (\ref{TransfLegendre}) defines a local diffeomorphism
iff $L$ is regular. In particular, $L$ is said to be
\emph{hyperregular} if $Leg_{L}$ is a global diffeomorphism.
\\
\linebreak \indent Let $c(t)=(q(t))$ be a smooth curve on
$\mathscr{C}$ and $\dot{c}(t)=(q(t),\dot{q}(t))$ its natural
prolongation to $T\mathscr{C}$. Along $\dot{c}$, we have
$p_{i}=\partial L/\partial \dot{q}^{i}$; in this context, $p_{i}$ is
called the \emph{momentum canonically conjugate to} $q^{i}$.
\\
\linebreak \indent Consider a hyperregular Lagrangian function
$L\in\mathcal{F}(T\mathscr{C})$ and define its associated
Hamiltonian $H:T^{*}\mathscr{C}\rightarrow\mathbb{R}$ by
\begin{eqnarray*}
H:=E_{L}\circ Leg_{L}^{-1}\,.
\end{eqnarray*}
Let $X_{H}$ be the corresponding Hamiltonian vector field. Then,
$(Leg_{L})_{*}\xi_{L}=X_{H}$. The integral curves of $\xi_{L}$ are
mapped by $Leg_{L}$ onto integral curves of $X_{H}$ satisfying the
Hamilton equations (\ref{Hameq}) corresponding to $H$. Furthermore,
these curves have the same projections on $\mathscr{C}$. The
Lagrangian and Hamiltonian formalisms are, thus, globa\-lly
equivalent in the hyperregular case,\footnote{This equivalence is
just local for regular functions.} being transformed one into the
other by the Legendre transformation.

\subsection{Dirac-Bergmann algorithm}

\indent If the Lagrangian function $L$ is singular (i.e., degenerate
or non-regular), $Leg_{L}$ does not define a local diffeomorphism.
Let us assume by hypothesis that the image
$\Gamma_{p}:=Leg_{L}(T\mathscr{C})\subset T^{*}\mathscr{C}$ is an
embedded submanifold of $T^{*}\mathscr{C}$, called the \emph{primary
constraint surface}, and that the rank of the Hessian matrix
$(\partial^{2}L/\partial v^{i}\partial v^{j})$ is constant and equal
to $R=n-M$ on $T\mathscr{C}$. Then, there are $M=n-R$ independent
equations, called the \emph{primary constraints}, that locally
describe the $(2n-M)$-dimensional surface $\Gamma_{p}$,
\begin{equation}\label{ligadprimarias}
\phi_{m}(q^{i},p_{i})=0\,,\,\,\, 1\le m\le M\,.
\end{equation}
These relations directly follow from the definition of the conjugate
momenta, in the sense that when the momenta $p_{i}$ are replaced by
their definitions $p_{i}=\partial L/\partial\dot{q}_{i}$ in terms of
genera\-lized coordinates and velocities the previous equations are
identically satisfied.
\\
\linebreak \indent Given the energy function $E_{L}$ associated with
$L$, the map $Leg_{L}$ projects a function $h$ on $\Gamma_{p}$ such
that $h(Leg_{L}(x))\equiv E_{L}(x)$, $\forall\,x\in T\mathscr{C}$.
If $H$ is an arbitrary extension of $h$ to $T^{*}\mathscr{C}$, all
the Hamiltonian functions of the form
$\tilde{H}=H+\sum_{m=1}^{M}u^{m}\phi_{m}$, with $u^{m}$ being
Lagrange multipliers, are weakly equal\footnote{A function $f$
defined in the neighborhood of $\Gamma_{p}$ is said to be
\emph{weakly zero} if its restriction on $\Gamma_{p}$ vanishes,
$f|_{\Gamma_{p}}=0$. This condition is usually denoted as $f\approx0$.} on $\Gamma_{p}$,
i.e., $\tilde{H}\approx H\approx h$. The corresponding equations of
motion are given by
\begin{eqnarray}
\dot{q}^{i}&=&\mathcal{L}_{X_{\tilde{H}}}(q^{i})=\{q^{i},\tilde{H}\}\,,\label{qn}\\
\dot{p}_{i}&=&\mathcal{L}_{X_{\tilde{H}}}(p_{i})=\{p_{i},\tilde{H}\}\,,\,\,\, 1\le i\le n\,,\\
\phi_{m}&=&0\,,\,\,\,1\le m\le M\,,
\end{eqnarray}
i.e.,
\begin{equation}\label{eqmovHp}
\dot{F}=\mathcal{L}_{X_{\tilde{H}}}(F)=\{F,\tilde{H}\}\,,\,\,\,\phi_{m}=0\,,\,\,\,1\le
m\le M\,,
\end{equation}
for any function $F\in\mathcal{F}(T^{*}\mathscr{C})$, where
$X_{\tilde{H}}$ is the Hamiltonian vector field of $\tilde{H}$. We
clearly see that there is an ambiguity in the description of the
dynamics, characterized by the multipliers $u^{m}$, $1\le m\le M$;
by using them, it is possible to define an invertible mapping from
the $2n$-dimensional phase space of velocities $T\mathscr{C}$ to the
$2n$-dimensional $\Gamma_{p}\times\{u^{m}\}$ space,
\begin{equation*}
q^{i}=q^{i}\,,\,\,\,p_{i}=\frac{\partial L}{\partial
\dot{q}^{i}}(q^{i'},\dot{q}^{i'})\,,\,\,\,u^{m}=u^{m}(q^{i},\dot{q}^{i})\,,\,\,\,
1\le i\le n,\,\,\, 1\le m\le M\,,
\end{equation*}
with inverse transformation
\begin{equation*}
q^{i}=q^{i}\,,\,\,\,\dot{q}^{i}=\frac{\partial H}{\partial
p_{i}}+u^{m}\frac{\partial \phi_{m}}{\partial
p_{i}}\,,\,\,\,\phi_{m}(q^{i},p_{i})=0\,,\,\,\,1\le i\le n,\,\,\,
1\le m\le M\,.
\end{equation*}
A basic consistency requirement for the dynamics is that the primary
constraints be preserved under the time evolution. That is, given an
initial condition $(q^{i},p_{i})$ in $\Gamma_{p}$, the dynamical
trajectory should remain there at later times. We thus impose
$X_{\tilde{H}}(\phi_{m})\approx 0$, $\forall\,m$, so that
$X_{\tilde{H}}$ is tangential to $\Gamma_{p}$,
\begin{equation}\label{Consistencia}
\{\phi_{m},H\}+\{\phi_{m},\phi_{m'}\}u^{m'}\approx 0\,,\,\,\,1\le
m\le M\,.
\end{equation}
For inadmissable Lagrangian functions (for instance, $L=\dot{q}-q$),
these relations will be inconsistent (in the previous example,
$H=q$, $\phi=p-1$, so that $1\approx 0$). The vanishing of
(\ref{Consistencia}) can yield two types of consequences: (i) some
of the arbitrary functions $u^{m}$ is determined or (ii) a new
independent constraint arises. The new constraints so obtained are
called \emph{secondary constraints}; they are consequence of the
definition of the momenta and of the equations of motion as well.
Again, the secondary constraints should be preserved under the
dynamics, so we must impose new consistency conditions. This
process, called the \emph{Dirac-Bergmann algorithm} $[4]$, is
iteratively applied until a final surface defined by primary and
secondary constraints is obtained, where consistent solutions exist.
A different classification of constraints is introduced at this
point, playing a central role in the theory of constrained dynamical
systems; namely, a constraint function is said to be of \emph{first
class} if its Poisson brackets with all the remaining constraints
weakly vanish. Otherwise, it is said to be of \emph{second class}.
In the next section we will focus on purely \emph{first class
systems}, where all primary and secondary constraints are of first
class, since this is the situation of interest in this thesis.

\section{First class constrained manifolds}

\indent Let $V$ be a finite-dimensional vector space,
$\omega\in\Lambda^{2}V$ a 2-form, and $K$ a subspace of $V$. The
subspace $K^{\perp}:=\{u\in V\,|\,\omega(u,v)=0,\,\forall\,v\in K\}$
is called the \emph{orthocomplement of $K$ in $V$ with respect to
$\omega$}. Given a vector $v\in V$, we define $v^{\perp}:=\{u\in
V\,|\,\omega(u,v)=0\}$. One has: (i)
$\mathrm{ker}\chi_{\omega}=V^{\perp}$, which implies
$\mathrm{corank}\,\omega=\mathrm{dim}V^{\perp}$ --see equation
(\ref{chi})--; (ii) $\mathrm{dim}V+\mathrm{dim}(V^{\perp}\cap
K)=\mathrm{dim}K+\mathrm{dim}K^{\perp}$; in particular, if $\omega$
is a symplectic form, then
$\mathrm{dim}V=\mathrm{dim}K+\mathrm{dim}K^{\perp}$.
\\
\indent Consider a symplectic vector space $(V,\omega)$. A subspace
$K\subset V$ is said to be \emph{isotropic}, resp.
\emph{coisotropic}, resp. \emph{Lagrangian}, resp. \emph{symplectic}
in $V$ if $K\subset K^{\perp}$, resp. $K^{\perp}\subset K$, resp.
$K$ is a maximal isotropic subspace\footnote{This amounts to
demanding $K=K^{\perp}$.} of $(V,\omega)$, resp. $K\cap
K^{\perp}=0$.
\\
\linebreak \indent Let $(M^{2n},\omega)$ be a symplectic manifold. A
submanifold $K\subset M^{2n}$ is called \emph{isotropic}, resp.
\emph{coisotropic}, resp. \emph{Lagrangian}, resp. \emph{symplectic}
in $(M^{2n},\omega)$ if, for each $x\in K$,
$T_{x}K\subset(T_{x}K)^{\perp}$, resp. $(T_{x}K)^{\perp}\subset
T_{x}K$, resp. $K$ is a maximal isotropic submanifold of $M^{2n}$,
resp. $(T_{x}K)\cap(T_{x}K)^{\perp}$. Here, $(T_{x}K)^{\perp}$
denotes the orthocomplement of $T_{x}K$ in $T_{x}M^{2n}$ with
respect to $\omega(x)\in\Lambda^{2}(T_{x}M^{2n})$. If $K$ is
isotropic, then $\mathrm{dim}K\le n$; if it is coisotropic,
$\mathrm{dim}K\ge n$, and if it is Lagrangian, $\mathrm{dim}K=n$.
\\
\indent A set of smooth functions
$f_{1},\ldots,f_{k}\in\mathcal{F}(M^{2n})$ is said to be
\emph{independent} if the co\-rresponding Hamiltonian vector fields
$X_{f_{1}},\ldots,X_{f_{k}}$ are linearly independent (or
equivalently, if the 1-forms
$\mathrm{d}f_{1},\ldots,\mathrm{d}f_{k}$ are linearly independent).
It is then possible to prove the following theorem.
\begin{theorem}\label{cosymplectic}
Let $K$ be a $(2n-k)$-dimensional submanifold of $M^{2n}$, locally
defined by the independent functions $f_{1}=\ldots=f_{k}=0$, $k\le
n$. $K$ is coisotropic iff $\{f_{i},f_{j}\}=0$ on $K$, for all $1\le
i,j\le k$. This implies that there exist functions ${t_{ij}}^{l}$ on
$M^{2n}$, $1\le i,j,l\le k$, called \emph{structure functions}, such
that $\{f_{i},f_{j}\}=\sum_{l=1}^{k}{t_{ij}}^{l}f_{l}$.
\end{theorem}
\noindent We can reformulate this result by saying that $K$ is
coisotropic iff the functions $f_{i}$, $1\le i\le k$, weakly
commute, $\{f_{i},f_{j}\}\approx0\,,\,\,\forall\,i,j$. A coisotropic
submanifold corresponds to a \emph{fist class constrained manifold}
in Dirac's terminology, the $f_{i}$ functions being \emph{first
class constraints} in this context. If $\varphi:K\rightarrow M^{2n}$
denotes the embedding, then the symplectic form $\omega$ induces a
2-form $\bar{\omega}$ on $K$, $\bar{\omega}=\varphi^{*}\omega$,
whose rank will be assumed to be constant along $K$. Let $X_{f_{i}}$
be the Hamiltonian vector fields corresponding to the functions
$f_{i}$; these fields are linearly independent and tangential to $K$
by definition. Note that, since the exterior derivative commutes
with the pull-back action, the induced 2-form is closed, although
possibly degenerate. The pair $(K,\bar{\omega})$ is said to define a
\emph{presymplectic} manifold. In fact, it is possible to show that
the induced 2-form $\bar{\omega}$ is, in this case, maximally
degenerate ($\mathrm{rank}\bar{\omega}=2n-2k$), with its kernel
generated by the $X_{f_{i}}$ vector fields.
\\
\linebreak \indent At each point $x\in K$, the vector fields
$X_{f_{i}}$ span a $k$-dimensional subspace $\mathcal{G}_{x}\subset
T_{x}(K)$, assumed to vary smoothly with $x$ in the sense that for
each $x\in K$ there exists an open neighborhood $U\ni x$, such that,
in $U$, $\mathcal{G}$ is generated by $C^{\infty}$ vector fields.
Denote by $\mathcal{G}$ the collection of subspaces
$\mathcal{G}_{x}$. According to the Frobenius integrability theorem,
$\mathcal{G}$ possesses integral submanifolds --i.e., through each
point $x\in K$ we can find an embedded submanifold $S$ such that the
tangent space to this submanifold at each $y\in S$ coincides with
$\mathcal{G}$-- iff $\mathcal{G}$ is involutive, i.e.,
$[Y_{1},Y_{2}]\in\mathcal{G}$, $\forall\,Y_{1},Y_{2}\in\mathcal{G}$.
This is precisely our case, since by virtue of the Jacobi identity,
$[X_{f_{i}},X_{f_{j}}]=-X_{\{f_{i},f_{j}\}}=
-\sum_{l=1}^{k}{t_{ij}}^{l}X_{f_{l}}$, with the structure functions
defined as in \emph{Theorem \ref{cosymplectic}}. Vector fields in
$\mathcal{G}$ are called \emph{constraints vector fields}. As Dirac
pointed out, motion along these directions corresponds, within the
context of the analysis of constrained dynamical systems, to
\emph{gauge transformations} of the physical system, i.e.,
transformations that do not alter the physical state of the system
$[5,6]$.
\\
\linebreak \indent One can recover a symplectic form from the
degenerate $\bar{\omega}$ by taking the space of orbits of the gauge
diffeomorphisms $\Gamma_{R}:=K/\mathcal{G}$ (this is possible
because $\mathcal{G}$ is integrable and
$\mathcal{L}_{X_{f_{i}}}\omega=0\Rightarrow
\mathcal{L}_{X_{f_{i}}}\bar{\omega}=0$), called the \emph{reduced
phase space}. As a different alternative, one can also proceed to
perform a \emph{gauge fixing process} by defining a global section
$\Gamma_{G}\subset K$ intersecting the gauge orbits on $K$ once and
only once (see the figure \emph{B.2}). The number of independent
gauge fixing conditions $C_{a}=0$ describing $\Gamma_{G}$ together
with $f_{i}=0$, $1\le i\le k$, must be equal to the number $k$ of
(independent) first class constraints. The type of intersections
mentioned above is locally guaranteed if $[6]$
\begin{figure}[t] \centering
\includegraphics[width=5.6cm]{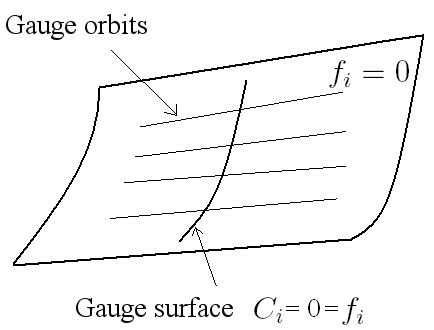}
\caption{Gauge fixing surface $C_{i}=0=f_{i}$ intersecting the gauge
orbits once and only once.}
\end{figure}
\begin{equation}\label{gf}
\mathrm{det}\big[\{C_{i},f_{j}\}\big]\neq 0
\end{equation}
on the gauge surface $\Gamma_{G}$. We may then restrict ourselves to
states lying on that surface, with the pull-back of the 2-form
$\bar{\omega}$ being nondegenerate. It should be taken into account,
however, that such a global cross-section need not always exists,
depending on the geometry of the constraint surface and of the gauge
orbits. This problem is usually referred to as the \emph{Gribov
obstruction}. In such a situation, even if the local condition (\ref{gf})
is fulfilled, the gauge surface would intersect some orbits (at
least) twice, or would not intersect some others.
\\
\linebreak \linebreak \noindent \textbf{References}
\begin{itemize}
\item [1] M. de Le\'on and P. R. Rodrigues, \emph{Methods of Differential Geometry in Analytical
Mechanics}, North Holland, Mathematics Studies 158, 1989.
\item [2] R. Abraham and J. E. Marsden, \emph{Foundations of Mechanics}, 2${^{\mathrm{a}}}$ ed., Benjamin, New York, 1978.
\item [3] P. R. Chernoff and J. E. Marsden, \emph{Properties of Infinite Dimensional Hamiltonian
Systems}, Springer-Verlag, 1974.
\item [4] M. J. Gotay, J.M. Nester, and G. Hinds, J. Math. Phys.
\textbf{19}, 2388 (1978).
\item [5] P. A. M. Dirac, \emph{Lectures on Quantum Mechanics},
Belfer Graduate School of Science Monograph, Yeshiva University, New
York, 1964; Proc. Roy. Soc. (London) A \textbf{246}, 333 (1958);
Phys. Rev. \textbf{114}, 924 (1959).
\item [6] M. Henneaux and C. Teitelboim, \emph{Quantization of Gauge
Systems}, Princeton University Press, 1994.
\item [7] N. M. J. Woodhouse, \emph{Geometric quantization}, Oxford University Press, USA; 2nd edition
(1997).
\end{itemize}


\chapter{Mathematical Structure of Physical Theories}\label{MSPT}


\pagestyle{fancy} \fancyhf{}

\fancyhead[LO]{\textsc{Appendix C. Mathematical Structure of
Physical Theories}} \fancyhead[RO,LE]{\thepage}
\renewcommand{\headrulewidth}{0.6pt}


\indent In this appendix, we briefly revise the algebraic
formulation of (classical and quantum) physical systems in terms of
$C^{*}$-algebras. In our opinion, in the case of atomic systems,
this description is actually better motivated than the usual
Dirac-von Neumann axiomatic structure of quantum theory $[1]$, which
becomes nearly inevitable in this context --except for the so-called
measurement problem and the reduction of the wave packet, concerning
the interaction between the quantum system and the measuring
apparatus, which will not be analyzed here. For a more detailed
study, the reader is strongly suggested to consult the bibliography
given at the end of this appendix and references therein.

\section{Observables and states}

\indent In any physical system it is necessary to properly
distinguish between the measuring instruments and the objects on
which the measurements are performed $[2]$. Denote by $Q$ the
measuring apparatus properly prepared into definite initial conditions and by $\omega$ a \emph{preparation state} of
the object under study.\footnote{It is still under discussion if the word
\emph{state} refers to an individual system or an ensamble. This
topic is beyond the scope of this appendix.} Suppose
that we perform $N$ replicated measurements of $\omega$ by the
instrument $Q$, in such a way that the \emph{measured value} $q\in
D_Q\subset\mathbb{R}$ is obtained $n(q)$ times. For simplicity,
$D_Q$ is assumed to be a discrete set. Of course, it is possible to
reformulate this analysis in terms of continuous random variables
and their associated probability density functions. Nevertheless, it
suffices to consider the discrete version in order to explain the
concepts of observables and states of a system in a simple way. The
foundations of experimental physics assume the existence of the
limit of the ratio $n(q)/N$ as $N\rightarrow+\infty$,
\begin{equation*}
P^{Q}_{\omega}(q)=\lim_{N\rightarrow+\infty}\frac{n(q)}{N}\,,
\end{equation*}
interpreted as the probability that the physical measurement of $Q$
has the value $q$ in the preparation state $\omega$. By definition,
this quantity satisfies the usual probabilistic properties
\begin{equation*}
P^{Q}_{\omega}(q)\in[0,1]\quad\mathrm{and}\quad\sum_{q}P^{Q}_{\omega}(q)=1\,,
\end{equation*}
where the sum must be extended over all possible measured values
$q$. Clearly, if for all preparation states $\omega$ and
experimental results $q$ one obtains the same probabilities
$P_{\omega}^{Q_1}(q)=P_{\omega}^{Q_2}(q)$ for two different
instruments $Q_1$ and $Q_2$, then they must be identified since they
measure the same physical quantity. This introduces an equivalence
relation between the measuring instruments; the set of all
equivalent classes is denoted by $\mathcal{O}$ and its elements are
again identified by the letter $Q$, being called the
\emph{observables} of the system. Similarly, two preparation states
$\omega_1$ and $\omega_2$ cannot be distinguished by any measurement
if the relation $P^{Q}_{\omega_1}(q)=P^{Q}_{\omega_2}(q)$ holds for
all $Q$ and $q$. Again, this defines an equivalence relation called
a \emph{state} of the system, also denoted by $\omega$.
\\
\linebreak \indent Given a real function
$f:\mathbb{R}\rightarrow\mathbb{R}$ and two observables
$Q,Q^{\prime}\in\mathcal{O}$, if for all states $\omega$ and
measured values $q^{\prime}$ the equation
\begin{equation*}
P_{\omega}^{Q^\prime}(q^\prime)=\sum_{q\,:\,f(q)=q^\prime}P^{Q}_{\omega}(q)
\end{equation*}
is satisfied, then the observable $Q^\prime$ is said to be a
function of $Q$, being denoted as $Q^\prime=f(Q)$. For instance,
$f(q)=c q$ defines a rescaling of the
apparatus by $c\in\mathbb{R}$. Note that $Q$ and $Q^{\prime}$ are then
\emph{simultaneously measurable} observables, since that by
measuring $Q$ the experimental value of $Q^{\prime}$ is known at the
same time. More generally, consider a finite set of simultaneously
measurable observables $\{Q_{k}\}_{k=1}^{n}$, all of them functions
of a single observable $Q\in\mathcal{O}$, i.e., $Q_{k}=f_{k}(Q)$,
$(k=1,\ldots,n)$. The \emph{joint probability} of getting the values
$\{q_{k}\}_{k=1}^{n}$ when these observables are measured for the
state $\omega$ is given by
\begin{equation}\label{jointprobability}
P_{\omega}^{Q}\left(\{q_{k}\}_{k=1}^{n};\{Q_{k}\}_{k=1}^{n}\right)=\sum_{q\,:\,f_{k}(q)=q_{k},\,k=1,\ldots,n}P^{Q}_{\omega}(q)\,.
\end{equation}
A (non necessarily finite) set of observables
$\mathcal{C}\subset\mathcal{O}$ is said to be a \emph{full system of
compa\-tible simultaneous measurable observables} if the following
three conditions are verified: (i) Any finite number of observables
belonging to $\mathcal{C}$ can be expressed as functions of a
(probably non-unique) observable $Q\in\mathcal{C}$; (ii) any
function $f(Q)$ of $Q\in\mathcal{C}$ is in $\mathcal{C}$; (iii) the
joint probability (\ref{jointprobability}) is independent of the
observable $Q$ in terms of which the $Q_{k}$ observables can be
written.
\\
\linebreak \indent The \emph{expectation value} of an observable
$Q\in\mathcal{O}$ in the state $\omega$, denoted $\omega(Q)$, is
defined as the average over the results of measurements
\begin{equation}\label{Meanvalue}
\omega(Q):=\sum_{q}qP_{\omega}^{Q}(q)\,.
\end{equation}
Obviously, the state $\omega$ is fully characterized by all the
expectation values $\omega(Q)$ as $Q$ varies over $\mathcal{O}$.
Taking into account that, for simultaneously measurable observables,
the relation
\begin{equation*}
\omega\left(\sum_{k=1}^{n}c_{k}Q_{k}\right)=\sum_{k=1}^{n}c_{k}\omega(Q_{k})
\end{equation*}
is verified, we conclude that any state of a physical system can be
interpreted as a \emph{real linear functional} on the set of
observables $\mathcal{O}$. In addition, this functional is
\emph{posi\-tive}, since that $\omega(Q)\ge0$ for any positive
observable $Q\in\mathcal{O}$, i.e., an observable for which all the
results of measurements are positive real numbers; $Q$ is then of
the form $Q=\tilde{Q}^{2}$, $\tilde{Q}\in\mathcal{O}$. Note that, by
definition of observables and states, $\omega(Q_1)=\omega(Q_2)$ for
all states $\omega$ implies $Q_1=Q_2$ (the states separate the
observables), and conversely $\omega_{1}(Q)=\omega_{2}(Q)$,
$\forall\,Q\in\mathcal{O}$, implies $\omega_1=\omega_2$ (the
observables separate the states).
\\
\linebreak \indent Finally, it is important to distinguish between
two disjoint classes of states, the so-called \emph{pure} and
\emph{mixed states}, the first of them being states that cannot be
expressed as nontrivial convex combinations of two different states,
i.e., $\omega=\lambda\omega_1+(1-\lambda)\omega_2$,
$\lambda\in\mathbb{R}$, is pure iff $\lambda\in\{0,1\}$ or
$\omega_1=\omega_2$ ($\lambda$ arbitrary). Any state that is not
pure is called a mixed state.

\section{Segal systems}

\indent From a purely operational point of view, and making use of
generic arguments regar\-ding the physical properties of states and
observables, Segal established the mathemati\-cal basis for the
description of any (classical or quantum) physical system $[3]$.
Concretely, the set of conditions that the observables in
$\mathcal{O}$ should satisfy can be enunciated as fo\-llows: (i)
$(\mathcal{O},\|\cdot\|)$, with the norm
$\|Q\|:=\sup_{\omega}|\omega(Q)|$, is a real Banach space (thus, it
is assumed in particular that $\mathcal{O}$ is linear); (ii) the
square $Q\mapsto Q^{2}$ is continuous in the norm; (iii)
$\|Q^2\|=\|Q\|^2$ and $\|Q_1^2-Q_2^2\|\le\max(\|Q_1\|^2,\|Q_2\|^2)$,
$\forall\,Q,Q_1,Q_2\in\mathcal{O}$. Note that, following the
operational description of the system, only bounded obser\-vables
($\|Q\|<+\infty$) are considered as basic, since any measurement of
an observable $Q$ must belong to a bounded set of real numbers,
given the intrinsic limitations of the measuring instruments.\\
\indent A Segal system is called \emph{special} if there exists a
$C^{*}$-algebra $\mathfrak{A}$ with identity $\mathbf{1}$
gene\-rated by (complex linear combination of elements of)
$\mathcal{O}$, with $\mathcal{O}$ identified as the subset of
$*$-invariant elements of $\mathfrak{A}$ (i.e., elements satisfying
$Q=Q^{*}$, usually referred to as \emph{self-adjoint}). The system
is called \emph{exceptional} otherwise. Since it is quite difficult
to construct concrete examples of this last class of Segal systems,
and it is not clear their physical usefulness either, we will focus
our attention in the special case only.
\\
\linebreak \indent A self-adjoint element $A\in\mathfrak{A}$ is said
to be \emph{positive} if $A=B^{2}$ for some self-adjoint
$B\in\mathfrak{A}$. It is possible to show that any positive element
of the $C^{*}$-algebra is of the form $A^{*}A$. Any
\emph{state}\footnote{It is possible to prove that the set of
positive linear functionals on a $C^{*}$-algebra $\mathfrak{A}$
keeps separating the elements of the algebra $[4]$. However, one can
think in the possibility that the states with physical
interpretation (\emph{physical states}) is smaller than the set of
all positive linear functional on $\mathfrak{A}$; in this case, the
set of physical states must separate the observables (and,
conversely, the observables separate the states).} is defined as a
linear functional $\omega:\mathfrak{A}\rightarrow\mathbb{C}$
satisfying: (i) $\omega$ is \emph{normalized}, in the sense that
$\omega(\mathbf{1})=1$; (ii) $\omega$ is \emph{positive} on positive
elements, $\omega(A^{*}A)\ge0$, $\forall\,A\in\mathfrak{A}$.
Positivity implies $\omega(A^{*})=\overline{\omega(A)}$, where the
bar denotes complex conjugation, and also that $\omega$ is
continuous, so that $\omega$ belongs to the dual $\mathfrak{A}^{*}$
of $\mathfrak{A}$.
\\
\linebreak \indent Let us now recall the abstract definition of a
$C^{*}$-algebra:

\begin{definit}
A set $\mathfrak{A}$ is called a $C^{*}$-algebra (with identity) if the following properties are satisfied:
\begin{enumerate}
\item $\mathfrak{A}$ is an associative algebra (with identity) with complex numbers $\mathbb{C}$ as the coe\-fficient field.
\item A bijection $*:A\in\mathfrak{A}\mapsto A^{*}\in\mathfrak{A}$ is defined on $\mathfrak{A}$,
satisfying
$(c_{1}A_1+c_{2}A_2)^{*}=\bar{c}_{1}A_{1}^{*}+\bar{c}_{2}A_{2}^{*}$,
$(A_{1}A_2)^{*}=A_{2}^{*}A_{1}^{*}$, and $(A^{*})^{*}=A$, for all
$A_{k}\in\mathfrak{A}$, $c_{k}\in\mathbb{C}$. Here, the bar denotes
complex conjugation. Such a mapping is called an \emph{involution},
and $\mathfrak{A}$ becomes a $*$-algebra.
\item A norm $\|\cdot\|$ is defined on $\mathfrak{A}$, with respect to which the product is continuous, $\|AB\|\le\|A\|\|B\|$, and $\mathfrak{A}$ is complete
respect to the metric topology defined by the norm, referred to as
the \emph{uniform topology}: A neighborhood basis of an element
$A\in\mathfrak{A}$ is given by the sets
$\,\mathscr{U}(A;\epsilon)=\{B\in\mathfrak{A}\,:\,\|B-A\|<\epsilon\}$,
$\epsilon>0$. Furthermore, the normed algebra is assumed to verify
$\|A\|=\|A^{*}\|$, $\forall\,A\in\mathfrak{A}$, so that
$\mathfrak{A}$ is a Banach $*$-algebra.
\item The norm verifies the so-called $C^{*}$\emph{-condition}: $\|A^{*}A\|=\|A\|^{2}$, for all $A\in\mathfrak{A}$.
\end{enumerate}
\end{definit}
\noindent Next, we will look for suitable realizations of this
abstract structure in order to facilitate concrete physical
calculations. For this purpose, taking advantage of the fact that
the set of all bounded linear operators on a Hilbert space
$\mathfrak{H}$, denoted $\mathscr{B}(\mathfrak{H})$, defines a
$C^{*}$-algebra with identity,\footnote{The sums and products of
elements of $\mathscr{B}(\mathfrak{H})$ are defined in the standard
manner, and the set is equipped with the operator norm
$\|A\|:=\sup\{\|A\Psi\|_{\mathfrak{H}}\,;\,\Psi\in\mathfrak{H}\,,\,\|\Psi\|_{\mathfrak{H}}=1\}$,
with $\|\cdot\|_{\mathfrak{H}}$ being the norm defined by the inner
product. The Hilbert space adjoint operation defines an involution
on $\mathscr{B}(\mathfrak{H})$, which becomes a $C^{*}$-algebra.} we
introduce the concept of \emph{representation} as follows:
\begin{definit}
A \emph{representation} $\pi$ of a $C^{*}$-algebra $\mathfrak{A}$
with identity in a Hilbert space $\mathfrak{H}$ is a
$*$-homomorphism of $\mathfrak{A}$ into the $C^{*}$-algebra
$\mathscr{B}(\mathfrak{H})$ of bounded linear operators in
$\mathfrak{H}$, i.e., a linear mapping
$\pi(c_{1}A_{1}+c_{2}A_{2})=c_{1}\pi(A_{1})+c_{2}\pi(A_{2})$,
$*$-preserving $\pi(A^{*})=(\pi(A))^{*}$, and multiplicative
$\pi(A_{1}A_{2})=\pi(A_{1})\pi(A_{2})$,
$\pi(\mathbf{1}_{\mathfrak{A}})=\mathbb{I}_{\mathfrak{H}}$, with
$\mathbb{I}_{\mathfrak{H}}$ being the identity operator. The
representation is called \emph{faithful} if it is injective
(i.e.,
$\mathrm{ker}(\pi)=\{\Psi\in\mathfrak{H}\,:\,\pi(A)\Psi=0\,,\,\forall\,A\in\mathfrak{A}\}=\{0\}$;
$\pi$ is called a $*$-isomorphism in this case). The representation
is \emph{irreducible} if $\{0\}$ and $\mathfrak{H}$ are the only
closed subspaces invariant under $\pi$; in this case, every vector
$\Psi\in\mathfrak{H}$ is \emph{cyclic}, i.e.,
$\pi(\mathfrak{A})\Psi:=\{\pi(A)\Psi\,;\,A\in\mathfrak{A}\}$ is
dense in $\mathfrak{H}$. Two representations of the same algebra
$\mathfrak{A}$,
$\pi_{i}:\mathfrak{A}\rightarrow\mathscr{B}(\mathfrak{H}_{i})$,
(i=1,2), are said to be \emph{unitarily equivalent} if there exists
a unitary transformation
$U:\mathfrak{H}_{1}\rightarrow\mathfrak{H}_{2}$ verifying
$U\pi_{1}(A)U^{-1}=\pi_{2}(A)$, $\forall\,A\in\mathfrak{A}$.
\end{definit}
\noindent Now, we can enunciate the following fundamental result
due to Gel'fand, Naimark, and Segal (see $[4]$ for a proof):
\begin{theorem}[GNS construction]\label{GNS}
Given a state $\omega$ over a $C^{*}$-algebra $\mathfrak{A}$ with
identity, there exist a Hilbert space $\mathfrak{H}_{\omega}$ and a
representation
$\pi_{\omega}:\mathfrak{A}\rightarrow\mathscr{B}(\mathfrak{H}_{\omega})$
such that:
\begin{enumerate}
\item $\mathfrak{H}_{\omega}$ contains a cyclic vector $\Psi_{\omega}$, i.e., $\overline{\pi(\mathfrak{A})\Psi_\omega}=\mathfrak{H}_{\omega}$, with the bar denoting closure.
\item $\omega(A)=\langle\Psi_{\omega}\,|\,\pi_{\omega}(A)\Psi_{\omega}\rangle_{\mathfrak{H}_{\omega}}$, for all $A\in\mathfrak{A}$, where $\langle\cdot|\cdot\rangle_{\mathfrak{H}_{\omega}}$ denotes the inner product in $\mathfrak{H}_{\omega}$.
\item Every other representation $\pi$ in a Hilbert space $\mathfrak{H}$ with a cyclic vector $\Psi$ such that $\omega(A)=\langle\Psi\,|\,\pi(A)\Psi\rangle_{\mathfrak{H}}$, $\forall\,A\in\mathfrak{A}$, is unitarily equivalent to $\pi_{\omega}$,
i.e., there exists a unitary transformation $U:\mathfrak{H}\rightarrow\mathfrak{H}_{\omega}$ satisfying
$U\pi(A)U^{-1}=\pi_{\omega}(A)$, $\forall\,A\in\mathfrak{A}$, and $U\Psi=\Psi_{\omega}$.
\end{enumerate}
The set $(\mathfrak{H}_{\omega},\pi_{\omega},\Psi_{\omega})$
satisfying these conditions is called the \emph{GNS triplet},
containing the \emph{cyclic representation space}
$\mathfrak{H}_{\omega}$, the \emph{cyclic representation}
$\pi_{\omega}$, and the \emph{cyclic vector} $\Psi_{\omega}$
associated with the state $\omega$.
\end{theorem}

\indent Note that every \emph{unit} vector $\Phi$ of the cyclic
representation space $\mathfrak{H}_{\omega}$ defines a state
$\omega_{\Phi}$ on $\mathfrak{A}$ through the formula
$\omega_{\Phi}(A):=\langle\Phi\,|\,\pi(A)\Phi\rangle_{\mathfrak{H}_{\omega}}$,
$\forall\,A\in\mathfrak{A}$; indeed, this is a positive normalized
linear functional on $\mathfrak{A}$. The GNS construction then
provides a mapping between states and Hilbert space vectors, usually
called \emph{state vectors} in this context. If $\Phi$ is a cyclic
vector, then according to \emph{point 3} of GNS theorem, the
representation $\pi_{\omega}$ is unitarily equivalent to the cyclic
representation defined by the state $\omega_{\Phi}$. It is possible
to prove the following result $[4]$:
\begin{theorem}
The cyclic representation $\pi_{\omega}$ is irreducible iff the state $\omega$ is pure.
\end{theorem}
\noindent It easily follows from the above theorem that, when dealing with
an irreducible GNS construction, any state vector $\Phi$ of
the cyclic representation space $\mathfrak{H}_{\omega}$ defines a
pure state $\omega_{\Phi}$ on $\mathfrak{A}$. In such a situation,
it is not possible to represent a mixed state by a state vector, but
rather by a \emph{density operator}: Given a
positive trace class operator $D$ on $\mathfrak{H}_{\omega}$
($\mathrm{Tr}(|D|)<+\infty$), with trace equal to one
($\mathrm{Tr}(D)=1$), the formula
\begin{equation}\label{Traceclass}
\omega_{D}(A):=\mathrm{Tr}(D\pi_{\omega}(A))\,,\,\,\,\forall\,A\in\mathfrak{A}\,,
\end{equation}
defines a state on $\mathfrak{A}$. The set
$\mathfrak{F}(\pi_{\omega})$ of all states of this form (called
\emph{normal states}) defines the \emph{folium of the
representation} $\pi_{\omega}$. Pure states are included in this
class if $D$ is a one-dimensional projection.
\\
\linebreak \indent  A state $\omega$ is \emph{faithful} if
$\omega(A^{*}A)>0$, $\forall\,A\neq0$, and then the corresponding
GNS representation is also faithful. In general, the GNS realization
of a $C^{*}$-algebra $\mathfrak{A}$ as a family of operators in a
Hilbert space may not be a $*$-isomorphism. Nevertheless, the
Gel'fand-Naimark theorem guarantees the existence of at least one
faithful representation $[4]$:
\begin{theorem}[Gel'fand-Naimark characterization of $C^{*}$-algebras]\label{GNtheorem}
A $C^{*}$-algebra is isomorphic to an algebra of bounded operators
in a Hilbert space.
\end{theorem}
\noindent This result encodes the Dirac-von Neumann quantum theory
axiom according to which the observables of any quantum system are
realized as bounded operators in a Hilbert space. Only for
\emph{abelian} (or \emph{commutative}) algebras (as characteristic
of classical systems) this representation is equivalent to a
description in terms of continuous functions, the states acting in
this case as probability measures (see next section) $[4]$:
\begin{theorem}[Gel'fand-Naimark characterization of abelian $C^{*}$-algebras]\label{GNtheoremAbel}
Let $\mathfrak{A}$ be an abelian $C^{*}$-algebra with identity. A
\emph{character} of $\,\mathfrak{A}$ is a nonzero linear map
$\omega:\mathfrak{A}\rightarrow\mathbb{C}$ such that
$\omega(AB)=\omega(A)\omega(B)$, $\forall\,A,B\in\mathfrak{A}$. The
\emph{Gel'fand spectrum} of $\mathfrak{A}$, denoted
$sp(\mathfrak{A})$, is defined as the set of all characters on
$\mathfrak{A}$. It is a subset of the dual $\mathfrak{A}^{*}$ of
$\mathfrak{A}$. It is proved that a state $\omega$ is pure iff it is
a character. The set $sp(\mathfrak{A})$, endowed with the weak*
topology\footnote{In the weak* topology, a neighborhood basis of an
element $\omega\in\mathfrak{A}^{*}$ is indexed by finite sets of
elements $A_{1},\ldots,A_{n}\in\mathfrak{A}$, and $\epsilon>0$; one
has the sets
$\mathscr{U}(\omega;A_{1},\ldots,A_{n};\epsilon)=\{\omega^{\prime}\in\mathfrak{A}\,:\,
|\omega^{\prime}(A_{i})-\omega(A_{i})|<\epsilon\,,\,\,i=1,\ldots,n\}$.}
inherited from the dual $\mathfrak{A}^{*}$ of $\mathfrak{A}$, is a
compact Hausdorff topological space. Moreover, $\mathfrak{A}$ is
isometrically isomorphic to the $C^{*}$-algebra of continuous
functions over $sp(\mathfrak{A})$.
\end{theorem}
\noindent Given a state $\omega$ (i.e., a normalized positive linear
functional) on the $C^{*}$-algebra $\mathfrak{A}$, the Riesz-Markov
representation theorem then ensures the existence of a unique
associated probability measure $\mu_{\omega}$ on $sp(\mathfrak{A})$
such that
\begin{equation*}
\omega(A)=\int_{sp(\mathfrak{A})}f_{A}\,\mathrm{d}\mu_{\omega}\,,\,\,\,\mu_{\omega}(sp(\mathfrak{A}))=\omega(\mathbf{1})=1\,,
\end{equation*}
where $f_{A}$ is the \emph{Gel'fand transform} of $A\in\mathfrak{A}$
assigned by the isomorphism.

\subsection{Classical systems}

Let us consider a classical system described by a phase space
$\Gamma$, that will be assumed to be compact in order to facilitate
subsequent discussions.\footnote{Note, however, that this excludes
phase spaces characterized by cotangent bundles. The possibility to
consider these cases will be discussed later.} This is the case if
the system under study is confined into a finite spatial region and
its energy is also bounded. The classical observables will belong to
a proper class of functions on $\Gamma$, for instance the continuous
real functions $\mathcal{O}_{c}=C(\Gamma;\mathbb{R})$. It is then
straightforward to define an \emph{abelian} $C^{*}$-algebra
$\mathfrak{A}_{c}$ with identity by considering the complex
continuous functions $C(\Gamma;\mathbb{C})$ (the algebraic product
being the pointwise composition of functions), where the identity
$\mathbf{1}$ is the unit function\footnote{For noncompact phase
spaces, the continuous functions are restricted to vanish at
infinity, so that $f=1$ is not an observable and the resulting
$C^{*}$-algebra is not unital. The absence of an identity can to a
large extent be avoided, however, by embedding this algebra into
another suitable one with identity $[5]$.} $f=1$, the $*$ bijection
is given by the standard complex conjugation
$\,\bar{}:\mathbb{C}\rightarrow\mathbb{C}$, and the norm for
elements $f\in\mathfrak{A}_{c}$ is defined as
$\|f\|:=\sup_{x\in\Gamma}|f(x)|$. The product is continuous in the
norm topology since $\|fg\|\le\|f\|\|g\|$, and the $C^{*}$-condition
$\|f^{*}f\|=\|f\|^{2}$ is obviously verified. $\mathcal{O}_{c}$
coincides with the class of functions satisfying $f=\bar{f}=f^{*}$,
and we clearly deal with a special Segal system.
\\
\linebreak \indent Given a state $\omega$ on the abelian
$C^{*}$-algebra $\mathfrak{A}_{c}$ of continuous functions on the
compact (Hausdorff) phase space $\Gamma$, the Riesz-Markov
representation theorem guarantees the existence of a unique
associated probability measure $\mu_{\omega}$ on $\Gamma$ such that
\begin{equation*}
\omega(f)=\int_{\Gamma}f(x)\,\mathrm{d}\mu_{\omega}(x)\,,\,\,\,\,f\in\mathcal{O}_{c}\,,
\end{equation*}
with $\mu_{\omega}(\Gamma)=\omega(\mathbf{1})=1$. Conversely, every
probability measure $\mu$ defines a state $\omega_{\mu}$ on
$\mathfrak{A}_{c}$ through the formula
$\omega_{\mu}(f)=\int_{\Gamma}f(x)\,\mathrm{d}\mu(x)$ and, thus, we
can identify the classical space of states with the space of
probability measures on $\Gamma$. In particular, pure states --those
that cannot be expressed as convex linear combinations of other
states-- correspond to singular $\delta$ measures, i.e., probability
measures concentrated on definite points $x_{0}\in\Gamma$, in such a
way that $\omega_{x_{0}}(f)=f(x_{0})$. Note that for this class of
states the mean square deviation or \emph{variance} relative to
$\omega_{x_0}$ of \emph{any} observable $f\in\mathcal{O}_{c}$,
\begin{equation*}
\Delta_{\omega_{x_0}}^{2}(f):=\omega_{x_0}(f^2)-\omega_{x_0}(f)^2\,,
\end{equation*}
is identically zero. This is the reason why these states are also
called \emph{dispersion free states}. The idealized nature of such
states is a consequence not only of the experimental impossibility
to determine with infinite precision the position and momentum of
particles, but also of the need to perform some type of statistical
description when the number of degrees of freedom of the system is
too large, typically $\sim 10^{23}$, owing to the unfeasibility of
setting out an initial value problem in this case. The realistic
states define, in this way, probability distributions on the
\emph{random variables} describing the observables of the system.
From a theoretical point of view, however, there is no obstruction
to closely approximate the idealized pure states, obtaining states
for which the dispersion of the configuration and momentum variables
are arbitrarily small. This fact lies on the assumption that the
algebra of observables is commutative, as will be clarified in the
next section.

\subsection{Quantum systems}

\indent The following theorem is easily shown to be true for any
noncommutative $C^{*}$-algebra $\mathfrak{A}$ (the result is trivial
for the abelian case):
\begin{theorem}[Heisenberg uncertainty relations]
Given two observables $A,B\in\mathfrak{A}$, ($A=A^{*}$, $B=B^{*}$),
the inequality
\begin{equation*}
\Delta_{\omega}(A)\cdot\Delta_{\omega}(B)\ge\frac{1}{2}|\omega([A,B])|
\end{equation*}
holds for any state $\omega$ on $\mathfrak{A}$, where
$\Delta_{\omega}(A)=(\omega(A^2)-\omega(A)^2)^{1/2}$ denotes the
variance of $A$ respect to $\omega$, and $[A,B]:=AB-BA$ is the
\emph{commutator} of the observables.
\begin{proof}
Define the observables $A^{\prime}:=A-\omega(A)\mathbf{1}$ and
$B^{\prime}:=B-\omega(B)\mathbf{1}$. Given the positi\-vity of
$(A^{\prime}-i\lambda B^{\prime})(A^{\prime}+i\lambda B^{\prime})$,
$\forall\,\lambda\in\mathbb{R}$, one has
$\omega({A^{\prime}}^2)+\lambda^2\omega({B^{\prime}}^2)+\lambda\omega(i[A^{\prime},B^{\prime}])\ge0$;
the positive-definiteness of this quadratic form in $\lambda$
requires
$4\omega({A^{\prime}}^2)\omega({B^{\prime}}^2)\ge|\omega(i[A^{\prime},B^{\prime}])|^2$,
so that
$\Delta_{\omega}(A)\cdot\Delta_{\omega}(B)\ge|\omega([A,B])|/2$.
Here, we have made use of the equivalent expression
$\Delta_{\omega}(A)^{2}=\omega((A-\omega(A)\mathbf{1})^2)$ and the
fact that $[A^{\prime},B^{\prime}]=[A,B]$.
\end{proof}
\end{theorem}
\noindent Using convincing physical arguments, Heisenberg showed
that it is not possible to measure the position $X$ of an atomic
particle without affecting its momentum $P$, in such a way that the
uncertainties of the components of these observables in definite
spatial directions are subject to verify the
inequalities\footnote{Here, subscript denotes the component of the
observables in the $k$-th spatial direction.}
$\Delta_{\omega}(X_{k})\cdot\Delta_{\omega}(P_{k^{\prime}})\ge(\hbar/2)\delta(k,k^{\prime})$,
for all physical states. According to the theorem proved above, it
is then expected that the commutator of these quantities satisfies
the \emph{Heisenberg commutation relations}\footnote{On the
contrary, it is possible to measure any two components of the
position (or the momentum) simultaneously, so that the corresponding
commutators must vanish,
$[X_{k},X_{k^{\prime}}]=0=[P_{k},P_{k^{\prime}}]$.}
\begin{equation*}
[X_{k},P_{k^{\prime}}]=i\hbar\delta(k,k^{\prime})\mathbf{1}\,.
\end{equation*}
Thus, the atomic particles are characterized by noncommutative
algebras. Only in the classical limit $\hbar\rightarrow 0$, where
the perturbing effects of the measurement processes can be ignored,
we recover the abelian algebras.
\\
\linebreak \indent In order to analyze the probabilistic
interpretation of quantum physics, take a \emph{normal} element
$A\in\mathfrak{A}$, (by definition, $A$ and $A^{*}$ are assumed to
commute) and construct the abelian $C^{*}$-algebra $\mathfrak{A}(A)$
generated by $\{\mathbf{1},A,A^{*}\}$. The Gel'fand spectrum
$sp(\mathfrak{A}(A))$ coincides in this case with the
spectrum\footnote{More generally, if $\mathfrak{A}$ is an abelian
$C^{*}$-algebra generated by
$\{\mathbf{1},A_{i},A_{i}^{*}\}_{i=1,\ldots,n}$ (the set assumed to
be algebraically independent), then the Gel'fand spectrum of
$\mathfrak{A}$ is given by the Cartesian product
$sp(\mathfrak{A})=\times_{i=1}^{n}\sigma(A_{i})$.} $\sigma(A)$ of
$A$, i.e.,
\begin{equation*}
sp(\mathfrak{A}(A))\equiv\sigma(A):=\{\lambda\in\mathbb{C}\,:\,(\lambda\mathbf{1}-A)\,\,\mathrm{does}\,\,\mathrm{not}\,\,\mathrm{have}\,\,\mathrm{a}\,\,\mathrm{two}\,\,\mathrm{sided}\,\,\mathrm{inverse}\,\,\mathrm{in}\,\,\mathfrak{A}\}\,.
\end{equation*}
According to \emph{Theorem \ref{GNtheoremAbel}} and the discussion
below it, given a generic state $\omega$ on $\mathfrak{A}(A)$ there
exists a probability measure $\mu_{\omega,A}$ supported on
$\sigma(A)$ such that
\begin{equation*}
\omega(B)=\int_{\sigma(A)}f_{B}(\lambda)\,\mathrm{d}\mu_{\omega,A}(\lambda)\,,\,\,\,\,B\in\mathfrak{A}(A)\,,
\end{equation*}
where $f_{B}$ denotes the Gel'fand transform of $B$. In particular,
\begin{equation*}
\omega(A)=\int_{\sigma(A)}\lambda\,\mathrm{d}\mu_{\omega,A}(\lambda)\,.
\end{equation*}
If the element $A$ is an observable ($A=A^{*}$), then it can be
shown that\footnote{For a positive element, $\sigma(A)$ is a subset
of the positive half-line,
$\sigma(A)\subseteq[0,\|A\|]\subset\mathbb{R}_{+}$.}
$\sigma(A)\subseteq[-\|A\|,\|A\|]\subset\mathbb{R}$, and the (Born)
probabilistic interpretation of measurements on $A$ is the
following: The possible measured values of the observable $A$ belong
to its spectrum $\sigma(A)$, and the probability that the observable
takes values within the Borel set
$\Delta\in\mathrm{Bor}(\mathbb{R})$ when the system is in the state
$\omega$ is given by $\mu_{\omega,A}(\Delta)$; compare this with
equation (\ref{Meanvalue}). In the GNS cons\-truction context,
$\mu_{\omega,A}(\Delta)=\langle\Psi_{\omega}\,|\,E^{\pi_{\omega}(A)}(\Delta)\Psi_{\omega}\rangle_{\mathfrak{H}_{\omega}}$,
where $E^{\pi_{\omega}(A)}$ denotes the unique spectral measure
associated with the (bounded and self-adjoint) operator
$\pi_{\omega}(A)$. The two main differences of noncommutative
algebras with respect to the abelian case are the following: (i) The
probability measure $\mu_{\omega,A}$ depends on the observable $A$;
as a consequence of this, the algebra of observables $\mathfrak{A}$
cannot be realized as an algebra of random variables on a single
probability space. This can be done for any abelian subalgebra of
$\mathfrak{A}$, but the probability spaces associated with
non-commuting observables will in fact be different; (ii) the
probability measure associated with pure states are not $\delta$
measures, i.e., the statistical interpretation cannot be avoided
even if restricting our considerations to pure states; the
statistical description is, thus, intrinsic to any quantum system.
\\
\linebreak \indent For a concrete example, let us consider the
simplest physical system consisting of a single spinless point
particle moving along the real line (this study can be easily
gene\-ralized to higher dimensions). The basic classical observables
are the position $X$ and the momentum $P$ of the particle.
Tentatively, one could propose as quantum algebra of observables
that generated by $\{\mathbf{1},X,P\}$ satisfying the Heisenberg
commutation relations (from now on we will consider units such that
$\hbar=1$)
\begin{equation}\label{HeisenbergCR}
[X,P]=i\mathbf{1}\,,\quad [X,X]=0=[P,P]\,.
\end{equation}
This \emph{Heisenberg algebra}, however, does not fall into the
Segal scheme, because it is a well-known fact that any $X$ and $P$
satisfying (\ref{HeisenbergCR}) cannot be realized as bounded
self-adjoint elements of a $C^{*}$-algebra.\footnote{Indeed, by
induction one gets $[X,P^{n}]=inP^{n-1}$, $n\in\mathbb{N}$, which
implies
$n\|P^{n-1}\|\le\|XP^{n}\|+\|P^{n}X\|\le2\|X\|\|P\|\|P^{n-1}\|$ for
some $C^{*}$-norm. Since $\|P^{n-1}\|\ne0$ --otherwise, $P=0$ and,
hence, (\ref{HeisenbergCR}) is not verified--, one finally obtains
$\|X\|\|P\|\ge n/2$, $\forall\,n\in\mathbb{N}$.} This is not a
surprise from the operational point of view: The technical
limitations of the measuring apparatuses imply that only bounded
functions of $X$ and $P$ can be measured. Of course, one could try
to obtain a representation of the Heisenberg algebra in terms of
unbounded self-adjoint operators $X$ and $P$ with dense domains in a
certain Hilbert space $\mathfrak{H}$, but this would not follow the
operational description given in this section. Nevertheless,
this problem can be easily overcome by introducing the so-called
\emph{Weyl algebra} --in fact, this will allow us to recover the
Heisenberg algebra in a definite sense at the end of this section.
Consider the bounded formal functions of $X$ and $P$, called
\emph{Weyl operators},
\begin{equation}\label{UalphaVbeta}
U(\alpha):=\exp(i\alpha X)\,,\quad V(\beta):=\exp(i\beta
P)\,,\quad\alpha,\beta\in\mathbb{R}\,.
\end{equation}
The Heisenberg commutation relations (\ref{HeisenbergCR}) are now
replaced by the \emph{Weyl relations}
\begin{eqnarray*}
&&U(\alpha)V(\beta)=V(\beta)U(\alpha)\exp(-i\alpha\beta)\,,\\
&&U(\alpha)U(\beta)=U(\alpha+\beta)\,,\quad
V(\alpha)V(\beta)=V(\alpha+\beta)\,,\nonumber
\end{eqnarray*}
formally obtained from (\ref{HeisenbergCR}) and (\ref{UalphaVbeta})
by applying the Baker-Campbell-Hausdorff formula
$\exp(A)\exp(B)=\exp(A+B+[A,B]/2)$, with $[A,B]$ assumed to be a
c-number. There is a natural $*$-operation on the Weyl algebra
generated by linear complex combinations and products of these
basics elements, as suggested by the self-adjointness of $X$ and
$P$,
\begin{equation*}
U(\alpha)^{*}=U(-\alpha)\,,\quad V(\beta)^{*}=V(-\beta)\,,
\end{equation*}
so that $U(\alpha)$ and $V(\beta)$ are unitary, i.e.,
$U(\alpha)^{*}U(\alpha)=U(\alpha)U(\alpha)^{*}=\mathbf{1}$, and
similarly for $V(\beta)$. It is possible to prove the existence of a
unique $C^{*}$-norm such that the completion of the Weyl algebra in
this norm is a (non commutative) $C^{*}$-algebra $[4]$, called the
Weyl $C^{*}$-algebra and denoted $\mathfrak{A}_{\mathrm{Weyl}}$.
According to the $C^{*}$-condition, we have
$\|U(\alpha)\|=\|V(\beta)\|=\|U(\alpha)V(\beta)\|=1$,
$\forall\,\alpha,\beta\in\mathbb{R}$. One faces now the problem of
finding suitable representations for this algebra. The following
theorem was rigorously proved for the first time by von Neumann
$[1]$:
\begin{theorem}[von Neumann]
All regular irreducible representations of the Weyl $C^{*}$-algebra
$\mathfrak{A}_{\mathrm{Weyl}}$ in a separable Hilbert space are
unitarily equivalent.
\end{theorem}
\noindent Here, a representation
$\pi:\mathfrak{A}_{\mathrm{Weyl}}\rightarrow\mathscr{B}(\mathfrak{H})$
into a separable Hilbert space $\mathfrak{H}$ is said to be
\emph{regular} if $\pi(U(\alpha))$ and $\pi(V(\beta))$ are
(one-parameter unitary groups) strongly continuous in the real
$\alpha$ and $\beta$ parameters, respectively. Thus, it suffices to
find one regular irreducible representation to univocally
characterize all the representations of this class; concretely, a
well-known solution is given by the \emph{Schr\"{o}dinger
representation} $\pi_{s}$ into the separable Hilbert space
$\mathfrak{H}_{s}=L^{2}(\mathbb{R},\mathrm{d}x)$, where for all pure
states ($L^{2}$-functions) $\Psi\in\mathfrak{H}_{s}$,
\begin{equation*}
\left(\pi_{s}(U(\alpha))\Psi\right)(x)=\exp(i\alpha
x)\Psi(x)\,,\quad
\left(\pi_{s}(V(\beta))\Psi\right)(x)=\Psi(x+\beta)\,.
\end{equation*}
It is easy to check the regularity and irreducibility of this
representation. Note that the strong continuity in the $\alpha$ and
$\beta$ parameters ensures, by virtue of Stone's theorem, the
existence of (unbounded) self-adjoint generators $X$ and $P$ with
dense domain in $\mathfrak{H}_{s}$. In particular, the Schwartz
space $\mathscr{S}(\mathbb{R})$ of smooth rapidly decreasing
functions in $\mathbb{R}$ is a common invariant dense domain of
essential self-adjointness for $X$ and $P$, where the Heisenberg
commutation relations (\ref{HeisenbergCR}) are satisfied. For all
$\Psi\in\mathscr{S}(\mathbb{R})$,
\begin{equation*}
\left(X\Psi\right)(x)=x\Psi(x)\,,\quad\left(P\Psi\right)(x)=-i\Psi^{\prime}(x)\,,
\end{equation*}
where the prime denotes derivative. Therefore, thanks to regularity,
the Heisenberg algebra can be recovered in a definite sense from the
Weyl $C^{*}$-algebra.

\section{Algebraic dynamics}

\indent Consider a $C^{*}$-algebra of observables, $\mathfrak{A}$.
An autonomous (algebraic) dynamical system is a triplet
$(\mathfrak{A},I\subseteq\mathbb{R},\alpha_{t})$, with $\alpha_{t}$,
$t\in I$, being a (weakly) continuous one-parameter group of
$*$-automorphisms (i.e., $*$-isomorphisms of $\mathfrak{A}$ into
itself). An (irreducible) representation of the algebra
$\pi:\mathfrak{A}\rightarrow\mathscr{B}(\mathfrak{H})$ is called
\emph{stable under time evolution} if $\pi$ and $\pi\circ\alpha_{t}$
(equivalently $\pi\circ\alpha^{-1}_{t}$) are unitarily equivalent
for all $t\in I$. In this case, there exists a unitary
operator-valued function $U(t):\mathfrak{H}\rightarrow\mathfrak{H}$,
called the \emph{time-evolution operator}, such that
\begin{equation}\label{A(t)}
\pi(\alpha_{t}(A))=U(t)^{-1}\pi(A)U(t)\,,\quad\forall\,A\in\mathfrak{A}\,.
\end{equation}
Furthermore, the weak continuity of $\alpha_{t}$ ensures the weak
continuity of $U(t)$, so that by Stone's theorem we can write
\begin{equation*}
U(t)=\exp(-itH)\,,\quad t\in\mathbb{R}\,,
\end{equation*}
where the generator $H$, called the \emph{Hamiltonian operator}, is
a self-adjoint operator with dense domain
$\mathscr{D}_{H}\subset\mathfrak{H}$. It is obtained as the strong
limit
$s\!-\!\lim_{t\rightarrow0}t^{-1}(U(t)-\mathbb{I})\Psi=-iH\Psi$,
$\Psi\in\mathscr{D}_{H}$. Note that the $H$ operator is unbounded in
general --so it does not belong to the $C^{*}$-algebra of physical
observables-- and its expression depends on the concrete
representation $\pi$. It is also important to realize that the
relations (\ref{A(t)}) determine $U(t)$ univocally modulo a complex
phase, so that the quantum Hamiltonian is unique up to an
(irrelevant) real term proportional to the identity. Let
$\Psi_{0}\in\mathscr{D}_{H}$ be a state vector of the stable
representation $\pi$, and let $\omega_{0}$ be the pure state defined
by it. Then, we have
\begin{equation*}
\omega_{0}(\alpha_{t}(A))=\langle\Psi_{0}\,|\,U(t)^{-1}\pi(A)U(t)\Psi_{0}\rangle_{\mathfrak{H}}=\langle
U(t)\Psi_{0}\,|\,\pi(A)U(t)\Psi_{0}\rangle_{\mathfrak{H}}=:\omega_{t}(A)\,,
\end{equation*}
where $\omega_{t}$ is the pure state defined by
$U(t)\Psi_{0}=:\Psi(t)\in\mathfrak{H}$. This is an algebraic
repre\-sentation of the evolution of states in the Schr\"{o}dinger
picture. $\mathscr{D}_{H}$ contains a dense subdomain
$\mathscr{D}\subset\mathscr{D}_{H}$ invariant under the action of
$U(t)$, in which the Hamiltonian is essentially self-adjoint. For
$\Psi_{0}\in\mathscr{D}$, differentiating the state vector $\Psi(t)$
with respect to the time parameter $t$, we obtain the
\emph{Schr\"{o}dinger equation}
\begin{equation*}
i\frac{\mathrm{d}}{\mathrm{d}t}\Psi(t)=H\Psi(t)\,,\quad \Psi(0)=\Psi_{0}\,.
\end{equation*}
The time evolution can be equivalently formulated in terms of
observables. Indeed, the $*$-automorphism $A\mapsto\alpha_{t}\circ
A$, $\forall\,A\in\mathfrak{A}$ with $A=A^{*}$, can be interpreted
as the algebraic representation of the evolution of observables in
the Heisenberg picture. Denoting
$A(t):=\pi(\alpha_{t}A)=U(t)^{-1}\pi(A)U(t)$ (see equation
(\ref{A(t)})), and differentiating this relation with respect to
$t$, one finally obtains the \emph{Heisenberg equation}
\begin{equation*}
\frac{\mathrm{d}}{\mathrm{d}t}A(t)=i[H,A(t)]\,,\quad A(0)=\pi(A)\,.
\end{equation*}
\\
\noindent \textbf{References}
\begin{itemize}
\item [1] J. Von Neumann, \emph{Mathematical Foundations of Quantum
Mechanics}, Princeton University Press (1964).
\item [2] H. Araki, \emph{Mathematical Theory of Quantum Fields}, The International Series of Monographs
on Physics, Vol. 101, Oxford University Press (1999).
\item [3] I. E. Segal, ``Postulates for General Quantum Mechanics'', The Annals of Mathematics \textbf{48}, 930-948
(1947); \emph{Mathematical Problems of Relativistic Physics},
Lectures in Applied Mathematics Series, American Mathematical
Society (1967).
\item [4] F. Strocchi, \emph{An Introduction to the Mathematical Structure of Quantum Mechanics. A Short Course
for Mathematicians}, Advanced Series in Mathematical Physics, Vol.
27, World Scientific (2005).
\item [5] O. Bratteli and D. W. Robinson, \emph{Operator Algebras and Quantum Statistical Mechanics I. $C^{*}$- and $W^*$-Algebras. Symmetry Groups. Decomposition of States},
2${^{\mathrm{a}}}$ ed., Springer (1997).
\item [6] C. Isham, \emph{Lectures on Quantum Theory: Mathematical and Structural Foundations}, World Scientific
(1995).
\item [7] G. W. Mackey, \emph{The Mathematical Foundations of Quantum
Mechanics}, W. A. Benjamin Inc., New York (1963).
\item [8] B. Simon, \emph{Quantum Mechanics for Hamiltonians Defined as Quadratic
Forms}, Princeton Univ. Press. Princeton, N. J. (1971).
\end{itemize}


\chapter{The Time-dependent Harmonic Oscillator}\label{AppendixTDHO}
\begin{flushright}
\small{\vspace*{-0.9cm} \textbf{D. G. Vergel and E. J. S.
Villase\~{n}or}\\\textbf{Annals of Physics 324, 1360-1385 (2009)}}
\end{flushright}


\pagestyle{fancy} \fancyhf{}

\fancyhead[LO]{\textsc{Appendix D. The Time-dependent Harmonic
Oscillator}} \fancyhead[RO,LE]{\thepage}
\renewcommand{\headrulewidth}{0.6pt}


\indent In this appendix, we reformulate the quantum theory of a
single one-dimensional time-dependent harmonic oscillator,
summarizing some basic results concerning the unitary implementation
of the dynamics. This is done by employing techniques different from
those used so far to derive the Feynman propagator. In particular,
we calculate the transition amplitudes for the usual harmonic
oscillator eigenstates and define suitable semiclassical states for
some physically relevant systems such as Gowdy-like oscillators.

\section{Properties of the TDHO equation}\label{TDHOproperties}

\indent We will review in this section some properties of the
classical equation of motion of a single harmonic oscillator with
time-dependent frequency, from now on referred to as the TDHO
equation, and its connection with the so-called Ermakov-Pinney
equation, which plays an auxiliary role in the calculation of
invariants for nonquadratic Hamiltonian systems. The TDHO equation
is given by
\begin{eqnarray}
\ddot{u}(t)+\kappa(t)u(t)=0\,,\quad t\in I=(t_-,t_+)\subseteq
\mathbb{R}\,,\label{TDHO_eq}
\end{eqnarray}
where $\kappa:I\rightarrow \mathbb{R}$ is a real-valued continuous
function and time-derivatives are denoted by dots. Given an initial
time $t_0\in I\,$, let $c_{t_0}$ and $s_{t_0}$ be the independent
solutions of (\ref{TDHO_eq}) such that
$c_{t_0}(t_0)=\dot{s}_{t_0}(t_0)=1$ and
$s_{t_0}(t_0)=\dot{c}_{t_0}(t_0)=0$. These can be written in terms
of any set of independent solutions to (\ref{TDHO_eq}), say $u_1$
and $u_2$, as
 \begin{eqnarray}\label{c&srelations}
c_{t_0}(t)
=\frac{\dot{u}_2(t_0)u_1(t)-\dot{u}_1(t_0)u_2(t)}{W(u_1,u_2)}\,,\quad
s_{t_0}(t) =\frac{u_1(t_0)u_2(t)-u_2(t_0)u_1(t)}{W(u_1,u_2)}\,,
\end{eqnarray}
where $(t_0,t)\in I\times I$ and
$W(u_1,u_2):=u_1\dot{u}_2-\dot{u}_1u_2$ denotes the
(time-independent) Wronskian of $u_1$ and $u_2$. In what follows, we
will use the notation $c(t,t_0):=c_{t_0}(t)$,
$\dot{c}(t,t_0):=\dot{c}_{t_0}(t)$, $s(t,t_0):=s_{t_0}(t)$, and
$\dot{s}(t,t_0):=\dot{s}_{t_0}(t)$. Note that the $s$ function
belongs to the class $C^2(I\times I)\,$, whereas $c(\cdot,t_0)\in
C^2(I)$ and $c(t,\cdot)\in C^1(I)\,$. As a concrete example, for the
time-independent harmonic oscillator (TIHO) with constant frequency
$\kappa(t)=\kappa_0\in \mathbb{R}$, we simply get ($\omega>0$)
\begin{eqnarray}\label{c&sMinkowski}
\kappa_0&=&\omega^2\,, \phantom{aaa}
c(t,t_0)=\cos((t-t_0)\omega)\,,\quad \phantom{a}
s(t,t_0)=\omega^{-1}\sin((t-t_0)\omega)\,;\\
\kappa_0&=&0\,, \phantom{aaaa}
c(t,t_0)=1\,,\quad\phantom{aaaaaaaaaaa}
s(t,t_0)=t-t_0\,;\label{c&sfree}\\
\kappa_0&=&-\omega^2\,,  \phantom{a,}
c(t,t_0)=\cosh((t-t_0)\omega)\,,\quad
s(t,t_0)=\omega^{-1}\sinh((t-t_0)\omega)\,.\label{c&stachy}
\end{eqnarray}
In fact, as well known from Sturm's theory, the $c$ and $s$
functions corresponding to arbitrary frequencies share several
properties with the usual cosine and sine functions. Firstly, their
Wronskian is normalized to unit, $W(c,s)=1$. Hence, if one of them
vanishes for some time $t=t_*$, then the other is automatically
different from zero at that instant. In view of this condition and
relations (\ref{c&srelations}), their time-derivatives satisfy
\begin{equation}\label{derivat}
\dot{s}(t,t_0)=c(t_0,t)\,,\quad
\dot{c}(t,t_0)=\frac{c(t,t_0)c(t_0,t)-1}{s(t,t_0)}\,,
\end{equation}
where the last equation must be understood as a limit for those
values of the time parameter $t_*$ such that $s(t_*,t_0)=0$. The odd
character of the sine function translates into the condition
$s(t_0,t)=-s(t,t_0)$. Finally, the well known formula for the sine
of a sum of angles can be generalized as
\begin{equation}\label{sum}
s(t_2,t_1)=c(t_1,t_0)s(t_2,t_0)-c(t_2,t_0)s(t_1,t_0)\,.
\end{equation}
\indent It is well known that solutions to the TDHO equation
(\ref{TDHO_eq}) are related to certain non-linear differential
equations. Here, we will restrict our attention to the so-called
Ermakov-Pinney (EP) equation (see $[1,2]$; the interested reader is
strongly suggested to consult the historical account of $[3]$ and
references therein). Let
$$
A=A^t= \left[
  \begin{array}{cc}
    a_{11} & a_{12} \\
    a_{12} & a_{22} \\
  \end{array}
\right]\in\mathrm{Mat}_{2\times2}(\mathbb{R})
$$
be a positive definite quadratic form with $\det A=1$. Then, the
(never vanishing) function $\rho:I\rightarrow (0,+\infty)$ defined
as
\begin{eqnarray}
\rho(t):=\sqrt{a_{11}c^2(t,t_0) +a_{22}s^2(t,t_0)+2a_{12}
s(t,t_0)c(t,t_0)}\label{rho_sc}
\end{eqnarray}
verifies the EP equation
\begin{eqnarray}
\ddot{\rho}(t)+\kappa(t)\rho(t)=\frac{1}{\rho^3(t)}\label{EP_eq}\,,\quad
t\in I\,.
\end{eqnarray}
According to (\ref{c&srelations}), the most general analytic
solution to (\ref{EP_eq}) can be written as $[4,5]$
\begin{equation}\label{rhoABC}
\rho(t)=\sqrt{b_{11}u_1^2(t)+b_{22}u_2^2(t)+2b_{12}u_1(t)u_2(t)}\,,
\end{equation}
where, as a consequence of (\ref{rho_sc}) and (\ref{EP_eq}), the
coefficients  $b_{11}$, $b_{12}$, $b_{22}\in\mathbb{R}$ satisfy
$W^2(u_1,u_2)=(b_{11}b_{22}-b_{12}^2)^{-1}>0$. Conversely, given
\emph{any} solution  to the EP equation it is possible to find the
general solution to the TDHO equation. Indeed, it is straightforward
to prove the following theorem.

\begin{theorem}\label{ThrEP}

Let $\rho$ be any solution to the EP equation (\ref{EP_eq}); then,
the $c$ and $s$ solutions to (\ref{TDHO_eq}) are given by
\begin{eqnarray}
c(t,t_0)&=&\frac{\rho(t)}{\rho(t_0)}\cos\left(\int_{t_0}^t\frac{\mathrm{d}\tau}{\rho^2(\tau)}\right)
-\rho(t)\dot{\rho}(t_0)\sin\left(\int_{t_0}^t\frac{\mathrm{d}\tau}{\rho^2(\tau)}\right)\,,\label{c_rho}\\
s(t,t_0)&=&\rho(t)\rho(t_0)\sin\left(\int_{t_0}^t\frac{\mathrm{d}\tau}{\rho^2(\tau)}\right)\,,\label{s_rho}\quad
(t,t_0)\in I\times I\,.
\end{eqnarray}
\end{theorem}
\begin{rem} \emph{By using (\ref{c_rho}) and
(\ref{s_rho}), it is possible to find other $\rho$-independent
objects. For example, the combination
$$
\frac{\rho(t_0)}{\rho(t)}\cos\left(\int_{t_0}^t\frac{\mathrm{d}\tau}{\rho^2(\tau)}\right)
+\rho(t_0)\dot{\rho}(t)\sin\left(\int_{t_0}^t\frac{\mathrm{d}\tau}{\rho^2(\tau)}\right)=c(t_0,t)=\dot{s}(t,t_0)
$$
and the zeros of $s(t,t_0)$, characterized by
$$
\int_{t_0}^t\frac{\mathrm{d}\tau}{\rho^2(\tau)}\equiv 0 \,\,\,
(\textrm{mod}\, \pi)\,,
$$
are independent of the particular solution $\rho$ to the EP
equation. These results will be profusely applied along this
appendix.}
\end{rem}

\section{Unitary quantum time evolution}\label{Unitaryoperator}

\subsection{General framework}

The \emph{canonical} phase space description of the classical system
under consideration consists of a nonautonomous Hamiltonian system
$(I\times\mathbf{P},\mathrm{d}t,\bm{\omega},H(t))$. Here,
$\mathbf{P}:=\mathbb{R}^2$ denotes the space of Cauchy data $(q,p)$
endowed with the usual symplectic structure
$\bm{\omega}((q_1,p_1),(q_2,p_2)):=p_1q_2-p_2q_1$,
$\forall\,(q_1,p_1), (q_2,p_2)\in\mathbf{P}$. The triplet
$(I\times\mathbf{P},\mathrm{d}t,\bm{\omega})$ then has the
mathematical structure of a cosymplectic vector space. The
time-dependent Hamiltonian $H(t):\mathbf{P}\rightarrow\mathbb{R}$,
$t\in I$, is given by
\begin{equation}\label{Hclass}
H(t,q,p):=\frac{1}{2} \Big(p^2+\kappa(t)q^2\Big)\,.
\end{equation}
The solution to the corresponding Hamilton equations with initial
Cauchy data $(q,p)$ at time $t_0$ can be written down as
\begin{eqnarray}
\left[\begin{array}{c}
                 q_H(t,t_0) \\
                 p_H(t,t_0) \\
               \end{array}
             \right]=\mathcal{T}_{(t,t_0)}\cdot\left[
               \begin{array}{c}
                 q \\
                 p \\
               \end{array}
             \right],\quad \mathcal{T}_{(t,t_0)}:=\left[
        \begin{array}{cc}
          c(t,t_0) & s(t,t_0) \\
       \dot{c}(t,t_0)& \dot{s}(t,t_0) \\
        \end{array}
      \right]. \label{Hamilton_class}
\end{eqnarray}
Note that the properties stated in \emph{section
\ref{TDHOproperties}} about the $c$ and $s$ solutions to the TDHO
equation (\ref{TDHO_eq}) guarantee that $\mathcal{T}_{(t,t_0)}\in
SL(2,\mathbb{R})=SP(1,\mathbb{R})$ for all $(t,t_0)\in I\times I\,$,
i.e., the classical time evolution is implemented by symplectic
transformations.
\\
\indent We now formulate the quantum theory of the TDHO by defining
the corresponding Weyl $C^*$-algebra of quantum observables
$\mathscr{W}(\mathbf{P})$ and choosing a suitable representation in
terms of self-adjoint operators in some separable Hilbert space
$\mathscr{H}$. As we pointed out in \emph{appendix \ref{MSPT}}, the
natural realization for this algebra is given by the well-known
Schr\"{o}dinger representation $(Q\psi)(q)=q\psi(q)$,
$(P\psi)(q)=-i\psi^\prime(q)$, $\psi\in L^2(\mathbb{R})$. Another
possibility is to represent the canonical commutation relations
(CCR) in the space $L^{2}(\mathbb{R},\mathrm{d}\mu_{\alpha})$ where,
given some $\alpha\in\mathbb{C}\setminus\{0\}$, $\mu_{\alpha}$
denotes the Gaussian probability measure
$$
\mathrm{d}\mu_{\alpha}=\frac{1}{\sqrt{2\pi}|\alpha|}\exp\left(-\frac{q^2}{2|\alpha|^2}\right)\,\mathrm{d}q\,.
$$
To each $\alpha$ there corresponds a family of unitary
transformations
$V_\alpha(\beta):L^2(\mathbb{R},\mathrm{d}q)\rightarrow
L^{2}(\mathbb{R},\mathrm{d}\mu_{\alpha})$ connecting the standard
Hilbert space with the new one in the form
\begin{eqnarray}\label{Valphabeta}
\Psi(q)=\big(V_\alpha(\beta)\psi\big)(q)=\big(\sqrt{2\pi}|\alpha|\big)^{1/2}\exp\big(-i\bar{\beta}q^2/(2\bar{\alpha})\big)\psi(q)\,,
\end{eqnarray}
where the complex numbers $\beta$ must satisfy
$\alpha\bar{\beta}-\beta\bar{\alpha}=i\,$. Note that the unitary
transformations $V_\alpha(\beta)$ map the `vacuum' state
$$\psi_0(q)=\big(\sqrt{2\pi}|\alpha|\big)^{-1/2}\exp\big(i\bar{\beta}q^2/(2\bar{\alpha})\big)\in
L^{2}(\mathbb{R},\mathrm{d}q)$$ onto the unit function
$\Psi_0(q)=(V_\alpha(\beta)\psi_0)(q)=1\in
L^{2}(\mathbb{R},\mathrm{d}\mu_\alpha)\,$. In these cases, the
position and momentum operators act on state vectors as
$$(Q\Psi)(q)=q\Psi(q)\quad \mathrm{and}\quad (P\Psi)(q)=-i\Psi'(q)+\frac{\bar{\beta}}{\bar{\alpha}}q\Psi(q)\,,$$
where, with the aim of simplifying the notation, $Q$ and $P$
respectively denote the transformed operators
$V_\alpha(\beta)QV_\alpha(\beta)^{-1}$ and
$V_\alpha(\beta)PV_\alpha(\beta)^{-1}$ with common dense domain
$V_\alpha(\beta)\mathscr{S}(\mathbb{R})\subset
L^{2}\big(\mathbb{R},\mathrm{d}\mu_\alpha\big)\,$.
\\
\indent Any regular irreducible representation
$\pi:\mathscr{W}(\mathbf{P})\rightarrow\mathscr{B}(\mathscr{H})$ is
stable under time evolution, i.e., there exists a unitary evolution
operator $U(t,t_0):\mathscr{H}\rightarrow\mathscr{H}$, $(t_0,t)\in
I\times I$, implementing the quantum dynamics. It is important to
notice at this point that, if the classical evolution has
singularities at the boundary of the interval $I$, they also occur
for the quantum dynamics, i.e., there is no resolution of classical
singularities. The Heisenberg equations for $Q$ and $P$ can be
solved just by the same expressions involved in the classical
solutions (\ref{Hamilton_class}), i.e.,
\begin{eqnarray}\label{QHPH}
&&\hspace*{-0.8cm}\left[
  \begin{array}{c}
    Q_H(t,t_0) \\
    P_H(t,t_0)
  \end{array}
\right]:=U^{-1}(t,t_0)\left[
  \begin{array}{c}
    Q \\
    P
  \end{array}
\right] U(t,t_0)=\left[
        \begin{array}{cc}
          c(t,t_0) & s(t,t_0) \\
          \dot{c}(t,t_0)& \dot{s}(t,t_0) \\
        \end{array}
      \right]\cdot \left[
  \begin{array}{c}
    Q \\
    P
  \end{array}
\right].
\end{eqnarray}
With more generality, given any well-behaved (analytic) classical
observable $F:\mathbf{P}\rightarrow\mathbb{R}$ for the TDHO, the
time evolution of its quantum counterpart in the Heisenberg picture
$F_{H}(t,t_0):=U^{-1}(t,t_0)F(Q,P)U(t,t_0)$ is simply given by
\begin{equation}
\hspace*{-3mm}F_H(t,t_0)\!=\!F(Q_H(t,t_0),P_H(t,t_0))
=F\big(c(t,t_0)Q+s(t,t_0)P,\dot{c}(t,t_0)Q+
\dot{s}(t,t_0)P\big)\,.\label{Heisen}
\end{equation}
Hence, the matrix elements $\langle
\Psi_2\,|\,U^{-1}(t_2,t_1)F(Q,P)U(t_2,t_1) \Psi_1 \rangle$,
$\Psi_1,\Psi_2\in\mathscr{H}$, can be computed without the explicit
knowledge of the unitary evolution operator. This is also the case
of the transition probabilities
$\mathrm{Prob}(\Psi_2,t_2\,|\,\Psi_1,t_1)=|\langle\Psi_2\,|\,U(t_2,t_1)\Psi_1
\rangle|^2$, as will be discussed in detail in \textit{subsection
\ref{TransInstab}}. The commutators of time-evolved observables can
be also calculated without the concrete expression of $U(t_2,t_1)$.
For instance, from (\ref{Heisen}) we easily obtain
$$\big[Q_{H}(t_1,t_0),Q_{H}(t_2,t_0)\big]=is(t_1,t_2)\mathbf{1}\,,$$
where we have used the relation (\ref{sum}) stated in \emph{section
\ref{TDHOproperties}}. As expected, the commutator given above is
proportional to the identity operator and independent of the choice
of the initial time $t_0\,$. Note, in contrast with the transition
probabilities, that the calculation of transition amplitudes of the
type $\langle \Psi_2\,|\, U(t_2,t_1)\Psi_1\rangle$ does require the
explicit knowledge of (the phase of) the evolution operator. This is
also the case of the (strong) derivatives of both $U(\cdot,t_0)$ and
$U(t,\cdot)\,$.
\\
\indent The dynamics of the quantum TDHO is governed by an
(unbounded) nonautonomous Hamiltonian operator
$H(t):\mathscr{H}\rightarrow\mathscr{H}$, $t\in I$, satisfying
\begin{equation}
\dot{U}(t,t_0)=-iH(t)U(t,t_0)\,.\label{Uequa}
\end{equation}
Given the quadratic nature of the classical Hamiltonian
(\ref{Hclass}), $H(t)$ must coincide with the operator directly
promoted from the classical function modulo a $t$-dependent real
term proportional to the identity $\mathbf{1}$ encoding the election
of $U(t,t_0)$. For a concrete representation of the CCR, we will
simply take
\begin{equation}\label{quantHamiltonian}
H(t):=\frac{1}{2}\left(P^{2}+\kappa(t)Q^{2}\right).
\end{equation}
This choice fixes $U(t,t_0)$ uniquely. The Hamiltonian
(\ref{quantHamiltonian}) is a self-adjoint operator with dense
domain $\mathscr{D}_{H(t)}$ --equal to $C_0^\infty(\mathbb{R})$ in
the standard Schr\"{o}dinger representation-- for each value of the
time parameter $t\,$. We will prove the following theorem in the
next subsections.

\begin{theorem}\label{ThrPropag} The action of the unitary TDHO evolution operator  $U(t,t_0)$ corresponding to
the Hamiltonian (\ref{quantHamiltonian}) on any state vector
$\psi\in\mathscr{S}(\mathbb{R})\subset
L^{2}(\mathbb{R},\mathrm{d}q)$ in the traditional Schr\"{o}dinger
representation is given by
$$
\big(U(t,t_0)\psi\big)(q)=\int_{\mathbb{R}}
K(q,t;q_0,t_0)\psi(q_0)\,\mathrm{d}q_0\,,
$$
where the propagator $K(q,t;q_0,t_0)$ depends on the times $t_0$ and
$t$ through the classical TDHO solutions $c$ and $s\,$. Explicitly,
\begin{eqnarray}
\hspace*{-2mm}K(q,t;q_0,t_0)=\frac{1}{\sqrt{2\pi
i}}\,s^{-1/2}(t,t_0)\exp\left(\frac{i}{2
s(t,t_0)}\Big(c(t_0,t)q^2+c(t,t_0)q_0^2-2qq_0\Big)\right)\!,&&\label{K_qq0a}
\end{eqnarray}
wherever $s(t,t_0)\neq 0\,$, and
\begin{eqnarray}
K(q,t;q_0,t_0) =
c^{-1/2}(t,t_0)\exp\left(i\frac{\dot{c}(t,t_0)}{2c(t,t_0)}\right)\label{K_qq0b}
\delta(q_0-q/c(t,t_0))
\end{eqnarray}
if $s(t,t_0)=0\,$.
\end{theorem}

\begin{rem}
\emph{Given a solution $u(t)$ to the TDHO equation (\ref{TDHO_eq})
which is positive in some interval $(t_0,t_0+\varepsilon)\subset I$,
$\varepsilon>0$, we define
$$
u^{\epsilon}(t,t_0):=\exp\big(i\epsilon\pi
\,\mathfrak{m}(u;t,t_0)\big)|u(t)|^{\epsilon}\,,\,\,\,\epsilon\in\mathbb{R}\,,\,\,\,t\in
I\,,
$$
where $\mathfrak{m}(u;t,t_0)\in \mathbb{Z}$ is the index function of
$u$, with  $\mathfrak{m}(u;t_0,t_0)=0\,$, in such a way that
$\mathfrak{m}(u;t_2,t_0)-\mathfrak{m}(u;t_1,t_0)$, $t_1<t_2$, gives
the number of zeros of $u(\cdot,t_0)$ in the interval $(t_1,t_2]\,$.
Finally, $\delta(q)$ denotes the Dirac delta distribution.}
\end{rem}
\begin{rem} \emph{Let $\vartheta:I\rightarrow \mathbb{R}$ be a real-valued continuous function and consider the Hamiltonian
$$H_1(t):=H(t)+\vartheta(t)\mathbf{1}\,$$
defined in terms of (\ref{quantHamiltonian}). The unitary evolution
$U_1(t,t_0)$ associated with $H_1(t)$ satisfying (\ref{QHPH}) gives
rise to the propagator
\begin{eqnarray}
K_1(q,t;q_0,t_0)=K(q,t:q_0,t_0)\exp\left(-i\int_{t_0}^t\vartheta(\tau)\mathrm{d}\tau\right).\label{K1}
\end{eqnarray}
Note that
$U^{-1}_1(t,t_0)\,\mathcal{O}\,U_1(t,t_0)=U^{-1}(t,t_0)\,\mathcal{O}\,U(t,t_0)$
for any quantum observable $\mathcal{O}$.}
\end{rem}
\begin{rem}
\emph{In the $L^{2}(\mathbb{R},\mathrm{d}\mu_\alpha)$-representation
defined by the unitary transformation $V_\alpha(\beta)$ (see
equation (\ref{Valphabeta})), the evolution is given by
$$
\big(U(t,t_0)\Psi\big)(q)=\int_{\mathbb{R}}
K_{\alpha\beta}(q,t;q_0,t_0)\Psi(q_0)\,\mathrm{d}\mu_\alpha(q_0)\,,
$$
where
\begin{eqnarray}\label{Kalphabeta}
K_{\alpha\beta}(q,t;q_0,t_0):=\sqrt{2\pi}|\alpha|\exp\left(
\frac{i\beta}{2\alpha}q_0^2-\frac{i\bar{\beta}}{2\bar{\alpha}}q^2\right)
K(q,t;q_0,t_0)\,.
\end{eqnarray}}
\end{rem}

\subsection{Constructing the evolution operator}\label{ConstructingU}

\indent  In order to calculate the unitary evolution operator
$U(t,t_0)$ we will perform a gene\-ralization of the method
developed in $[6]$ that will clarify the appearance of the auxiliary
Ermakov-Pinney solution (\ref{rhoABC}) in this context, and will
allow us also to warn the reader about other problematic choices
that have appeared before in the related literature. We first
introduce on $\mathscr{H}$ the (one-parameter family of) unitary
operators
$$D(x):=\exp\left(-\frac{i}{2} x Q^{2}\right),\,\,\,x\in \mathbb{R}\,,$$
generating a displacement of the momentum operator, $D(x) P
D^{-1}(x)=P+xQ$ (the position operator being unaffected by them),
and define the unitary squeeze operators
$${S}(y):=\exp\left(\frac{i}{2}y\big({Q}{P}+{P}{Q}\big)\right),\,\,\,y\in\mathbb{R}\,,$$
scaling both the position and momentum operator respectively as
$S(y){Q}S^{-1}(y)=e^{y}{Q}$ and $S(y){P}{S}^{-1}(y)=e^{-y} P$. Let
$\Psi(t)\in\mathscr{D}_{H(t)}$, $t\in I$, be a solution to the
Schr\"{o}dinger equation, i.e., $i\dot{\Psi}(t)=H(t)\Psi(t)$, and
let $x,y\in C^1(I)\,$. We now introduce the unitary operators
$$T(t)=T(t;x,y):=S(y(t))D(x(t))\,,$$
where the functions $x$ and $y$ remain arbitrary at this stage. Let
us consider the time evolution for the transformed state vector
$$
\Phi(t)=\Phi(t;x,y):=T(t;x,y)\Psi(t)\,,
$$
given by
\begin{eqnarray*}
i\dot{\Phi}(t)&=&\Big(T(t)H(t)T^{-1}(t)-iT(t)\dot{T}(t)\Big)\Phi(t)\\
&=&\frac{1}{2}\Big(e^{-2y(t)}P^2+(x(t)-\dot{y}(t))(QP+PQ)
+e^{2y(t)}(x^2(t)+\kappa(t)+\dot{x}(t))Q^2\Big)\Phi(t)\,.
\end{eqnarray*}
We note at this point that it is possible to get a notable
simplification of the previous expression just by imposing
\begin{equation}\label{naturalconditions}
x(t)=\dot{y}(t)\quad\mathrm{and}\quad
x^2(t)+\kappa(t)+\dot{x}(t)=\exp(-4y(t))\,.
\end{equation}
The most natural way to achieve this is to choose
$$
y(t):=\log\rho(t)\quad \textrm{and, hence,}\quad
x(t)=\dot{\rho}(t)/\rho(t)\,,
$$
with $\rho$ being \emph{any} solution to the auxiliary EP equation
(\ref{EP_eq}) introduced in \emph{section \ref{TDHOproperties}}. In
this way, the state vector
$\Phi(t;\dot{\rho}/\rho,\log\rho)=:\Phi_\rho(t)$ satisfies the
differential equation
$$i\dot{\Phi}_{\rho}(t)=\frac{1}{2\rho^2(t)}\big(P^2+Q^2\big)\Phi_\rho(t)\,.$$
Solving this equation and going back to the original state vector
$\Psi(t)$, we finally obtain the unitary evolution operator for the
system. We can then enunciate the following theorem.

\begin{theorem}\label{ThrU}
The time evolution operator $U(t,t_0)$ for the quantum TDHO whose
dynamics is governed by the Hamiltonian (\ref{quantHamiltonian}) is
given by a composition of unitary operators
$$ U(t,t_0)=T_\rho^{-1}(t)R_\rho(t,t_0)T_\rho(t_0)\,,$$
where
\begin{eqnarray}
R_\rho(t,t_0):=\exp\left(-\frac{i}{2}\int_{t_0}^t\frac{\mathrm{d}\tau}{\rho^2(\tau)}\big(P^2+Q^2\big)\right),\label{U1}
\end{eqnarray}
and $T_\rho(t)=S_\rho(t)D_\rho(t)$, with
\begin{eqnarray}
D_\rho(t):=\exp\left(-\frac{i}{2}\frac{\dot{\rho}(t)}{\rho(t)}Q^{2}\right)\quad
\mathrm{and}\quad
S_\rho(t):=\exp\left(\frac{i}{2}\log\rho(t)\big({Q}{P}+{P}{Q}\big)\right).\label{U2}
\end{eqnarray}
\end{theorem}
\begin{rem} \emph{Note that instead of introducing $\rho$, we could have used
other choices for the $x$ and $y$ functions. In these cases,
conditions (\ref{naturalconditions}) may not hold and the
expressions of the evolution operator would differ from the one
obtained here. For instance, one can select $x(t)=\dot{u}(t)/u(t)$
and $y(t)=\log u(t)$ as in $[6]$, with $u(t)$ being any solution to
the TDHO equation, but this choice is problematic because the set
$\{t\in I\,|\,u(t)=0 \}$ may be non-empty and, hence, the resulting
formula for the unitary operator is generally not well-defined for
all values of the time parameter $t$. This is the reason why the
election of the Ermakov-Pinney solution is especially convenient in
this context --recall that $\rho$ is a positive function. It follows
from the above argument that the appearance of this solution is
nearly unavoidable in this context.}
\end{rem}

\indent Note that the eigenstates of the $R_\rho(t,t_0)$ operator
(\ref{U1}) are given by those of the Hamiltonian operator
corresponding to a quantum harmonic oscillator with unit frequency
$\sqrt{\kappa(t)}=1$,
\begin{equation}\label{H0}
H_0:=\frac{1}{2}\big(P^2+Q^2\big)\,.
\end{equation}
This fact will be shown particularly useful to calculate the Feynman
propagator. It is also important to point out that the procedure
employed in this section is implicitly based upon the transformation
of the so-called Lewis invariant $[7]$
\begin{equation}\label{Lewis}
I_\rho(t):=\frac{1}{2}\left(\frac{Q^{2}}{\rho^2(t)}+\big(\rho(t)P-\dot{\rho}(t)Q\big)^{2}\right)\,,\quad
\dot{I}_{\rho_H}=0\,,
\end{equation}
into an explicitly time-independent quantity --although in order to
obtain the unitary operator it has not been necessary to use it. In
this case, we simply have
\begin{equation}\label{TIT-1}
T_\rho(t)I_\rho(t) T_\rho^{-1}(t)=H_0\,.
\end{equation}
The Lewis invariant is often used to generate exact solutions to the
Schr\"{o}dinger equation, and turns out to be especially useful to
construct semiclassical states for these systems, as will be
discussed later.

\subsection{Propagator formula}

We finally proceed to derive the Feynman propagator for the quantum
TDHO corresponding to the Hamiltonian (\ref{quantHamiltonian}). In
the previous subsection, we have written down the evolution operator
for this system explicitly in closed form in terms of the position
and momentum operators (see \emph{Theorem \ref{ThrU}}). It is given
by the product of the unitary operators (\ref{U1}) and (\ref{U2}).
We calculate now the action of these factors on test functions
$\psi\in\mathscr{S}(\mathbb{R})\subset L^2(\mathbb{R},\mathrm{d}q)$
in the standard Schr\"{o}dinger representation. First, it is
straightforward to see that
\begin{eqnarray*}
\big(T_\rho(t)\psi\big)(q)&=&\sqrt{\rho(t)}\exp\left( - \frac{i}{2}\dot{\rho}(t)\rho(t)q^2\right)\psi(\rho(t)q)=\int_\mathbb{R} K_\rho^+(q,t;q_0) \psi(q_0)\,\mathrm{d}q_0\,,\\
\big(T^{-1}_\rho(t)\psi\big)(q)&=&\frac{1}{\sqrt{\rho(t)}}\exp\left(
\frac{i}{2}\frac{\dot{\rho}(t)}{\rho(t)}q^2\right)\psi(q/\rho(t))=\int_\mathbb{R}
K_\rho^-(q,t;q_0) \psi(q_0)\,\mathrm{d}q_0\,,
\end{eqnarray*}
where we have introduced the distributions
\begin{eqnarray}
K_\rho^+(q,t;q_0)&:=&\sqrt{\rho(t)}\exp\left( - \frac{i}{2}\dot{\rho}(t)\rho(t)q^2\right)\delta(q_0-\rho(t)q)\,,\label{K-}\\
K_\rho^-(q,t;q_0)&:=&\frac{1}{\sqrt{\rho(t)}}\exp\left(
\frac{i}{2}\frac{\dot{\rho}(t)}{\rho(t)}q^2\right)\delta(q_0-q/\rho(t))\,.\label{K+}
\end{eqnarray}
The propagator for $R_\rho(t,t_0)$, satisfying
$$
\big(R_\rho(t,t_0)\psi\big)(q)=\int_\mathbb{R}
K_\rho^0(q,t;q_0,t_0)\psi(q_0)\, \mathrm{d}q_0\,,
$$
can be easily derived from the one corresponding to the TIHO with
unit frequency. As is well known $[8,9]$, the Green function $K^0$
for the Hamiltonian (\ref{H0}) is given by the Feynman-Soriau
formulae
\begin{eqnarray*}
K^0(q,\upsilon;q_0,0)&=&\frac{1}{\sqrt{2\pi
i}}\sin^{-1/2}(\upsilon,0)\exp\left(\frac{i}{2\sin\upsilon}\Big((q^2+q_0^2)\cos\upsilon-2qq_0\Big)\right),
\end{eqnarray*}
whenever $\upsilon\not\equiv 0 \,\,\,(\mathrm{mod}\,\pi)$, and
\begin{eqnarray*}
K^0(q,\upsilon
;q_0,0)&=&\cos^{-1/2}(\upsilon,0)\exp\left(-\frac{i\sin
\upsilon}{2\cos\upsilon}\right)\delta(q_0-q/\cos\upsilon)\,,\quad
\upsilon\equiv 0 \,\,\,(\mathrm{mod}\,\pi)\,,
\end{eqnarray*}
where the so-called Maslov correction factor $[9]$, which allows the
calculation of the propagator beyond the caustics $\{\upsilon\in
\mathbb{R}\,:\,\sin(\upsilon)=0\}=\{\pi k\,:\, k\in \mathbb{Z}\}\,$,
has been conveniently absorbed into the definition of
$\sin^{1/2}(\upsilon,0)$ and $\cos^{1/2}(\upsilon,0)$ given in the
formulation of \emph{Theorem \ref{ThrPropag}}. In view of
(\ref{U1}), we simply get
\begin{eqnarray}
K_\rho^0(q,t;q_0,t_0)=K^0\left(q,\int_{t_0}^t\frac{\mathrm{d}\tau}{\rho^2(\tau)}\,;q_0,0\right).\label{Kr}
\end{eqnarray}
Therefore,
$$
\big(U(t,t_0)\Psi\big)(q)=\big(T^{-1}_\rho(t)
R_\rho(t,t_0)T_\rho(t_0)\Psi\big)(q)=\int_\mathbb{R}
K(q,t;q_0,t_0)\Psi(q_0)\, \mathrm{d}q_0\,,
$$
where
\begin{eqnarray}
K(q,t;q_0,t_0)&=&\int_{\mathbb{R}^2}
K_\rho^-(q,t;q_2)K_\rho^0(q_2,t;q_1,t_0)K_\rho^+(q_1,t_0;q_0)\,\mathrm{d}q_1\,\mathrm{d}q_2\,.\label{K}
\end{eqnarray}
By combining (\ref{K-})-(\ref{K}) with (\ref{c_rho}) and
(\ref{s_rho}), we find the formula for the propagator (\ref{K_qq0a})
enunciated in \emph{Theorem \ref{ThrPropag}} expressed in terms of
the $c$ and $s$ solutions to the classical TDHO equations
(\ref{TDHO_eq}). As expected, the propagator --and hence the
evolution operator itself-- does not depend on the particular
solution $\rho$ to the EP equation (\ref{EP_eq}) chosen to factorize
$U(t,t_0)$. Taking the appropriate limits one obtains, after
straightforward calculations, the propagator evaluated at caustics
(\ref{K_qq0b}). The resulting expressions are in agreement with
those obtained by other authors (see, for example, $[9,10,11,12]$,
where more complicated path integration techniques are often
employed), though in our case they have been achieved within a
different scheme, based essentially on the previous obtention of a
closed expression for the evolution operator. Finally, a direct
calculation shows that the propagator $K(q,t;q_0,t_0)$, viewed as a
function of $(q,t)$, formally satisfies the evolution equation
$$
i\partial_t K =-\frac{1}{2}\partial_q^2
K+\frac{1}{2}q^2\kappa(t)K\,.
$$

\subsection{Transition amplitudes and vacuum
instability}\label{TransInstab}

\indent The exact expressions for the Green functions (\ref{K_qq0a})
and (\ref{K_qq0b}) can be used to \emph{exactly} compute both
transition amplitudes and probabilities. Here, we will restrict
ourselves to the class of normalized states $\phi_n^\omega$ defined
in $L^2(\mathbb{R},\mathrm{d}q)$ as
\begin{equation}\label{phin}
\phi^\omega_n(q):=\frac{\omega^{1/4}}{\sqrt{2^n
n!\sqrt{\pi}}}\exp\Big(-\omega q^2/2\Big)
H_n(\sqrt{\omega}q)\,,\,\,\, \omega>0\,,\,\,\, n\in\mathbb{N}_0\,,
\end{equation}
with $H_n(z)$ denoting the $n$-th Hermite polynomial in the variable
$z\,$. For any fixed value $\omega$, the set
$(\phi_n^\omega\,:\,n\in\mathbb{N}_0)$ defines the usual orthonormal
basis of $L^2(\mathbb{R})$ constituted by the eigenvectors of the
quantum Hamiltonian (\ref{quantHamiltonian}) corresponding to a TIHO
of constant frequency $\sqrt{\kappa(t)}=\omega$. Since the
$\phi_n^\omega$ states are complete, the corresponding transition
amplitudes and probabilities for other states are readily
obtainable. By using the generating function for Hermite
polynomials,
$$
\exp\big(2\sqrt{\omega}qx-x^2\big)=\sum_{n=0}^\infty
H_n(\sqrt{\omega}q)\frac{x^n}{n!}\,,
$$
it is clear that
\begin{equation}\label{transition}
\langle\phi^{\omega_2}_{n_2}\,|\,U(t_2,t_1)\phi^{\omega_1}_{n_1}\rangle=\frac{1}{\pi}\left(\frac{n_1!n_2!\sqrt{\omega_1\omega_2}}
{2^{n_1+n_2+1} i
}\right)^{1/2}s^{-1/2}(t_2,t_1)\,[x_1^{n_1}x_2^{n_2}]\,I(x_1,x_2;\Lambda(t_1,t_2;\omega_1,\omega_2))\,,
\end{equation}
where $[x_1^{n_1}x_2^{n_2}]f(x_1,x_2)$ denotes the complex
coefficient appearing in the $x_1^{n_1}x_2^{n_2}$-term of the Taylor
expansion of the function $f$. Here, for any matrix
$\Lambda\in\mathrm{Mat}_{2\times2}(\mathbb{C})$, we define
\begin{eqnarray*}
I(x_1,x_2;\Lambda)&:=&\exp\big(-(x^2_1+x^2_2)\big)\int_{\mathbb{R}^2}\exp\left(-\frac{1}{2} \vec{q}\,^{t}\Lambda\,\vec{q}+2\vec{x}\,^{t}\,\mathrm{diag}(\sqrt{\omega_1},\sqrt{\omega_2}) \,  \vec{q}\right) \,\mathrm{d}^2\vec{q}\\
&=&\frac{2\pi}{\sqrt{\det\,\Lambda}}\exp\bigg(
\vec{x}\,^{t}\Big(2\mathrm{diag}(\sqrt{\omega_1},\sqrt{\omega_2})\Lambda^{-1}\mathrm{diag}(\sqrt{\omega_1},\sqrt{\omega_2})-\mathbb{I}\Big)
\vec{x} \bigg)\,,
\end{eqnarray*}
whenever $\mathrm{Re}[\Lambda]\geq 0\,$ and ${\det}\,\Lambda\ne0\,$.
In this formula, $\vec{x}$ denotes the column vector with first and
second components given by $x_1$ and $x_2$, respectively; we define
$\vec{q}$ similarly. In our case,
\begin{eqnarray*}
 \Lambda(t_1,t_2;\omega_1,\omega_2):=\left[
             \begin{array}{cc}
               \omega_1-i\displaystyle\frac{c(t_2,t_1)}{s(t_2,t_1)} &\displaystyle \frac{i}{s(t_2,t_1)} \\
              \displaystyle \frac{i}{s(t_2,t_1)} & \omega_2-i\displaystyle\frac{c(t_1,t_2)}{s(t_2,t_1)}\\
             \end{array}
           \right],
\end{eqnarray*}
with
$$\det\Lambda(t_1,t_2;\omega_1,\omega_2)=
\left(\omega_1\omega_2-\frac{\dot{c}(t_2,t_1)}{s(t_2,t_1)}\right)-i\left(\frac{\omega_1c(t_1,t_2)+\omega_2
c(t_2,t_1)}{s(t_2,t_1)}\right).
$$
Here, $\mathrm{Re}[\Lambda(t_0,t;\omega_1,\omega_2)]\ge0$ and
$\det\Lambda(t_0,t;\omega_1,\omega_2)\neq0$ for all $(t_0,t)\in
I\times I$ and $\omega_1,\omega_2\in(0,+\infty)\,$. The Taylor
expansion of $I(x_1,x_2;\Lambda)$ can be efficiently computed by
applying the following lemma, that trivially follows from the
multinomial formula.

\begin{lema}
Let
$$B=B^t=\left[\begin{array}{cc}b_{11} & b_{12}\\ b_{12} & b_{22}\end{array}\right]\in \mathrm{Mat}_{2\times 2}(\mathbb{C})\,.
$$
Then, using the notation introduced above, we have
\begin{eqnarray*}
[x_1^{n_1}x_2^{n_2}]\exp\bigg(\vec{x}\,^t B
\vec{x}\bigg)=b_{11}^{(n_1-n_2)/2}(2b_{12})^{n_2} \sum_{m\in
\Delta(n_1,n_2)} \frac{(b_{11}b_{22})^m
(4b^2_{12})^{-m}}{m!(m+(n_1-n_2)/2)!(n_2-2m)!}\,,
\end{eqnarray*}
whenever $n_1$ and $n_2$ have the same parity, and vanishes
otherwise. Here, $\Delta(n_1,n_2):=\big(m\in \mathbb{N}_0\,:\,
\max\{0,(n_2-n_1)/2\}\leq m\leq \lfloor n_2/2\rfloor\,\big)\,$,
where $\lfloor x\rfloor$ denotes the largest integer less than or
equal to $x\in\mathbb{R}\,$. In particular, taking $n_1=0\,$, we get
\begin{equation}\label{b22}
[x_1^{0}x_2^{n_2}]\exp\bigg(\vec{x}\,^t B
\vec{x}\bigg)=\frac{b_{22}^{n_2/2}}{(n_2/2)!}\quad\mathrm{for}\quad
n_2\equiv 0 \,\,\,(\mathrm{mod}\, 2)\,,
\end{equation}
and vanishes if $n_2$ is an odd number.
\end{lema}
\noindent \textbf{Remarks.} Note that the TDHO quantum dynamics is
invariant under parity inversion $\mathbf{P}$ and states
$\phi_n^\omega$ satisfy $\mathbf{P}\phi_n^\omega=
(-1)^n\phi^\omega_n$. Hence,
$\langle\phi^{\omega_2}_{n_2}\,|\,U(t_2,t_1)\phi^{\omega_1}_{n_1}\rangle=0$
if $n_1$ and $n_2$ have different parity.
\\
\linebreak \indent As a concrete example, in the case of a TIHO with
constant frequency $\omega=\omega_1=\omega_2$, we identify
$$
B=2\,\mathrm{diag}(\sqrt{\omega},\sqrt{\omega})\,
\Lambda^{-1}(t_1,t_2;\omega,\omega)\,
\mathrm{diag}(\sqrt{\omega},\sqrt{\omega})-\mathbb{I}
=\exp\big(-i\omega(t_2-t_1)\big) \left[\begin{array}{cc} 0 & 1\\ 1 &
0
\\\end{array}\right]
$$
and, hence,
$$
I(x_1,x_2;\Lambda(t_1,t_2;\omega,\omega))=\sum_{n=0}^\infty\frac{2^n}{n!}\exp\big(-i\omega
n (t_2-t_1)\big) x_1^nx_2^n\,.
$$
This is in perfect agreement with
$$
\langle\phi^\omega_{n_2}\,|\,U(t_2,t_1)\phi^\omega_{n_1}\rangle=\exp\big(-i\omega(n_1+1/2)(t_2-t_1)\big)\delta(n_1,n_2)\,,
$$
where $\delta(n_1,n_2)$ denotes the Kronecker delta. For arbitrary
time-dependent frequencies the formula (\ref{transition}), when
restricted to the same initial and final frequencies
$\omega_1=\omega_2$, coincides with the one given in $[13]$ written
in terms of associated Legendre functions.
\\
\indent We conclude this section with the analysis of the
instability of the vacuum state $\phi_0^\omega$ due to the
nonautonomous nature of the Hamiltonian (\ref{quantHamiltonian}).
This can be easily derived from the formulae (\ref{transition}) and
(\ref{b22}).

\begin{theorem}\label{th_vacuum}
The quantum time evolution of the vacuum state $\phi_0^\omega$ is
generally given by a superposition of states
$U(t,t_0)\phi_0^\omega=\sum_{n\in\mathbb{N}_0}\langle\phi_{2n}^\omega\,|\,U(t,t_0)\phi_0^\omega\rangle\phi_{2n}^\omega\,,$
where the probability amplitudes
$\langle\phi_{2n}^\omega\,|\,U(t,t_0)\phi_0^\omega\rangle$ are given
by
\begin{equation}\label{2n0}
\langle
\phi_{2n}^\omega\,|\,U(t,t_0)\phi_0^\omega\rangle=\frac{\sqrt{(2n)!}}{2^n
n!}\big(2\omega
(\Lambda^{-1}(t_0,t;\omega,\omega))_{22}-1\big)^n\langle\phi_0^\omega\,|\,U(t,t_0)\phi_0^\omega\rangle\,,\,\,\,n\in\mathbb{N}\,,
\end{equation}
in terms of the the expectation value
$$\langle\phi_0^\omega\,|\,U(t,t_0)\phi_0^\omega\rangle=\sqrt{\frac{2\omega}{\det\Lambda(t_0,t;\omega,\omega)}}\exp(-i\pi/4)\,s^{-1/2}(t,t_0)\,,$$
with
$$(\Lambda^{-1}(t_0,t;\omega,\omega))_{22}=\frac{\omega s^2(t_2,t_1)-i
s(t_2,t_1)c(t_2,t_1)}{1+\omega^2s^2(t_2,t_1)-c(t_2,t_1)c(t_1,t_2)-i\omega
s(t_2,t_1)\big(c(t_2,t_1)+c(t_1,t_2)\big)}\,.$$
\end{theorem}
\noindent \textbf{Remarks.} Consider the usual annihilation and
creation operators
\begin{equation}\label{aa*usual}
a_\omega:=\frac{1}{\sqrt{2}}\big(\sqrt{\omega}Q+iP/\sqrt{\omega}\big)\quad
\mathrm{and} \quad
a_\omega^*:=\frac{1}{\sqrt{2}}\big(\sqrt{\omega}Q-iP/\sqrt{\omega}\big)
\,,
\end{equation}
with $[a_\omega,a_\omega^*]=\mathbf{1}$ and
$[a_\omega,a_\omega]=0=[a_\omega^*,a_\omega^*]$, such that
$a_\omega^*\phi_n^\omega=\sqrt{n+1}\phi_{n+1}^\omega$ and
$a_\omega\phi_n^\omega=\sqrt{n}\phi_{n-1}^\omega$,
$\forall\,n\in\mathbb{N}$, with $a_\omega\phi_0^\omega=0\,$. The
evolution of these operators in the Heisenberg picture can be
obtained directly from (\ref{QHPH}),
\begin{eqnarray}\label{BogolTransf}
U^{-1}(t,t_0)\,a_\omega\,U(t,t_0)&=&A_\omega(t,t_0)a_\omega+B_\omega(t,t_0)a_\omega^{*}\,,\\
U^{-1}(t,t_0)\,a_\omega^*\,U(t,t_0)&=&\bar{B}_\omega(t,t_0)a_\omega+\bar{A}_\omega(t,t_0)a_\omega^{*}\,,\nonumber
\end{eqnarray}
where $A_\omega(t,t_0)$ and $B_\omega(t,t_0)$ are the Bogoliubov
coefficients
\begin{eqnarray}
A_\omega(t,t_0)&:=&\frac{1}{2}\Big(c(t,t_0)+\dot{s}(t,t_0)+i\big(\omega^{-1}\dot{c}(t,t_0)-\omega
s(t,t_0)\big)\Big)\,,\label{BogolCoeff1}\\
B_\omega(t,t_0)&:=&\frac{1}{2}\Big(c(t,t_0)-\dot{s}(t,t_0)+i\big(\omega^{-1}\dot{c}(t,t_0)+\omega
s(t,t_0)\big)\Big)\,,\label{BogolCoeff2}
\end{eqnarray}
satisfying $A_\omega(t,t_0)=\bar{A}_\omega(t_0,t)$,
$B_\omega(t,t_0)=-B_\omega(t_0,t)$, and
$|A_\omega(t,t_0)|^2-|B_\omega(t,t_0)|^2=1\,$, $\forall\,(t,t_0)\in
I\times I\,$. Note, in particular, that $A_\omega(t,t_0)$ never
vanishes. For example, for the TIHO of constant frequency $\omega>0$
we have $B_\omega(t,t_0)=0$ and
$A_\omega(t,t_0)=\exp(-i(t-t_0)\omega)$. A straightforward
calculation yields (see also $[14]$)
\begin{equation}\label{vacuumevol}
U(t,t_0)\phi_0^\omega
=\langle\phi_0^\omega\,|\,U(t,t_0)\phi_0^\omega\rangle\exp\left(-\frac{1}{2}\frac{B_\omega(t_0,t)}{A_\omega(t_0,t)}a_\omega^{*2}\right)\phi_0^\omega\,,
\end{equation}
This formula is in perfect agreement with the transitions
(\ref{2n0}). Indeed, it is straightforward to check that
$$2\omega(\Lambda^{-1}(t_0,t;\omega,\omega))_{22}-1=-B_\omega(t_0,t)/A_\omega(t_0,t)\,.$$
Since $\det\Lambda(t_0,t;\omega,\omega)=-2i\omega
s^{-1}(t,t_0)A_\omega(t_0,t)$, the expectation value
$\langle\phi_0^\omega\,|\,U(t,t_0)\phi_0^\omega\rangle$ can be
rewritten as
\begin{equation}\label{Up0U}
\langle\phi_0^\omega\,|\,U(t,t_0)\phi_0^\omega\rangle=\frac{1}{\sqrt{|A_\omega(t_0,t)|}}\exp(i\sigma(t,t_0))\,,
\end{equation}
where the phase $\sigma(t,t_0)\in C^1(I\times I)$ comes from a
careful calculation of the principal argument. For a TIHO with
constant frequency $\omega>0$, we have
$\sigma(t,t_0)=(t_0-t)\omega/2$ for all $t,t_0\in\mathbb{R}\,$.
Given an arbitrary squared frequency $\kappa(t)$, the phase
$\sigma(t,t_0)$ evaluated at times $t$ close to $t_0$ is simply
given by
\begin{equation}\label{sigma}
\sigma(t,t_0)=-\frac{1}{2}\arctan\left(\frac{\omega
s(t,t_0)-\omega^{-1}\dot{c}(t,t_0)}{c(t,t_0)+\dot{s}(t,t_0)}\right).
\end{equation}
The $\sigma$ phase can be conveniently canceled through a suitable
redefinition of the Hamiltonian (\ref{quantHamiltonian}) just in the
case when $\dot{\sigma}(t,t_0)$ is independent of $t_0\,$. In that
situation, by identifying $\vartheta(t)=\dot{\sigma}(t,t_0)$ in
equation (\ref{K1}), we have that the redefined evolution operator
satisfies
$\langle\phi_0^\omega\,|\,U_1(t,t_0)\phi_0^\omega\rangle=1/\sqrt{|A_\omega(t_0,t)|}\,$.
In the TIHO case, we get $\vartheta(t)=-\omega/2$ (this amounts to
considering normal order). In general, it is not possible to proceed
in this way in all cases when dealing with arbitrary time-dependent
frequencies. In any case, the $\sigma$ phase is irrelevant for the
calculation of transition probabilities. In particular, given
$\Psi_1,\Psi_2\in\mathscr{H}$ with
$\Psi_1=F_1(a_\omega,a_\omega^*)\,\phi_0^\omega\,$, where $F_1$ is
some analytic function, we have
\begin{eqnarray*}
&&\hspace*{-0.8cm}\mathrm{Prob}(\Psi_2,t_2\,|\,\Psi_1,t_1)=|\langle\Psi_2\,|\,U(t_2,t_1)\Psi_1\rangle|^2\\
&&\hspace*{1cm}=\frac{|\langle\Psi_2\,|\,F_1(a_{\omega
H}(t_1,t_2),a_{\omega H}^{*}(t_1,t_2))\exp\big(-B_\omega(t_1
,t_2)/(2A_\omega(t_1,t_2))a_\omega^{*2}\big)\phi_0^\omega\rangle|^2}{|A_\omega(t_1,t_2)|}\,,
\end{eqnarray*}
where the time dependence only appears through the Bogoliubov
coefficients (\ref{BogolCoeff1}) and (\ref{BogolCoeff2}). Finally,
it is important to point out that the transformations
(\ref{BogolTransf}) and the evolution of the vacuum state
(\ref{vacuumevol}) fully characterize the quantum time evolution of
the TDHO. By using these relations, we can easily compute the action
of $U(t,t_0)$ on any basic vector
$\phi_n^\omega=(1/\sqrt{n!})\,a_\omega^{*n}\phi_0^\omega\,$.

\section{Semiclassical states}\label{SemiclStates}

\indent In this section, we will look for states that behave
semiclassically under the dynamics defined by the quantum
Hamiltonian (\ref{quantHamiltonian}). We will base our study on the
concrete factorization of the evolution operator defined in
\emph{Theorem \ref{ThrU}}. To achieve this goal, note that the
eigenvalue problem for the Lewis invariant (\ref{Lewis}) can be
exactly solved. Indeed, let us fix $t_0\in I$ and let
$(\phi_n\,:\,n\in\mathbb{N}_0)$ be the eigenstates (\ref{phin}) of
the auxiliary  Hamiltonian $H_0$ (\ref{H0}) corresponding to unit
frequency $\omega=1$. According to relation (\ref{TIT-1}), the
initial states $\psi^{\rho}_n(t_0):=T_\rho^{-1}(t_0)\phi_n$,
$$\psi^{\rho}_n(t_0,q)=\left(\frac{1}{2^n n!\sqrt{\pi}\rho(t_0)}\right)^{1/2}\exp\left(\frac{i}{2}\left(\frac{\dot{\rho}(t_0)}{\rho(t_0)}
+\frac{i}{\rho^2(t_0)}\right)q^2\right)H_n(q/\rho(t_0))\in
L^2(\mathbb{R},\mathrm{d}q)\,,$$ labeled both  by $\rho$ and the
integers $n\in\mathbb{N}_0$, are eigenstates of $I_\rho(t_0)$ with
eigenvalues equal to $n+1/2$. Consider now the initial pure state
\begin{equation}\label{semiclass}
\Phi_{\rho}^{(z)}(t_0):=T_\rho^{-1}(t_0)\Phi^{(z)}
=e^{-|z|^{2}/2}\sum_{n=0}^{\infty}\frac{z^n}{\sqrt{n!}}\psi^\rho_n(t_0)\,,\,\,\,\,z\in\mathbb{C}\,,
\end{equation}
with
$\Phi^{(z)}:=e^{-|z|^2/2}\sum_{n=0}^{\infty}\frac{z^n}{\sqrt{n!}}\phi_n$
being the well-known coherent states for the Hamiltonian $H_0\,$.
Let us take the annihilation and creation operators $a$ and $a^*$
for unit frequency $\omega=1$ defined in (\ref{aa*usual}). The
superposition (\ref{semiclass}) is a normalized eigenvector of the
(time-dependent) annihilation operator
\begin{equation}\label{arho1D}
a_\rho(t_0):=T_\rho^{-1}(t_0)\,a\,T_\rho(t_0)
=\frac{1}{\sqrt{2}}\big(Q/\rho(t_0)+i(\rho(t_0)P-\dot{\rho}(t_0)Q)\big)\,,
\end{equation}
in the sense that
$a_\rho(t_0)\Phi^{(z)}_\rho(t_0)=z\Phi^{(z)}_\rho(t_0)\,$. This
operator, together with the associated creation operator
$$a^*_\rho(t_0):=T_\rho^{-1}(t_0)\,a^*\,T_\rho(t_0)
=\frac{1}{\sqrt{2}}\big(Q/\rho(t_0)-i(\rho(t_0)P-\dot{\rho}(t_0)Q)\big)\,,$$
satisfies the Heisenberg algebra,
$[a_\rho(t_0),a^*_\rho(t_0)]=\mathbf{1}$,
$[a_\rho(t_0),a_\rho(t_0)]=0=[a^*_\rho(t_0),a^*_\rho(t_0)]$, for
each initial value of the time parameter $t_0\,$. In particular, the
Lewis invariant (\ref{Lewis}) may be expressed in terms of these
operators as
$I_\rho(t_0)=a^*_\rho(t_0)a_\rho(t_0)+(1/2)\mathbf{1}\,$. Through
unitary time evolution, we get
\begin{equation}\label{Phit}
\Phi_\rho^{(z)}(t,t_0):=U(t,t_0)\Phi_{\rho}^{(z)}(t_0)
=\exp\left(-\frac{i}{2}\int_{t_0}^{t}\frac{\mathrm{d}\tau}{\rho^{2}(\tau)}\right)\Phi_{\rho}^{(z_\rho(t,t_0))}(t)\,,
\end{equation}
where we have denoted
$$
z_\rho(t,t_0):=\exp\left(-i\int_{t_0}^{t}\frac{\mathrm{d}\tau}{\rho^2(\tau)}\right)\,z\,,\quad
z\in\mathbb{C}\,.$$ We want to remark that the time-dependent phase
appearing in (\ref{Phit}) is necessary for these states to verify
the Schr\"{o}dinger equation. In our case, they coincide with those
defined in equation (4.6) of reference $[15]$. We conclude that the
family of states (\ref{semiclass}) is closed under the dynamics.
Moreover, the following theorem can be used to justify that these
states can be considered as semiclassical under certain assumptions.

\begin{theorem}\label{th_semiclass}
Let $z=x+iy\in \mathbb{C}$ and $t_0\in I$. The position and momentum
expectation values in the state
$\Phi_\rho^{(z)}(t,t_0)=U(t,t_0)\Phi_{\rho}^{(z)}(t_0)$ satisfy
\begin{eqnarray*}
q_H(t,t_0)&=&\big\langle\Phi_{\rho}^{(z)}(t,t_0)\,\big|\,Q\,\Phi_{\rho}^{(z)}(t,t_0)\big\rangle
=\sqrt{2}\rho(t)\,\mathrm{Re}\big[z_\rho(t,t_0)\big]\,,\\
p_H(t,t_0)&=&\big\langle\Phi_{\rho}^{(z)}(t,t_0)\,\big|\,P\Phi_{\rho}^{(z)}(t,t_0)\big\rangle
=\sqrt{2}\,\mathrm{Re}\big[\big(\dot{\rho}(t)-i/\rho(t)\big)z_\rho(t,t_0)\big]\,,
\end{eqnarray*}
where $(q_H,p_H)$ is the classical solution (\ref{Hamilton_class})
determined by the Cauchy data $(q,p)=\big(\sqrt{2}\rho(t_0)
x,\sqrt{2}( \dot{\rho}(t_0)x+y/\rho(t_0))\big)$ at time $t_0\,$.
Moreover, the mean square deviations of the position and momentum
operators with respect to the evolved state
$\Phi_\rho^{(z)}(t,t_0)$,
\begin{eqnarray}\label{varianceQP}
\Delta_{\Phi_{\rho}^{(z)}(t,t_0)}Q=\frac{1}{\sqrt{2}}\,\rho(t)\,,\quad
\Delta_{\Phi_{\rho}^{(z)}(t,t_0)}P=\frac{1}{\sqrt{2}}\,\big|\dot{\rho}(t)-i\rho^{-1}(t)\big|\,,
\end{eqnarray}
are independent of both $t_0$ and the Cauchy data defined by $z\,$.
\end{theorem}

\begin{rem} \emph{Given any observable $\mathcal{O}$,  its uncertainty in the
state $\Psi\in\mathscr{D}_\mathcal{O}$ is defined as $\Delta_\Psi
\mathcal{O}:=\big(\langle\Psi\,|\,\mathcal{O}^2\Psi\rangle-\langle\Psi\,|\,\mathcal{O}\Psi\rangle^2\big)^{1/2}$.
Note that, in general, the elements of the family of states under
consideration are neither standard coherent states nor squeezed
states. For instance, for the free particle (\ref{c&sfree}) one can
choose $\rho(t)=\sqrt{1+(t-t_0)^2}$ and, hence,
$\Delta_{\Phi_{\rho}^{(z)}(t,t_0)}Q\sim t/\sqrt{2}$ for large values
of $t\,$; similar results occur for other elections of $\rho\,$.
Nevertheless, it is obvious that we will obtain good semiclassical
states for a system whenever the solution $\rho$ to the auxiliary EP
equation (\ref{EP_eq}) has a suitable behavior, for instance, if
$\rho$ is periodic in time or is simply a bounded function. We will
analyze some clarifying examples in this respect.}
\end{rem}

\begin{figure}
  \includegraphics[width=16cm]{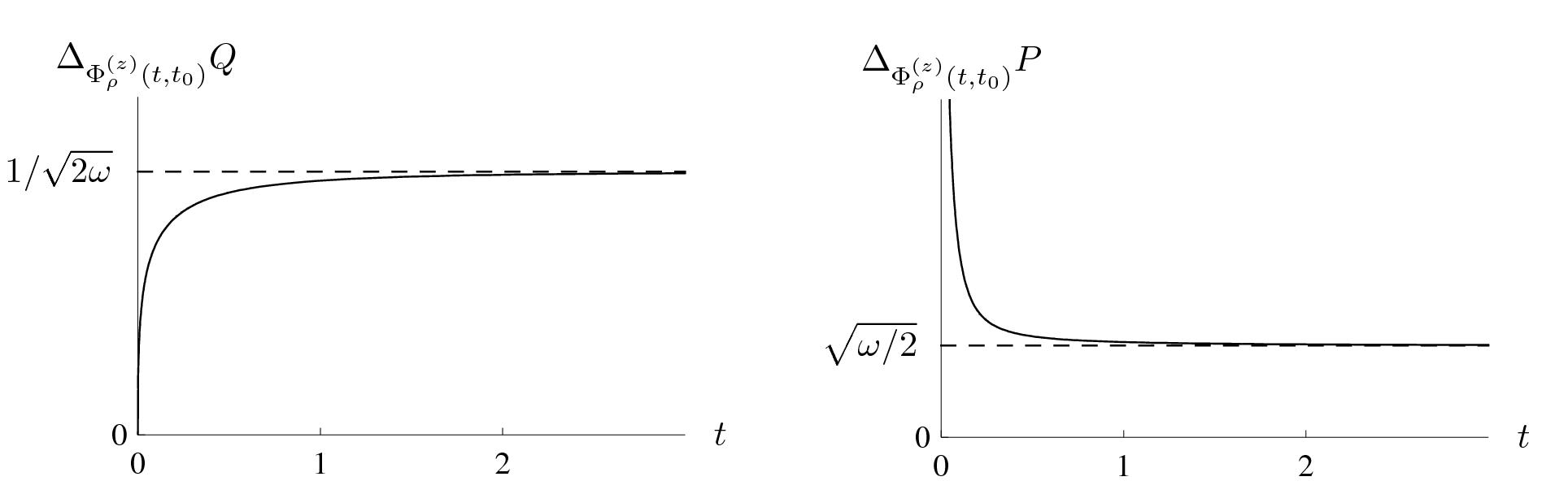}\\
  \caption{Variances of the position and momentum operators for the 3-torus Gowdy-type oscillator. Here,
  $\rho(t)=\sqrt{\pi t\big(J_0^2(\omega t)+Y_0^2(\omega t)\big)/2}$. The $\Phi^{(z)}_\rho(t,t_0)$ are states of minimum uncertainty
  for times $t$ far from the singularity at $t=0$\,.}\label{GrGowdyT3}
\end{figure}

\begin{figure}
  \includegraphics[width=16cm]{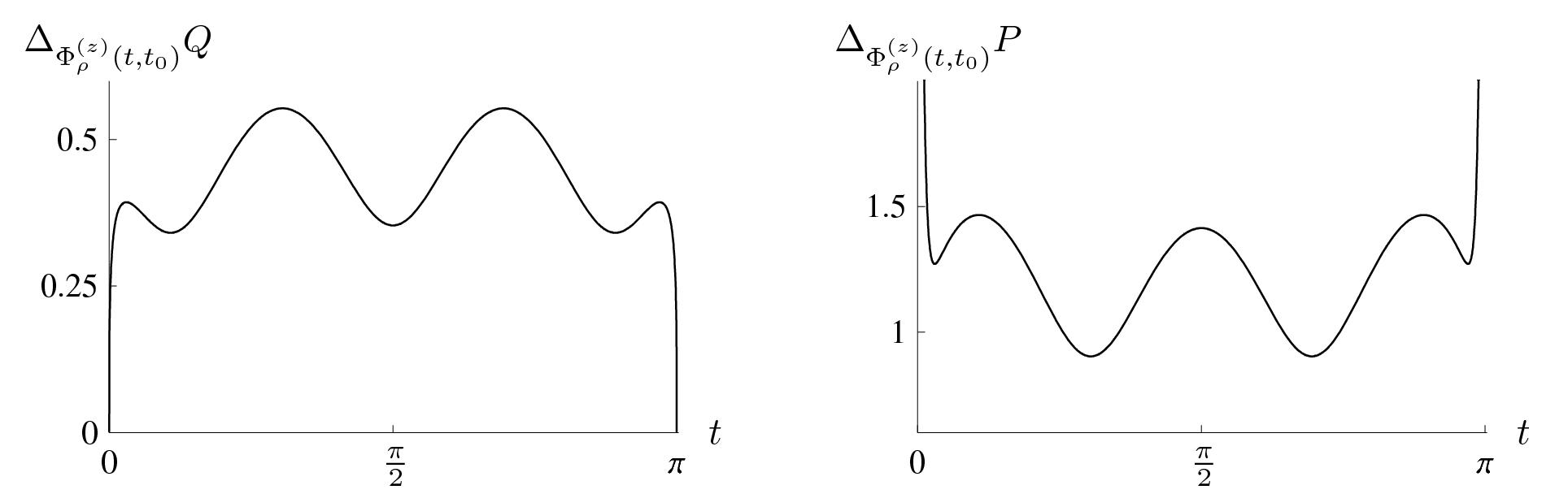}\\
  \caption{Variances of the position and momentum operators for the 3-handle and 3-sphere Gowdy-type oscillators.
  Here, we take the solution $\rho(t)=\sqrt{\sin t}\big(\mathscr{P}^2_{(\omega^\prime-1)/2}(\cos t)+\mathscr{Q}^2_{(\omega^\prime-1)/2}(\cos t)\big)^{1/2}$ to the auxiliar
  Ermakov-Pinney equation. In particular, graphics correspond to $\omega^{\prime}=5\,$.}\label{GrGowdyS1S2&S3}
\end{figure}

\noindent \textbf{Example 1} (Vertically driven pendulum)\textbf{.}
Consider the vertically driven pendulum $[16]$, i.e., the motion of
a physical pendulum whose supporting point oscillates in the
vertical direction. In the small angles regime, it is described by
the Mathieu equation in its canonical form $[17]$
$$\ddot{u}(t)+\kappa(t;a,b)u(t)=0\,,\quad \kappa(t;a,b):=a-2b\cos(2t)\,,\quad a,b\in\mathbb{R}\,.$$
The general solution to this equation is a real linear combination
of the so-called Mathieu cosine and sine functions $[18,19]$,
denoted respectively as $Ce(t;a,b)$ and $Se(t;a,b)$. Given a nonzero
$b$ value, it is a well-known fact that the Mathieu cosine and sine
functions are periodic in the time parameter $t$ only for certain
(countable number of) values of the $a$ parameter, called
\emph{characteristic values}. The procedure to calculate these
characteristic values for even or odd Mathieu functions with
\emph{characteristic exponent}\footnote{All Mathieu functions have
the form $\exp(irt)F(t)$, where $r$ is the characteristic exponent
and the function $F(t)$ has period $2\pi$.} $r\in\mathbb{Z}$ and
parameter $b$ can be efficiently implemented in a computer. In this
case, solutions to the EP equation (\ref{EP_eq}) inherit the
periodic behavior from the Mathieu solutions, in such a way that one
obtains well-behaved semiclassical states for which the average
position and momentum follow the classical trajectories, whereas the
corresponding uncertainties vary periodically in time. Note that,
for small values of the $b$ parameter, we have
$Ce(t;a,b)\sim\cos(\sqrt{a}t)$ and $Se(t;a,b)\sim\sin(\sqrt{a}t)$,
and the system closely approximates the TIHO with squared frequency
given by the $a$ parameter.
\\
\linebreak \noindent \textbf{Example 2} ($\mathbb{T}^{3}$ Gowdy-like
oscillator)\textbf{.} Consider the TDHO equation
$$\ddot{u}(t)+\kappa(t;\omega)u(t)=0\,,\quad \kappa(t;\omega):=\omega^2+\frac{1}{4t^2}\,,\,\,\,\omega\in\mathbb{R}\,,\,\,\,t\in(0,+\infty)\,.$$
This equation is satisfied for each mode of the scalar fields
encoding the information about the gravitational local degrees of
freedom of the $\mathbb{T}^3$ Gowdy models. In terms of the zero
Bessel functions of first and second kind $[19]$, denoted $J_0$ and
$Y_0$ respectively, the $c$ and $s$ solutions introduced in
\emph{section \ref{TDHOproperties}} are given by
\begin{eqnarray}
c(t,t_0)\!\!&=&\!\!\frac{\pi}{4}\Big(\sqrt{\frac{t}{t_0}}Y_0(\omega
t_0)\!-\!2\omega \sqrt{t_0t}Y_1(\omega t_0)\Big) J_0(\omega t)\\
&-&\frac{\pi}{4}\Big(\sqrt{\frac{t}{t_0}}J_0(\omega t_0)-2\omega
\sqrt{t_0t}J_1(\omega t_0)\Big) Y_0(\omega t),\nonumber\\
s(t,t_0)\!\!&=&\!\!-\frac{\pi}{2}\sqrt{t_0t}Y_0(\omega
t_0)J_0(\omega t)+\frac{\pi}{2}\sqrt{t_0t}J_0(\omega t_0)Y_0(\omega
t)\,.\label{c&s3torus}
\end{eqnarray}
Note that the squared frequency is a sum of a positive constant
$\omega^2$ plus a decreasing function of time, so that the system
approaches a time-independent oscillator as $t$ tends to infinity.
In \emph{figure \ref{GrGowdyT3}}, we show states
$\Phi_{\rho}^{(z)}(t,t_0)$ that behave as coherent states for large
values of the time parameter. The classical equation of motion has a
singularity at $t=0$ which translates into the vanishing of the
uncertainty of the position operator --and, hence, into the
divergence of the variance for the conjugate momentum-- at that
instant of time.
\\
\indent There are other interesting effects due to the classical
singularity. Let us consider again the study of transition
amplitudes developed in \emph{subsection \ref{TransInstab}} and take
$\omega_1=\omega_2=\omega$. We proceed to analyze the behavior of
the (unique) state $\Psi(t_2,t_1)$ that evolves to the vacuum state
$\phi_0^\omega$ at time $t_2$ when used as Cauchy data in
$t_1<t_2\,$, i.e.,
$$
U(t_2,t_1)\Psi(t_2,t_1)=\phi^\omega_0\Leftrightarrow
\Psi(t_2,t_1)=U(t_1,t_2)\phi_0^\omega\,.
$$
The transition amplitudes $\langle
\phi^\omega_{2n}\,|\,\Psi(t_2,t_1)\rangle = \langle
\phi^\omega_{2n}\,|\,U(t_1,t_2)\phi_0^\omega\rangle$,
$n\in\mathbb{N}_0$, can be computed by using (\ref{2n0}). We
recognize two regions of interest in the time domain,
\begin{eqnarray*}
T_{0+}:=\{(t_1,t_2)\,\,|\,\,0<t_1\ll\omega^{-1}\ll t_2\}
\quad\mathrm{and}\quad T_{++}:=\{(t_1,t_2)\,\,|\,\,\omega^{-1}\ll
t_1< t_2\}\,.
\end{eqnarray*}
In $T_{++}$, the asymptotic behavior of the Bessel functions for
large values of the time parameter $[19]$ leads the system to behave
as a TIHO of constant frequency $\omega$, with $\Psi(t_2,t_1)\sim
\phi_0^\omega\,$. On the other hand, in the region $T_{0+}\,$, the
proximity of $t_1$ to the classical singularity manifests itself in
the fact that the wave function takes the form $\Psi(t_2,t_1)\sim
0\,$. Note that this behavior is in conflict with the unitary
evolution of the system, which implies $\|\Psi(t_2,t_1)\|=1\,$.
\\
\linebreak \noindent \textbf{Example 3} Let us consider now a
harmonic oscillator of the type analyzed in \emph{Chapters
\ref{ChapterI}} and \emph{\ref{ChapterII}} corresponding to the
$\mathbb{S}^1\times\mathbb{S}^2$ and $\mathbb{S}^3$ Gowdy models.
Here, the modes satisfy equations of motion of the form
$$\ddot{u}(t)+\kappa(t;\omega)u(t)=0\,,\quad \kappa(t;\omega):=\omega^2+\frac{1}{4}(1+\csc^2 t)\,,\,\,\,\omega\in\mathbb{R}\,,\,\,\,t\in(0,\pi)\,.$$
In this case, in terms of first and second class Legendre functions
$[19]$ $\mathscr{P}_x$ and $\mathscr{Q}_x$, $x\in\mathbb{R}\,$, we
have
\begin{eqnarray}\label{c&sremaintopol}
&&\hspace*{-1.3cm}c(t,t_0)=\frac{1}{2}\sqrt{\sin t/\sin
t_0}\mathscr{Q}_{\frac{\omega^\prime-1}{2}}(\cos
t)\big((1+\omega^{\prime})\mathscr{P}_{\frac{1+\omega^\prime}{2}}(\cos
t_0)
-\omega^\prime\mathscr{P}_{\frac{\omega^\prime-1}{2}}(\cos t_0)\big)\nonumber\\
&&\hspace*{0.0cm}-\frac{1}{2}\sqrt{\sin t/\sin
t_0}\mathscr{P}_{\frac{\omega^\prime-1}{2}}(\cos
t)\big((1+\omega^{\prime})\mathscr{Q}_{\frac{1+\omega^\prime}{2}}(\cos
t_0)-\omega^\prime\mathscr{Q}_{\frac{\omega^\prime-1}{2}}(\cos
t_0)\big)\,,\nonumber\\
&&\hspace*{-1.3cm}s(t,t_0)=\sqrt{\sin t\sin
t_0}\big(\mathscr{Q}_{\frac{\omega^\prime-1}{2}}(\cos
t_0)\mathscr{P}_{\frac{\omega^\prime-1}{2}}(\cos
t)-\mathscr{P}_{\frac{\omega^\prime-1}{2}}(\cos
t_0)\mathscr{Q}_{\frac{\omega^\prime-1}{2}}(\cos t)\big)\,,
\end{eqnarray}
where $\omega^\prime:=\sqrt{1+4\omega^2}\,$. In \emph{figure
\ref{GrGowdyS1S2&S3}}, we show the behavior of states
$\Phi_{\rho}^{(z)}(t,t_0)$ for which the uncertainties of the
position and momentum operators have an oscillatory behavior far
enough from the singularities occurring at $t=0$ and $t=\pi$.
Although $\rho$ does not vary periodically, the function remains
bounded and, thus, the $\Phi_{\rho}^{(z)}(t,t_0)$ states can be used
to perform a semiclassical study of these models. Finally, one may
proceed as in the 3-torus case in order to analyze the way the
classical singularities affect the quantum behavior of the systems,
obtaining similar results.

\bigskip

\noindent \textbf{References}
\begin{itemize}
\item [1] V. P. Ermakov, Univ. Izv. Kiev. \textbf{20},1 (1880); see Appl. Anal. Discrete Math. \textbf{2}, 123-145 (2008) for an English translation of Ermakov's original paper.
\item [2] E. Pinney, Proc. Amer. Math. Soc. \textbf{1}, 681 (1950).
\item [3] P. G. L. Leach and K. Andriopoulos, Appl. Anal. Discrete Math. \textbf{2}, 146-157 (2008).
\item [4] H. R. Lewis, Jr., J. Math. Phys. \textbf{9}, 1976 (1968).
\item [5] M. Lutzky, Phys. Lett. A \textbf{68}, 3 (1978).
\item [6] M. Fern\'andez Guasti and H. Moya-Cessa, J. Phys. A: Math. Gen. \textbf{36}, 2069-2076 (2003).
\item [7] H. R. Lewis, Jr., Phys. Rev. Lett. \textbf{18}, 510, 636 (1967).
\item [8] B. Felsager, \emph{Geometry, Particles and Fields}, Odense University Press, (1981).
\item [9] J. Rezende, J. Math. Phys. \textbf{25}, 3264-3269 (1984).
\item [10] D. C. Khandekar and S. V. Lawande, J. Math. Phys. \textbf{16}, 384-388 (1973).
\item [11] A. K. Dhara and S. V. Lawande, Phys. Rev. A \textbf{30}, 560-567 (1984).
\item [12] B. K. Cheng and F. T. Chan, J. Phys. A: Math. Gen. \textbf{20}, 3771-3775 (1987).
\item [13] L. F. Landovitz, A.M. Levine, and W.M. Schreiber, J. Math. Phys. \textbf{21}, 2159-2163 (1980).
\item [14] K. Pilch and N. Warner, Class. Quant. Grav. \textbf{4}, 1183 (1987).
\item [15] J. G. Hartley and J. R. Ray, Phys. Rev. D \textbf{25}, 382-386 (1982).
\item [16] J. V. Jos\'e and E. J. Saletan, \emph{Classical Dynamics: A Contemporary Approach}, Cambridge University Press, (1998).
\item [17] E. Mathieu, Journal of Mathematiques Pures et Appliquees \textbf{13}, 137–203 (1868).
\item [18] N. W. McLachlan, \emph{Theory and Application of Mathieu Functions}, Dover, (1965).
\item [19] M. Abramowitz and I. A. Stegun, \emph{Handbook of Mathematical Functions with Formulas, Graphs, and Mathematical Tables}, Dover, (1964).
\end{itemize}


\newpage
\thispagestyle{plain} \mbox{}

\chapter{Fock Spaces}\label{AppendixFock}


\pagestyle{fancy} \fancyhf{}

\fancyhead[LO]{\textsc{Appendix E. Fock spaces}}
\fancyhead[RO,LE]{\thepage}
\renewcommand{\headrulewidth}{0.6pt}


\indent Consider a quantum mechanical system consisting of (possibly
a variable number of) identical particles. Their
indistinguishability at the atomic level is reflected by the
so-called \emph{symmetrization principle}, according to which the
pure states of the system under consideration must be either
completely symmetric or antisymmetric under the exchange of any two
particles. Starting from such a general assumption, it is then
possible to prove in the context of the quantum theory of local
fields the so-called \emph{spin-statistics theorem}, which divides
particles into two groups: The pure states of a system of identical
particles are symmetric (antisymmetric) if they have integer
(half-integer) spin. Particles which exhibit symmetric states
(bosons) satisfy the Bose-Einstein statistics, whereas particles
with antisymmetric states (fermions) obey the Fermi-Dirac
statistics. In this appendix, we construct the bosonic/fermionic
Fock spaces used for the description of these systems, as well as
the representation of the corresponding canonical
commutation/anticommutation relations.

\section{Symmetric and antisymmetric Fock spaces}

\indent Let $\mathscr{H}$ be a (separable) one-particle Hilbert
space. Denote by $\mathscr{H}^{\otimes n}$, $n\in\mathbb{N}$, the
$n$-fold Hilbert tensor product of $\mathscr{H}$, i.e., the Cauchy
completion of the pre-Hilbert space of finite linear combinations of
elements of the form $x_{1}\otimes\cdots\otimes x_{n}$,
$x_{1},\ldots,x_{n}\in\mathscr{H}$, with respect to the inner
product
\begin{equation*}
\langle x_{1}\otimes\cdots\otimes
x_{n}\,|\,y_{1}\otimes\cdots\otimes
y_{n}\rangle_{\mathscr{H}^{\otimes n}}:=\langle
x_{1}\,|\,y_{1}\rangle_{\mathscr{H}}\cdots\langle
x_{n}\,|\,y_{n}\rangle_{\mathscr{H}}\,,\,\,\,\,
x_{1},\ldots,y_{n}\in\mathscr{H}\,.
\end{equation*}
If $\{\varphi_{k}\,:\,k\in\mathbb{N}\}$ is an orthonormal basis in
$\mathscr{H}$, then
$\{\varphi_{k_{1}}\otimes\varphi_{k_{2}}\otimes\cdots\otimes
\varphi_{k_{n}}\,:\,k_i\in\mathbb{N}\}$ is an orthonormal basis in
$\mathscr{H}^{\otimes n}$. We define the \emph{Fock space} over
$\mathscr{H}$ by the direct sum
\begin{equation*}
\mathscr{F}(\mathscr{H}):=\bigoplus_{n=0}^{\infty}\mathscr{H}^{\otimes
n}=\mathbb{C}\oplus\mathscr{H}\oplus\mathscr{H}^{\otimes
2}\oplus\cdots,
\end{equation*}
where we have set $\mathscr{H}^{\otimes 0}:=\mathbb{C}$. This is the
state space for an indeterminate number of particles of the same
species. It is separable if $\mathscr{H}$ is. Note that each element
$\psi\in\mathscr{F}(\mathscr{H})$ can be identified with a sequence
of vectors $\psi=\{\psi^{(n)}\}_{n\ge0}$, with
$\psi^{(0)}\in\mathbb{C}$ and $\psi^{(n)}\in\mathscr{H}^{\otimes
n}$, $n\ge1$. The inner product of any two vectors
$\varphi,\psi\in\mathscr{F}(\mathscr{H})$ is then given by
\begin{eqnarray*}
\langle
\varphi\,|\,\psi\rangle_{\mathscr{F}(\mathscr{H})}=\bar{\varphi}^{(0)}\psi^{(0)}+\sum_{n=1}^{\infty}\langle
\varphi^{(n)}\,|\,\psi^{(n)}\rangle_{\mathscr{H}^{\otimes n}}\,.
\end{eqnarray*}
\noindent Obviously, it is not $\mathscr{F}(\mathscr{H})$ itself but
the closed subspaces of symmetric/antisymmetric tensor products
which are frequently used in quantum field theory. Denote by
$\Pi_{n}$ the permutation group of $n$ elements. For each
$\sigma\in\Pi_{n}$, we define the operator $T_{\sigma}$ on basis
elements of $\mathscr{H}^{\otimes n}$ by
\begin{equation*}
{T}_{\sigma}\left({\varphi_{k}}_{1}\otimes\cdots\otimes
{\varphi_{k}}_{n}\right):={\varphi_{k}}_{\sigma(1)}\otimes\cdots\otimes
{\varphi_{k}}_{\sigma(n)}\,.
\end{equation*}
${T}_{\sigma}$ extends by linearity to a unitary operator on
$\mathscr{H}^{\otimes n}$. The map
$\sigma\in\Pi_{n}\mapsto{T}_{\sigma}\in\mathscr{B}(\mathscr{H}^{\otimes
n})$ then defines a unitary representation of $\Pi_{n}$ on
$\mathscr{H}^{\otimes n}$, with
${T}_{\sigma_{1}\sigma_{2}}={T}_{\sigma_{1}}{T}_{\sigma_{2}}$ and
${T}_{\sigma}^{\dag}={T}_{\sigma}^{-1}={T}_{\sigma^{-1}}$. A tensor
$u\in\mathscr{H}^{\otimes n}$ is called \emph{symmetric} if
${T}_{\sigma}u=u$, $\forall\,\sigma\in\Pi_{n}$, and
\emph{antisymmetric} if ${T}_{\sigma}u=\mathrm{sgn}(\sigma)u$,
$\forall\,\sigma\in\Pi_{n}$, where
$\mathrm{sgn}:\Pi_{n}\rightarrow\{-1,1\}$ takes the values $+1$ or
$-1$ depending on whether $\sigma$ is an even or odd permutation,
respectively. It is easy to show that the operators
\begin{eqnarray*}
\mathscr{P}_{+}^{(n)}&:=&\frac{1}{n!}\sum_{\sigma\in\Pi_{n}}{T}_{\sigma}\,,\\
\mathscr{P}_{-}^{(n)}&:=&\frac{1}{n!}\sum_{\sigma\in\Pi_{n}}\mathrm{sgn}(\sigma){T}_{\sigma}\,,
\end{eqnarray*}
are orthogonal projections on $\mathscr{H}^{\otimes n}$. The range
of $\mathscr{P}_{+}^{(n)}$ (resp. $\mathscr{P}_{-}^{(n)}$) is called
the $n$-fold symmetric (resp. antisymmetric) tensor product of
$\mathscr{H}$. We now define the \emph{bosonic} or \emph{symmetric
Fock space} over $\mathscr{H}$ by
\begin{equation*}
\mathscr{F}_{+}(\mathscr{H})=\mathscr{P}_{+}\mathscr{F}(\mathscr{H}):=\bigoplus_{n=0}^{\infty}\mathscr{P}_{+}^{(n)}\left(\mathscr{H}^{\otimes
n}\right),
\end{equation*}
and, similarly, the \emph{fermionic} or \emph{antisymmetric Fock
space} over $\mathscr{H}$ as
\begin{equation*}
\mathscr{F}_{-}(\mathscr{H})=\mathscr{P}_{-}\mathscr{F}(\mathscr{H}):=\bigoplus_{n=0}^{\infty}\mathscr{P}_{-}^{(n)}\left(\mathscr{H}^{\otimes
n}\right),
\end{equation*}
with
$\mathscr{P}_{\pm}:\mathscr{F}(\mathscr{H})\rightarrow\mathscr{F}_{\pm}(\mathscr{H})$
being called the \emph{symmetrization} (the $+$ sign) and
\emph{antisymmetrization} (the $-$ sign) \emph{projections}.
Finally, we introduce the so-called \emph{vacuum state},
\begin{equation*}
\Psi_{0}:=\left(1,0,0,\ldots\right),
\end{equation*}
which represents the state of the system with no particles. Note
that the state so defined is normalized.
\\
\indent Let $\{\varphi_k\}$ be an orthonormal basis of
$\mathscr{H}$. With the aim of constructing an orthonormal basis for
the Fock spaces $\mathscr{F}_{\pm}(\mathscr{H})$, it suffices to
consider the states
\begin{eqnarray*}
&&|n_{1},n_{2},\ldots\rangle_{\pm}:=\Theta_{\pm}\mathscr{P}_{\pm}(\varphi_1^{\otimes
n_1}\otimes\varphi_2^{\otimes n_2}\otimes
\ldots)\,,\,\,\,\sum_{k}n_{k}=N<+\infty\,,\\
&&\Theta_+=\left(\frac{(\sum_k n_k)!}{\prod_k
(n_k!)}\right)^{1/2}\,, \quad
\Theta_-=\left(\frac{1}{\prod_k(n_k!)}\right)^{1/2},
\end{eqnarray*}
for finite nonzero sequences $\{n_{k}\in\mathbb{N}_0\}$. The
$|n_{1},n_{2},\ldots\rangle_{\pm}$ vector describes an assembly of
$N$ identical bosonic (the $+$ sign) or fermionic (the $-$ sign)
particles, in which the state $\varphi_{k}$ is occupied by $n_{k}$
particles. The integers $n_{k}$, $k\in\mathbb{N}$, are called the
\emph{occupation numbers} of these states. Note that
$\mathscr{P}_{-}(\varphi_1^{\otimes 2}\otimes\cdots)=0$ in the
antisymmetric case, so that two identical fermions cannot occupy the
same quantum state at the same time. This last result is referred to
as the \emph{Pauli exclusion principle}, according to which
$n_{k}\in\{0,1\}$, $\forall\,k$, for fermionic systems. This is
precisely the main qualitative difference between fermion and boson
particles; the absence of the Pauli principle for the latter
particles implies that there is no bound on the number of bosons
that can occupy a state.

\section{Creation and annihilation operators}

\indent Let $N$ be the self-adjoint \emph{number operator} on
$\mathscr{F}(\mathscr{H})$,
\begin{equation*}
N\psi=\{n\psi^{(n)}\}_{n\ge0}\,,\,\,\,\psi\in\mathscr{D}_{N}\,,
\end{equation*}
with dense domain
$\mathscr{D}_{N}=\{\psi=\{\psi^{(n)}\}_{n\ge0}\,|\,\sum_{n\ge0}n^{2}\|\psi^{(n)}\|^{2}<+\infty\}$.
For each vector $f\in\mathscr{H}$ we define the \emph{creation} and
\emph{annihilation operators} on $\mathscr{F}(\mathscr{H})$,
respectively $C(f)$ and $A(f)$, by initially setting
$A(f)\psi^{(0)}=0$, $C(f)\psi^{(0)}=f$, $f\in\mathscr{H}$, and
\begin{eqnarray*}
C(f)(f_{1}\otimes f_{2}\otimes\cdots\otimes f_{n}) & = &
(n+1)^{1/2}f\otimes f_{1}\otimes\cdots\otimes f_{n}\,,\\
A(f)(f_{1}\otimes f_{2}\otimes\cdots\otimes f_{n}) & = &
n^{1/2}\langle
f\,|\,f_{1}\rangle_{\mathscr{H}}f_{2}\otimes\cdots\otimes f_{n}\,.
\end{eqnarray*}
Extension by linearity leads to densely defined operators. Given
$\psi^{(n)}\in\mathscr{H}^{\otimes n}$, it is straightforward to
check that
\begin{equation}\label{CfAf}
\|C(f)\psi^{(n)}\|\le (n+1)^{1/2}\|
f\|\|\psi^{(n)}\|\,\,\,\,\,\mathrm{and}\,\,\,\,
\|A(f)\psi^{(n)}\|\le n^{1/2}\| f\|\|\psi^{(n)}\|\,,
\end{equation}
so that $C(f)$ and $A(f)$ have well-defined extensions to the domain
of the $N^{1/2}$ operator, satisfying
\begin{equation}\label{relac1}
\|C(f)\psi\|,\|A(f)\psi\|\le\|
f\|\|(N+1)^{1/2}\psi\|\,,\,\,\,\psi\in\mathscr{D}_{N^{1/2}}\,.
\end{equation}
The adjoint relation
\begin{equation}\label{relac2}
\langle C(f)\varphi\,|\,\psi\rangle_{\mathscr{F}(\mathscr{H})}=\langle\varphi\,|\,A(f)\psi\rangle_{\mathscr{F}(\mathscr{H})}
\end{equation}
holds for all $\varphi,\psi\in\mathscr{D}_{N^{1/2}}$. We then define
the creation and annihilation operators on the
symmetric/antisymmetric Fock spaces $\mathscr{F}_{\pm}(\mathscr{H})$
by
\begin{equation*}
C_{\pm}(f)=\mathscr{P}_{\pm}C(f)\mathscr{P}_{\pm}\,,\quad
A_{\pm}(f)=\mathscr{P}_{\pm}A(f)\mathscr{P}_{\pm}\,,
\end{equation*}
verifying, by virtue of relations (\ref{relac1}) and (\ref{relac2}),
\begin{equation*}
\langle C_{\pm}(f)\varphi\,|\,\psi\rangle_{\mathscr{F}(\mathscr{H})}=\langle\varphi\,|\,A_{\pm}(f)\psi\rangle_{\mathscr{F}(\mathscr{H})}\,,\quad
\| C_{\pm}(f)\psi\|\le\|f\|\|(N+1)^{1/2}\psi\|\,,
\end{equation*}
for all $\varphi,\psi\in\mathscr{D}_{N^{1/2}}$. Moreover, since
$A(f)$ leaves the subspaces
$\mathscr{F}_{\pm}(\mathscr{H})\prec\mathscr{F}(\mathscr{H})$
invariant, i.e., $[A(f),\mathscr{P}_{\pm}]=0$, $f\in\mathscr{H}$, we
get
\begin{equation*}
C_{\pm}(f)=\mathscr{P}_{\pm}C(f)\,,\quad
A_{\pm}(f)=A(f)\mathscr{P}_{\pm}\,.
\end{equation*}
Note that the maps $f\mapsto C_{\pm}(f)$ and $f\mapsto A_{\pm}(f)$,
$f\in\mathscr{H}$, are linear and antilinear, respectively. Finally,
it is straightforward to calculate the \emph{canonical commutation
relations}
\begin{equation}\label{CCR}
[{A}_{+}(f),{A}_{+}(g)]=0=[{C}_{+}(f),{C}_{+}(g)]\,,\quad
[{A}_{+}(f),{C}_{+}(g)]=\langle
f\,|\,g\rangle_{\mathscr{H}}\mathbb{I}\,,\quad f,g\in\mathscr{H}\,,
\end{equation}
as well as the \emph{canonical anticommutation relations}
\begin{equation}\label{CAR}
\{{A}_{-}(f),{A}_{-}(g)\}=0=\{{C}_{-}(f),{C}_{-}(g)\}\,,\quad
\{{A}_{-}(f),{C}_{-}(g)\}=\langle
f\,|\,g\rangle_{\mathscr{H}}\mathbb{I}\,,\quad f,g\in\mathscr{H}\,,
\end{equation}
where we have used the notation $\{{A},{B}\}:=AB+BA$. The fact that
the occupation numbers can vary over all $\mathbb{N}$ for bosonic
particles is reflected by the unboundedness of the creation and
annihilation operators. On the contrary, these operators have
bounded extensions in the antisymmetric case as a consequence of
Pauli principle.
\\
\linebreak \linebreak \noindent \textbf{References}
\begin{itemize}
\item [1] O. Bratteli and D. W. Robinson, \emph{Operator Algebras and Quantum Statistical Mechanics II. Equilibrium States. Models in Quantum Statistical
Mechanics}, 2${^{\mathrm{a}}}$ ed., Springer (1997).
\item [2] M. Reed and B. Simon, \emph{Methods of Modern Mathematical Physics I: Functional Analysis; II: Fourier Analysis. Self-Adjointness}, Academic Press, New York (1975).
\end{itemize}

\end{document}